\documentclass[11pt, a4paper, oneside]{Thesis} 

\graphicspath{{Pictures/}} 

\usepackage[square,sort,compress,numbers]{natbib}
\usepackage{cite}
\usepackage{amsmath,latexsym} 
\usepackage{float}
\usepackage{caption}
\usepackage[T1]{fontenc}
\usepackage{lmodern} 
\usepackage{microtype} 
\usepackage{hyperref}
\usepackage{subcaption} 
\usepackage{graphicx} 
\usepackage{lscape}
\usepackage{enumitem}
\usepackage{physics}
\hypersetup{colorlinks,citecolor=red}
\hypersetup{colorlinks=true,urlcolor=black}
\title{\ttitle} 
\DeclareUnicodeCharacter{2212}{-}
\DeclareUnicodeCharacter{2212}{+}

\begin{document}

\setstretch{1.3} 

\fancyhead{} 
\rhead{\thepage} 
\lhead{} 

%

\thesistitle{Cosmological and Dynamical  Aspects of Teleparallel Gravity and its Extension}
\documenttype{\textbf{THESIS}}
\supervisor{Prof. BIVUDUTTA MISHRA}
\supervisorposition{Professor}
\supervisorinstitute{BITS-Pilani, Hyderabad Campus}
\examiner{}
\degree{Ph.D. Research Scholar}
\coursecode{\textbf{DOCTOR OF PHILOSOPHY}}
\coursename{\textbf{Thesis}}
\authors{\textbf{Lokesh Kumar Duchaniya}}
\IDNumber{\textbf{2020PHXF0478H}}
\addresses{}
\subject{}
\keywords{}
\university{\texorpdfstring{\href{http://www.bits-pilani.ac.in/} 
                {Birla Institute of Technology and Science, Pilani}} 
                {Birla Institute of Technology and Science, Pilani}}
\UNIVERSITY{\texorpdfstring{\href{http://www.bits-pilani.ac.in/} 
                {\textbf{BIRLA INSTITUTE OF TECHNOLOGY AND SCIENCE, PILANI}}} 
                {BIRLA INSTITUTE OF TECHNOLOGY AND SCIENCE, PILANI}}


\department{\texorpdfstring{\href{http://www.bits-pilani.ac.in/pilani/Mathematics/Mathematics} 
                {Mathematics}} 
                {Mathematics}}
\DEPARTMENT{\texorpdfstring{\href{http://www.bits-pilani.ac.in/pilani/Mathematics/Mathematics} 
                {Mathematics}} 
                {Mathematics}}
\group{\texorpdfstring{\href{Research Group Web Site URL Here (include http://)}
                {Research Group Name}} 
                {Research Group Name}}
\GROUP{\texorpdfstring{\href{Research Group Web Site URL Here (include http://)}
                {RESEARCH GROUP NAME (IN BLOCK CAPITALS)}}
                {RESEARCH GROUP NAME (IN BLOCK CAPITALS)}}
\faculty{\texorpdfstring{\href{Faculty Web Site URL Here (include http://)}
                {Faculty Name}}
                {Faculty Name}}
\FACULTY{\texorpdfstring{\href{Faculty Web Site URL Here (include http://)}
                {FACULTY NAME (IN BLOCK CAPITALS)}}
                {FACULTY NAME (IN BLOCK CAPITALS)}}

\maketitle

\clearpage
\pagestyle{empty}
\pagenumbering{gobble}
\frontmatter
\Certificate

\Declaration

\setstretch{1.3} 

\pagestyle{empty} 

\Dedicatory{\bf \begin{LARGE}
To   
\end{LARGE} 
\\
\vspace{1.5cm}
 My loving Parents\\
 \vspace{1cm}
 }

\addtocontents{toc}{\vspace{2em}}

\begin{acknowledgements}

I am grateful to many people for their help and support while writing this thesis and during my doctorate studies.

I extend my deepest gratitude to my supervisor, \textbf{Prof. Bivudutta Mishra}, Professor of Mathematics at BITS-Pilani, Hyderabad Campus, for his invaluable guidance throughout my research endeavors. His mentorship has provided a thorough understanding of general relativity and cosmology and imparts essential life values that transcend academia. Collaborating with him has been a profound honor, and I deeply appreciate his support, expertise, and patience during this journey.

I am deeply grateful to my DAC members, \textbf{Prof. Pradyumn Kumar Sahoo} and \textbf{Prof. Nirman Ganguly}, for their insightful inquiries, valuable critiques and continuous encouragement throughout this study.

I would like to express my deepest gratitude to the Head of the Department, the DRC Convener and the entire staff of the Department of Mathematics at BITS-Pilani, Hyderabad Campus, for their invaluable support, guidance and encouragement throughout my research. I also thank the Associate Dean of AGRSD BITS-Pilani, Hyderabad Campus.

I am thankful to the University Grants Commission (UGC) for their financial assistance throughout my PhD journey.    

I would like to extend my sincere gratitude to \textbf{Prof. Jackson Levi Said}, Professor, Department of Physics, University of Malta, Msida, MSD 2080, Malta, for allowing me to collaborate with him. I have learned a great deal from our work together and I deeply appreciate his constant support, guidance and willingness to clarify my doubts throughout the collaboration. 

I would like to thank my co-authors, {\bf Prof. Sunil Kumar Tripathy, Prof. Giovanni Otalora, Prof. Manuel Gonzalez-Espinoza, Prof. I.V. Fomin and Prof. S.V. Chevron}, for their invaluable contributions and collaboration on various projects. 

Special thanks go to my Cosmic Researchers group Dr. Amar, Dr. Siddheshwar, Dr. Santosh, Shubham, Priyobarta, Shivam and my friends.

In my personal life, I am especially thankful to my parents, {\bf Mrs. Chameli Duchaniya} and {\bf Mr. Jagdish Duchaniya}, whose unwavering love and guidance have shaped my journey. I also extend my heartfelt appreciation to my brother {\bf Rakesh} and my sisters {\bf Krishna, Manju}, and {\bf Suman} for their constant support, encouragement, and care. Their presence has been a strength and inspiration in everything I do.      \end{acknowledgements}

\clearpage

\begin{abstract}  
This thesis investigates the characteristics of modified teleparallel gravity models that incorporate a scalar field and a trace of the energy-momentum tensor, with particular attention on their cosmological effects, particularly about late-time cosmic acceleration. 

Teleparallel gravity, as an alternative framework to General Relativity, defines gravity through torsion instead of curvature. Telleparallel gravity, along with the scalar field may explore a wider variety of cosmological scenarios. Our investigation delves into the late-time acceleration of the cosmos. It examines the various phases of the Universe through dynamical system analysis within the framework of modified teleparallel gravity theories. 

Chapter~\ref{Chapter1} offers an in-depth exploration of fundamental formulations and the mathematical framework of $f(T)$, $f(T, \mathcal{T})$, $f(T, \phi)$ and Horndeski teleparallel gravity and the quintessence dark energy model. Additionally, it delves into the theory of dynamical systems and their applications while also presenting the cosmological datasets employed in our analysis across various gravity theories discussed in Chapters~[\ref{Chapter2}-\ref{Chapter6}].  

In Chapter~\ref{Chapter2}, the analysis of $f(T)$ gravity cosmological models is conducted at both background and perturbation levels. Three distinct functional forms of $f(T)$ are examined, with corresponding cosmological parameters derived from dynamical variables. Critical points for each model are identified along with their necessary existence condition. Each model is analyzed individually, with stability discussed through eigenvalues and phase portraits, revealing at least one stable node in each case. Additionally, the evolution plots of cosmological parameters show the accelerating behavior of the cosmological models. 

In Chapter~\ref{Chapter3}, two problems are discussed. The first focuses on stability analysis and the dynamical system approach to understanding the viability of models through critical point dynamics. The second one examines the $f(T, \mathcal{T})$ cosmological model using various data set combinations, comparing results from Pantheon+ (without SH0ES) and Pantheon+\&SH0ES (with SH0ES). We have also incorporated BAO data and $H_0$ priors. Also, we have calculated the $\chi^2_{min}$ for various combinations of the data set to evaluate the model against the $\Lambda$CDM model. The best-fit values for cosmological parameters are determined and the findings suggest that the selected models adequately represent the late-time cosmic dynamics of the Universe.

Chapter~\ref{Chapter4} examined the dynamical system analysis in the $f(T, \phi)$ gravity. It presents two models featuring different functional forms of the torsion scalar, identifying critical points for each model and analyzing their stability and corresponding cosmology. It also includes graphical representations of the EoS parameter and density parameters for radiation, Matter and dark energy sectors. According to the behavior of the cosmological parameters, we conclude that both models show the late-time cosmic acceleration of the Universe.

In Chapter~\ref{Chapter5}, the study explores cosmological observations in quintessence dark energy models using three different potential functions \( V(\phi) \). The analysis focuses on CC, Pantheon+\&SH0ES and BAOs datasets, comparing the results of these models with the standard \( \Lambda \)CDM model and presenting the findings through a whisker plot. Additionally, we computed the $\chi^2_{min}$ for different combinations of the dataset to assess the model against the $\Lambda$CDM model.

In Chapter~\ref{Chapter6}, focusing on an exponential form of the coupled scalar field $F(\phi)$ and three different forms of $V(\phi)$ to study the phase space analysis in modified Galileon cosmology is explored. We derive the critical points of the autonomous system, as well as their stability criteria and cosmological characteristics. It also examines various phases and the late-time cosmic acceleration of the Universe, with results analyzed about the Hubble rate \( H(z) \) and Supernovae Ia cosmological data sets.

Chapter~\ref{Chapter7}, summarizes the results and discussion of the research presented in the thesis. 
\end{abstract}

\clearpage


\addtocontents{toc}{\vspace{1em}}
\tableofcontents 
\addtocontents{toc}{\vspace{1em}}
\lhead{\emph{List of Tables}}
\listoftables 
\addtocontents{toc}{\vspace{1em}}
\lhead{\emph{List of Figures}}
\listoffigures 
\addtocontents{toc}{\vspace{1em}}
 


\setstretch{1.5}
\lhead{\emph{List of Symbols and Abbreviations}}
\listofsymbols{ll}{
\begin{tabular}{cp{0.5\textwidth}}
$\mathcal{L}_{m}$ & : Matter Lagrangian\\
$\mathcal{L}_{r}$ & : Radiation Lagrangian\\
$H(z)=\frac{\dot a}{a}$& : Hubble parameter\\
$a(t)$ & : Scale factor\\
$q$ & : Deceleration parameter\\
$g_{\mu\nu}$ &: Metric tensor\\
 $g$ &: Determinant of $g_{\mu\nu}$  \\ 
$G_{\mu \nu} \equiv R_{\mu \nu} - \frac{1}{2} g_{\mu \nu} R$ & : Einstein tensor\\
$\Lambda$ & : Cosmological constant\\
 $\hat\Gamma^{\sigma}_{\mu \nu}$  &: Teleparallel  Weitzenb\"{o}ck connection \\ 
$R^{k}_{\sigma \mu \nu}$ &: Riemann tensor \\
$R_{\mu \nu}$ &: Ricci tensor \\
$R$ &: Ricci scalar \\
$S_{m}, S_{r}$ & : Matter, radiation action\\
$T^{\alpha}_{\ \ \mu \nu}$ & : Torsion tensor\\
$T$ & : Torsion scalar \\
$S_{\rho}^{~~\mu \nu}$ & : Superpotential\\
$K^{\mu \nu}_{~~~~\rho}$ & : Contortion tensor\\
$e^{A}_{\,\,\,\,\mu}$ &: Tetrad\\
$E^{\,\,\,\, \mu}_{A}$&: Inverse tetrad\\
$e$ & : Determinant of tetrad\\
$(-\,+\,+\,+)$ &: Metric signature\\
$\rho_{m}$, $\rho_{DE}$ & : Energy density of dark matter and dark energy \\
$p_{m}$, $p_{DE}$&: Pressure of dark matter and dark energy \\
\end{tabular}
}

\clearpage

\noindent
\begin{center}
\begin{tabular}{cp{0.5\textwidth}}
$G$ &: Gravitational constant and  G= $6.67 \times 10^{-11} N m^2 kg^{-2}$\\
$\omega_{DE}$&: Dark energy sector equation-of-state parameter\\
$\omega_{tot}$&: Total equation-of-state parameter\\
$\Omega_r$, $\Omega_m$, $\Omega_{\rm DE}$ &: Standard density parameters for radiation, dark matter and dark energy\\
  $\lor$ &: Or \\
  $\land$ &: And\\
  \textbf{GR}&: \textbf{G}eneral \textbf{R}elativity\\
\textbf{TEGR}&: \textbf{T}eleparallel \textbf{E}quivalent of \textbf{G}eneral \textbf{R}elativity\\
\textbf{DE}&: \textbf{D}ark \textbf{E}nergy\\
\textbf{DM}&: \textbf{D}ark \textbf{M}atter\\
\textbf{EoS}&: \textbf{E}quation \textbf{o}f \textbf{S}tate\\
\textbf{$\Lambda$CDM}&: \textbf{L}ambda-\textbf{C}old-\textbf{D}ark-\textbf{M}atter\\
\textbf{FLRW}&: \textbf{F}riedmann \textbf{L}ema\^{i}tre \textbf{R}obertson \textbf{W}alker \\
\textbf{CMB}&: \textbf{C}osmic \textbf{M}icrowave \textbf{B}ackground \\
\textbf{CC}&: \textbf{C}osmic \textbf{C}hronometers \\
\textbf{SNIa}&: \textbf{S}upernovae Type Ia \\
\textbf{BAO}&: \textbf{B}aryon \textbf{A}coustic \textbf{O}scillation \\
\textbf{TRGB}&: \textbf{T}ip of the \textbf{R}ed \textbf{G}iant \textbf{B}ranch \\
\textbf{MCMC}&: \textbf{M}arkov  \textbf{C}hain \textbf{M}onte \textbf{C}arlo\\
\textbf{AIC}&: \textbf{A}kaike  \textbf{I}nformation \textbf{C}riterion\\
\textbf{BIC}&: \textbf{B}ayesian  \textbf{I}nformation \textbf{C}riterion\\
\textbf{PN$^+$}&:\textbf{Pantheon+}\\
$D_A$&: \textbf{A}ngular  \textbf{D}iameter \textbf{D}istance\\
$D_L$&: \textbf{L}uminosity \textbf{D}istance\\
$D_M$&: \textbf{C}omoving \textbf{D}istance\\
\end{tabular}
\end{center}


%
%


\clearpage 

\lhead{\emph{Glossary}} 

 


\mainmatter 
\setstretch{1.2} 
\pagestyle{fancy} 



\chapter{Introduction} 
\label{Chapter1}

\lhead{Chapter 1. \emph{Introduction}} 


\newpage 
\section{Cosmology Overview}\label{ch1cosmologicaloverview}
People have been intrigued by profound questions throughout the ages: What initiated the existence of the Universe? What principles dictate its functions? Will it persist indefinitely, or does it have a conclusion? Is the Universe boundless? These timeless enigmas led to the emergence of cosmology, the scientific field focused on exploring the Universe. Cosmology aims to reveal the narrative of the cosmos, from its beginnings in the remote past to its active development and ultimate destiny. By reconstructing the history of the Universe, cosmology tackles essential inquiries, offering a basis for comprehending present phenomena and predicting future changes.
\subsection{Cosmic Principle}\label{CosmicPrinciple}
The Copernican principle suggests that Earth is not in a special or central position in the Universe, indicating that our location is typical and unremarkable. This principle indicates that the Universe is largely homogeneous and isotropic when observed on large scales. Homogeneity refers to the concept that every location in the Universe has the same properties. Meanwhile, isotropy signifies no favored direction, implying that the Universe appears consistent in every direction. It suggests that the laws of physics remain uniform throughout the cosmos, without any favored location or direction.
\subsection{Hubble's observation}\label{Hubbleobservation}
Hubble's discovery marked a major advancement in cosmology, which greatly improved our comprehension of the structure and dynamics of the Universe. In 1929, Edwin Hubble carried out detailed surveys of galaxies, uncovering a relationship between their redshift and their distance from the Earth specifically, that galaxies are moving away from us at speeds proportional to how far away they are. Hubble's Law describes the relationship between the recession speed of galaxies and their distance, indicated as \( v = H_0 d \), where \( v \) is the recession speed, \( H_0 \) is the Hubble constant and  \( d \) is the distance to the galaxy. Hubble noted that the light emitted by distant galaxies is redshift, meaning the wavelength becomes longer as they recede due to the Doppler effect. The observation that galaxies are distancing themselves from one another resulted in the inference that the Universe is undergoing expansion. This served as crucial evidence in favor of the Big Bang theory. The movement of galaxies away from each other implies that, at some point in the past, the Universe was significantly smaller and more concentrated before it started to expand from a single point, an event termed the Big Bang. 
\subsection{Distance and time measurements in cosmology}\label{timemeasurements}
The parallax technique directly measures distances to astronomical objects within roughly 100 parsecs. Indirect methods must be relied upon for estimating distances to extragalactic or more distant objects. When a distant object emits a photon with a certain wavelength $\lambda_{em}$, it is detected later due to the causality limit. The expansion of the Universe causes this photon to experience redshift, making it observed at a longer wavelength as,
\begin{equation}\label{timewavelength}
\lambda_{0}= \frac{a_0}{a_{em}}\lambda_{em}\,.    
\end{equation}
In this scenario, \( a_{em} \) and \( a_{0} \) denote the scale factors at the moment of photon emission and at the current time of observation, respectively. The redshift for an astronomical object can be expressed as,
\begin{equation}\label{redshiftequation}
z_{obj}=\frac{a_0}{a_{obj}}-1\,.    
\end{equation}
The variable $a_{obj}$ signifies the scale factor when the light from that object reaches to observer. The cosmological expansion history, denoted as $a(t)$, allows for an accurate connection between redshift, the distance of an object and the time of its light emission. In cosmology, distances and times are typically expressed in $z$. The main distance measures used in this area are the angular diameter and luminosity distance. These measurements are essential for interpreting the dynamics of the expanding Universe and gaining insight into the characteristics of far-off astronomical bodies.

In cosmology, the luminosity distance measures distances to astronomical objects based on their observed brightness. The $D_L$ is the distance an object would be visible if the Universe were stable, non-expanding and  Euclidean. In mathematical terms, it connects the intrinsic  $D_L$ of an astronomical entity to its observed flux $F$ via the equation,
\begin{equation}\label{luminosityobservedflux}
D_L = \sqrt{\frac{L}{4\pi F}}\,,    
\end{equation}
where $L$ represents the inherent brightness and $F$ denotes the detected flux. 

In an expanding Universe, $D_L$ considers the impact of redshift $z$, which occurs when light from an object is stretched to greater wavelengths due to cosmic expansion. For an object at redshift $z$, the luminosity distance can be indicated as,
\begin{equation}\label{luminosityredshift}
D_L= (1+z)D_M\,,    
\end{equation}
where $D_M$ considers the geometry and expansion of the Universe, the luminosity distance depends on the cosmological model, including $H_0$ tension, matter density parameter $\Omega_{m}$ and the curvature of the Universe. In a flat Universe, the comoving distance is given by,
\begin{equation}\label{comovingdistance}
D_M= c \int_0^z \frac{dz'}{H(z')}\,,    
\end{equation}
Where $H(z')$ refers to a particular redshift $z'$, while $c$ represents the speed of light. As a result, the formula for luminosity distance can be expressed as follows,
\begin{equation}\label{luminositydistance1}
 D_L = (1+z) \, c \int_0^z \frac{dz'}{H(z')}\,.    
\end{equation}
Indirect measurements can be used to establish the luminosities of various astronomical objects. For instance, SNIa exhibits a narrow range in total luminosity. Gamma-ray bursts (GRBs) can be precisely calibrated, allowing their luminosity to be accurately determined. The flow of these and other standardizable candles at various redshifts can be evaluated to obtain cosmological parameters using Eq.~\eqref{luminositydistance1}.

The angular diameter distance is a cosmological concept that connects the actual size of an object to its apparent angular size as observed from the Earth. This idea is essential for determining distances within the expanding Universe and understanding its structure. The  $D_A$ is defined as,
\begin{equation}\label{angulardistance}
D_A = \frac{D}{\theta}\,,    
\end{equation}
where $D$ indicates the actual physical size of the object, while $\theta$ refers to the angle at which the object appears. The angular diameter distance is related to $D_L$ through the redshift $z$,
\begin{equation}\label{angularluminosity}
D_A = \frac{D_L}{(1+z)^2}\,,    
\end{equation}
This relationship arises due to the luminosity distance changes with the square of the redshift factor, which stems from the combined effects of light dilution and time dilation in an expanding Universe.
\subsection{FLRW cosmology}\label{FLRWcosmology}
 The dynamics of the Universe are governed by the Einstein field equations, which are derived from the variational principle applied to the Einstein-Hilbert action. A common approach to solving these equations is to leverage general symmetries inherent in the problem. The FLRW metric \cite{Liddle2000metric, Weinberg1971metric, Padmanabhan2000metric} is the standard framework used in cosmology, which is predicated on the assumptions of homogeneity and isotropy throughout the Universe at all points. This metric encapsulates the geometry of a cosmos that is uniform and isotropic on large scales. The FLRW metric is given as 
  \begin{align}
     d\textit{s}^2 = -N(t)^2 dt^2 +a^2 (t) \left(\frac{1}{ 1 - k r^2} dr^2+r^2 d\theta^ 2+r^2 sin^ 2\theta d\phi^2 \right)\,,\label{FLRWW}
\end{align}
where $N(t)$  denotes the lapse function and $a(t)$  is the scale factor, which encapsulates the evolution of the Universe. The determination of the scale factor is based on the Einstein field equations, which are influenced by the matter content of the Universe. The constant \( k \) in the metric describes the geometric characteristics of the spatial portion of spacetime, where \( k = 1 \) indicates a closed Universe, \( k = 0 \) signifies a flat Universe and  \( k = -1 \) represents an open Universe. This thesis employs the flat version of the metric with the $diag$[ -, +, +, +], which is represented as,
\begin{equation}
    ds^2 = -N(t)^2dt^2+a(t)^2(dx^2+dy^2+dz^2)\,.\label{FLATFLRW}
\end{equation}
\subsection{The mathematical foundations of the Universe epochs through the Einstein field equation}\label{EinsteinsFieldEquation}
The Einstein field equations establish a mathematical link between geometry and matter. The Einstein field equations can be expressed as
\begin{align}
G_{\mu \nu} = 8\,\pi\,G\,T_{\mu\nu} \,,\label{GRFE}
\end{align}
where \(G_{\mu \nu} \equiv R_{\mu\nu} - \frac{1}{2}R g_{\mu\nu}\) is the Einstein tensor and  \(T_{\mu \nu}\) is the energy-momentum tensor.

In the Newtonian framework, the gravitational field is directly proportional to the mass distribution within the system. In the context of GR, mass is one of several factors contributing to the curvature of the spacetime. The energy-momentum tensor encompasses all possible sources of energy that creates curvature in spacetime, facilitating a thorough representation of the density and flow of 4-momentum in a specific system. Within the framework of GR, a perfect fluid is an idealized fluid that lacks heat conduction and viscosity. Its thermodynamic state is completely characterized by two primary parameters: mass density $(\rho)$ and pressure $(p)$. This conceptualization streamlines the examination of fluid dynamics and spacetime curvature, making it a useful theoretical model in physics and cosmology for investigating various astrophysical events. The energy-momentum tensor can be written as,
\begin{equation}
T_{\mu \nu} = \left(\rho + p\right)u_{\mu}u_{\nu} + p \, g_{\mu \nu}\,,\label{stressenergytensor}
\end{equation}
where $u_{\mu}$ is the four-velocity of the fluid. For a comoving observer, the 4-velocity is expressed as $\vec{u} = (1, 0, 0, 0)$, simplifying the energy-momentum tensor as,
\[
T_{\mu\nu} =
\begin{pmatrix}
\rho & 0 & 0 & 0 \\
0 & p\,a^2(t) & 0 & 0 \\
0 & 0 & p\,a^2(t) & 0 \\
0 & 0 & 0 & p\,a^2(t)
\end{pmatrix}
\]
As \( p \to 0 \), the perfect fluid model approaches the behavior of dust-like matter. An ideal fluid acts as an appropriate approximation for the components of the Universe during earlier phases, particularly when radiation is the primary constituent. Within the framework of the flat FLRW metric~\eqref{FLATFLRW}, the Friedmann equation can be represented as follows,
\begin{align}
    3H^2=8\pi G \, \rho -\frac{k}{a^2} \,,\label{FriedmanEQ1}\\
    3H^2+2\dot{H}=-8\pi G \, p + \frac{k}{a^2}\,,\label{FriedmanEQ2}
\end{align}
where \( H \) is the Hubble parameter and it can be defined in terms of scale factor by \( H = \frac{\dot{a}}{a} \), where \( \dot{a} \) indicates the time derivative of the scale factor \( a(t) \). The continuity equation can be defined as,
\begin{align}
\dot{\rho}+3H\left(\rho+p\right)=0\,.\label{inConservationEq}
\end{align} 
By adjusting Eqs.~\eqref{FriedmanEQ1} and \eqref{FriedmanEQ2}, the acceleration equation is derived, which can follow the following expression,
\begin{equation}\label{accelerationequation}
\frac{\ddot{a}}{a}=-\frac{4 \pi G}{3} (\rho+3p) \,.    
\end{equation}
Cosmologists use the deceleration parameter \( q = -\frac{\ddot{a}a}{\dot{a}^2} \) to analyze cosmic dynamics. A negative value of $q$ indicates that the Universe is expanding at an accelerated rate. In contrast, a positive value of $q$ suggests a deceleration expansion phase of the Universe. Recent observational data, which is discussed in the next section, strongly supports that our Universe is undergoing the accelerating expansion phase. One can write Eq.~\eqref{FriedmanEQ1} in terms of density parameter as,
\begin{equation}\label{densitycurvature}
\Omega(t)-1 = \frac{k}{(aH)^2}\,,    
\end{equation}
where $\Omega(t)=\frac{\rho(t)}{\rho_{c}(t)}$ is the energy density parameter and $\rho_{c}(t)=\frac{3 H^2}{8\pi G}$ is the critical density.
\[
\begin{aligned}
\text{If } \rho(t) > \rho_c(t), \, \Omega(t) > 1 & \implies k = +1, \text{ the Universe is closed.} \\
\text{If } \rho(t) < \rho_c(t), \, \Omega(t) < 1 & \implies k = -1, \text{ the Universe is open.} \\
\text{If } \rho(t) = \rho_c(t), \, \Omega(t) = 1 & \implies k = 0, \text{ the Universe is flat.}
\end{aligned}
\]
Recent observations suggest that the Universe is almost a spatially flat geometry with its energy density approximately equal to the critical density. The relationship between energy and the scale factor can be established from the continuity equation as
\begin{align}
    \rho \propto a^{-3(1+\omega)}\,.
\end{align}
The symbol $\omega$ represents the EoS parameter, which characterizes the link between $\rho$ and $p$. This relationship is expressed as $p = \omega \rho$, where $\omega$ determines the state of the material, reflecting its thermodynamic properties. Based on the values of the EoS parameters, the link between energy density and the scale factor as 
\[
\begin{aligned}
\text{Radiation dominated phase:}\,\,\, \omega_{r}=\frac{1}{3},\,\,\,\, \, \rho \propto a^{-4}\,, \\
\text{Matter dominated phase:}\,\,\, \omega_{m}=0,\,\,\,\, \, \rho \propto a^{-3}\,, \\
\text{DE dominated phase:}\,\,\, \omega_{de}=-1, \,\,\,\, \, \rho \propto a^{0}\,.  
\end{aligned}
\]

Furthermore, the total EoS (\(\omega_{tot}\)), the DE EoS (\(\omega_{DE}\)) and  the deceleration parameter (\(q\)) can be articulated as follows,
\begin{eqnarray}
\omega_{tot.}&=&-1-\frac{2\dot{H}}{3H^{2}}\equiv \frac{p_m+p_r+p_{DE}}{\rho_{m}+\rho_{r}+\rho_{DE}}, \label{ch1_Eosparameter}\\  
\omega_{DE}&=&\frac{p_{DE}}{\rho_{DE}}, \label{ch1_darkenergyEosparameter}\\ 
q&=&-1-\frac{\dot{H}}{H^{2}}.\label{ch1_deceleration}
\end{eqnarray}

From Eq.~\eqref{ch1_Eosparameter} and Eq.~\eqref{ch1_deceleration}, the connection between the total EoS and the deceleration parameters as follows:
\begin{eqnarray}\label{relationdeceEOS}
 q=\frac{1}{2}(1+3\omega_{tot})\,.   
\end{eqnarray}
The density parameters can be written as,
\begin{equation}\label{ch1_densityparametesr}
\Omega_{m}=\frac{8 \pi G \rho_{m} }{3 H^{2}}\,, \hspace{0.8cm} \Omega_{r}=\frac{8 \pi G \rho_{r} }{3 H^{2}}\,, \hspace{0.8cm} \Omega_{DE}=\frac{8 \pi G \rho_{DE} }{3 H^{2}}\,,  
\end{equation}
satisfying
\begin{equation}\label{ch1_constraints}
\Omega_{m}+\Omega_{r}+\Omega_{DE}=1.    
\end{equation}
\section{Discovery of accelerated expansion of the Universe}\label{DiscoveryofAccelerated}
According to recent observational data \cite{Riess:1998cb, SupernovaCosmologyProject:1998vns}, it appears probable that the Universe is experiencing accelerated expansion. The question arises regarding the source of the energy that would be required to sustain this potential accelerated expansion. One possible explanation is that a type of energy referred to as ``DE''
 \cite{Hinshaw:2013, DIVALENTINO:2016, Peebles:2002gy, Padmanabhan_2003, Weinberg_1989} could be the driving force behind it. To account for the mysterious energy density permeating space, Einstein introduced the cosmological constant $\Lambda$ into the Einstein field equations. This modification was intended to facilitate understanding of this unknown energy source within the framework of GR. While GR is commonly recognized as a precise representation of gravity for substantial masses within the scale of the solar system \cite{misner1973gravitation, Capozziello:2011et, Faraoni:2008mf}. This theory necessitates the inclusion of a cosmological constant $\Lambda$ to explain the accelerated expansion of the Universe. This cosmological constant can be described as a fluid that exhibits negative pressure, offering a basis for interpreting violations of particular energy conditions and the subsequent impact of a repulsive gravitational force within the framework of GR. Nevertheless, because of the insufficient theoretical comprehension of the cosmological constant $\Lambda$ \cite{sahni2000case}, scientists are investigating different interpretations of the accelerated expansion of the Universe during later epochs. Two main strategies have emerged to tackle the aforementioned challenges in cosmology. The first strategy involves investigating a dynamical cosmological constant or more generally, DE, through the lens of GR \cite{Amendola_2000_cup, Copeland:2006wr, Tsujikawa:2013fta}. This approach seeks to elucidate the role of evolving DE in the accelerated expansion of the Universe within the established framework of GR. The second strategy is based on the formulation of innovative models that incorporate more intricate dynamics at cosmological scales, necessitating modifications to the fundamental principles of gravitational theory \cite{DeFelice:2010aj, Capozziello:2011et, Cai:2015emx}. This approach aims to extend our understanding of the influence of gravity on cosmic evolution beyond the confines of traditional theories.

In GR, the Levi-Civita connection is employed and is characterized by its symmetry and absence of torsion. An alternative approach is to replace this connection with the Weitzenb$\ddot{o}$ck connection \cite{Weitzenbock1923}, defined by its skew-symmetry, absence of curvature and the presence of a non-zero torsion tensor. An alternative method includes broadening the action of the corresponding torsional formulation of GR, known as the TEGR \cite{Maluf:1994ji, Aldrovandi:2013wha, deAndrade:2000nv, Pereira:2019woq, Bahamonde:2021gfp, 2023Symm15291C}. The modified gravity theory that results in second-order equations within four-dimensional space-time can be derived from the TEGR. In the framework of TEGR, the metric tensor is replaced by a dynamic variable known as the tetrad. In this context, the dynamic components consist of four linearly independent tetrad fields that establish the orthogonal bases for the tangent space at each point in space-time. Furthermore, the torsion tensor is derived from the product of the first derivatives of the tetrad fields. This thesis sets out to study the simple modification of TEGR non-minimal coupling with a trace of the energy-momentum tensor. This study will also emphasize the scalar field coupling function, which offers a wider range of viable cosmic models to explore cosmic evolution. The investigation centers on the coupling function of the scalar field, which broadens the spectrum of plausible cosmological models and deepens the analysis of theories concerning cosmic evolution.   
While GR modifications are often utilized to tackle theoretical challenges, propositions for alterations to teleparallel gravity have emerged only recently. This raises questions about whether modified teleparallel gravity theories could effectively address or lessen cosmological problems such as DE, DM and the $H_0$ tension. Compared to traditional modifications, do modified teleparallel gravity theories provide any advantage in cosmology? Do these theories present an opportunity to explore the historical development of the Universe? This thesis will introduce various modified teleparallel theories to tackle these questions. 
\section{{\texorpdfstring{$H_0$}{H0}} tension in the {\texorpdfstring{$\Lambda$}{Lambda}}CDM model}\label{H0tensionflatCDMmodel}
The standard $\Lambda$CDM can be define as 
\begin{equation}\label{hubbleLCDMintro}
H_{\Lambda CDM}= H_{0}\sqrt{(1+z)^3 \Omega_{m}+(1+z)^4 \Omega_{r}+(1-\Omega_{m}-\Omega_{r})} \,,  
\end{equation}

where $H_{0}$ is the Hubble constant. The $\Lambda$CDM model that best corresponds with observational evidence and clarifies the evolution of the Universe is commonly known as the standard model of cosmology, often called the Benchmark model \cite{ryden2017cosmology}. Although the \(\Lambda\)CDM framework effectively accounts for cosmological evolution during both the early and late epochs at both background and perturbation levels, recent years have revealed some potential inconsistencies with certain data sets, such as the \(H_0\) tension \cite{DiValentino:2021izs, Brout:2022pan, Bernal:2016gxb, Benisty:2021cmq}. The \(H_0\) tension refers to the discrepancies noted in the measurements of the Hubble constant from various observational methods, which challenge the standard \(\Lambda\)CDM model \cite{Di_Valentino_2017, Vagnozzi_2020newph, Abdalla:2022yfr, Moresco_2022_H0, Aghanim:2018eyx, DiValentino:2021izs}. Recent observations of the \(H_0\), reveal a significant difference between values inferred from early Universe data, such as the CMB measurements from the Planck satellite \cite{plank2013} and those obtained from local distance indicators like Cepheid variables and SNIa \cite{Riess:2021jrx}. Specifically, the SH0ES Team, led by Riess et al. \cite{Riess:2021jrx}, reported a Hubble constant of \(73.30 \pm 1.04 \, \text{km s}^{-1} \, \text{Mpc}^{-1}\) based on data from SNIa. In contrast, the H0LiCOW Collaboration \cite{wang_H0LiCOWmnras} derived a value of \(73.3^{+1.7}_{-1.8} \, \text{km s}^{-1} \, \text{Mpc}^{-1}\) using strong gravitational lensing techniques applied to quasars. Conversely, Freedman et al. \cite{Freedman_2019TRGB} provided a lower estimate of \(69.8 \pm 1.9 \, \text{km s}^{-1} \, \text{Mpc}^{-1}\), determined through the TRGB as a distance marker. In the context of the early Universe, the Planck Collaboration \cite{Aghanim:2018eyx} reported \(H_0 = 67.4 \pm 0.5 \, \text{km s}^{-1} \, \text{Mpc}^{-1}\) and  Abbott et al. \cite{Abbott_2018mnras} proposed a comparable value of \(67.2^{+1.2}_{-1.0} \, \text{km s}^{-1} \, \text{Mpc}^{-1}\). The \(H_0\) tension, presents major challenges and fascinating questions for modern cosmology.
\section{Theoretical constructs addressing cosmic acceleration
}\label{Theoreticalconstructsaddressing}
This section will delve into the motivations and foundational principles behind investigating modified gravity as a potential approach to address certain challenges that GR or the TEGR fail to resolve adequately. Theoretical frameworks in cosmology can be distinctly classified into two primary categories:

I) Modified gravity models:
The framework offers an alternative viewpoint on gravitational principles that diverge from GR. A key aspect of these models is their ability to examine cosmological evolution without depending on exotic forms of matter.

II) DE models:
The modified matter model is a framework in the Einstein field equations that incorporates an energy-momentum tensor with a component representing exotic matter characterized by negative pressure. This framework helps explain the cosmological evolution of the Universe.

\section{ Modified gravity model}\label{modifiedgravitymodel}
A variety of modified gravity models. Some examples of these include $f(T)$ gravity \cite{Ferraro:2006jd, Krssak:2015oua}, $f(T, \mathcal{T})$ gravity \cite{Harko_2014a}, $f(T,\phi)$ gravity  \cite{Gonzalez-Espinoza:2021mwr},  scalar-tensor theories \cite{Amendola_2000_cup, Copeland:2006wr, Tsujikawa:2013fta} and  Galileon gravity \cite{Nicolis:2008in}, among others.
\subsection{Mathematical formalism of TEGR and its modifications} \label{mathTEGRItsmodification}
After Einstein developed the field equations of GR, alternative interpretations emerged, such as the TEGR. The equations of motion in  TEGR are aligned with those in GR, with the only difference being a total derivative term in their action. This results in the two theories being indistinguishable through experimental means. In contrast to GR, TEGR uses force equations rather than the geodesic equation to describe particle movement influenced by gravity. In TEGR, the tetrad is the main dynamic variable instead of the metric used in GR. Einstein introduced these concepts in 1928, building on mathematical insights from Weitzenb$\ddot{o}$ck's research in 1923. For further details on TEGR, refer to Ref.~\cite{Aldrovandi:2013wha}.

The concept of gravity is based on a manifold, with each point having a tangent space. This tangent space is modeled as Minkowski spacetime, characterized by the metric $\eta_{AB} = \text{diag} [-1, +1, +1, +1]$. In this thesis, Greek indices such as $\alpha, \beta, \gamma,...$ represent spacetime indices, while Latin indices like $A, B, C,...$ denote tangent space indices. The tetrad is symbolized as $e^{A}_{~\mu}$, with its inverse denoted as $E^{~\mu}_{A}$. For a flat FLRW spacetime, it will be represented in the following way,
\begin{align}\label{FLRWTETRAD}
e^{A}_{~\mu}=(1,a(t),a(t),a(t))\,.
\end{align}
The orthogonality conditions for tetrads can described as follows:
\begin{align}
 e^{A}_{\,\,\,\,\mu} E_{B}^{\,\,\,\,\mu}=\delta^A_B\,,\hspace{2cm} e^{A}_{\,\,\,\,\mu} E_{A}^{\,\,\,\,\nu}=\delta^{\nu}_{\mu}\,. 
\end{align}
The relationship between Minkowski space-time and the metric tensor can be articulated as follows,
\begin{align}
    g_{\mu\nu} = e^A_{~\mu}\, e^B_{~\nu} \,\,\eta_{AB}\,, \quad
    g^{\mu\nu} = E^{~\mu}_{A}\,E^{~\nu}_{B}\,\,\eta^{AB} \,.
\end{align}
In GR, the Levi-Civita connection is utilized, which is characterized by its torsion-free property. But, in TEGR, the Weitzenb$\ddot{o}$ck connection is employed instead of the Levi-Civita connection. The Weitzenb$\ddot{o}$ck connection can be expressed as follows,
\begin{align}\label{witzenbockconnection}
    \hat\Gamma^{\sigma}_{\,\,\,\,\nu\mu} := E_{A}^{\,\,\,\,\sigma}\left(\partial_{\mu}e^{A}_{\,\,\,\,\nu} + \omega^{A}_{\,\,\,\,B\mu} e^{B}_{\,\,\, \nu}\right)\,,
\end{align}
The symbol \(\omega^{A}_{\,\,\,\, B\mu}\) denotes the spin connection, which incorporates the inertial effects found in the tetrad formalism. Another aspect of the spin connection is its antisymmetry in the first two indices $(\omega_{AB\mu} = -\omega_{BA\mu})$. This connection \cite{Aldrovandi:2013wha, Krssak:2015oua} can be defined as,
\begin{align}
        \omega^{A}_{\, \, \, B\mu}=-\Lambda_{\,B}^{C}\partial_{\mu} \Lambda_{\,C}^{A} \,,
    \end{align}
The notation $\Lambda_{\,C}^{A}$ denotes the Lorentz metric. The connection coefficients $\Gamma^{\mu}_{\, \, \,\nu\sigma}$ are introduced primarily to ensure the covariance of the covariant derivative. Nonetheless, the consequences of these concepts reach further than just maintaining mathematical validity. They are crucial in defining the geometric characteristics of spacetime. Specifically, the connection describes the trajectories of freely falling particles, while the metric delineate the causal structure of manifold \cite{Cai:2015emx, Capozziello:2011et}. To differentiate among various geometric configurations, one employs the Riemann curvature tensor, the torsion tensor and the principle of metric compatibility, all of which are intrinsically linked to the characteristics of the connection. The Riemann tensor \cite{Krssak:2015oua} can be defined as,
\begin{equation}
    R^\theta_{\ \sigma\mu\nu} = \partial_\mu \Gamma^\theta_{\nu\sigma} - \partial_\nu \Gamma^\theta_{\mu\sigma} + \Gamma^\theta_{\mu\alpha} \Gamma^\alpha_{\nu\sigma} - \Gamma^\theta_{\nu\alpha} \Gamma^\alpha_{\mu\sigma}\,,\label{Riemanntensor}
\end{equation} 
and the term $\Gamma^{\mu}_{\, \, \,\nu\sigma}$ (Levi-Civita connection) can be defined as,
 \begin{equation}
    \Gamma^\mu_{\nu\sigma} = \frac{1}{2} g^{\mu\alpha} \left( \partial_\nu g_{\alpha\sigma} + \partial_\sigma g_{\alpha\nu} - \partial_\alpha g_{\nu\sigma} \right)\,,
\end{equation}
It provides an idea of curvature, which can be seen as a method to gauge variation in direction when a vector is transported along a closed curve. Therefore, a connection is defined as flat (i.e., without curvature) when  
$R^\theta_{\ \sigma\mu\nu} = 0.$  In the Weitzenb$\ddot{o}$ck gauge, the spin connection preserves local Lorentz invariance, and its components vanish identically. This relationship enables the characterization of \cite{Hayashi:1979qx} as an antisymmetric operator related to the Riemann tensor, which becomes null for the teleparallel connection.
 In this thesis, the work related to the zero spin connection ($\omega^{A}_{\, \, \, B\mu}=0$) or the non-covariant teleparallel framework. For the non-covariant case, the torsion tensor can be described as, 
\begin{equation}
    T^\theta_{~\mu\nu} = \hat\Gamma^\theta_{\nu\mu} - \hat\Gamma^\theta_{\mu\nu}\,.
\end{equation}
The torsion tensor is covariant under local Lorentz transformations and diffeomorphisms. The torsion scalar \cite{Krssak:2018ywd, Bahamonde:2021gfp} is derived from contractions of the torsion tensor, which can be written as,
\begin{equation}\label{torsionscalar}
T \equiv \frac{1}{4} T^{\rho \mu \nu} T_{\rho \mu \nu}+\frac{1}{2} T^{\rho \mu \nu} T_{\nu \mu \rho}-T_{\rho \mu}^{~~\rho} T^{\nu \mu}_{~~\nu}.
\end{equation}
The field equations describe the relationship between curvature geometry and matter. They can be derived using the principle of least action and a Lagrangian density, a concept introduced by Hilbert \cite{Hilbert:1915tx}. The Lagrangian incorporates a scalar and given that curvature represents gravitational effects, it follows that this scalar must be derived from the Riemann curvature tensor. In GR, the Ricci scalar \( R = R^{\mu}_{~\mu} \) represents this relationship, where \( R_{\mu \nu} = R^{\alpha}_{~\mu \alpha \nu} \) is referred to as the Ricci tensor. The matter Lagrangian \( \mathcal{L}_{\text{m}} \) incorporates the effects of matter fields and is added to the total Lagrangian. This results in an overall action that combines both gravitational and matter aspects. The total action can be defined as,
\begin{equation}\label{einsteinaction}
    S_{EH} = \frac{1}{2} \int d^4x \, \sqrt{-g} \, R + \int d^4x \, \sqrt{-g} \, \mathcal{L}_{\text{m}} + \int d^4x \, \sqrt{-g} \, \mathcal{L}_{\text{r}} \,.
\end{equation}
By applying variations of the action \eqref{einsteinaction} concerning the metric and lapse function, the Einstein field equations are derived as detailed in Eqs.~(\ref{FriedmanEQ1}, \ref{FriedmanEQ2}).

The action formula of the TEGR theory is described as,
\begin{equation}\label{TEGRaction}
    S_{TEGR} = \frac{1}{2} \int d^4x \, e \, T + \int d^4x \, e \, \mathcal{L}_{\text{m}}+  \int d^4x \, e \, \mathcal{L}_{\text{r}}\,.
\end{equation}
By varying TEGR action concerning the tetrad field \cite{Aldrovandi:2013wha}, one can derive the field equations for TEGR as,
\begin{equation}
    e^{-1} \partial_{\nu} (e S_{A}^{~~\mu\nu})+\frac{1}{4} E^{~\mu}_{A} T -T^{B}_{~~\nu A} S_{B}^{~~\nu\mu}+\omega^{B}_{~~A\nu} S_{B}^{~~\nu\mu}=\frac{1}{2} T^{~~\mu}_{A}\,.
\end{equation}
The superpotential term \( S_{\rho}^{~~\mu \nu} \) and the contortion tensor \( K^{\mu \nu}_{~~~\rho} \) can be expressed as,
\begin{align}
 S_{\rho}^{~~\mu \nu} &\equiv \frac{1}{2}(K^{\mu \nu}_{~~~\rho} + \delta^{\mu}_{\rho} T^{\alpha \nu}_{~~~\alpha} - \delta^{\nu}_{\rho} T^{\alpha \mu}_{~~~\alpha}) \,, \label{superpotentialterm}\\
 K^{\mu \nu}_{~~~\rho} &\equiv \frac{1}{2}(T^{\nu \mu}_{~~~\rho} + T_{\rho}^{~~\mu \nu} - T^{\mu \nu}_{~~~\rho}) \,. \label{contorsiontensor}
\end{align}
The equivalence between TEGR and GR arises from the relationship between quantities associated with curvature (Riemann tensor) and those related to torsion (Torsion tensor). This relationship is established by integrating the definition of the Riemann tensor as shown in Eq.~\eqref{Riemanntensor}, with the expression provided by the contortion tensor in Eq.~\eqref{contorsiontensor}. Ref. \cite{Aldrovandi:2013wha} illustrated that this integration simplifies the teleparallel field equations to those of GR. The continued development of the TEGR framework leads to an alternative theory of gravity, which will be discussed in the next section.
\subsection{\texorpdfstring{$f(T)$}{} gravity}
The first simple extension of the TEGR Lagrangian incorporates a general function of the torsion scalar $f(T)$, which can be regarded as the torsional counterpart of $f(R)$ gravity, a modification of GR that involves curvature. The gravitational action associated with $f(T)$ gravity can be expressed as,
\begin{equation} \label{fTaction}
    S_{f(T)} = \frac{1}{2} \int d^4x \, e \, f(T) + \int d^4x \, e \, \mathcal{L}_{m}+\int d^4x \, e \, \mathcal{L}_{r}\,.
\end{equation}
The general field equations of  $f(T)$ gravity given by \cite{Ferraro:2006jd, Linder:2010py, Bengochea:2008gz, Tamanini:2012hg, Krssak:2015oua, Cai:2015emx,  Farrugia:2016xcw},
\begin{equation}
   f_{TT} S_{A}^{~~\mu\nu} \partial_{\nu} T+\frac{1}{4} E_{A}^{~~\mu} f +f_{T} e^{-1} \partial_{\nu} (e S_{A}^{~~\nu \mu})-f_{T} T^{B}_{~~\nu A} S_{B}^{~~\nu\mu}+f_{T}\omega^{B}_{~~A\nu}\, S_{B}^{~~\nu\mu}=\frac{1}{2} T_{A}^{~~\mu}\,.\label{f(T)FE}
\end{equation}
The $f(T)$ gravity field equations related to the tetrad field are of the second order, in contrast to the fourth-order equations found in $f(R)$ gravity. It is crucial to consider the local Lorentz invariance of the theory, which has been explored in numerous studies (see Ref. \cite{Cai:2015emx}). The $f(T)$ gravity is recognized for violating this invariance, which is attributed to the non-local Lorentz invariance of the torsion scalar. In TEGR, this issue is not a concern since the field equations still maintain local Lorentz invariance. However, in the case of $f(T)$ gravity, the equations would cease to exhibit local Lorentz invariance \cite{Li:2010cg, Krssak:2015oua}. 
\subsection{\texorpdfstring{$f(T, \mathcal{T})$}{} gravity}
The next teleparallel extension examines a coupling between matter and gravity that involves an arbitrary function of the torsion scalar $T$ and the trace of the stress-energy tensor $\mathcal{T}$, a framework initially proposed in Ref. \cite{Harko_2014a}. The action for $f(T, \mathcal{T})$ gravity can be expressed as,
\begin{equation}\label{fTTaction}
    S_{f(T, \mathcal{T})} = \frac{1}{2} \int d^4x \, e \, f(T, \mathcal{T}) + \int d^4x \, e \, \mathcal{L}_{m}+ \int d^4x \, e \, \mathcal{L}_{r}\,.
\end{equation}
The general field equations of  $f(T, \mathcal{T})$ gravity written as,
\begin{equation}
 \frac{1}{4}\, E^{\,\rho}_{A} \, f + f_{T}\, [e^{-1}\partial_{\sigma}(e \, S^{\,\,\,\rho \sigma}_{A})-T^{B}_{\,\, \nu A} \, S^{\,\,\,\nu \rho}_{B}+ \omega^{B}_{\,\, A\nu}\, S_{B}^{\,\,\,\nu\rho}]+ S_{A}^{\,\,\,\sigma\rho} \partial_{\sigma} \, f_{T}+ \frac{f_{\mathcal{T}}}{2} \, (T^{\,\,\rho}_{A}+p\, E^{\,\rho}_{A})= \frac{1}{2} T^{\,\,\rho}_{A}  \,.\label{f(TT)FE}
\end{equation}
The theory of matter coupling \cite{Harko_2014galaxies} has been widely explored in the literature, particularly its curvature analog, which displays characteristics distinct from previously mentioned behaviors. Including such coupling broadens the potential interactions between gravity and matter, potentially offering additional understanding of DM and DE without introducing new exotic types of matter. Research on $f(T, \mathcal{T})$ gravity has been explored across various contexts, highlighting its potential as a robust alternative framework for both early and late-time cosmology \cite{Harko_2014a}. Notable studies have examined the growth factor for subhorizon perturbations in the late Universe \cite{Jackson2016a}. In contrast, others have focused on reconstructing the gravitational action associated with the $\Lambda$CDM model, including a preliminary stability analysis \cite{Junior_2016}. Additionally, investigations into the Tolman-Oppenheimer-Volkoff equation within the context of quark star systems have been conducted, further elucidating the implications of modified $f(T, \mathcal{T})$ gravity models \cite{Pace_2017}.  
\subsection{ Scalar tensor theories}\label{Scalartensortheorie} 
The Brans-Dicke theory \cite{Tsujikawa_2010brance, Brans:1961mach} represents the most fundamental formulation of scalar-tensor theories. A scalar field is associated with the Ricci scalar \( R \) and  the Lagrangian density can be represented in the following way,
\begin{align}\label{brancedickelagrangian}
    \mathcal{L}= \frac{\phi R}{2}- \frac{\omega_{BD}}{2 \phi} (\nabla \phi)^2\,.
\end{align}
In this context, $\omega_{BD}$ denotes the Brans-Dicke parameter, while $\phi$ signifies the scalar field inherent to the Brans-Dicke theory of gravity. In the weak field approximation, the theory of Brans-Dicke aligns with the predictions of GR as $\omega_{BD}$ approaches infinity. However, it has been shown that the theory does not align with the principles of GR in the nonlinear regime \cite{Banerjee:1997brance}. When an additional potential \( U(\phi) \) is considered and  \( \omega_{BD} = 0 \), the generalized Brans-Dicke theory is essentially comparable to \( f(R) \) gravity when expressed in the metric formulation \cite{Hanlon:1972}. By applying a conformal transformation to the action of the generalized Brans-Dicke theory, it can be shown to be equivalent to a coupled quintessence scenario \cite{Amendola_2000_cup}. Banerjee and Pavon \cite{Banerjee:2011accequi} demonstrated that the Brans-Dicke theory alone can accelerate expansion without needing exotic matter. Nevertheless, a limitation is that it does not provide a transition from a decelerated phase to an accelerated one.
\subsection{\texorpdfstring{$f(T, \phi)$}{} gravity}
This section will explore a further extension of the TEGR involving the scalar field $\phi$, known as $f(T,\phi)$ gravity. Just like the modifications that have been applied in GR with the scalar field \cite{Capozziello:2011et}, similar adaptations have been implemented in the TEGR \cite{Cai:2015emx, Bahamonde:2017ize}. It is crucial to understand that while TEGR and GR share dynamic equivalence, a scalar-torsion theory with non-minimal coupling does not have the same equivalence as its curvature-based version (scalar-tensor theory). A scalar-torsion theory incorporating a non-minimal coupling term, specifically $\xi \phi^2 T$, where $\phi$ represents the dynamic scalar field, was originally applied to explore DE in Ref. \cite{Geng:2011aj, Geng_2012}. This framework can be expanded by considering a more general non-minimally coupled scalar field represented as $F(\phi)G(T)$. Such generalizations are viewed as an extension of $f(T)$ gravity within the gravitational sector \cite{Ferraro:2008ey, Linder:2010py}. Motivated by the study of scaling solutions highlighted in Ref. \cite{Uzan1999, Amendola1999}, the $f(T,\phi)$ gravity framework \cite{Gonzalez-Espinoza:2021mwr, Gonzalezreconstruction2021, Gonzalez-Espinoza:2020jss} gains significance in the context of the TEGR.

The gravitational action for $f(T,\phi)$ is \cite{Gonzalez-Espinoza:2021mwr},
\begin{equation}\label{ActionEqf(Tphi)}
S =\int d^{4}xe[f(T,\phi)+P(\phi)X]+ \int d^4x \, e \, \mathcal{L}_{m}+ \int d^4x \, e \, \mathcal{L}_{r} \,,
\end{equation}
where \( X = -\frac{1}{2} \partial_{\mu} \phi \partial^{\mu} \phi \) is  the kinetic term of the scalar field \( \phi \). The total action \( S \) encompasses contributions from the matter lagrangian $\mathcal{L}_{m}$ and radiation lagrangian $\mathcal{L}_{r}$ sectors.

By varying this action for the tetrads, corresponding field equations are derived as
\begin{equation}
    f_{T}\, G_{\mu\nu} + S_{\mu\nu}^{\ \ \ \theta}\, \partial_{\theta} \, f_{T} + \frac{1}{4} g_{\mu\nu} \left( f - T\, f_{T} \right) + \frac{P}{4} \left( g_{\mu\nu}\, X + \partial_\mu \phi \, \partial_\nu \phi \right) = -\frac{1}{4} T_{\mu\nu},
\end{equation}
where \( G^{\mu}_{~\nu} = E_A^{\ \mu} G^A_{\ \nu} \) is the Einstein tensor \cite{Aldrovandi:2013wha} and it can be established in the following way,
\begin{equation}
    G^{\mu}_{\ A} \equiv e^{-1} \partial_\nu \left( e E^{\,\sigma}_{A} S_{\sigma}^{\,\,\,\mu\nu} \right) - E^{\,\sigma}_{\, A} T^{\alpha}_{\ \theta\sigma} S^{\,\,\,\theta\mu}_{\alpha}
    + E^{\,\alpha}_{\ B} S^{\,\,\,\theta\mu}_{\ \alpha} \omega^{B}_{\,\, A\theta} + \frac{1}{4} E^{\,\mu}_{A} T \,.
\end{equation}
\subsection{Teleparallel Framework of the Horndeski theory}\label{Teleparallel FrameworkofHorndeskitheory}
Horndeski's framework provides a comprehensive method for describing gravitational interactions in four-dimensional spacetime \cite{Horndeski:1974wa}. According to Lovelock's theorem \cite{Lovelock:1971yv}, second-order theories restrict the types of Lagrangian terms that can be used. In this framework, the Teleparallel equivalent of Horndeski's theory \cite{Bahamonde:2019shr} can be expressed through a specific Lagrangian formulation as
\begin{equation}\label{horndeskilagrangian}
\mathcal{L}=\sum_{i=2}^5 \mathcal{L}_i \,.    
\end{equation}
This theory represents the most comprehensive framework utilizing a single scalar field, characterized by scalar invariants limited to quadratic expressions in the contractions of the torsion tensor. Consequently, it yields second-order field equations about the derivatives of both the tetrad and the scalar field. The Lagrangians \(\mathcal{L}_{i}\) for \(i=2,3,4,5\) \cite{Horndeski:1974wa} correspond to well-defined action within this context defined as,
\begin{align}
 \mathcal{L}_{2} & :=G_{2}(\phi,X)\,, \quad
    \mathcal{L}_{3} :=G_{3}(\phi,X)\Box\phi\,,\label{eq:LagrHorn1}\\
    \mathcal{L}_{4} & :=G_{4}(\phi,X)\left(-T+B\right)+G_{4,X}(\phi,X)\left(\left(\Box\phi\right)^{2}-\phi_{;\mu\nu}\phi^{;\mu\nu}\right)\,,\\
     \mathcal{L}_{5} & :=G_{5}(\phi,X)G_{\mu\nu}\phi^{;\mu\nu}-\frac{1}{6}G_{5,X}(\phi,X)\bigg(\left(\Box\phi\right)^{3}+2 \phi_{;\mu}^{~~\nu} \phi_{;\nu}^{~~\alpha}\phi_{;\alpha}^{~~\mu} -3\phi_{;\mu\nu}\phi^{;\mu\nu}\,\Box\phi\bigg)\,.\label{eq:LagrHorn5}  \end{align}
In this context, the boundary term \( B \) is related to the Ricci scalar \( R \) and the torsion scalar \( T \) through the relation \( R = -T + B \). Here, $\phi$ represents the scalar field and  $X$ indicates the kinetic energy. Additionally, $G_{i}$, for $i=2,3,4,5$, is a function that depends on the scalar field $\phi$ and the kinetic energy $X$. Furthermore, $G_{i, X}$ and $G_{i,\phi}$ correspond to the partial derivatives of $G_{i}$ with respect to $X$ and $\phi$, respectively. The entire action can be written as, 
\begin{align}\label{Horndeski_action_Lagran}
    \mathcal{S} = \frac{1}{2} \int \mathrm{d}^4 x e\, \mathcal{L}+\int \mathrm{d}^4 x e\,(\mathcal{L}_m+\mathcal{L}_r) \,,
\end{align}   
\section{DE models}\label{darkenergymodel}
To tackle the problem associated with the cosmological constant, it is important to concentrate on a type of DE that changes as time progresses. Numerous scalar field DE models have been explored in the literature, including Quintessence \cite{Amendola_2000_cup, Tsujikawa:2013fta, Copeland:2006wr, Bassett:2005xm, Cai:2009zp, Bahamonde:2017ize, Otalora:2013dsa}, Phantom \cite{Hoyel:1948phan, Francisco:2021phantom} and k-essence \cite{Armendariz_Picon_2000:kessence, Chervon_2019prd, Fomin_2022jcap} and   Tracker models \cite{Ivaylo:1999:trackor}, etc.. These models seek to explain the accelerated expansion of the Universe and often incorporate varying equations of state and scalar field dynamics to address both observational data and theoretical considerations. The next section focuses on the quintessence model among the various scalar field models.
\subsection{Quintessence scalar field model}\label{quintessencemodel}
Quintessence serves as a dynamic substitute for the cosmological constant. A standard scalar field $\phi$ is minimally coupled to gravity in DE models involving quintessence. This scalar field exhibits negative pressure and gradually moves down its potential. The corresponding action for the quintessence scalar field \cite{Bartolo_1999, Faraoni_2000, Geng:2011aj, Geng_2012} can be expressed as,
\begin{equation}\label{actionformulaquintessence}
S = \int d^4x \, \sqrt{-g} \left(\frac{R}{2 \kappa^2}-\frac{1}{2} g^{\mu \nu} \partial_\mu \phi \, \partial_\nu \phi - V(\phi) \right) + S_m + S_r\,. \end{equation}
In this context, \( S_m \) and \( S_r \) represent the contributions from matter and radiation, respectively. The term \( V(\phi) \) denotes the potential function associated with the quintessence scalar field. The energy-momentum tensor corresponding to the quintessence scalar field can be articulated as,
\begin{align}
T_{\mu \nu}= \partial_\mu \phi \, \partial_\nu \phi 
- g_{\mu \nu} \left[\frac{1}{2} g^{\gamma \delta} \partial_\gamma \phi \, \partial_\delta \phi + V(\phi)\right] \,.  
\end{align}
In a flat FLRW metric \eqref{FLATFLRW}, the energy density and pressure associated with the scalar field are described as,
\begin{align}
 \rho_{\phi}= \frac{1}{2} \dot{\phi}^2 + V(\phi) \,, \label{rhophi}\\
 p_{\phi}= \frac{1}{2} \dot{\phi}^2 - V(\phi) \,. \label{pphi}
\end{align}
In this framework, the Einstein field equations can be expressed in the following form,
\begin{align}
  H^2&= \frac{8 \pi G}{3}\left(\rho_m+\rho_r+\dot{\phi}^2 + V(\phi)\right) \,, \label{quinfriedmann1}\\
  2 \dot{H}&=-\kappa^2 \left(\dot{\phi}^2+\rho_m+\frac{4}{3}\rho_r\right)\,.  \label{quinfriedmann2} 
\end{align}
By varying the action concerning $\phi$, leads to Klein-Gorden equation,
\begin{align}\label{kleingordanequation}
\ddot{\phi}+ 3 H \dot{\phi}+ \frac{dV}{d\phi} =0 \,.    
\end{align} 
\section{Dynamical systems technique}\label{dynamicalsystemtechnique}
This section provides a concise overview of the fundamental principles of dynamical systems analysis \cite{Perko2001, James:2007, Wainwright_Ellis_1997, coley2003dynamical}, highlighting the key aspects applied in our study.

Differential equations indicate the links between functions and their derivatives. This idea was first uncovered by Newton in the mid-seventeenth century. He applied it in his gravitational theory and found a solution for two-body systems, such as the sun's and earth's motion. A differential equation involving only one independent variable is known as an ordinary differential equation (ODE). We will focus solely on ODEs, which are pertinent to our analysis. Consider an ODE presented in the following manner,
\begin{align}\label{nsyatem}
\dot{X}=f(X)\,,    
\end{align}
where $X=(X_1, X_2, X_3, ...., X_n) \in \mathbb{R}^n, \quad f : \mathbb{R}^n \to \mathbb{R}^n$ and  $\dot{X}=\frac{dX}{dt}$. When a set of differential equations has no explicit time dependence, it is an autonomous system. Our discussion will primarily concentrate on autonomous systems. A system that is not autonomous can be viewed as autonomous by introducing time as a new variable, meaning $X_{n+1}=t$ and $\dot{X}_{n+1}=1$. This approach will expand the dimensions of the system by one. Although $f(X)$ is generally nonlinear, the discussion begins with linear systems, which provide a foundation for understanding nonlinear behavior. A linear differential equation can be formulated as,
\begin{align}\label{linergeneralsystem}
    \dot{X}= f(X)= A\,X
\end{align}
where $A$ stands for an $n \times n$ matrix. Beginning with the initial state \(X(0) = X_0\), the solution to the ordinary differential equation \eqref{linergeneralsystem} can be represented as \(X =X_0 e^{At}\). The term \(e^{At}\) denotes an \(n \times n\) matrix and can be computed using its Taylor series expansion. Various techniques are available for computing the exponential of a matrix based on its eigenvalues (for more details, see Ref. \cite{Perko2001}). In n-dimensional space, solutions appear as curves in phase space, where these paths are known as phase trajectories. To grasp the qualitative dynamics of a system, it is essential to pinpoint its critical points. Critical points are locations where the solutions remain constant. In mathematical language, critical points refer to the shared solutions of the equation $f(X)=0$. Depending on their stability, critical points can be classified as unstable, stable or saddle points. To evaluate the stability of a critical point, one can introduce slight disturbances to the system away from that point. If the system returns to the critical point, it is classified as a stable critical point; conversely, it is considered an unstable fixed point if it does not return. If the direction of the disturbance influences the behavior, then the critical point is identified as a saddle point.

If all eigenvalues of matrix \( A \) are negative, the critical point is stable. If all eigenvalues are positive, the critical point is either a repeller or instability. A critical point is a saddle point when there is a mix of positive and negative eigenvalues. 

When eigenvalues are imaginary of the form $\lambda = a \pm ib$, the behavior of the phase portrait near the critical point is determined by the real part $a$. Specifically, the trajectories spiral inward for $a < 0$, indicating stability. In contrast, for $a > 0$, they spiral outward suggesting instability. When \( a = 0 \), the critical point acts like a center. 

Critical points are classified into two categories: hyperbolic and non-hyperbolic. Hyperbolic critical points occur when \( Re(\lambda) \neq 0 \) (where $Re(\lambda)$ is the real part of the eigenvalue of $\lambda$), while non-hyperbolic critical points do not meet this condition.

The behavior of nonlinear systems in phase space is explored. To begin, a nonlinear system of differential equations is represented as follows:
\begin{align} \label{nonlinersystemgeneral}
    \dot{X} = f(X)\,.
\end{align}
Let \( f: E \rightarrow \mathbb{R}^n \), where \( E \) is an open set within \( \mathbb{R}^n \). In non-linear dynamical systems, the governing differential equations cannot be represented in matrix form, unlike what is achievable with linear systems. Nevertheless, it is possible to linearize the non-linear system in the neighborhood of a hyperbolic fixed point, making it easier to analyze using the linear approximation method. The liner system can be written as $\dot{X}=AX$. Where $A=Df(X)$ denotes the Jacobian matrix associated with the system. The stability of a stationary point can be assessed by examining the eigenvalues of the Jacobian matrix, which is expressed as \( A = Df(X) = \left( \frac{\partial f_i}{\partial X_j} \right) \). Our analysis used the center manifold theory to assess the stability of non-hyperbolic critical points. The following section explains the center manifold theory approach.
\subsection{Center manifold theory (CMT)}\label{Centralmanifoltheory}
CMT is a distinct area within the theory of dynamical systems that focuses on examining the behavior of systems near their fixed points. Perko \cite{Perko2001} provided a comprehensive explanation of the essential mathematical structure of CMT. Traditional linear stability theory often struggles to accurately assess the stability of critical points when the associated eigenvalues are zero. Conversely, CMT facilitates a stability evaluation by effectively reducing the dimensionality of the system near these points. When a system transitions through a critical point, its dynamics are dictated by the invariant local center manifold, denoted as $W_c$. This central manifold $W_c$ corresponds to eigenvalues that have zero real parts and the behaviors that unfold within this manifold encapsulate the critical characteristics of the system near equilibrium \cite{wiggins2003introduction,carr1981applications, Bahamonde:2017ize}. \\
 Consider the dynamical system characterized by the following equations,
 \begin{align} \label{CMT1}
\tau'= F(\tau)\,,
\end{align}
where $\tau= (\mu, \nu)$. A geometric space is regarded as a center manifold for this system can be locally represented as,
\begin{align} \label{invariantmanifold}
W_c = \{ (\mu, \nu) \in \mathbb{R} \times \mathbb{R} : \nu = h(\mu), |\mu| < \delta, h(0) = 0, \nabla h(0) = 0 \}\,,
\end{align}
for sufficiently small values of $\delta$, the function $h(\mu)$ behaves smoothly on $\mathbb{R}$. The subsequent steps examine the CMT:
\begin{enumerate}
    \item \textbf{Coordinate shift:} Initially, the locations of the non-hyperbolic critical points are shifted to the origin, resulting in a revised set of autonomous equations expressed in the newly transformed coordinate system.
    \item \textbf{Reformulation of the Dynamical System:} As illustrated below, the revised dynamical system is provided in a standard format, facilitating further analysis.
    \begin{align} \label{ReformulationoftheDynamicalSystem}
\tau' =
\begin{pmatrix}
\mu' \\ 
\nu'
\end{pmatrix}
=
\begin{pmatrix}
A\mu \\ 
B\nu
\end{pmatrix}
+
\begin{pmatrix}
\phi(\mu, \nu) \\ 
\psi(\mu, \nu)
\end{pmatrix}\,,
\end{align}
the functions \( \phi \) and \( \psi \) meet the following requirements,
\begin{align}
 \phi(0, 0) = 0, \quad \nabla \phi(0, 0) = 0\,, \quad  \psi(0, 0) = 0, \quad \nabla \psi(0, 0) = 0\,. \nonumber   
\end{align}

Using $\nabla$ to represent the gradient operator, it is important to note that $A$ and $B$ are square matrices that have the following properties: 
\begin{itemize} 
\item The eigenvalues of $A$ have zero as their real parts. 
\item The eigenvalues of $B$ have negative real parts. 
\end{itemize}
  \item \textbf{Finding the Function $h(\mu)$:} Subsequently, a function \( h(\mu) \) is established, often employing a series expansion that incorporates \( \mu^2 \) term. This function \( h(\mu) \) satisfies the following quasilinear partial differential equation,
  \begin{align} \label{approximationfunction}
\mathcal{N}(h(\mu))=\nabla h(\mu)[A\mu+\phi(\mu,h(\mu))] -Bh(\mu)-\psi(\mu,h(\mu))=0 \,,
\end{align}
under the conditions $h(0) = 0$ and $\nabla h(0) = 0$.
   \item \textbf{Center Manifold Dynamics:} Utilizing the approximate solution for $h(\mu)$ obtained from Eq.~\eqref{approximationfunction}, the dynamics of the original system restricted to the center manifold can be represented in the following manner,
   \begin{equation} \label{dynamicsofcentermanifoldequation}
   \mu' = A\mu + \phi(\mu, h(\mu))\,,
   \end{equation}
   for $\mu \in \mathbb{R}$ is sufficiently small.
   \item \textbf{Final State of the Reduced System:} The equation $\mu' = A\mu + \phi(\mu, h(\mu))$ can be reduced to the form $\mu' = k \mu^n$, where $k$ represents a constant and $n$ denotes a positive integer, specifically about the term with the lowest order in the series expansion.
   \begin{itemize}
       \item If $k<0$ and $n$ is an odd integer, the system exhibits stability, indicating that the original system is also stable.
       \item The reduced and the original systems will display instability in all other circumstances.
   \end{itemize}
\end{enumerate}


\section{Application of dynamical systems technique in cosmological studies}
The foundation of contemporary cosmological models is built on non-linear differential equations, which pose significant challenges in deriving exact solutions. A fruitful strategy for analyzing these non-linear systems is to employ a dynamical systems framework, which facilitates understanding their qualitative dynamics. This approach typically involves introducing normalized, dimensionless variables and a dimensionless time parameter, reformulating the system into an autonomous form. These newly defined variables are directly linked to measurable physical quantities and demonstrate stable behavior. By pinpointing the fixed point of the system and assessing its stability characteristics, it becomes possible to qualitatively investigate the gravitational origins of the Universe and its potential end state. Notably, heteroclinic trajectories connect unstable fixed points to stable ones; thus, unstable fixed points may correspond to the early eras of the Universe, while stable fixed points could signify their ultimate destiny (late-time cosmic phenomena). This analysis clarifies the dynamic evolution of cosmic systems and enhances our understanding of cosmological phenomena.
This kind of analysis is well-established in GR and cosmology. Nearly all significant models in GR and cosmology have been examined through dynamical systems analysis. Some noteworthy examples include modified gravity, scalar-tensor theories, Bianchi-type models and  non-minimally coupled scalar field models. Several cosmological systems in modified theories of gravity may be seen in Refs. \cite{Amendola_2000_cup, Geng:2011aj, Otalora:2013dsa, Otalora:2013tba, roy2018dynamical, Franco:2021compl, Gonzalez-Espinoza:2020jss, Kadam:2022lgq, dutta2018jcap, Awad_2018, Narawade_2022a, Khyllep2023, Agrawal_2023, Gonzalez-Espinoza:2020jss, Patil_2023,  Lohakare_2023_40, Agrawal_2024, Kadam_2024ttg, Gonzalez_Espinoza_2024pha, Alfredo:2024kantow, Chervon_2023GC, Chervon_2020Uni, Vishwakarma_2024}.
\section{Observational Cosmology} \label{ObservationalCosmology}
This section provides detailed summaries of the diverse cosmological data sets and the methodologies used to derive optimal estimates of the cosmological parameters. An MCMC \cite{Foreman_Mackey_2013} approach was employed to determine the maximum likelihood estimates of these parameters, leveraging Bayesian statistics to inform the model fitting process.
\subsection{MCMC technique} \label{MCMCapproachsection}
MCMC method refers to algorithms designed to sample from probability distributions that are challenging to calculate directly. It merges the concepts of Markov chains with Monte Carlo integration, rendering it an effective tool in fields such as statistics, machine learning, physics and cosmology. There are MCMC applications, such as Bayesian Inference, To estimate posterior distributions for complex models. Cosmology: To constrain cosmological parameters by analyzing observational data. Machine Learning: For probabilistic models such as Bayesian networks and neural networks. Physics and Chemistry: To model systems with many interacting particles. Through this approach, we have compared our chosen cosmological model with the cosmological observation data sets and obtained the best-fit values for the model parameter.
\subsection{Bayesian Inference} \label{BayesianInference}
Bayesian inference is a statistical method based on Bayes' Theorem. It allows for updating the probability of a hypothesis as new evidence or data is acquired. The theorem relates a parameter $\Theta$ to observed data $D$. Bayes' Theorem is expressed as, 
\begin{align}
    P(\Theta | D) = \frac{P(D | \Theta) P(\Theta)}{P(D)}, 
\end{align}
where \(P(\Theta | D)\): The posterior distribution, \(P(D | \Theta)\): The likelihood function, \(P(\Theta)\): The prior distribution and  \(P(D)\): The evidence or marginal likelihood,  \begin{align}
        P(D) = \int P(D \mid \Theta) P(\Theta) \, d\Theta.
\end{align}

In our study, we have used the CC, PN$^+$\&SH0ES and  BAO data sets. The following section delves into the mathematical formulation associated with these data sets.
\subsection{CC data set} \label{CCdatasetsintro}
The primary concept of the CC data set is to utilize passive galaxy evolution to estimate the age of the Universe across various redshifts. The ages of these galaxies can be inferred from spectroscopic data, particularly from absorption features in the spectra, which offer insights into the stellar population and its age. We utilized 31 data points for the Hubble parameter obtained through the CC method. This approach directly assesses the Hubble function over a spectrum of redshifts up to \( z \leq 2 \). The advantage of CC data is its capability to determine the age difference between two galaxies that evolve passively and formed at the same time but exhibit a slight variation in redshift, which permits the calculation of \( \frac{\Delta z}{\Delta t} \). Refs \cite{Jimenez_2002, Zhang_2014hz, Jimenez_2003cmb, Moresco_2016hubb, Simon_2005prd, M_Moresco_2012JCAP, Daniel_Stern_2010jcap, Moresco_2015mnras} serve as the foundation for the CC data points. The corresponding \( \chi^{2}_{H} \) follow as,
\begin{equation}\label{chisqure_hz}
\chi^2_{H}(\Theta) = \sum_{i=1}^{31} \frac{ \left( H(z_i, \Theta) - H_{\text{obs}}(z_i) \right)^2 }{ \sigma^2_{H}(z_i) } \,.  
\end{equation}
The Hubble parameters can be classified into two groups: \(H(z_i, \Theta)\) represents the theoretical values of the Hubble parameter at a specific redshift \(z_i\), while \(H_{\text{obs}}(z_i)\) denotes the observed values of the Hubble parameter at \(z_i\), accompanied by an observational uncertainty indicated by \(\sigma_{H}(z_i)\).
\subsection{SNIa data set } \label{pantheonplusintro}
The SNIa data set is one of the most important observational data sets in cosmology. It provides insights into the rate of expansion of the Universe, helps refine cosmological models and is central to our understanding of DE and the evolution of the Universe. SNIa are a subclass of supernovae (exploding stars) used as standard candles for measuring cosmic distances. Their consistent luminosity makes them reliable indicators for determining distances to far-off galaxies and they play a vital role in estimating cosmological parameters such as the Hubble constant, the matter density and the EoS of DE. The data set utilized for our MCMC analysis includes through collection of SNIa, comprising 1701 data points that offer relative luminosity distances within the redshift range of $0.01 < z < 2.3$ \cite{Brout_2022panplus, Riess:2021jrx, Scolnic_2022panplus}. The distance modulus function is the difference between the apparent magnitude $m$ and the absolute magnitude $M$. For a specific redshift $z_i$, the functional representation of the distance modulus $\mu(z_i, \Theta)$ can be articulated as follows,
\begin{equation}\label{modulus_function}
\mu(z_i, \Theta) = m - M = 5 \log_{10} \left[ D_L(z_i, \Theta) \right] + 25\,,    
\end{equation}
the luminosity distance $D_L(z_i, \Theta)$ can be formulated as
\begin{equation}\label{luminosity_distance}
D_L(z_i, \Theta) = c(1 + z_i) \int_0^{z_i} \frac{dz'}{H(z', \Theta)}\,,  
\end{equation}
where $c$ is the speed of light. To find the $\chi^2_{\text{SN}}$ value with the PN$^{+}$\&SH0ES  compilation, which includes 1701 Supernovae data points,  the following formula \cite{Conley_2010Apjs} is derived,
\begin{equation}\label{chisquare_pantheon}
 \chi^2_{\text{SN}} = \left(\Delta\mu(z_i, \Theta)\right)^{T} C_{SN}^{-1} \left(\Delta\mu(z_i, \Theta)\right)\,.
\end{equation}
In this context, the symbol $C_{SN}$ signifies the covariance matrix that includes systematic and statistical uncertainties inherent in the measurements. Furthermore, $\Delta\mu(z_i, \Theta) = \mu(z_i, \Theta) - \mu(z_i)_{\text{obs}}$ represents the difference between the predicted and observed distance modulus at the specific redshift $z_i$. 
\subsection{BAO data set } \label{baointro}
BAO represents imprints in the large-scale structure of the Universe, arising from the interplay of sound waves within the photon-baryon plasma before recombination. These oscillations reflect the acoustic pressure waves propagating through the primordial medium, leading to characteristic density fluctuations that serve as a cosmic ruler, helping delineate the scale of structure formation in the Universe. A composite BAO data set is examined, focusing on various discrete data points. This thesis utilizes a BAO data set that includes observations from the Six-degree Field Galaxy Survey at an effective redshift of \( z_{\text{eff}} = 0.106 \) \cite{Beutler_2011baosixdegree}, the BOSS DR11 quasar Lyman-alpha measurements at \( z_{\text{eff}} = 2.4 \) \cite{du_Mas_des_Bourboux_2017} and the SDSS Main Galaxy Sample at \( z_{\text{eff}} = 0.15 \) \cite{Ross_2015}. Furthermore, measurements of $H(z)$ are included and angular diameter distances derived from the SDSS-IV eBOSS DR14 quasar survey at effective redshifts \( z_{\text{eff}} = \{0.98, 1.23, 1.52, 1.94\} \) \cite{Zhao_2018sdss_IV}, along with the agreed-upon BAO measurements of the Hubble parameter and the associated comoving angular diameter distances from SDSS-III BOSS DR12 at \( z_{\text{eff}} = \{0.38, 0.51, 0.61\} \) \cite{Alam_2017sdss_III}. Our examination considers the complete covariance matrix for the BAO data sets. To assess the BAO data set about the cosmological model, it is essential to compute the volume-average distance \( D_V(z) \), the Hubble distance \( D_H(z) \) and  the comoving angular diameter distance \( D_M(z) \) 
\begin{eqnarray}\label{BAOdistancesintro}
 D_V(z) = \left[(1 + z)^2 D_A^2(z) \frac{z}{H(z)}\right]^{1/3}\,, \quad D_H(z) = \frac{c}{H(z)}\,,\quad D_M(z) = (1 + z)D_A(z)\,,
\end{eqnarray}
where the angular diameter distance $D_A(z) = (1+z)^{-2} D_L(z)$. To incorporate the findings from BAO into MCMC analyses, it is necessary to consider the relevant combinations of parameters.
\begin{eqnarray} \label{baoparametercombination}
\mathcal{F}(z_i) &= \bigg\{D_A(z_i) \bigg( \frac{r_{s, \text{fid}}(z_d)}{r_s(z_d)} \bigg), H(z_i) \bigg( \frac{r_s(z_d)}{r_{s, \text{fid}}(z_d)} \bigg), \frac{r_s(z_d)}{D_V(z_i)},\nonumber \\& D_M(z_i) \bigg( \frac{r_{s, \text{fid}}(z_d)}{r_s(z_d)} \bigg), \frac{D_V(z_i)}{r_s(z_d)}, D_H(z_i) \bigg\}\,.
\end{eqnarray}

Where \( r_s(z_d) \) denotes the sound horizon during the drag epoch, \( r_{s, \text{fid}}(z_d) \) refers to the fiducial sound horizon. To achieve this, the comoving sound horizon \( r_s(z) \) is determined at the redshift \( z_d \approx 1059.94 \) \cite{Aghanim:2018eyx}, which marks the end of the baryon drag epoch.
\begin{eqnarray}\label{sound_horizon}
r_s(z) = \int_{z}^{\infty} \frac{c_s(\tilde{z})}{H(\tilde{z})} d\tilde{z} = \frac{1}{\sqrt{3}} \int_{0}^{1/(1+z)}\frac{da}{a^2 H(a) \sqrt{1 + \left[\frac{3\Omega_{b,0}}{4\Omega_{\gamma,0}}\right] a}}\,,
\end{eqnarray}
The following values are applied: \(\Omega_{b,0} = 0.02242\) \cite{Aghanim:2018eyx}, \(T_0 = 2.7255 \, \text{K}\) \cite{Fixsen_2009temcmb}, along with a reference value of \(r_{s, \text{fid}}(z_d) = 147.78 \, \text{Mpc}\). The relevant calculation for \(\chi^{2}_{\text{BAO}}\) is provided by \cite{Conley_2010Apjs},
\begin{equation}\label{chisquare_BAO}
\chi^2_{\text{BAO}}(\Theta) = \left(\Delta \mathcal{F}(z_i, \Theta)\right)^T C^{-1}_{\text{BAO}} \Delta \mathcal{F}(z_i, \Theta)\,,
\end{equation}
The symbol \(C_{\text{BAO}}\) represents the covariance matrix associated with the selected BAO data, while \(\Delta \mathcal{F}(z_i, \Theta)\) signifies the difference between the theoretical and observed values of \(\mathcal{F}\) at a given redshift \(z_i\).
\subsection{Akaike and Bayesian Information Criteria} \label{AICBICIntro}
The AIC and BIC are commonly used statistical methods for selecting models, particularly when evaluating competing models for a specific data set. Both criteria balance how well a model fits the data with its complexity to avoid overfitting. The AIC, based on information theory assesses the fitting of the model while imposing a penalty for the number of free parameters used. A lower AIC value indicates a model that is favored. The BIC, which is based on Bayesian probability also penalizes complexity but places a greater emphasis on the number of observations, making it more rigorous as the data set size grows. The BIC is often preferred when the simplicity of the model is important or when dealing with large sample sizes. Whereas AIC tends to support more complex models in smaller data sets BIC typically leans towards simpler models due to its more stringent penalty. The \(\Lambda\)CDM model serves as a cornerstone for understanding cosmic evolution. In evaluating alternative cosmological models against \(\Lambda\)CDM, the AIC and  BIC are essential metrics for model selection. The AIC is formulated as follows,
\begin{equation}\label{AIC}
\text{AIC} = -2 \ln L_{\text{max}} + 2k\,,   
\end{equation}
where \( k \) represents the number of model parameters and \( L_{\text{max}} \) denotes the highest value of the likelihood function. The BIC can be expressed as,
\begin{equation}\label{BIC}
\text{BIC} = -2 \ln L_{\text{max}} + k \ln \mu \,, 
\end{equation}
where $\chi^{2}_{min}=-2 \ln L_{\text{max}} $. The term $\mu$ represents the size of the observational data set combination. Furthermore, a lower value for the $\Delta$AIC or $\Delta$BIC indicates that the selected cosmological model resembles the standard $\Lambda$CDM model. This analysis assists in assessing whether it is necessary to introduce novel cosmic phenomena or alter the established model. The definitions of $\Delta$AIC or $\Delta$BIC can be expressed as follows,
\begin{eqnarray}
\Delta \text{AIC} &= \text{AIC}_{\text{model}} - \text{AIC}_{\Lambda \text{CDM}} \,,\label{deltaAIC}\\ 
\Delta \text{BIC} &= \text{BIC}_{\text{model}} - \text{BIC}_{\Lambda \text{CDM}} \,. \label{deltaBIC}
\end{eqnarray}
\section{Present study}\label{Currentstudy}
In this thesis, we conducted a dynamical systems analysis to assess the stability of various cosmological models across different gravity theories. Utilizing cosmological data sets, including CC, PN$^+$\&SH0ES and  BAO, we determined the best-fit values for the parameters of these models. Following this, we generated plots depicting the evolution of the cosmological parameters such as the EoS, deceleration parameter, energy density etc. By examining these parameters, we explored the distinct phases of the Universe, focusing particularly on late-time cosmological phenomena. 

Through a dynamical systems analysis, we have investigated the various phases of the Universe by examining the behavior of critical points in autonomous systems. Our analysis encompasses DE-dominated, matter-dominated, radiation-dominated and stiff-matter solutions. We found that matter-dominated and radiation-dominated solutions exhibit unstable characteristics, whereas the DE-dominated solution demonstrates stability. This stability is indicative of late-time cosmic acceleration of the Universe. Additionally, we conducted a comparative analysis results of our model against the standard $\Lambda$CDM framework. 

 The following chapter will delve deeper into these methodologies and their outcomes.


\chapter{Attractor behaviour of $f(T)$ modified gravity and the cosmic acceleration} 
\label{Chapter2} 

\lhead{Chapter 2. \emph{Attractor behaviour of $f(T)$ modified gravity and the cosmic acceleration}} 

\vspace{5cm}
*The work in this chapter is covered by the following publication:\\

\textbf{L K Duchaniya}, Kanika Gandhi and B Mishra, "Attractor behaviour of $f(T)$ modified gravity and the cosmic acceleration", \textit{Physics of the Dark Universe}, \textbf{44}, 101461 (2024).

\clearpage
\section{Introduction} \label{ch2_SEC-I}
 This chapter will explore the dynamical system analysis in $f(T)$ at the background and perturbation levels. Through this analysis, we will investigate the late-time cosmic acceleration of the Universe and examine the various phases of the Universe, including DM and DE dominance. In the dynamical system, an attractor is a region in phase space toward which trajectories naturally evolve. In cosmology, this implies that the evolution of the Universe dictated by dynamical variables inevitably converges to a specific state, independent of initial conditions. Additional research on the late-time cosmic expansion concerns related to $f(T)$ gravity can be found in \cite{Dent_2011a, Cai:2015emx, Briffa_2022a, Wu_2010a, Hohmann_2017a, Mirza_2017, Duchaniya_2022}. 
\section {\texorpdfstring{$f(T)$}{f(T)} gravity field equations} \label{ch2_SEC-II}
 In this work, we explore the framework of $f(T)$ gravity, focusing on its action as delineated by Eq.~\eqref{fTaction}. We utilize the flat FLRW metric, specified in Eq.~\eqref{FLATFLRW}, along with the corresponding tetrad field outlined in Eq.~\eqref{FLRWTETRAD}. By varying the action given in Eq.~\eqref{fTaction} for the tetrad field, we derive the field equations governing $f(T)$ gravity,
 \begin{equation}
3H^2 = 8\pi G \rho_m-\frac{f}{2}+Tf_T \label{ch2_7},
\end{equation}
\begin{equation}
\dot{H} = -\frac{4\pi G(\rho_{m}+p_{m})}{1+f_T+2Tf_{TT}}. \label{ch2_8} 
\end{equation}
The Hubble parameter $H\equiv\frac{\dot{a}}{a}$ with an over dot denotes the derivative for cosmic time $t$. The matter-energy density and pressure, respectively $\rho_{m}$ and $p_{m}$ and the total energy-momentum tensor is comprised of the matter sector. Now the field equations of $f(T)$ gravity in the DE sector pressure ($p_{DE}$) and energy density ($\rho_{DE}$) can be defined as,
\begin{equation}
    \rho_{DE} \equiv \frac{1}{16 \pi G}\left[-f+2Tf_{T}\right], \label{ch2_9}
\end{equation}
\begin{equation}
    p_{DE} \equiv -\frac{1}{16 \pi G}\left[\frac{-f+Tf_{T}-2T^{2}f_{TT}}{1+f_{T}+2Tf_{TT}}\right].\label{ch2_10}
\end{equation}
The EoS parameter of the DE sector ($\omega_{DE}$) can be obtained as, 
\begin{equation}\label{ch2_11}
\omega_{DE}=-1+\frac{\left(f_T+2Tf_{TT}\right)\left(-f+T+2Tf_T\right)}{(1+f_{T}+2Tf_{TT})(-f+2Tf_{T})}.
\end{equation}

We are intending to study the interacting cosmology by performing the dynamical system analysis. From the expressions of Eq.~\eqref{ch2_9}--Eq.~\eqref{ch2_11}, we can see that the functional form of $f(T)$ is required for further study. Therefore, we shall consider three distinct forms of $f(T)$ in the following section.
\section{Phase space analysis} \label{ch2_SEC-III}
Here, we shall set up the dynamical system of the background and perturbed equations. To do this, the equation governing the growth of matter perturbations on sub-horizon scales can be invoked in the form \cite{Gannouji_2009a, Anagnostopoulos_2019a} 
\begin{equation}\label{ch2_23}
 \Ddot{\delta}_m+2 H \dot{\delta}_{m}=4 \pi G_{eff} \rho_{m} \delta_{m} \,, 
\end{equation} 

where $\delta_{m}=\frac{\delta \rho_{m}}{\rho_{m}}$ is the matter over density and the $G_{eff}(a)=G Y(a)$ is the effective Newton constant, with $G$ being the gravitational constant. Usually, $G_{eff}(a)$ is changeable, but the shape of $Y(a)$ is set from the basic theory of gravity. Here, we keep the general perturbation method for $f(T)$ cosmology, if we have the form of $G_{eff}(a)$ or $Y(a)$ for the $f(T)$ gravity. So, we take the form of $Y(a)$ as in Ref. \cite{Zheng_2011},
\begin{equation}\label{ch2_24}
Y(a)= \frac{G_{eff}(a)}{G} =\frac{1}{1+f_{T}}\,,   
\end{equation}
From Eqs.~(\ref{ch2_23}--\ref{ch2_24}), we have the following 
\begin{equation}\label{ch2_25}
  \Ddot{\delta}_{m}+2 H \dot{\delta}_{m}= \frac{4 \pi G \rho_{m} \delta_{m}}{1+f_{T}}\,.  
\end{equation}

Referring Eq.~\eqref{ch2_7}, Eq.~\eqref{ch2_8} and Eq.~\eqref{ch2_25}, initially we set up the dynamical variables of the background and perturbed equations for a general function of $f(T)$ as, 
\begin{equation} \label{ch2_26}
   x = -\frac{f}{6H^{2}},
    \hspace{1cm}
    y  = -2f_{T},
    \hspace{1cm}
    \sigma = \frac{d (ln \delta)}{d (lna)}\,. 
\end{equation}

Here, the variables $x$ and $y$ are related to describe the background evolution of the Universe whereas $\sigma$ quantifies the expansion of matter perturbations. So, when the matter density contrast is positive ($\sigma>0$), it means the matter perturbations are getting bigger, whereas for ($\sigma<0$), it gets smaller. We write Eq.~ \eqref{ch2_7} in term of dynamical variables as,
\begin{equation}\label{ch2_27}
 x+y+\Omega_{m}=1\,.   
\end{equation}

To note here, the dynamical variables $x$  and $y$ define the DE density parameter ($\Omega_{DE}$). Also, Eq.~\eqref{ch2_8}, in terms of dynamical variables can be,
\begin{equation}\label{ch2_28}
\frac{\dot{H}}{H^{2}} =  \frac{3(x+y-1)}{2(1+f_{T}+2Tf_{TT})}\,. 
\end{equation}

Using $x$, $y$ and $\sigma$ as phase space variables, we can perform $3D$ dimensional phase space analysis. In terms of the dynamical variables [Eq.~\eqref{ch2_26}], the cosmological equations can be written as an autonomous system as below,
\begin{eqnarray}
 \frac{dx}{dN}&=& -\frac{\dot{H}}{H^{2}}(y+2x)\,, \label{ch2_29} \\ 
 \frac{dy}{dN}&=& -4 \frac{\dot{H}}{H^{2}}(T f_{TT})\,, \label{ch2_30}\\
 \frac{d\sigma}{dN}&=& -\sigma(\sigma+2)-\frac{3(x+y-1)}{(2-y)}-\frac{\dot{H}}{H^{2}} \sigma \,,\label{ch2_31}
\end{eqnarray}

where $N=lna$. The EoS parameters and the deceleration parameter in terms of dimensionless variables are,
\begin{eqnarray}
 \omega_{DE}&=&\frac{-2x-y+4Tf_{TT}}{2(x+y)(1+f_{T}+2Tf_{TT})} \,,\label{ch2_32} \\
\omega_{tot}&=& -1-\frac{(x+y-1)}{(1+f_{T}+2Tf_{TT})} \,, \label{ch2_33}  \\
q&=&-1-\frac{3(x+y-1)}{2(1+f_{T}+2Tf_{TT})} \,.  \label{ch2_34}
\end{eqnarray}

The critical points of the system [Eq.~\ref{ch2_29}--Eq.~\ref{ch2_31}] will be obtained to determine the dynamic growth of the system and the stability of these critical points will be examined. From the physical point of view, it is well known that the stable point ($\sigma>0$) implies continuous growth of matter perturbations and also we can say that the system is not stable for matter perturbations. However, a stable point having $\sigma<0$ denotes the reduction in matter perturbation. When $\sigma=0$ is at a stable point, it is considered that the changes in matter perturbation are always the same. To solve the dynamical system [Eq.~\ref{ch2_29}--Eq.~\ref{ch2_31}], we need to choose the form of $f(T)$. In the following sections, we will look more closely at three forms, which may explain some interesting features of the Universe. 
\subsection{Model-I}
We choose the logarithmic form of $f(T)$ \cite{Zhang2011_jacp, bamba2011eos},
\begin{equation}\label{ch2_2A}
f(T) = \beta T \ln\left(\frac{T}{T_{0}}\right) 
 \end{equation}
 where $\beta$ is an arbitrary model parameter. The autonomous system [Eq.~\ref{ch2_29}--Eq.~\ref{ch2_31}] become, 
\begin{eqnarray}
\frac{dx}{dN} &=& -\frac{3(x+y-1)(2x+y)}{(2+4\beta -y)} \,,\label{ch2_35}\\
 \frac{dy}{dN} &=& -\frac{12\beta  (x+y-1)}{(2+4\beta -y)} \,,\label{ch2_36}
 \end{eqnarray}
 \begin{equation}
 \frac{d\sigma}{dN}= -\sigma(\sigma+2) - \frac{3(x+y-1)}{(2-y)} 
 -\frac{3\sigma (x+y-1)}{(2+4\beta -y)} \,. \label{ch2_37}
\end{equation}
The corresponding EoS and deceleration parameter in terms of dynamical variables become, 
 \begin{eqnarray}
 \omega_{DE}&=&\frac{-4 \beta +2 x+y}{(x+y) (-4 \beta +y-2)} \,, \label{ch2_38}\\
\omega_{tot}&=& \frac{4 \beta +2 x+y}{-4 \beta +y-2} \label{ch2_39} \,, \\
q&=&-1+\frac{3 (x+y-1)}{-4 \beta +y-2} \,.\label{ch2_40}
\end{eqnarray}
We can find the critical points by applying the criteria $\frac{dx}{dN}=0$, $\frac{dy}{dN}=0$ and $\frac{d\sigma}{dN}=0$ to the autonomous dynamical system [Eq.~\ref{ch2_35}--Eq.~\ref{ch2_37}]. Four critical points are obtained and are given in Table~\ref{ch2_TABLE-I} with its corresponding cosmology. In Table~\ref{ch2_TABLE-II}, we have derived eigenvalues of the Jacobian matrix, where  $\lambda_{1}$,  $\lambda_{2}$ and  $\lambda_{3}$ indicate eigenvalues of the Jacobian matrix at both background and perturbation levels.
\begin{table}[H]
    \caption{Critical points of Model-I} 
    \centering 
    \begin{tabular}{|c|c|c|c|c|c|c|c|c|c|} 
    \hline\hline 
    C.P. & $x_{c}$ & $y_{c}$ & $\sigma_{c}$  &$\omega_{DE}$ &$\omega_{tot}$&$q$&$\Omega_{DE}$&$\Omega_{m}$& Exists for \\ [0.5ex] 
    \hline\hline 
    $A_{1}$  & $x$ & $-2x$ & $1$ &$0$&$0$&$\frac{1}{2}$&$-x$&$1+x$ & Always \\
    \hline
    $A_{2}$ &$x$ & $-2x$ & $-\frac{3}{2}$&$0$&$0$&$\frac{1}{2}$&$-x$&$1+x$& Always \\
    \hline
    $A_{3}$ & $x$ & $1-x$ & $-2$  &$-1+\frac{8\beta}{1+4\beta+x}$&$-1$&$-1$&$1$&$0$&Always \\
    \hline
    $A_{4}$ & $x$ & $1-x$ & $0$  &$-1+\frac{8\beta}{1+4\beta+x}$&$-1$&$-1$&$1$&$0$& Always\\
    [1ex] 
    \hline 
    \end{tabular}
    \label{ch2_TABLE-I}
\end{table}

\begin{table}[H]
    \caption{Eigenvalues and stability condition. } 
    \centering 
    \begin{tabular}{|c|c|c|c|c|} 
    \hline\hline 
    C.P. & Stability Conditions  & $\lambda_{1}$ & $\lambda_{2} $  &$\lambda_{3}$  \\ [0.5ex] 
    \hline\hline 
    $A_{1}$  & Saddle Unstable  & $0$ & $-\frac{5}{2}$ &$3$ \\
    \hline
    $A_{2}$  & Node Unstable  & $0$ & $\frac{5}{2}$ &$3$ \\
    \hline
    $A_{3}$  & Saddle Unstable & $0$ & $-3$ &$2$ \\
    \hline
    $A_{4}$  & Node Stable & $0$ & $-3$ &$-2$\\
    [1ex] 
    \hline 
    \end{tabular}
    \label{ch2_TABLE-II}
\end{table}

{\bf Summary of the critical points (Model-I):}
 \begin{itemize}
 \item\textbf{$A_{1}$:}  This critical point represents a matter-dominated scaling solution at the background level. The density parameter for the matter phase is  $\Omega_{m}=1+x$. The total and DE sector EoS parameters are respectively $\omega_{tot}=0$ and  $\omega_{DE}=0$. At the background level, a positive value of the deceleration parameter $q=\frac{1}{2}$, indicates the decelerated phase of the Universe. We derive $\sigma=1$, at the perturbation level. The positive value of $\sigma$ suggests the growth in matter perturbation. From linear stability theory, both positive and negative eigenvalues indicate unstable saddle behavior. So, this point may be the best way to explain how structures formed when matter dominated, both at the background and perturbation levels.
 
\item\textbf{$A_{2}$:}  At the background level, this critical point is also related to matter dominated phase of the Universe and both the EoS parameters vanish. The deceleration parameter is $q=\frac{1}{2}$ shows the decelerating phase of the Universe. Similar to point $A_{1}$, the critical point $A_{2}$ does not exhibit any late time acceleration for any physically accepted value of $q$ and $\omega_{tot}$. Also, we get $\sigma=-\frac{3}{2}$, at the perturbation level, which indicates the decay in matter perturbation. The corresponding eigenvalues of the Jacobian matrix suggest node unstable behavior.

\item\textbf{$A_{3}$:} This critical point is absolutely DE dominated solution $\Omega_{DE}=1$, with  $\omega_{tot}=-1$ and $q=-1$ at the background level. The negative value of the deceleration parameter shows the accelerating phase of the Universe and $\omega_{tot}=-1$ behaves as a cosmological constant. At the perturbation level, we have derived $\sigma=-2$, which implies the decay in matter perturbation. The eigenvalues of this critical point show saddle unstable behavior. So, at the perturbation level, this critical point does not exhibit late time acceleration of the Universe.

\item\textbf{$A_{4}$:} This critical point again implies the de-Sitter phase solution of the Universe. The corresponding scaling solution of the EoS and the deceleration parameters are $\omega_{tot}=-1$ and $q=-1$ respectively at the background level. Similar to critical point $A_{3}$, this point also indicates late time acceleration of the Universe. At the perturbation level, we have accomplished $\sigma=0$. This suggests that the matter perturbation is unchanged. The eigenvalues of this critical point are negative real part and zero. Here, the curve is one dimensional with one vanishing eigenvalue and hence it is hyperbolic \cite{Coley:1999,aulbach1984}. This critical point shows stable node behavior. In addition, this critical point shows the late time acceleration phase of the Universe at both background and perturbation levels. 
\end{itemize}

Two matter-dominated critical points ($A_1, A_2$) and two DE-dominated critical points ($A_3, A_4$) are obtained for the considered logarithmic form of $f(T)$. Both the matter-dominated critical points are unstable. The saddle instability, a defined growth rate in a matter perturbation, is represented by critical points $A_1$. The unstable node, represented by critical point $A_2$, represents the decay in matter perturbation. Critical point $A_3$ shows the accelerating behavior of the Universe only at the background level, whereas the critical point $A_4$ shows similar behavior both at the background and perturbation levels, however, it is stable.     

\begin{figure}[H]
\centering
\includegraphics[width=80mm]{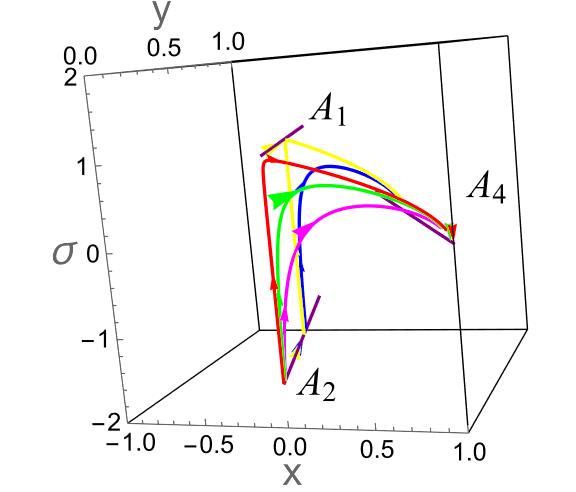}
\caption{3D phase portrait for Model-I.} \label{ch2_FigF}
\end{figure}

\begin{figure}[H]
\centering
\includegraphics[width=70mm]{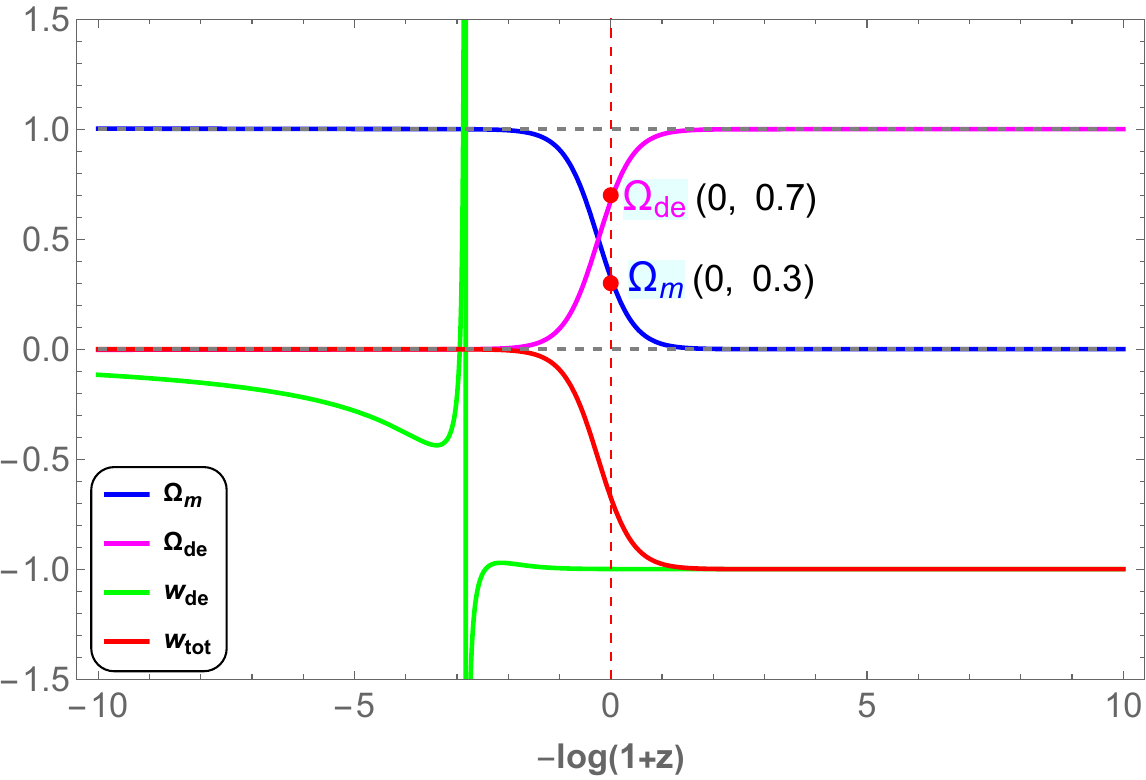}
\includegraphics[width=70mm]{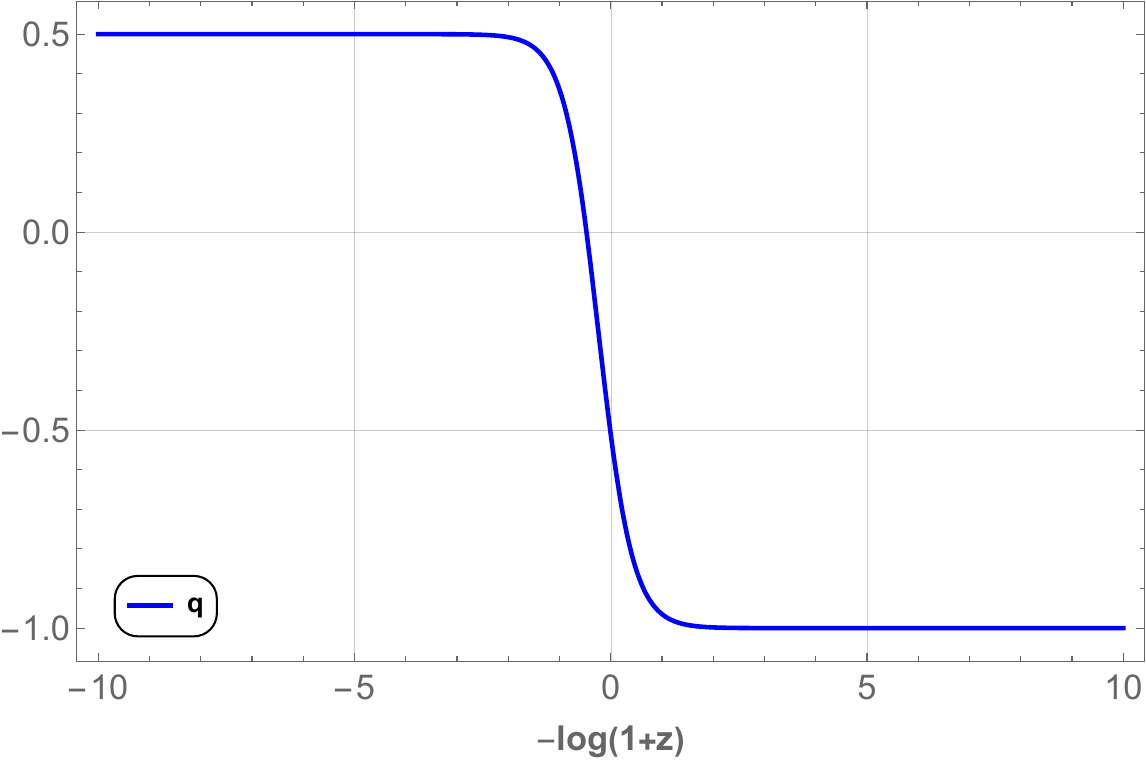}
\caption{Evolution of density parameters (Left panel) and the deceleration parameter (Right panel) for Model-I. The initial conditions $x = 10^{-2}$, $y =10^{-6} $ and $\beta = 0.0001$. The vertical dashed red line denotes the present time.} \label{ch2_FigA}
\end{figure}

Fig.~\ref{ch2_FigF} displays the phase portrait in $3D$ space. The selected trajectory moved from matter-dominated to dark-energy-dominated critical points. From Fig.~\ref{ch2_FigF}, we can observe the transition of the trajectory like $A_{2}$ (node unstable) \textrightarrow  $A_{1}$ (saddle unstable)  \textrightarrow  $A_{4}$ (node stable). Fig.~\ref{ch2_FigA} (Left panel), represents the evolutionary history of density and the EoS parameters. In the Left panel, the Universe transits from a matter to an acceleration era at late times. The current density parameters for the matter and DE sectors are $\Omega_m \approx0.3$ and $\Omega_{DE} \approx0.7$, respectively. The total EoS parameter starts with a matter-dominated era ($\omega_{tot}=0$) and approaches the dark-energy sector ($\omega_{tot.}=-1$) at late-time. Also, the DE EoS parameter goes to $-1$ at the late phase of the evolution. The present value of the DE EoS parameter $\omega_{DE}=-1$, matches with the current observational range $\omega_{DE}= -1.028 \pm 0.032 $ \cite{Aghanim:2018eyx}. In the Right panel, the deceleration parameter shows transient behavior from deceleration to the acceleration phase of the Universe.  The transition point from deceleration to acceleration is $z=0.59$ and the present value of the deceleration parameter is $q_0=-0.57$ \cite{Camarena:2020prr}.
\subsection{Model-II}
We consider the power law form of $f(T)$  \cite{Bengochea:2008gz} as,
\begin{equation} \label{ch2_2B}
f(T) = f_{0} (-T)^{m}, 
\end{equation}
where the arbitrary constants $f_{0}$ and $m$ are the model parameters and for this choice of $f(T)$, the dynamical system can be presented as, 
\begin{eqnarray}
\frac{dx}{dN}&=& -\frac{3(x+y-1)(2x+y)}{(2+(1-2m)y)} \,,\label{ch2_41}\\
 \frac{dy}{dN}&=& \frac{6y(m-1)(x+y-1)}{(2+(1-2m)y)} \,,\label{ch2_42}\\
 \frac{d\sigma}{dN}&=& -\sigma(\sigma+2) - \frac{3(x+y-1)}{(2-y)}-\frac{3 \sigma(x+y-1)}{(2+(1-2m)y)} \,.\label{ch2_43a}    
\end{eqnarray}
Also, the corresponding EoS and deceleration parameters are respectively,
\begin{eqnarray}
 \omega_{DE}&=&\frac{(2 m-1) y+2 x}{((2 m-1) y-2) (x+y)} \,, \label{ch2_44}\\
\omega_{tot}&=&-1+ \frac{2 (x+y-1)}{(2 m-1) y-2} \,,\label{ch2_451} \\
q&=&-1+\frac{3 (x+y-1)}{(2 m-1) y-2} \,.\label{ch2_46}
\end{eqnarray} 
Using the same approach as in Model--I, the critical points of the autonomous dynamical system [Eq.~\ref{ch2_41}--Eq.~\ref{ch2_43a}] are summarized in Table~\ref{ch2_TABLE-III}. The eigenvalues of the Jacobian matrix are presented in Table~\ref{ch2_TABLE-IV}.
 \begin{table}[H]
 \renewcommand{\arraystretch}{0.8}
    \caption{Critical points of Model--II } 
    \centering 
    \begin{tabular}{|c|c|c|c|c|c|c|c|c|c|} 
    \hline\hline 
    C.P. & $x_{c}$ & $y_{c}$ & $\sigma_{c}$  &$\omega_{DE}$ &$\omega_{tot}$&$q$&$\Omega_{DE}$&$\Omega_{m}$& Exists for \\ [0.5ex] 
    \hline\hline 
    $B_{1}$  & $0$ & $0$ & $1$ &$-$&$0$&$\frac{1}{2}$&$0$&$1$ & Always \\
    \hline
    $B_{2}$ &$0$ & $0$ & $-\frac{3}{2}$&$-$&$0$&$\frac{1}{2}$&$0$&$1$& Always \\
    \hline
    $B_{3}$ & $x$ & $1-x$ & $-2$  &$\frac{x (2 m-3)-2 m+1}{x (2 m-1)-2 m+3}$&$-1$&$-1$&$1$&$0$&Always \\
    \hline
    $B_{4}$ & $x$ & $1-x$ & $0$  &$\frac{x (2 m-3)-2 m+1}{x (2 m-1)-2 m+3}$&$-1$&$-1$&$1$&$0$& Always\\
    [1ex] 
    \hline 
    \end{tabular}
    \label{ch2_TABLE-III}
\end{table}
\begin{table}[H]
\renewcommand{\arraystretch}{0.6}
    \caption{Eigenvalues and stability condition. } 
    \centering 
    \begin{tabular}{|c|c|c|c|c|} 
    \hline\hline 
    C.P. & \begin{tabular}{@{}c@{}}Stability\\ Conditions\end{tabular}    & $\lambda_{1}$ & $\lambda_{2} $  &$\lambda_{3}$  \\ [0.5ex] 
    \hline\hline 
    $B_{1}$  & Unstable  & $3$ & $-\frac{5}{2}$ &$-3(m-1)$ \\
    \hline
    $B_{2}$  &  Unstable  & $3$ & $\frac{5}{2}$ &$-3(m-1)$ \\
    \hline
    $B_{3}$  & Unstable & $0$ & $2$ &$-\frac{3(3-2m+2x-x^{2}+2mx^{2})}{(1+x)(3-2m-x+2mx)}$ \\
    \hline
    $B_{4}$  & \begin{tabular}{@{}c@{}}Stable for\\ $\left(\left.x\right|m\right)\in \mathbb{R}$\end{tabular}    & $0$ & $-2$ &$-\frac{3(3-2m+2x-x^{2}+2mx^{2})}{(1+x)(3-2m-x+2mx)}$\\
    [1ex] 
    \hline 
    \end{tabular}
    \label{ch2_TABLE-IV}
\end{table}

{\bf Summary of the critical points (\bf Model--II):}
 \begin{itemize}
\item\textbf{$B_{1}$:} The solution of this critical point represents the matter phase of the Universe. From Table~\ref{ch2_TABLE-III},  $\Omega_{m}=1$, the critical point exists for all values of the free model parameter. The total  EoS and deceleration parameters imply that this solution has no late-time acceleration at the background level. The value of $\omega_{DE}$ is undefined for this critical point.  At the perturbation level, we find $\sigma=1$ and the positive value of $\sigma$  implies a growth factor in matter perturbation. This critical point shows saddle unstable behavior.  

\item\textbf{$B_{2}$:} At the background level, this critical point is similar to critical point $B_{1}$. We have discovered $\sigma=-\frac{3}{2}$ at the perturbation level, which denotes a decline in matter perturbation. This critical point exhibits node unstable behavior for ($1>m$) and saddle unstable for ($m>1$).    

\item\textbf{$B_{3}$:} This critical point is having DE-dominated solution with $\Omega_{DE}=1$. With $\omega_{tot}=-1$ and $q=-1$, the values of EoS and deceleration parameters show late time acceleration of the Universe at the background level. But at the perturbation level, we find $\sigma=-2$. The negative value of the perturbed variable indicates decay in matter perturbation. The eigenvalues of the Jacobian matrix imply unstable behavior for any value of $x$ and $m$. However, this critical point does not exhibit accelerating behavior at the perturbation level. It only indicates acceleration at the background level. 

\item\textbf{$B_{4}$:} At the background level both $B_{3}$  and $B_4$ are having similar behaviour. At the perturbation level, we obtain $\sigma=0$, which indicates that the perturbation of matter is unchanged. It shows the stable behavior for any choice of $m$ and $x$ \cite{Coley:1999,aulbach1984} and exhibits late time acceleration at both perturbation and background level. 
 \end{itemize} 
 
For the power law form of $f(T)$, two matter-dominated critical points ($B_1, B_2$) and two DE-dominated critical points ($B_3, B_4$) are obtained. Among these, one stable critical point $B_{4}$, which represents accelerated expansion and DE-dominated phase of the Universe. The second DE-dominated critical point $B_{3}$ shows saddle instability, decay in matter perturbation and accelerated expansion of the Universe only at the background level. The two unstable matter-dominated critical points, $B_{1}$ and $B_{2}$ show the decelerated phase of the Universe. 

\begin{figure}[H]
\centering
\includegraphics[width=80mm]{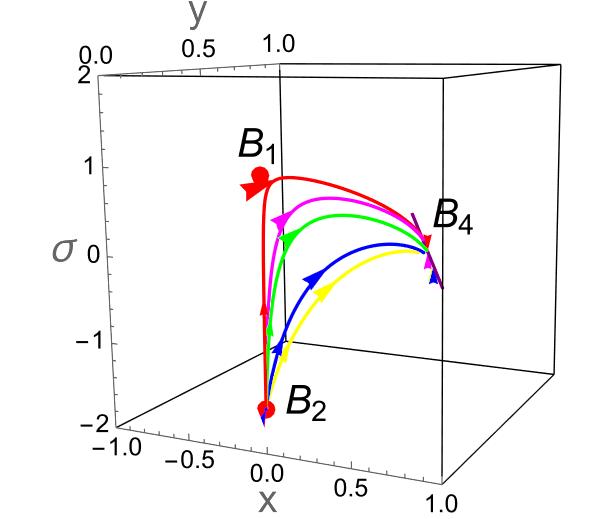}
\caption{3D phase portrait for Model-II.} \label{ch2_FigE}
\end{figure}

Fig.~\ref{ch2_FigE} shows the critical points in 3D space. The trajectories show a path from the matter-dominated unstable critical points $B_{1}$ and $B_{2}$ to the stable DE-dominated critical point $B_{4}$ ($B_{2}$\textrightarrow $B_{1}$\textrightarrow  $B_{4}$).
\begin{figure}[H]
\centering
\includegraphics[width=70mm]{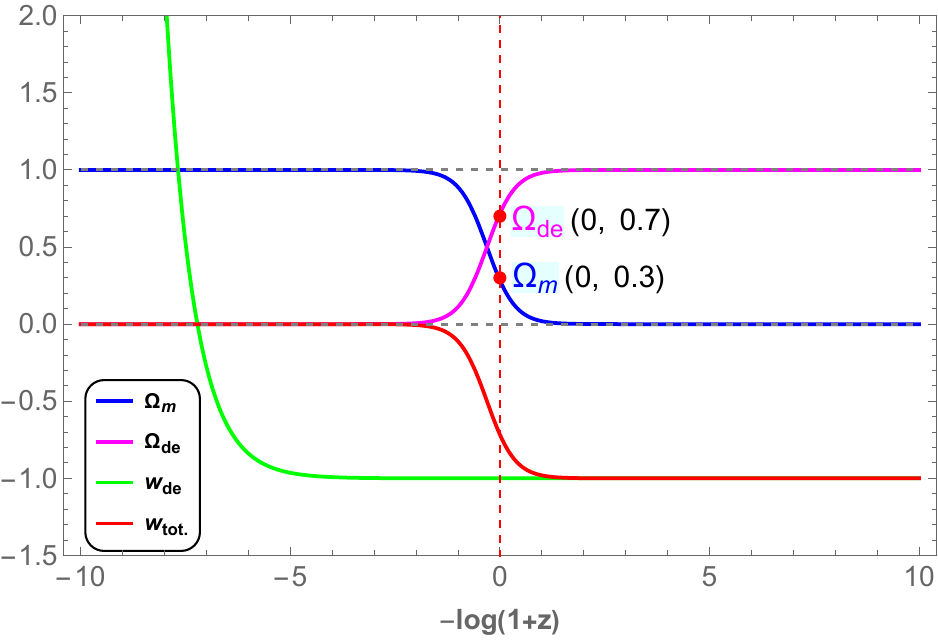}
\includegraphics[width=70mm]{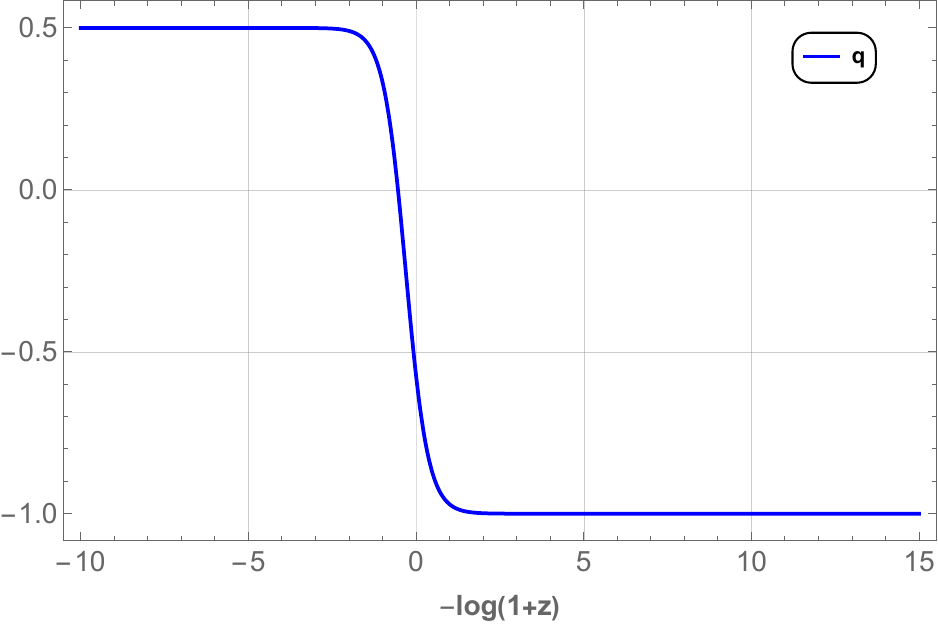}
\caption{Evolution of density parameters (Left panel) and the deceleration parameter (Right panel) for Model-II. The initial conditions $x = 10^{-3}$, $y =10^{-6} $ and $m = 0.5$. The vertical dashed red line denotes the present time.} \label{ch2_FigC}
 \end{figure}
 Fig.~\ref{ch2_FigC} (Left panel) shows the evolution of the density and EoS parameters for redshift $z$. The initial condition values are calibrated to eventually give present time (at $z=0$) values of both density parameters. The graph is per the expected behavior of the Universe to transition from a matter-dominated phase to a DE-dominated phase. Consequently, Fig.~\ref{ch2_FigC} (Right panel) shows transitions from decelerated to accelerated expansion at $z=0.64$ and the present value of the deceleration parameter is noted as $q_{0}=-0.62$ \cite{Camarena:2020prr}. As shown in the Left panel, the value of DE EoS parameter $\omega_{DE}=-1$ is also under the current observational value \cite{Aghanim:2018eyx}.
\subsection{Model-III}
Finally, we consider the form of $f(T)$ as the combined form of logarithmic and power law \cite{Mirza_2017} as, 
\begin{equation}\label{ch2_421}
 f(T)=\alpha (-T)^{n}  \ln\left(\frac{T}{T_{0}}\right),   
\end{equation}
where $\alpha$ and $n$ are arbitrary constants. For this $f(T)$ the autonomous system becomes, 
\begin{eqnarray}
\frac{dx}{dN}&=&-\frac{3 (x+y-1) (2 x+y)}{-4 n (n x+y)+y+2} \,,\label{ch2_43}\\
\frac{dy}{dN}&=&-\frac{6 (x+y-1) (y-2 n (n x+y))}{-4 n (n x+y)+y+2} \,,\label{ch2_441}
\end{eqnarray}
\begin{equation}
\frac{d\sigma}{dN}=-\frac{3 \sigma  (x+y-1)}{-4 n (n x+y)+y+2}-\sigma  (\sigma +2)+\frac{3 (x+y-1)}{y-2} \,.\label{ch2_45}
\end{equation}
The EoS parameters and deceleration parameter would be,
\begin{eqnarray}
\omega_{DE} &=& \frac{y-2 \left(2 n^2 x+2 n y+x\right)}{(x+y) (-4 n (n x+y)+y+2)} \,,\label{ch2_461}\\
\omega_{tot} &=&-1 -\frac{2 (x+y-1)}{-4 n (n x+y)+y+2} \,, \label{ch2_47}\\
q &=&-1 -\frac{3 (x+y-1)}{-4 n (n x+y)+y+2} \,.\label{ch2_48}
\end{eqnarray}
In a similar approach, the critical points of the autonomous dynamical system represented by Eqs.~(\ref{ch2_43}--\ref{ch2_45}) are calculated and are presented in Table~\ref{ch2_TABLE-V}. The corresponding eigenvalues of the Jacobian matrix are listed in Table~\ref{ch2_TABLE-VI}.
\begin{table}[H]
\renewcommand{\arraystretch}{0.8}
\caption{Critical points of Model-III } 
\centering 
\begin{tabular}{|c|c|c|c|c|c|c|c|c|c|} 
\hline\hline 
C.P. & $x_{c}$ & $y_{c}$ & $\sigma_{c}$  &$\omega_{DE}$ &$\omega_{tot}$&$q$&$\Omega_{DE}$&$\Omega_{m}$& Exists for \\ [0.5ex] 
\hline\hline 
\hline
$C_{1}$ & $0$ & $0$ & $1$  &$-$&$0$&$\frac{1}{2}$&$0$&$1$&Always \\
\hline
$C_{2}$ & $0$ & $0$ & $-\frac{3}{2}$  &$-$&$0$&$\frac{1}{2}$&$0$&$1$& Always\\
\hline
$C_{3}$ &$x$ & $1-x$ & $-2$&$\frac{4n-1+x(3+4n(n-1)}{4n-3+x(1-2n)^{2}}$&$-1$&$-1$&$1$&$0$& Always \\
\hline
$C_{4}$  & $x$ & $1-x$ & $0$ &$\frac{4n-1+x(3+4n(n-1)}{4n-3+x(1-2n)^{2}}$&$-1$&$-1$&$1$&$0$ & Always \\
[1ex] 
\hline 
\end{tabular}
\label{ch2_TABLE-V}
\end{table}
     
\begin{table}[H]
 \renewcommand{\arraystretch}{0.8}
\caption{Eigenvalues and stability condition. } 
\centering 
\begin{tabular}{|c|c|c|c|c|} 
\hline\hline 
C.P. &  \begin{tabular}{@{}c@{}}Stability\\ Conditions\end{tabular}   & $\lambda_{1}$ & $\lambda_{2} $  &$\lambda_{3}$  \\ [0.5ex] 
\hline\hline 
\hline
$C_{1}$  & Unstable & $-\frac{5}{2}$ & $-3(n-1)$ &$-3(n-1)$ \\
\hline
$C_{2}$  & Unstable & $\frac{5}{2}$ & $-3(n-1)$ &$-3(n-1)$\\
\hline
$C_{3}$  &  Saddle Unstable  & $0$ & $-3$ &$2$ \\
\hline
$C_{4}$  & Stable   & $0$ & $-3$ &$-2$ \\
[1ex] 
\hline 
\end{tabular}
\label{ch2_TABLE-VI}
\end{table}

{\bf Summary of the critical points (Model-III):}
\begin{itemize} 
\item\textbf{$C_{1}$:} The eigenvalues of this critical point show saddle unstable behavior for ($1>n$). The value of the DE EoS parameter ($\omega_{DE}$) is undefined and the total EoS parameter is, $\omega_{tot}=0$. Also, the deceleration parameter, $q=\frac{1}{2}$ indicates the decelerated expansion of the Universe at the background level. The positive value of $\sigma=1$ shows the increase in matter perturbation and consequently non-decelerated expansion of the Universe at the perturbation level. This critical point resembles matter dominated phase as $\Omega_{m}=1$.

\item\textbf{$C_{2}$:} This critical point shows similar behaviour as that of $C_{1}$ except the value of $\sigma$. Here, we obtain, $\sigma=-\frac{3}{2}$ which suggests a decay in matter perturbation and decelerated expansion of the Universe at both background and perturbation levels. This point indicates saddle unstable behavior for ($n>1$) and node unstable for ($1>n$). 

\item\textbf{$C_{3}$:} This critical point indicates DE dominated solution with $\Omega_{DE}=1$ and $\Omega_{m}=0$. Since the values of the deceleration parameter and total EoS parameter are respectively, $q=-1$ and $\omega_{tot}=-1$. This critical point shows the accelerated expansion of the Universe at the background level, but not at the perturbation level, as $\sigma = -2$ which resembles decay of matter perturbation. The eigenvalues show saddle unstable behavior as the sign of non-zero eigenvalues is positive.

\item\textbf{$C_{4}$:} This is the only critical point that shows stable behavior because the non-zero eigenvalues are negative. Since $\sigma = 0$, we observe that the matter perturbation remains constant. This also resembles an entirely DE-dominated phase and late-time accelerated expansion of the Universe at both, background and perturbation levels.
\end{itemize}

For this combined form of $f(T)$, we examine two matter-dominated critical points ($C_1, C_2$) and two DE-dominated critical points ($C_3, C_4$) like the other two models. Both the matter-dominated critical points $C_1$ and $C_2$ are unstable and indicate the decelerated phase of the Universe. One of the DE-dominated critical points $C_3$ shows saddle instability and accelerated expansion of the Universe at the background level but not at the perturbation level. The other DE-dominated critical point $C_4$ is stable, showing the accelerated expansion of the Universe at both levels and no change in matter perturbation.

\begin{figure}[H]
\centering
\includegraphics[width=80mm]{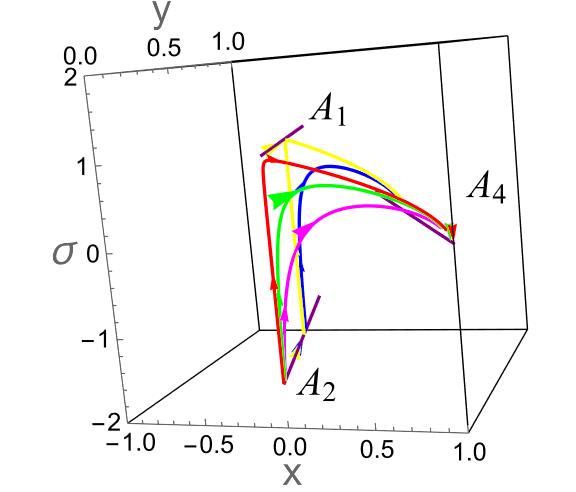}
\caption{3D phase portrait for Model-III.} \label{ch2_FigG}
\end{figure}
 \begin{figure}[H]
\centering
\includegraphics[width=70mm]{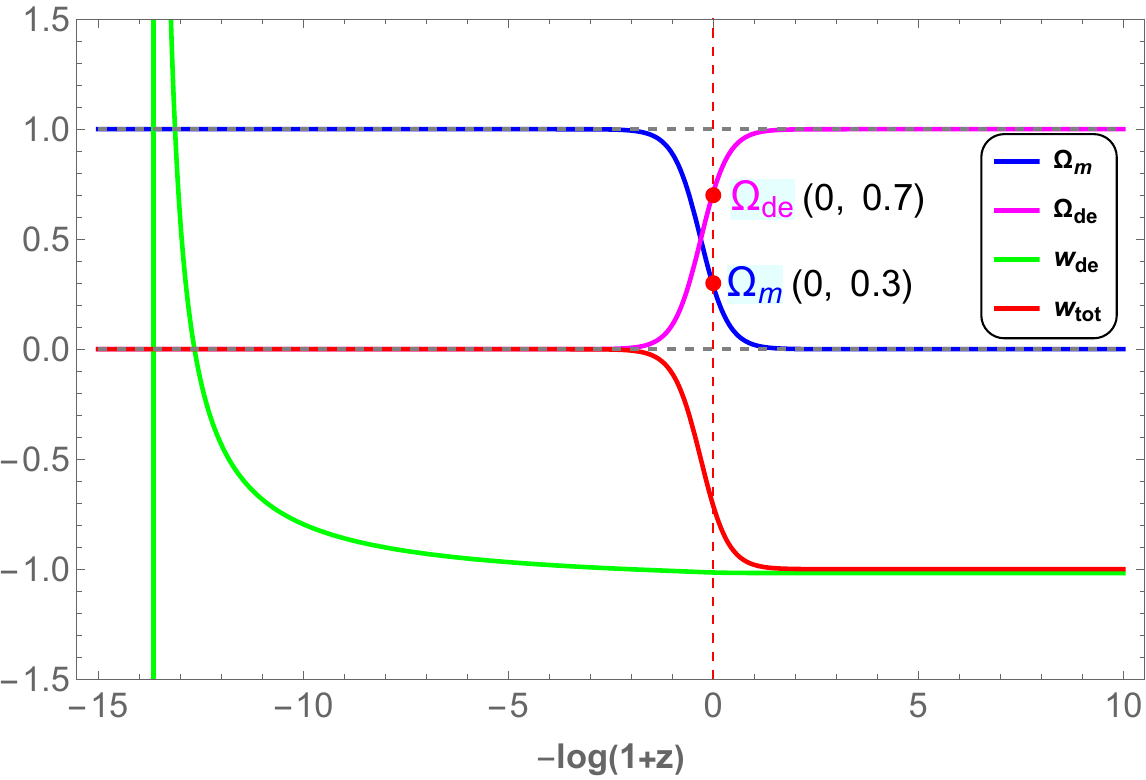}
\includegraphics[width=70mm]{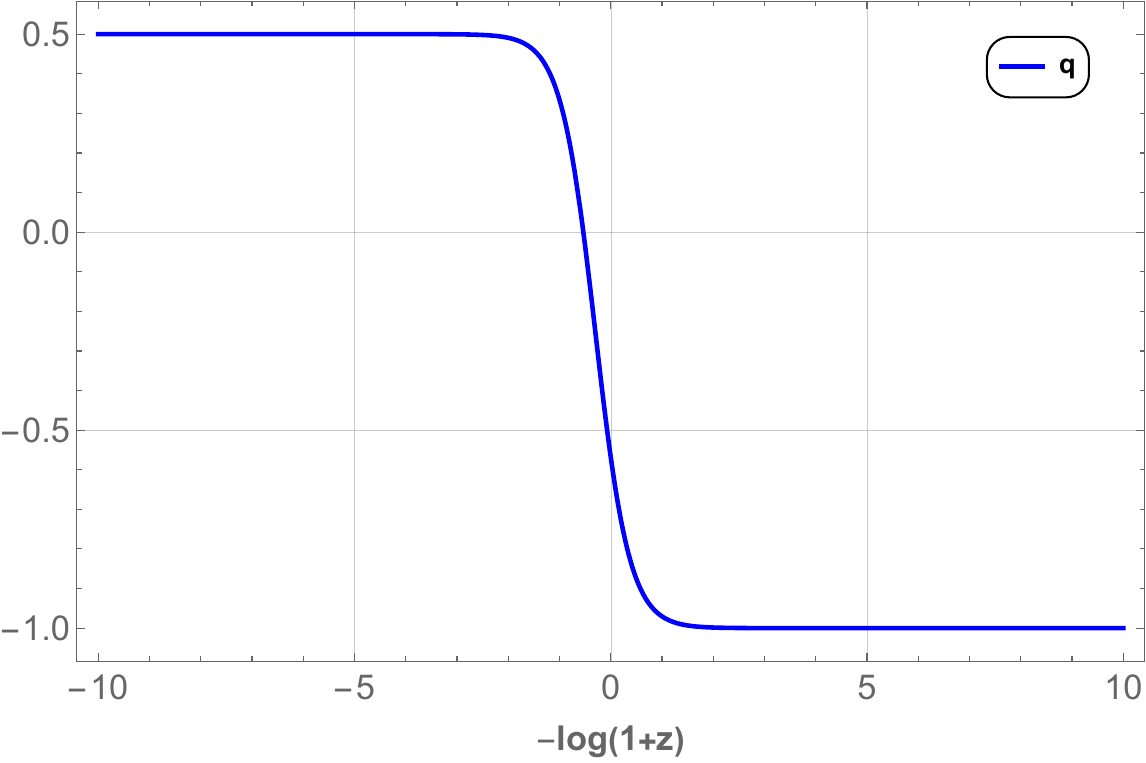}
\caption{Evolution of density parameters (Left panel) and the deceleration parameter (Right panel) for Model-III. The initial conditions $x = 10^{-2}$, $y =10^{-6} $ and $n = 0.03$. The vertical dashed red line denotes the present time.} \label{ch2_FigH}
 \end{figure}

Fig.~\ref{ch2_FigG} shows the critical points and the trajectories, $C_{2}$\textrightarrow $C_{1}$\textrightarrow  $C_{4}$ in $3D$ space. The trajectories show paths from matter to DE-dominated critical points. Fig.~\ref{ch2_FigH} shows the evolution of all the background parameters for redshift. It follows our observation that matter-dominated critical points represent the decelerated phase and DE-dominated critical points represent the accelerated expansion phase of the Universe. The initial condition values are set such that they give present time values of both, matter and DE density parameters as shown in Fig.~\ref{ch2_FigH} (Left panel). It also shows that both DE and total EoS parameters go to $-1$ at the late time phase of the Universe. Fig.~\ref{ch2_FigH} (Right panel) shows a transition from decelerated to the accelerated expansion of the Universe at redshift $z=0.68$. The present value of the deceleration parameter is obtained as, $q_{0}=-0.56$ \cite{Camarena:2020prr}.

\section{Conclusion} \label{ch2_SEC-IV}
Dynamical system analysis is a useful approach to analyze the qualitative behavior of the Universe. In this approach, we deal with the non-linear differential equation in terms of dynamical variables. This concept describes the evolution of the Universe through the critical points of autonomous systems. Taking these things into consideration, we examined dynamical system analysis in $f(T)$ gravity both at the background and perturbation levels in this study. In this approach, we have described general dynamical autonomous systems [Eq.~\eqref{ch2_29}--Eq.~\eqref{ch2_31}] in the teleparallel framework, to be specific in $f(T)$ gravity. The dynamical variables $x$ and $y$ represent the background behavior of the Universe, whereas the dynamical variable $\sigma$ defines the perturbed level of the Universe, i.e. the growth and decay in matter perturbation. In the defined autonomous systems, the functional form of $f(T)$ is incorporated; three distinct forms of $f(T)$ are proposed leading to three different models.

In Model--I, we have taken the logarithmic form of $f(T)$, which is displayed in Eq.~\eqref{ch2_2A}. For this form of $f(T)$, we have obtained four critical points, which are defined matter and the dark-energy era of the Universe at both background and perturbation levels. The critical points $A_{1}$ and $A_{2}$ describe the matter-dominated era, but  $A_{1}$ and $A_{2}$ show a growth rate and decay in the matter part respectively. The critical points $A_{3}$ and $A_{4}$ represent the dark-energy era of the Universe, but at the background level, the critical point $A_{3}$ shows accelerated expansion of the Universe and at perturbation level shows the decay in the matter part. But, the critical point $A_{4}$ shows accelerated expansion of the Universe at both levels. Also, it is showing stable node behavior. In Model--II, we have considered the power-law form of $f(T)$, which is presented in Eq.~\eqref{ch2_2B} and four critical points have been obtained. The behavior of critical points for this model is similar to that of Model--I despite the different form of $f(T)$. Critical points $B_{1}$ and $B_{2}$ determined the matter phase of the Universe and critical points $B_{3}$ and $B_{4}$ show the DE phase of the Universe but here, only critical point $B_{4}$ describes late time acceleration of the Universe at both levels. When we combine both the models in another model as in Eq.~\eqref{ch2_421}, the same four critical points are obtained. The qualitative behavior of this model is similar to the first two models both at the background and perturbation levels. We wish to mention here that the cosmological perturbation has been studied to check the stability of the cosmological models in $f(T)$ gravity in Refs. \cite{Coley_2307_12930, Coley_2310_14378}. The investigation was basically on the class of Einstein teleparallel geometries that would have a four-dimensional Lie algebra of affine connection. Further, the explicit form of $f(T)$ has been obtained for each parameter value. We have considered three such forms of $f(T)$ to show the late-time cosmic acceleration of the Universe through the dynamical system analysis. 

The cosmological behavior of the Universe through the density parameters of matter and DE, EoS parameters and deceleration parameters are shown. From the behavior of the deceleration parameter, it has been observed that in all models, the Universe shows early deceleration to late time acceleration with the transition noted respectively as:  $z=0.59$, $z=0.64$ and $z=0.68$. Also, the present values are: $q_0=-0.57$, $q_0=-0.62$ and $q_0=-0.56$. All three models present the same dark EoS parameter value, which is $\omega_{DE}=-1$. In all three models, the density parameters for matter and DE are $\Omega_{m} \approx 0.3$ and $\Omega_{DE} \approx 0.7$, consistent with recent cosmology observations. In 3D space, we have drawn phase space trajectories for all models. The trajectory moves from a matter-dominated (unstable) to a DE-dominated (stable) phase. Finally, we conclude that the dynamical stability analysis can be used extensively to analyze the cosmological behavior of the Universe.

\chapter{Dynamical system and observational evaluation of \texorpdfstring{$f (T, \mathcal{T})$}{f(T, T) gravity models} gravity models} 

\label{Chapter3} 

\lhead{Chapter 3. \emph{Dynamical system and observational evaluation of \texorpdfstring{$f (T, \mathcal{T})$}{f(T, T) gravity models} gravity models}} 

\vspace{10 cm}
*The work in this chapter is covered by the following publications:\\

 \textbf{L K Duchaniya}, Santosh V Lohakare and B. Mishra, ``Cosmological models in $f (T, \mathcal{T})$ gravity and the dynamical system analysis", \textit{Physics of the Dark Universe}, \textbf{43}, 101402 (2024).

  \textbf{L K Duchaniya} and B. Mishra, ``Late Time Phenomena in $f(T,\mathcal{T})$ Gravity Framework: Role of $H_0$ Priors", (Accepted for publication in European Physical Journal C).

\clearpage
\section{Introduction} \label{ch3_SEC-I}
In Chapter \ref{Chapter2}, we have analyzed the dynamical system within the $f(T)$ gravity framework. In this chapter, we have built upon our previous studies and examined the dynamical system analysis in $f(T, \mathcal{T})$, where $T$ denotes the torsion scalar and $\mathcal{T}$ signifies the trace of the energy-momentum tensor. The purpose of exploring $f(T, \mathcal{T})$ gravity is to investigate how the trace of the energy-momentum tensor $\mathcal{T}$ interacts with the torsion scalar $T$. Furthermore, in this chapter, we explored observational cosmology within the context of $f(T, \mathcal{T})$ gravity in the second phase. The data set employed includes Hubble, SNIa and BAO with $H_0$ priors based on R21 and TRGB. We calculated the one-dimensional parameter distribution to obtain the posterior distribution for each parameter. We derived the two-dimensional parameter distributions to uncover the covariance relationships among pairs of parameters. Our analysis resulted in the generation of MCMC corner plots at 1$\sigma$ and 2$\sigma$ confidence levels. This investigation assessed the behavior of the $f(T, \mathcal{T})$ cosmological model using various combinations of data sets. We also compared the outcomes of this model between the PN$^+$ (without SH0ES) and the PN$^+$\&SH0ES (with SH0ES) data sets. In addition, we integrated BAO data along with $H_0$ priors into our analysis. Moreover, we evaluated the $\chi^2_{min}$ value for each combination of data sets to examine the selected model against the standard $\Lambda$CDM model. After identifying the best-fit values for each combination of the data set cosmological and model parameters, we illustrated the cosmological background parameters. Some investigations on $f(T, \mathcal{T})$ gravity can be referred \cite{Harko_2014a, Momeni_2014,Jackson2016a,Junior_2016,Pace_2017, Nassur_2015a}.     
\section{Mathematical formalism} \label{ch3_SEC-II}
In this study, we investigate the $f(T, \mathcal{T})$ gravity framework, concentrating on its action as described by Eq.~\eqref{fTTaction}. We employ the flat FLRW metric, which is detailed in Eq.~\eqref{FLATFLRW}, together with the relevant tetrad field specified in Eq.~\eqref{FLRWTETRAD}. By applying variations to the action expressed in Eq.~\eqref{fTaction} concerning the tetrad field, we obtain the field equations that govern $f(T, \mathcal{T})$ gravity, leading to the following formulations,
\begin{eqnarray}
3H^2&=&8\pi G \rho_m-\frac{1}{2}(f+12 H^{2}f_T)+f_{\mathcal{T}}(\rho_m+p_m), \label{ch3_9}\\
\dot{H}&=&-4\pi G(\rho_{m}+p_{m})-\dot{H}(f_{T}-12 H^{2}f_{TT})-H(\dot{\rho}_{m}-3\dot{p}_{m}) f_{T \mathcal{T}}-f_{\mathcal{T}}\left(\frac{\rho_m+p_m}{2}\right)\,.\label{ch3_10} 
\end{eqnarray}
An over dot on the Hubble parameter $H$ denotes the ordinary derivative with respect to time $t$ and the trace of the energy-momentum tensor, $\mathcal{T}=\rho_{m}-3 p_{m}$. Here, $p_{m}$ represents the pressure of matter, whereas the equivalent energy density terms are represented by $\rho_{m}$. It is noteworthy to mention here that the matter sectors make up the overall energy-momentum tensor. Further, the Friedmann Eqs.~(\ref{ch3_9}--\ref{ch3_10}) are given as,
\begin{eqnarray}
3 H^{2}&=& 8 \pi G(\rho_{m}+\rho_{DE}), \label{ch3_11}\\
-\dot{2H}&=&8 \pi G(\rho_{m}+p_{m}+\rho_{DE}+p_{DE}). \label{ch3_12}
\end{eqnarray}

From Eqs.~(\ref{ch3_9}--\ref{ch3_12}), the expression of energy density ($\rho_{DE}$) and pressure ($p_{DE}$) for the DE sector can be obtained,  

\begin{eqnarray}
\rho_{DE}&\equiv&-\frac{1}{16 \pi G}\left[f+12 H^{2}f_{T}-2 f_{\mathcal{T}}(\rho_m+p_m)\right], \label{ch3_13} \\
p_{DE}&\equiv& (\rho_m+p_m)\left[\frac{1+\frac{f_{\mathcal{T}}}{8 \pi G}}{1+f_{T}-12 H^{2}f_{TT}+H \dfrac{d \rho_{m}}{dH}(1-3 c^{2}_{s})f_{T \mathcal{T}}}-1\right]+  \nonumber\\&& \frac{1}{16 \pi G}[f+12 H^{2}f_{T}-2 f_{\mathcal{T}}(\rho_m+p_m)].\label{ch3_14} 
\end{eqnarray}
In the next section, we will investigate the cosmic evolution of the Universe through the dynamical system analysis for two different forms of $f(T, \mathcal{T})$.
\section{The Dynamical system framework} \label{ch3_SEC-III}
 We introduce the following dynamical variables,
\begin{eqnarray}\label{ch3_23}
x&=&-\frac{f}{6 H^{2}}, \hspace{0.8cm} y=-2f_{T}, \hspace{0.8cm} u = \frac{\rho_{m} f_{\mathcal{T} }}{3 H^{2}}.   
\end{eqnarray}
With these dimensionless variables, the first Friedmann Eq. \eqref{ch3_9} becomes
\begin{equation}\label{ch3_24}
\Omega_{m}+x+y+u=1.    
\end{equation}
The density parameter at various stages of the evolutionary history of the Universe can be expressed in the dynamical system variable, 
\begin{eqnarray}
\Omega_{DE}&=& x+y+u ,\label{ch3_25} 
\end{eqnarray}
Also, Eq. \eqref{ch3_10} can be rewritten in terms of dynamical variables as
\begin{equation}\label{ch3_28}
\frac{\dot{H}}{H^{2}}= \frac{-3+3x+3y+6 \rho_{m} f_{T \mathcal{T}}}{(2-y+24 H^{2} f_{TT})}.   
\end{equation}
From Eq. \eqref{ch3_23}, we construct the set of autonomous differential equations 
\begin{eqnarray}
\frac{dx}{dN}&=&\frac{3 u}{2}-(y+2 x)\frac{\dot{H}}{H^{2}}, \label{ch3_29}\\
\frac{dy}{dN}&=& 24 \dot{H} f_{TT}+6 \rho_{m} f_{T \mathcal{T}}, \label{ch3_30}\\
\frac{du}{dN}&=& -\frac{\rho_{m}^{2}f_{\mathcal{T} \mathcal{T}} }{H^{2}}-4\rho_{m} f_{\mathcal{T} T}  \frac{\dot{H}}{H^{2}}-3u-2u\frac{\dot{H}}{H^{2}}\,. \label{ch3_31}
\end{eqnarray}

We derive the EoS parameter and deceleration parameter in terms of dimensionless variables as,
\begin{eqnarray}
\omega_{DE}&=&\frac{-6 H^2+6x H^2  +6y H^2 +12 H^2 \rho_{m} f_{T\mathcal{T}}+6 H^2(1-\frac{y}{2}+2 T f_{TT})}{(1-\frac{y}{2}+2 T f_{TT})(-6x H^2-6y H^2-2 f_{\mathcal{T}}\rho_{m})}, \label{ch3_34}\\
\omega_{tot}&=& -1-\frac{-1+x+y+2 \rho_{m} f_{T \mathcal{T}}}{(1-\frac{y}{2}+12 H^{2} f_{TT})}, \label{ch3_35}\\
q&=& -1-\frac{-3+3x+3y+6 \rho_{m} f_{T \mathcal{T}}}{(2-y+24 H^{2} f_{TT})}.\label{ch3_36}
\end{eqnarray}

To solve the autonomous system of differential Eqs.~(\ref{ch3_29}--\ref{ch3_31}), $f_{TT}$ and $f_{T \mathcal{T}}$ to be expressed either as a dynamical variable or in the form of dimensionless variables. This may be achieved by choosing some specific form of $f(T,\mathcal{T})$. Here, we have considered two forms of $f(T,\mathcal{T})$ and discussed the dynamical system analysis of two models.
\subsection{Model-I} \label{ch3_SEC-A}
Considering the dynamical system with the above dimensionless variables, we need to determine whether the modified gravity works as a model for the Universe. We consider the form of $f(T,\mathcal{T})$ \cite{Harko_2014a} as, 
\begin{equation}\label{ch3_37}
f(T,\mathcal{T})=\alpha T^{n} \mathcal{T}+\Lambda,   
\end{equation}
so that
\begin{eqnarray}
f_{T}= \alpha n T^{n-1} \mathcal{T} = -\frac{y}{2},\quad f_{TT}= \alpha n(n-1) T^{n-2}\mathcal{T} = -\frac{y(n-1)}{2T}, \quad f_{T \mathcal{T}}=\alpha n T^{n-1}.
\end{eqnarray}
Where the model parameters $\alpha$, $\Lambda$ and $n$ are constant \cite{Harko_2014a}. For this choice of $f(T,\mathcal{T})$, we have obtained the relation between dynamical variables $u=\frac{y}{2}$. The dynamical system is reduced to only two dynamical variables, $x$ and $y$. Subsequently, the autonomous system Eqs.~(\ref{ch3_29}--\ref{ch3_31}) can be written respectively as,
\begin{eqnarray}
\frac{dx}{dN}&=&\frac{(2 x+y) (3 (n-2) y-6 x+6)}{(2-4 n) y+4}+\frac{3 y}{4}, \label{ch3_39}\\
\frac{dy}{dN}&=&-y \left(\frac{3 (n-1) ((n-2) y-2 x+2)}{(2 n-1) y-2}-3\right). \label{ch3_40}
\end{eqnarray}

We can express the DE EoS parameter, total EoS parameter and the deceleration parameter with respect to the dynamical variables as,  
\begin{eqnarray}
\omega_{DE}&=&\frac{4 x-6 (n-1) y}{((2 n-1) y-2) (2 x+3 y)} , \label{ch3_44}\\
\omega_{tot}&=& -\frac{(2 x + 3 y - 3 n y)}{(2 + y - 2 n y)}, \label{ch3_45}\\
q&=& \frac{7 n y-6 x-8 y+2}{-4 n y+2 y+4}.\label{ch3_46}
\end{eqnarray}

We solve the combined equations as described in Eqs.~(\ref{ch3_39}--\ref{ch3_40}) to obtain the critical points to analyze the dynamical features of the autonomous system. Subsequently, we obtain the stability condition and the cosmology to describe the evolutionary phase of the Universe. The existence of critical points and their cosmological parameters are given in Table \ref{ch3_TABLE-I}. For the system Eqs.~(\ref{ch3_39}--\ref{ch3_40}), we have obtained three critical points and in the following section, we will go through the characteristics of each critical point and its possible connection with the evolutionary phase of the Universe. In this work, the eigenvalues are denoted by $\lambda_1$ and $\lambda_2$.

\begin{table}[H]
   \setlength{\tabcolsep}{2pt} 
    \caption{ The critical points and background parameters of the dynamical system. } 
    \centering 
    \begin{tabular}{|c|c|c|c|c|c|c|c|c|} 
    \hline 
    C.P. & $x_{c}$ & $y_{c}$ & $q$ & $\omega_{tot}$ & $\omega_{DE}$ & $\Omega_{DE}$ & $\Omega_{m}$ & Exists for \\ [0.5ex] 
    \hline\hline 
    $A_{1}$  & $0$ & $0$ & $\frac{1}{2}$ & $0$ & undefined & $0$ & $1$ & Always\\
    \hline
    $A_{2}$ & $1$ &$0$ & $-1$ & $-1$ & $-1$ &  $1$ & $0$& Always\\
     \hline
    $A_{3}$ & $\frac{3 n-n^2}{n^2-6 n+3}$ & $ \frac{2}{3}\left(n- \frac{9n-3n^{2}}{3-6n+n^{2}}+ \frac{3n^{2}-n^{3}}{3-6n+n^{2}}\right)$ & $\frac{1+2n}{2-2n}$ & $\frac{n}{1-n}$ & $\frac{3+n(6-n)}{(n-1)(n+3)}$ & $\frac{-n^2-3 n}{n^2-6 n+3}$ & $\frac{2 n^2-3n+3}{n^2-6 n+3}$& \begin{tabular}{@{}c@{}} $n^2-6 n+3\neq 0$, \\ $5 n^2-8 n+3\neq 0$\end{tabular} \\
    [1ex] 
    \hline 
    \end{tabular}
    \label{ch3_TABLE-I}
\end{table}
\begin{itemize}
\item \textbf{Critical Point $A_{1}$}: The critical point $A_{1}$ exists for all values of the free parameters and the density parameter, $\Omega_{m}=1$. The value of the total EoS parameter vanishes and therefore, it is the same as that of the EoS parameter of the matter-dominated phase and this critical point always exists. The DE EoS could not be determined for the critical point. Since the deceleration parameter is positive, there is no sign of cosmic acceleration. The respective eigenvalues of the Jacobian matrix for this critical point are defined below. The presence of both positive and negative sign eigenvalues means the critical point shows unstable node behavior if $n<0$. If $n>0$, then it shows unstable saddle behavior according to the linear stability theory.
 \begin{align*}
 \{\lambda_1=3, \quad \lambda_2=-3n\}\,. \nonumber    
\end{align*}   
\end{itemize}
\begin{itemize}
\item \textbf{Critical Points $A_{2}$:} The critical point $A_2$ appears in the DE era and the density parameter provides $\Omega_{DE}=1$. The $\omega_{tot}=\omega_{DE}=-1$ and the deceleration parameter $q=-1$ indicates the late time cosmic acceleration of the Universe. Also, the Hubble rate is constant for these critical points, i.e., $\dot{H}=0$; therefore, the expansion keeps accelerating close to the critical point. The signature of the eigenvalues of this critical point is negative, which means this critical point shows stable node behavior for any choice of model parameters. 
\begin{align*}
 \{\lambda_1=-3,\quad\lambda_2=-3\}\,. \nonumber    
\end{align*}
\end{itemize}
\begin{itemize}
\item \textbf{Critical Point $A_{3}$:} The dominance of different eras of the Universe for this critical point depends on the different range of model parameter $n$. We have presented the background parameters value in Table \ref{ch3_TABLE-I}, which depend on the parameter $n$. The value of background parameters at $n=0$ is the same as the critical point $A_{1}$. That means this critical also shows the matter-dominated phase of the Universe at $n=0$. According to the observations, the deceleration parameter shows the accelerated phase when $q<0$ and the decelerated phase when $q>0$ of the Universe. For this critical point, we have obtained the range of the model parameter $n$ to study different phases of the Universe. At $n<-\frac{1}{2}\lor n>1$, the deceleration parameter shows the accelerated phase of the Universe. For this range of $n$, the total EoS parameter satisfies the condition $\omega_{tot}<-\frac{1}{3}$. The total EoS parameter shows the phantom and quintessence regions of the Universe for $n>1$ and $n<-\frac{1}{2}$, respectively. The behavior of the eigenvalues depends on the model parameter $n$. This critical shows stable behavior for the range of model parameter $1<n\leq 2.14$. The signature of both eigenvalues is negative for this range. This critical point shows the overall dynamics of the Universe for different choices of model parameter $n$,
\begin{eqnarray*}
 \left\{\lambda_1=\frac{3 \left(2 n^3-9 n^2-\sqrt{A}+3\right)}{2 (n-1) (5 n-3)}, \quad\lambda_2=\frac{3 \left(2 n^3-9 n^2+\sqrt{A}+3\right)}{2 (n-1) (5 n-3)}\right\}   
\end{eqnarray*}
Where $A =4 n^6-36 n^5+101 n^4-120 n^3+78 n^2-36 n+9 $. 
\end{itemize} 

\begin{figure}[hbt!] 
    \centering
    \includegraphics[width=50mm]{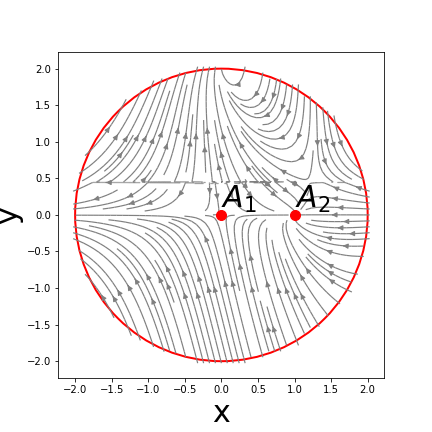}
    \includegraphics[width=50mm]{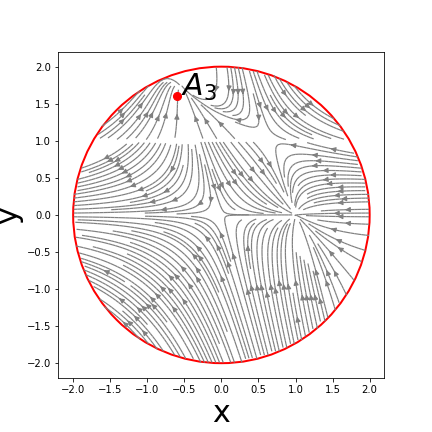}
    \includegraphics[width=50mm]{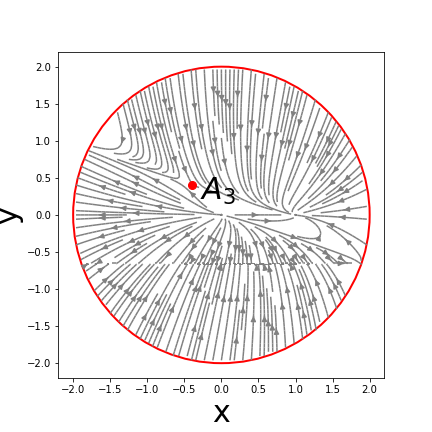}
    \caption{ $2$D Phase portrait for the dynamical system of {\bf Model-I}, (i) Left panel: Red dot denotes the critical point $A_{1}(0,0)$ and $A_{2}(1,0)$; (ii) Middle panel: Red dot denotes the critical point $A_{3}(-0.6,1.6)$ for the model parameter $n=1.5$; (iii)  Right panel: Red dot denotes the critical points $A_{3}(-\frac{2}{5}, \frac{2}{5})$ for the model parameter $n=-1$.} \label{ch3_Fig1}
\end{figure}

The phase portrait is an effective way to visualize the dynamics of the system as it shows the typical paths in the state space. The phase space of the critical points can be determined by setting the proper value for the parameters. Fig.~\ref{ch3_Fig1} depicts the $2D$ phase space of the dynamical system specified in Eqs.~(\ref{ch3_39}, \ref{ch3_40}). The stability of the model can be described through the phase portrait. According to the phase space diagram (Fig.~\ref{ch3_Fig1}), the trajectories of the critical point $A_{1}$ are moving away from the critical point, which indicates the saddle or unstable behavior. The trajectories show the attracting behavior towards the critical point, $A_{2}$ and hence show the stability. In addition, this critical point appears in the de-Sitter phase and may imply the current accelerated phase of the Universe. The trajectories for the critical point $A_{3}$ are moving towards the critical point $A_{3}$ for the stability range of model parameter $n<-\frac{1}{2}\lor n>1$, which is shown in Fig.~\ref{ch3_Fig1} (Middle-panel) for $n=1.5$ and outside the stability condition of the model parameter $n=-1$, the trajectories for the critical point $A_{3}$ are moving away to the critical point $A_{3}$, which means that outside the stability condition, the critical point shows unstable behavior, which is presented in Fig.~\ref{ch3_Fig1} (Right-panel).

We have plotted the background dynamical parameter plot for the two different regions, such as Phantom and Quintessence of the Universe. These regions are described from the total equation of the state parameter value for different choices of $n$. The Phantom region is defined for the range $n>1$ and the Quintessence region shows for the range $n<-\frac{1}{2}$. Here, we have drawn a plot for a specific choice of model parameter $n$; this particular choice of $n$ satisfies the condition of the Phantom and Quintessence regions.
\begin{figure}[H]    
\centering
      \includegraphics[width=70mm]{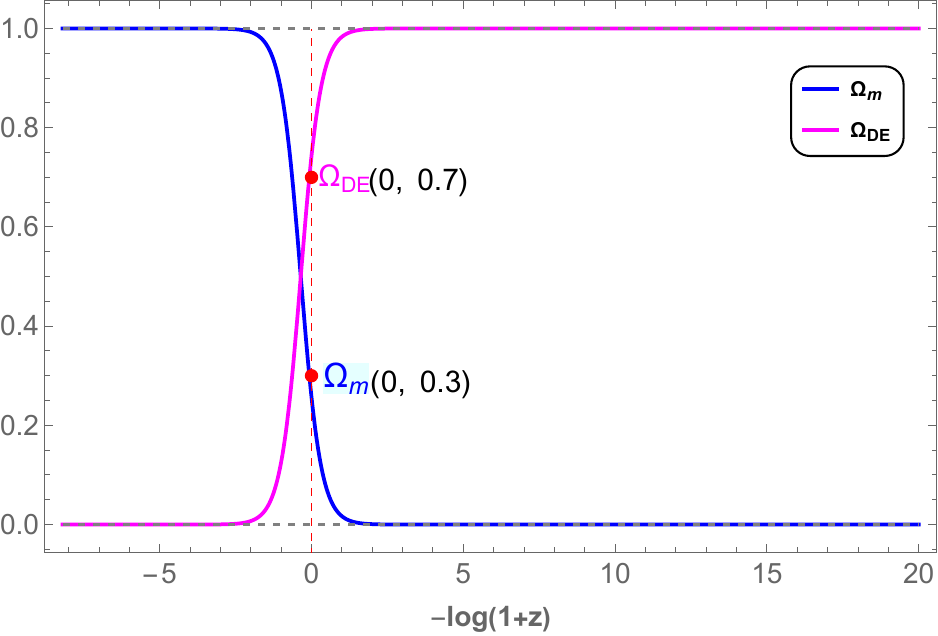}
      \includegraphics[width=70mm]{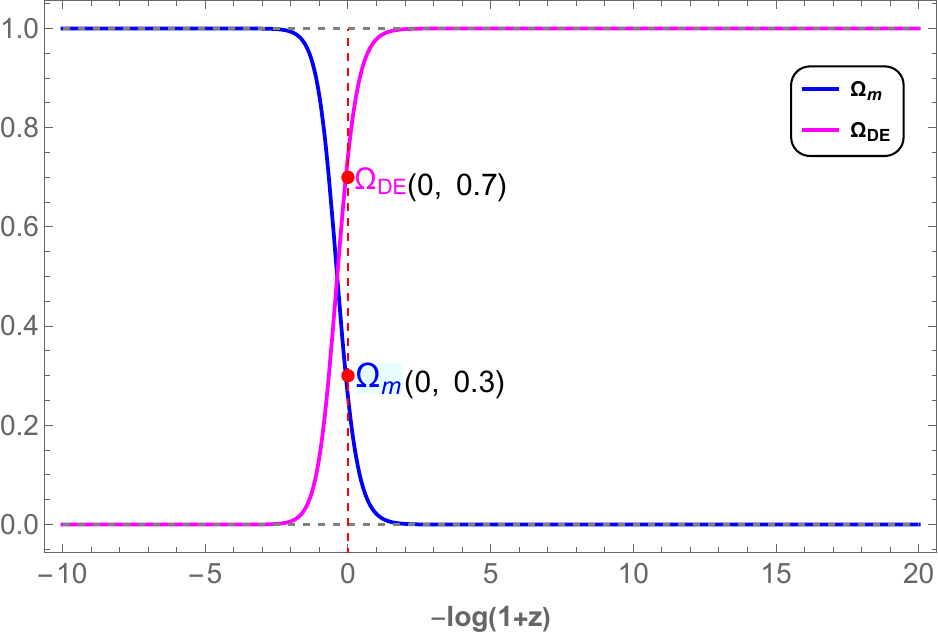}
    \caption{Evolution of density parameters for {\bf Model--I}. The initial conditions $x = 10^{-9}$, $y =10^{-8} $ and (Left panel n=1.5 Phantom region) and (Right panel n=-1 Quintessence region). The vertical dashed red line denotes the present time.} \label{ch3_Fig2}
\end{figure}
 
 \begin{figure}[H]
      \centering
      \includegraphics[width=70mm]{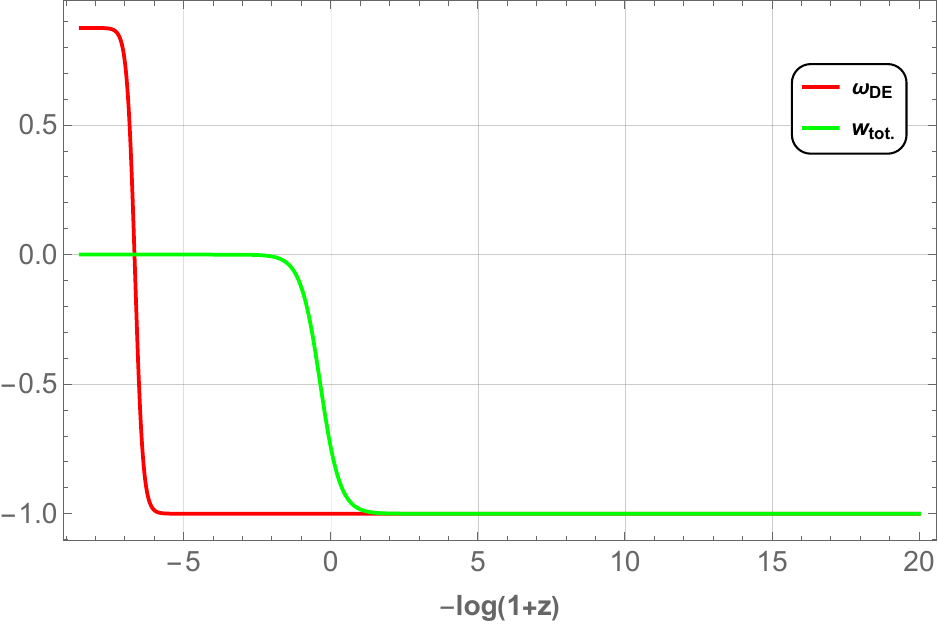}
      \includegraphics[width=70mm]{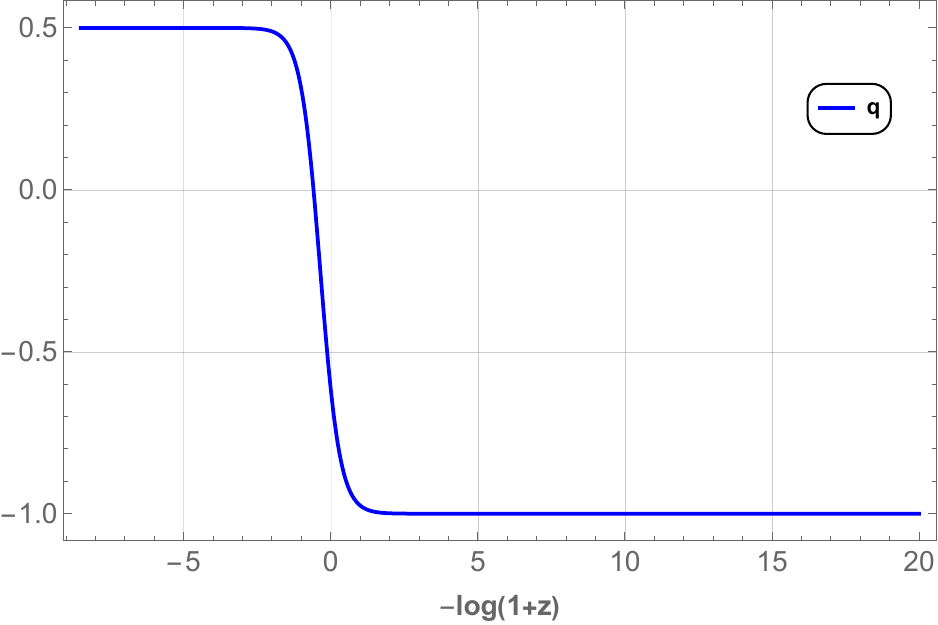}
      \caption{(Left panel) Evolution of total EoS parameter (Green) and DE EoS parameter (Red); (Right panel) Evolution of deceleration parameter (blue). The initial conditions are same as in Fig.~\ref{ch3_Fig2} and $n=1.5$ (Phantom region) .} \label{ch3_Fig3}
  \end{figure}

    \begin{figure}[H]
      \centering
      \includegraphics[width=70mm]{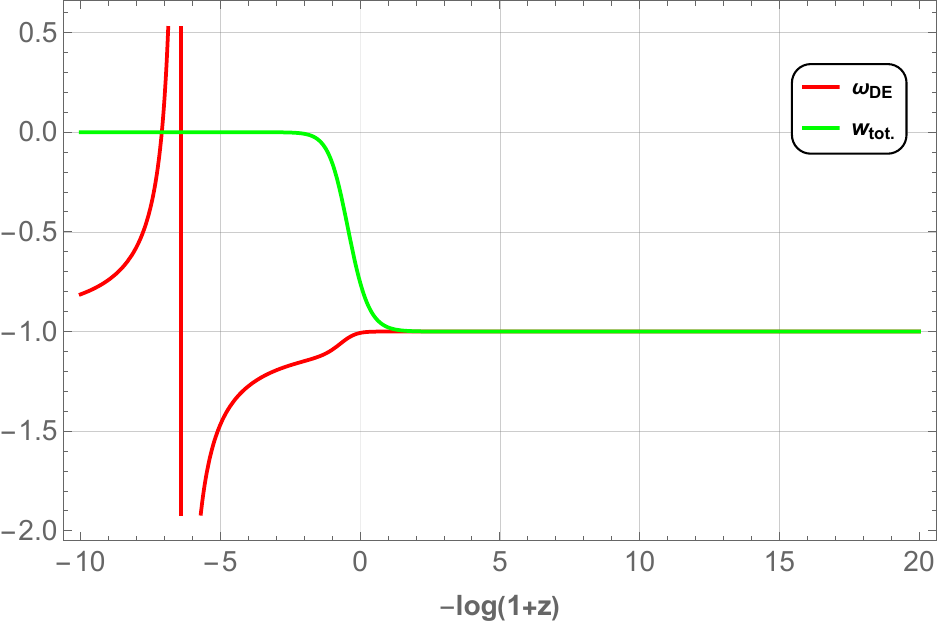}
      \includegraphics[width=70mm]{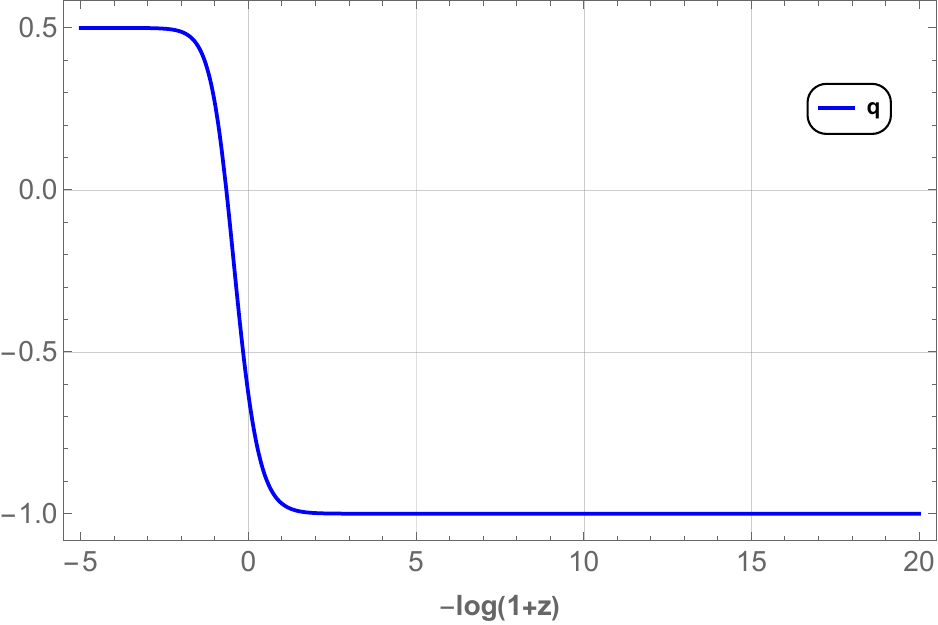}
      \caption{(Left panel) Evolution of total EoS parameter (Green) and DE EoS parameter (Red); (Right panel) Evolution of deceleration parameter (blue). The initial conditions are the same as in Fig.~\ref{ch3_Fig2} and  $n=-1$ (Quintessence region).} \label{ch3_Fig4}
  \end{figure}
For both the Phantom and Quintessence regions, the evolutionary behavior of the density parameters (Fig.~\ref{ch3_Fig2}) in redshift $\left(N=ln(\frac{1}{1+z})\right)$ illustrates the domination of matter in the early Universe and later approaches the DE sector. At present (vertical red dashed line), as the observation revealed, DM and DE predominate. DE makes up around $0.7$ of the total energy, with the remaining $0.3$ being DM. For both regions, We obtain $\Omega_{DE}\approx0.7$ and $\Omega_{m}\approx0.3$. We observe from Fig.~\ref{ch3_Fig3} (Left panel) that the total EoS parameter begins with the Phantom region of the Universe, falls to $0$ during the period when matter predominated and finally approaches $-1$ as the role of DE becomes more significant. We also notice the DE EoS parameter and, at present, $\omega_{DE}\approx -1$, which is compatible with the present Planck Collaboration result [$\omega_{DE}(z=0)= -1.028 \pm 0.032$ \cite{Aghanim:2018eyx}].  In Fig.~\ref{ch3_Fig3} (Right panel), the deceleration parameter shows a transition from deceleration to acceleration with the transition at $z=0.62$, which is consistent with the observational constraint $z_{trans.}=0.7679^{+0.1831}_{-0.1829}$ \cite{PhysRevD.90.044016a}. The present value of the deceleration parameter can be obtained as $q(z=0) \approx -0.59$, consistent with the visualized cosmological observations \cite{Camarena:2020prr}. The total EoS parameter in Fig.~\ref{ch3_Fig4} (Left panel) starts with a matter value of $0$ and moves towards $-1$ when DE plays a role. We also see the EoS parameter for DE, currently $\omega_{DE}\approx -1$, which agrees with the current Planck Collaboration conclusion. The deceleration parameter in Fig.~\ref{ch3_Fig4} (Right panel) exhibits a transition from deceleration to acceleration at $z=0.58$, which follows the observational constraint. The calculated current value of the deceleration parameter is $q(z=0) \approx -0.63$, following the depicted cosmological data. 
\subsection{Model-II} \label{ch3_SEC-B}
For further investigation of the dynamical system, we consider another form of $f(T,\mathcal{T})$ \cite{Harko_2014a} as
\begin{equation}\label{ch3_47}
f(T,\mathcal{T})= \gamma T^{2}+ \delta \mathcal{T},   
\end{equation}
where $\gamma$ and $\delta$ are free parameters. The framework of the model in terms of dynamical variables is as follows,
\begin{eqnarray}\label{ch3_48}
f_{T}=2 \gamma T  \equiv -\frac{y}{2}, \hspace{0.5cm} f_{TT}&=&2\gamma, \hspace{0.5cm} f_{\mathcal{T}}=\delta, \hspace{0.5cm} f_{\mathcal{T}\mathcal{T}}=0, \hspace{0.5cm} f_{T \mathcal{T}}=0.
\end{eqnarray}
We have discovered the relationship between the dynamical variables $y=-4(x+\frac{u}{2})$ for this choice of $f(T,\mathcal{T})$. Only the dynamic variables $x$ and $y$ remain in the simplified dynamic system. The autonomous system is therefore represented by the Eqs.~ (\ref{ch3_29}--\ref{ch3_31}) as follows,
\begin{eqnarray}
\frac{dx}{dN}&=&\frac{3 (x+y-1) (4 (\delta +1) x+\delta  (-y)+2 (\delta +y))}{2 (\delta +1) (3 y-2)}, \label{ch3_49}\\
\frac{dy}{dN}&=&-\frac{6 y (x+y-1)}{3 y-2}. \label{ch3_50}
\end{eqnarray}
In the form of a dimensionless variable, the EoS parameter and deceleration parameter may be represented as,
\begin{eqnarray}
\omega_{DE}&=&\frac{(\delta +1) (2 x-y)}{(3 y-2) (\delta +x+y)} , \label{ch3_54}\\
\omega_{tot}&=& \frac{2 x-y}{3 y-2}, \label{ch3_55}\\
q&=& \frac{1-3 x}{2-3 y}.\label{ch3_56}
\end{eqnarray}

In order to perform the dynamical analysis, the critical points of the system to be obtained from Eqs.~ (\ref{ch3_49}, \ref{ch3_50}). The nature and stability of each point to be established by perturbing the system around these critical points using the eigenvalues of the underlying perturbation matrix. The two critical points obtained are given in Table~\ref{ch3_TABLE-IV} along with its existence condition(s) and the corresponding value of the EoS parameters, deceleration parameter and density parameters.

\begin{table}[H]
    \caption{ The critical points and background parameters of the dynamical system. } 
    \centering 
    \begin{tabular}{|c|c|c|c|c|c|c|c|c|} 
    \hline\hline 
    C.P. & $x_{c}$ & $y_{c}$ & $q$ & $\omega_{tot}$ & $\omega_{DE}$ & $\Omega_{DE}$ & $\Omega_{m}$ & Exists for \\ [0.5ex] 
    \hline\hline 
    $B_{1}$  & $x$ & $1-x$ & $-1$ & $-1$ & $-1$ & $\frac{1-3x}{2}$ & $\frac{1+3x}{2}$ & $-\delta +3 \delta  x+3 x-1\neq 0$\\
    \hline
    $B_{2}$ & $-\frac{\delta }{2 (\delta +1)}$ &$0$ & $\frac{5 \delta +2}{4 \delta +4}$ & $\frac{\delta }{2 \delta +2}$ & $\frac{\delta +1}{2 \delta +1}$ &  $\frac{\delta }{2 (\delta +1)}$ & $\frac{\delta +2 }{2 (\delta +1)}$& $\delta +1\neq 0$\\
    [1ex] 
    \hline 
    \end{tabular}
    \label{ch3_TABLE-IV}
\end{table}
\begin{itemize}
 \item {\bf Critical Point $B_{1}$:} The total EoS and DE sector EoS parameter and the deceleration parameter assume the value $-1$, which indicates an accelerating de-Sitter phase of the Universe. The value of the dynamical parameters shows that against the critical points, it shows $\Lambda$CDM-like behavior. The solution of density parameters for the critical point $B_{1}$ is $\Omega_{DE}=\frac{1-3x}{2 }$ and  $\Omega_{m}=\frac{1+3x}{2}$. For $x=-\frac{1}{3}$, the density parameters reduced in the fully dominated DE sector $\Omega_{DE}=1$. The first eigenvalue of this critical point is zero and others depend on the model parameter $\delta$ and dynamical variable $x$. We know that the linear stability theory fails to tell about stability when a zero eigenvalue is presented along with positive eigenvalues. But in our case the second eigenvalue is negative when $\delta$ and $x$ satisfy the condition $x\in \mathbb{R}\land \left(\delta <-1\lor \delta >-\frac{2}{3}\right)$. For this condition, the eigenvalues of this critical point are negative real part and zero. The dimension of the set of eigenvalues for non-hyperbolic critical points equals the number of vanishing eigenvalues \cite{Coley:1999,aulbach1984}. As a result, the set of eigenvalues is normally hyperbolic and the critical point associated with it is stable but cannot be a global attractor. In our case, the dimension of the set of eigenvalues is one and only one eigenvalue vanishes. That means the dimension of a set of eigenvalues equals the number of vanishing eigenvalues. This critical point is consistent with recent observations and can explain the current acceleration of the Universe. The behavior of this critical point is a stable node.
 
 \begin{align*}
 \left\{\lambda_1=0, \quad \lambda_2=\frac{3 (3 \delta -9 \delta  x-6 x+2)}{2 (\delta +1) (3 x-1)}\right\}. \nonumber    
\end{align*}
\end{itemize}
\begin{itemize}
\item {\bf Critical Point $B_{2}$:} The critical point $B_{2}$ is the origin of the phase space, which corresponds to a Universe where matter predominates, $\Omega_{m}=1$ for $\delta=0$. Since $\omega_{tot}=0$ = $\omega_{m}$ and the total EoS overlaps with the matter EoS, no acceleration occurs for physically permissible $\omega_{m}$ values. The phase space trajectory and the eigenvalues show the unstable behavior of the critical point for any choice of model parameter $\delta$. 
 \begin{align*}
 \left\{\lambda_1=-\frac{3 (3 \delta +2)}{2 (\delta +1)}, \quad \lambda_2=\frac{3 (3 \delta +2)}{2 (\delta +1)}\right\}. \nonumber    
\end{align*}
\end{itemize}

\begin{figure}[hbt!]
    \centering
    \includegraphics[width=70mm]{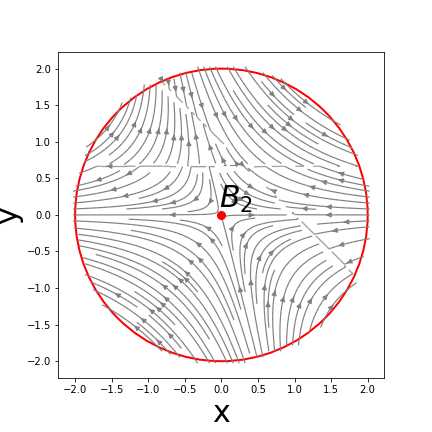}
    \includegraphics[width=70mm]{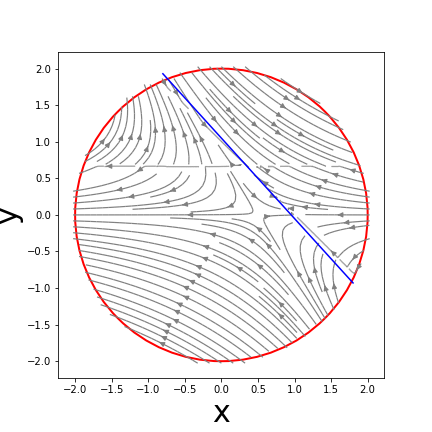}
     \caption{ $2D$ Phase portrait for the dynamical system of {\bf Model-II}, (i) Left panel: Red dot denotes the critical point  $B_{2}$;  (ii) Right panel: blue line denotes the critical points $B_{1}$.} \label{ch3_Fig5}
\end{figure}

 Further, to establish the result on the stability of the critical points, the $2D$ phase space trajectory of the dynamical system Eqs.~(\ref{ch3_49}, \ref{ch3_50}) are shown in Fig.~\ref{ch3_Fig5}. The direction of the trajectory describes the stability behavior of the critical point. The trajectories of the phase space are moving away from the critical point $B_{2}$, which shows the saddle or unstable behavior. In comparison, the trajectory exhibits attractive behavior for the curve of critical points $B_{1}$. Moreover, the phase portrait shows that $B_{1}$ is a global attractor.

\begin{figure}[H]
     \centering
     \includegraphics[width=70mm]{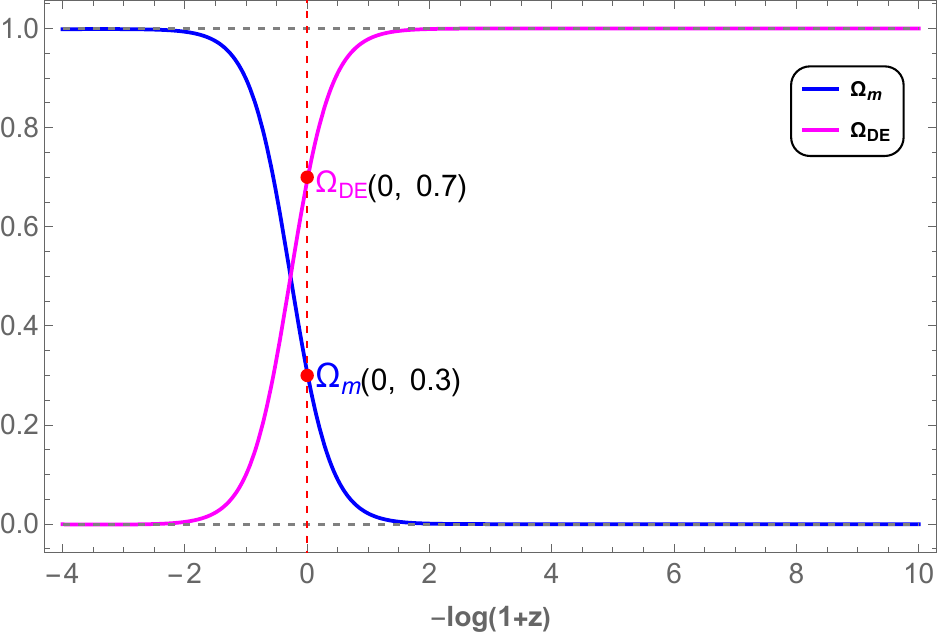}
     \caption{Evolution of density parameters for {\bf Model-II}. The initial conditions $x = 10^{-4}$, $y =10^{-5} $, $\delta=0.002$. The vertical dashed red line denotes the present time.}\label{ch3_Fig6}
\end{figure}
\begin{figure}[H]
     \centering
     \includegraphics[width=60mm]{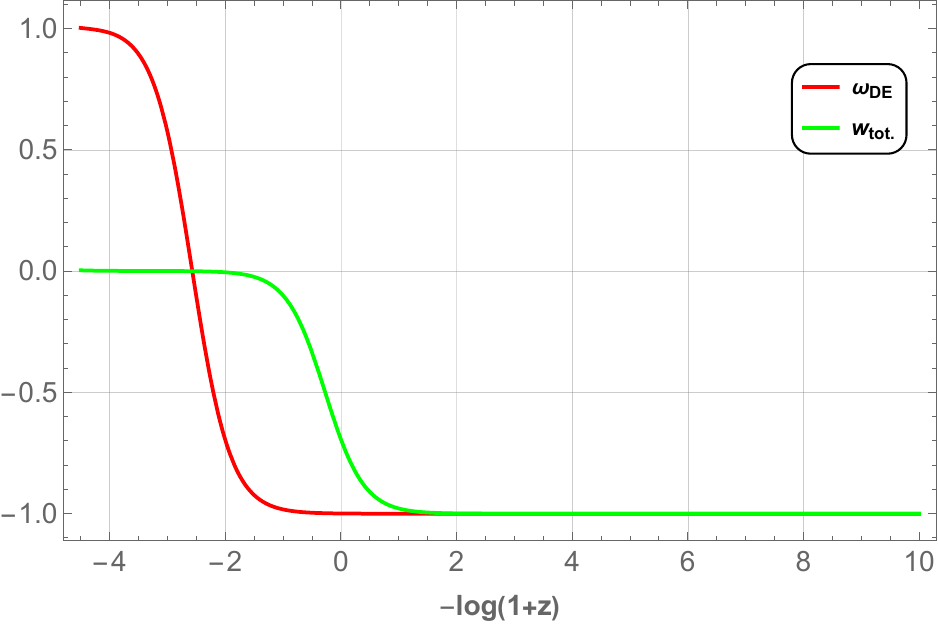}
     \includegraphics[width=60mm]{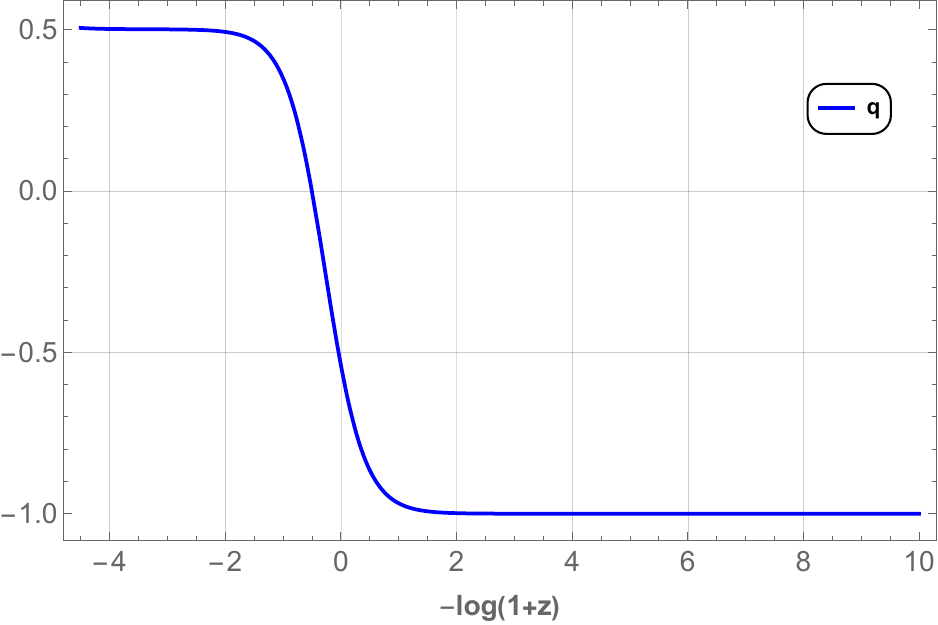}
      \caption{(Left panel) Evolution of total EoS parameter (Green line) and DE EoS parameter (Red line); (Right panel) Evolution of deceleration (Blue line). The initial conditions are same as in Fig.~\ref{ch3_Fig6}.}  \label{ch3_Fig7}
\end{figure}

Fig.~\ref{ch3_Fig6} shows how matter predominates early on and then gradually transitions into the de-Sitter phase of the Universe. The present value for matter and DE density parameters are respectively $\Omega_{m}\approx0.3$ and $\Omega_{DE}\approx0.7$. The behavior of the EoS parameter has been given in Fig.~\ref{ch3_Fig7}. The total EoS parameter starts with a matter value of  $0$ and finally approaches $-1$ as the role of DE becomes more substantial [Fig.~\ref{ch3_Fig7} (Left panel)]. The DE EoS parameter approaches $-1$ and the present value as suggested in Planck Collaboration is, $\omega_{DE}= -1.028 \pm 0.032 $ \cite{Aghanim:2018eyx}. The deceleration parameter shows transient behavior from deceleration to acceleration at the point $z=0.51$ with the present value noted as $q(z=0) \approx -0.56$ [Fig.~\ref{ch3_Fig7}] (Right panel)] \cite{Camarena:2020prr}.
\section{Cosmological Observation} \label{ch3_cosmological observation}
In this section, we present and evaluate the results following the methodology described in Sec.~\ref{ObservationalCosmology}, using the cosmological observations data set. We showcase contour plots of the constrained parameters and their respective $1\sigma$ and $2\sigma$ uncertainties, accompanied by tables summarizing the final results. In all tables and posterior plots, we include results for the Hubble constant $H_0$, the current matter density parameter $\Omega_{m0}$ and the parameters of the models. This will enable us to examine how the independent data sets and cosmological models influence the Hubble tension. We will briefly highlight the most significant findings, emphasizing the PN$^+$ (without SH0ES) and PN$^+$\&SH0ES (with SH0ES) distinctions. We have also considered the recent local measurement from SH0ES which gives $H_0 = 73.04 \pm 1.04 \, \text{km} \, \text{s}^{-1} \text{Mpc}^{-1}$ (R21) \cite{Riess:2021jrx}. The measurement using the TRGB as a standard candle with \( H_0 = 69.8 \pm 1.9 \, \text{km} \, \text{s}^{-1} \, \text{Mpc}^{-1} \) \cite{Freedman_2019TRGB} along with data set combination. In the Friedmann equations [Eq.\eqref{ch3_9}--Eq.\eqref{ch3_10}], we need to incorporate some functional form of \( f(T, \mathcal{T}) \). In this setting, we have considered the following form of \( f(T, \mathcal{T}) \) \cite{Harko_2014a}
\begin{equation} \label{modelconsidered}
f(T,\mathcal{T}) = \alpha T^{n} \mathcal{T} + \Lambda \,,  \end{equation}
where \( \alpha \neq 0\), \( n \neq 0 \) and \( \Lambda \) are arbitrary constants. At present, one can express the Friedmann Eq.~\eqref{ch3_9} as,
\begin{equation}\label{alphapresenttime}
\alpha = \frac{2-2\Omega_{m0}+\frac{\Lambda}{3 H_{0}^2}}{(1+2n)\Omega_{m0} (-6 H_{0}^2)^{-n}}\,,    
\end{equation}
where \( H_0 \) and \( \Omega_{m0} \) respectively represent the Hubble parameter and matter density parameter at present time. From Eq.~\eqref{alphapresenttime}, it can be inferred that the model parameter \( \alpha \) depends on other parameters such as \{\( H_0 \), \( \Omega_{m0} \), \( n \) and \( \Lambda \)\}. By defining the dimensionless Hubble parameter \( E(z)= \frac{H(z)}{H_0} \), the Friedmann Eq.~\eqref{ch3_9} for this model can be reformulated as
\begin{eqnarray}\label{modelHz}
E^{2}(z)= (1+z)^3 \Omega_{m0}-\frac{\Lambda}{6 H_{0}^2}+\bigg(1-\Omega_{m0}+ \frac{\Lambda}{6 H_{0}^2}\bigg) (1+z)^3 E^{2n}(z)\,.  
\end{eqnarray}
To ensure that the term \( \frac{\Lambda}{6 H_{0}^2} \) remains dimensionless, we define the parameter as \( \Lambda =H_{0}^2 \). Now, Eq.~\eqref{modelHz} becomes,
\begin{eqnarray}\label{conmodelhz}
E^{2}(z)= (1+z)^3 \Omega_{m0}-\frac{1}{6}+\left(1-\Omega_{m0}+ \frac{1}{6}\right) (1+z)^3 E^{2n}(z)\,.     
\end{eqnarray}
Eq.~\eqref{conmodelhz} is an implicit formulation for \(E(z)\). Considering that analytical solutions for Eq.~\eqref{conmodelhz} are impractical, we have adhered to the numerical methods to compute the parameters. The methodology used as described in Sec.-\ref{ObservationalCosmology} and the results obtained are described below.

\begin{figure}[H]
     \centering
     \begin{subfigure}[b]{0.49\textwidth}
         \centering
         \includegraphics[width=\linewidth]{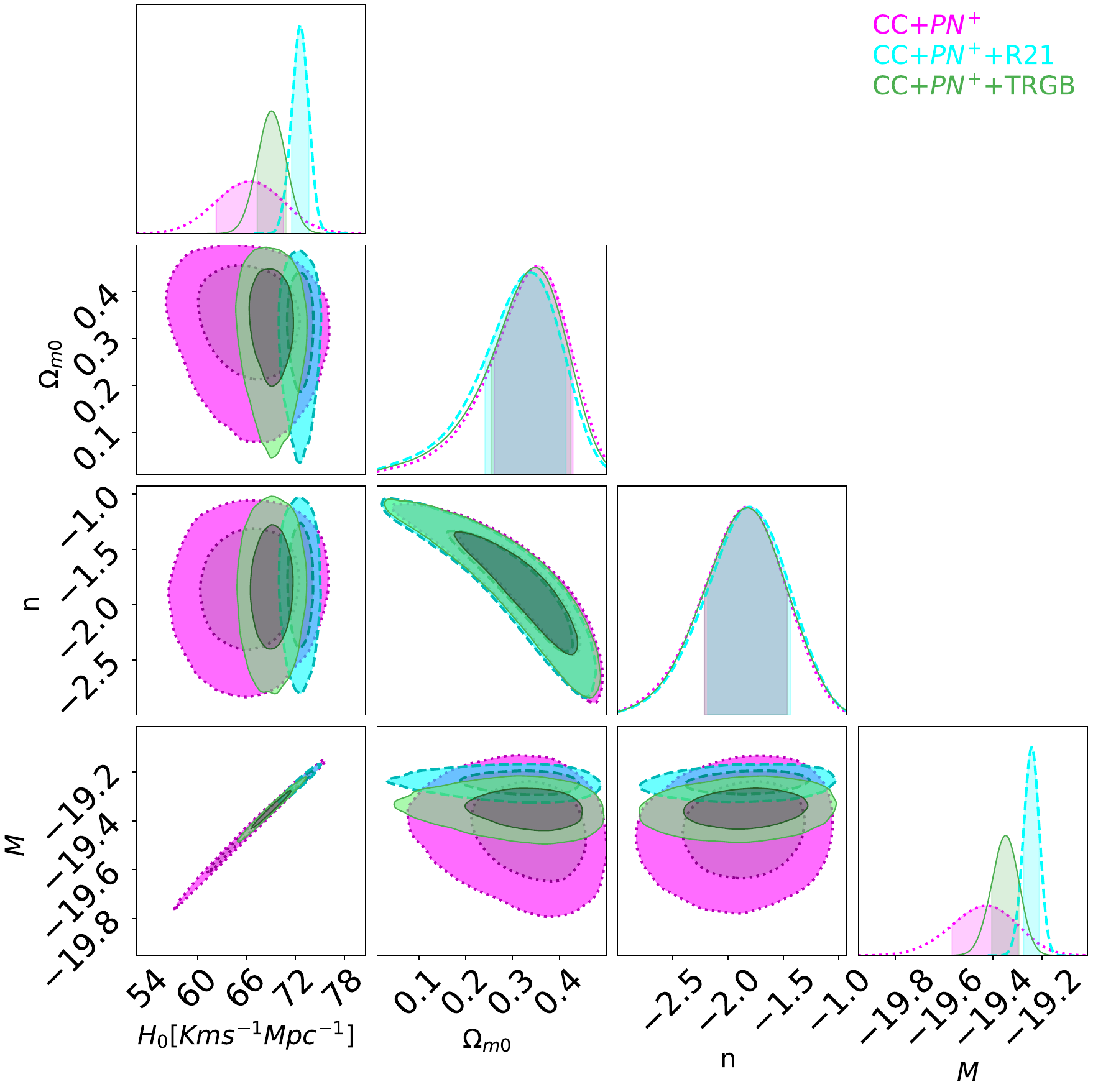}
          \caption{}
         \label{fig:CCMCMCplus}
     \end{subfigure}
     \hfill
     \begin{subfigure}[b]{0.49\textwidth}
         \centering
         \includegraphics[width=\linewidth]{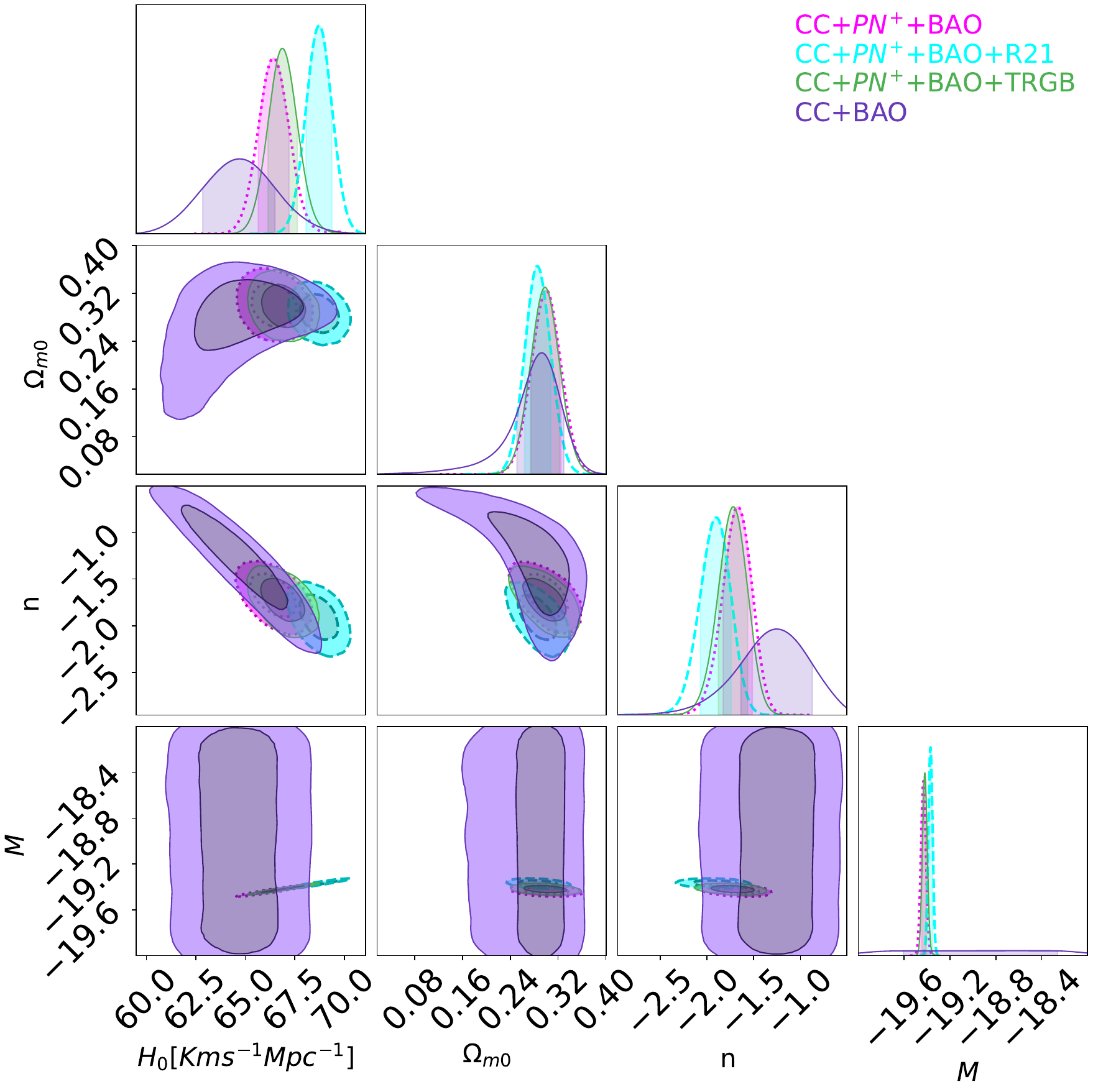} 
         \caption{}
         \label{fig:BAOMCMCplus}
     \end{subfigure}
\caption{The contour plot of $1\sigma$ and $2\sigma$ uncertainty regions and posterior distribution for the model parameters with the combination of data sets (a) CC, PN$^{+}$ (b) CC, PN$^{+}$ and BAO. The $H_0$ priors are: TRGB (Green) and R21 (Cyan).} 
\label{plusmodelMCMC}
\end{figure}

In Fig.~\ref{plusmodelMCMC}, we have presented the $1\sigma$ and $2\sigma$ confidence levels along with the posterior distributions for the parameters $H_0$, $\Omega_{m0}$, $n$ and $M$ using CC, PN$^{+}$ (without SH0ES) and BAO data sets in addition to the R21 and TRGB priors. It displays the marginalized posterior distributions for different combinations of parameters. The inner contours indicate the 68\% confidence level while the outer contours represent the 95\% confidence level. This visual representation facilitates a thorough evaluation of parameter uncertainties and correlations. The most noticeable aspect of the findings is the influence these priors exert on $H_0$ values, as it is clear that these priors generally lead to an increase in the Hubble constant value when contrasted with the scenario of no priors. Including priors decreases the estimated value of $\Omega_{m0}$, albeit to a lesser degree than the impact observed on $H_0$. This reduction is expected, as the priors effectively adjust the value of $H_0$. Table~\ref{tab:model_outputsmodelplus} summarizes the results derived from Fig.~\ref{plusmodelMCMC}. Notably, the highest value of $H_0$ is obtained from the data set combination CC+PN$^{+}$+R21, while the lowest is observed for CC+BAO. The diminished $H_0$ in the CC+BAO combination can be attributed to incorporating the BAO data set, which originates from early Universe measurements. Furthermore, the nuisance parameter $M$ remains unconstrained in the CC+BAO analysis due to the absence of the PN$^+$ data set. The determined value of \( H_0 \) for the CC+PN$^{+}$+R21 aligns with the elevated \( H_0 \) value reported by the SH0ES team (R22), which states \( H_{0} = 73.30\pm{1.04} \, \text{km s}^{-1} \text{ Mpc}^{-1} \) \cite{Riess:2021jrx}.

\begin{figure}[H]
     \centering
     \begin{subfigure}[b]{0.49\textwidth}
         \centering
         \includegraphics[width=\linewidth]{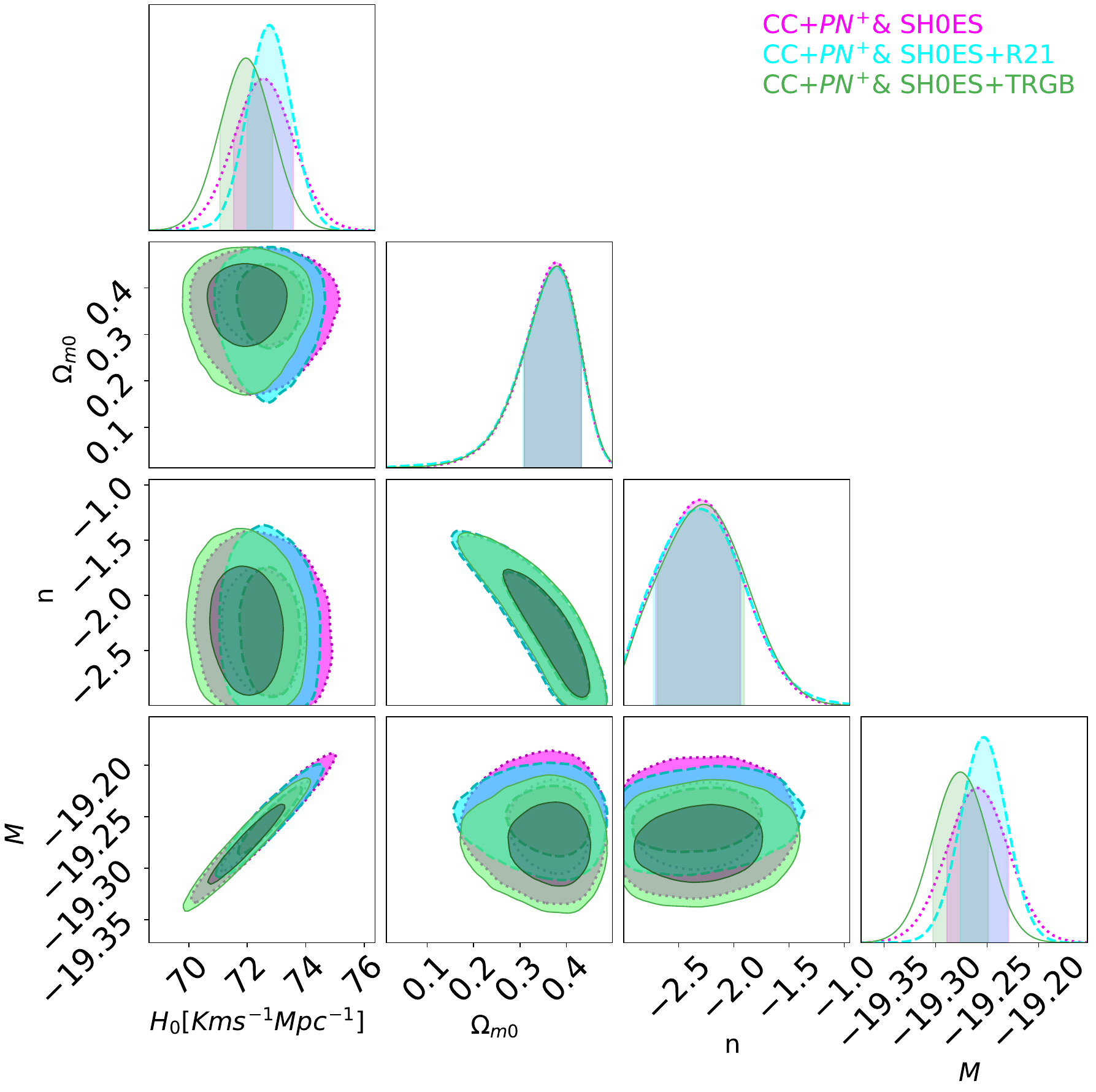}
          \caption{}
         \label{fig:CCMCMC}
     \end{subfigure}
     \hfill
     \begin{subfigure}[b]{0.49\textwidth}
         \centering
         \includegraphics[width=\linewidth]{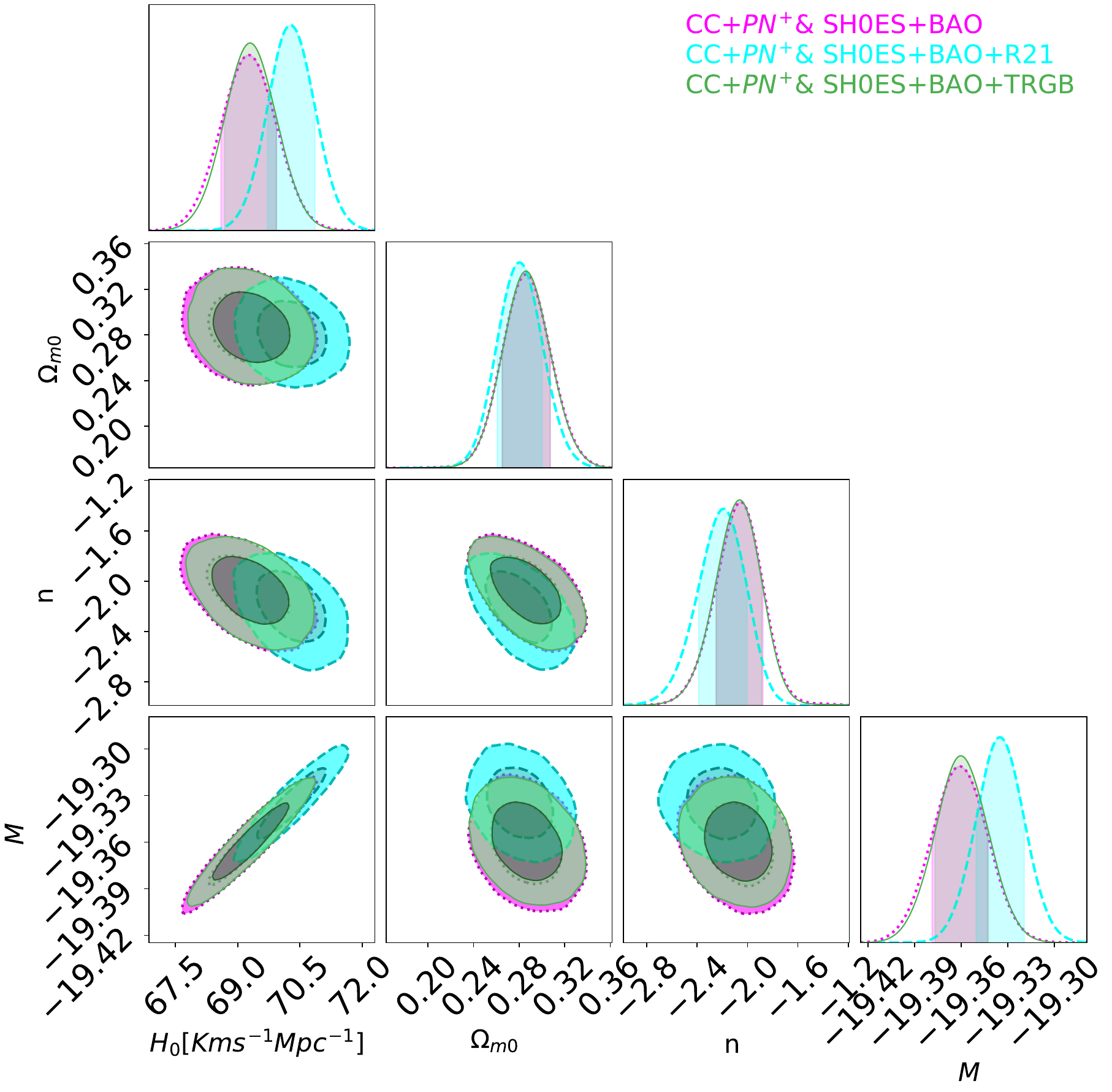} 
         \caption{}
         \label{fig:BAOMCMC}
     \end{subfigure}
\caption{The contour plot of $1\sigma$ and $2\sigma$ uncertainty regions and posterior distribution for the model parameters with the combination of data sets (a) CC, $PN^{+}\& SH0ES$ (b) CC, $PN^{+}\& SH0ES$ and BAO. The $H_0$ priors are: TRGB (Green) and R21 (Cyan).} 
\label{modelMCMC}
\end{figure}

In Fig.~\ref{modelMCMC}, we display the $1\sigma$ and $2\sigma$ confidence intervals alongside the posterior distributions for the parameters $H_0$, $\Omega_{m0}$, $n$ and $M$ utilizing the CC, PN$^{+}$\&SH0ES (with SH0ES) and BAO data set, in addition to the R21 and TRGB priors. In Table~\ref{tab:model_outputsmodel}, it is observed that the highest \( H_{0} \) value is noted for the combination of data sets CC+$PN^{+}\& SH0ES$+R21, which is \( H_{0} = 72.74^{+0.77}_{-0.74} \, \text{km s}^{-1} \text{ Mpc}^{-1} \). The addition of SH0ES data points alongside the PN$^+$ data set raises the \( H_{0} \) value in comparison to using only the PN$^{+}$ data set (without SH0ES). Integrating the BAO data set results in a lower \( H_0 \) value than the combination of the PN$^{+}$\& SH0ES data sets. The value of \(H_0\) for inclusion with the BAO data set is consistent with the Planck Collaboration \cite{Aghanim:2018eyx}, which reports a Hubble constant of \(67.4 \pm 0.5 \, \text{km s}^{-1} \, \text{Mpc}^{-1}\). In contrast, Aboot et al. \cite{Abbott_2018mnras} propose a value of \(67.2^{+1.2}_{-1.0} \, \text{km s}^{-1} \, \text{Mpc}^{-1}\). In this work, we have analyzed the differences between the PN$^+$ (without SH0ES) and PN$^{+}$\& SH0ES (with SH0ES) data set, also incorporating the BAO data set and $H_0$ priors. The findings can be observed in Tables-\ref{tab:model_outputsmodel} and Table~\ref{tab:model_outputsmodelplus}. From the results, we concluded that including the SH0ES data points with the PN$^+$ data set raises the $H_0$ value compared to PN$^+$, which in turn caused adjustments in $\Omega_{m0}$ and $n$ due to the change in the $H_0$ value. Our results for the Hubble constant \(H_0\) value from the data set combination PN\(^{+}\) (without SH0ES) and PN\(^{+}\)\& SH0ES (with SH0ES), along with the BAO data set combination, align with the findings presented by Brout et al.\cite{Brout_2022panplus} for the PN\(^{+}\) (without SH0ES) and PN\(^{+}\)\& SH0ES (with SH0ES).
           
We have also computed the AIC and BIC values, providing a statistical foundation for selecting the appropriate model. Results related to the \(\Lambda\)CDM model can be found in the Appendix, particularly in Table~\ref{tab:model_outputsLCDM}. Lower values of \(\Delta\)AIC and \(\Delta\)BIC suggest that the model using the chosen data sets closely resembles the \(\Lambda\)CDM model. In Tables~\ref{tab:model_outputAICBICplus} and \ref{tab:model_outputAICBIC}, we present the statistical results including $\chi^{2}_{\text{min}}$, $\Delta$AIC and $\Delta$BIC for PN$^+$ (without SH0ES) and PN$^+$\& SH0ES (with SH0ES), respectively. In this analysis, the values of \(\Delta \text{AIC}\) and \(\Delta \text{BIC}\) for the data set combinations for  CC+PN$^{+}$ and CC+$PN^{+}\& SH0ES$ that include \( H_{0} \) priors are notably lower than those for the BAO data set combinations that also incorporate \( H_{0} \) priors. The study indicates that our model for data set combination of CC+PN$^{+}$ and CC+$PN^{+}\& SH0ES$ performs more effectively than when BAO data is included in combination with the standard $\Lambda$CDM model. In fact, for the scenario where the statistical distance deviates most from $\Lambda$CDM, specifically for the inclusion of the BAO data set, the statistical criteria yield a negative result, suggesting a slight preference for the model. In contrast, the other scenarios slightly inclined towards the standard cosmological model. The lower value of the  \(\Delta \text{AIC}\) and \(\Delta \text{BIC}\) values for the data set combination with \( H_{0} \) priors signify that the particular data combination aligns more closely with the standard \(\Lambda\)CDM model.

\renewcommand{\arraystretch}{1.2} 
\begin{table}[H]
    \centering
    \caption{The best-fit values of the parameters explored by MCMC analysis. The first column enumerates a combination of data sets with the \(H_0\) priors. The second column presents the constrained \(H_0\) values. The third column contains the constrained \(\Omega_{m0}\) values. The fourth and fifth columns represent the optimal \(n\) and \(M\) values respectively. {\bf(Without SH0ES data points)}}
    \label{tab:model_outputsmodelplus}
      \begin{tabular}{cccccc}
        \hline
		Data set & $H_0[\text{km s}^{-1} \, \text{Mpc}^{-1}]$ & $\Omega_{m0}$  & n & M \\     
		\hline
		CC+PN$^{+}$ & $66.4\pm 4.1$ & $0.354^{+0.074}_{-0.094}$ & $-1.82^{+0.36}_{-0.39}$ & $-19.43^{+0.13}_{-0.14}$\\ 
		CC+PN$^{+}$+R21 & $72.6^{+1.1}_{-1.0}$ & $0.338^{+0.076}_{-0.096}$ & $-1.80^{+0.37}_{-0.38}$ & $-19.242^{+0.032}_{-0.033}$ \\ 
		CC+PN$^{+}$+TRGB & $69.1\pm 1.8$ & $0.347^{+0.077}_{-0.093}$ & $-1.82^{+0.36}_{-0.39}$ & $-19.348^{+0.055}_{-0.057}$ \\ 
		\cline{1-5}
       CC+PN$^{+}$+BAO & $66.43\pm 0.78$ & $0.300\pm 0.025$ & $-1.67^{+0.15}_{-0.16}$ & $-19.431^{+0.022}_{-0.023}$ \\ 
		CC+PN$^{+}$+BAO+R21 & $68.73^{+0.64}_{-0.67}$ & $0.285^{+0.023}_{-0.021}$ & $-1.90^{+0.16}_{-0.17}$ & $-19.370^{+0.019}_{-0.018}$ \\ 
		CC+PN$^{+}$+BAO+TRGB &$66.88^{+0.75}_{-0.74}$ & $0.298^{+0.025}_{-0.024}$ & $-1.72^{+0.15}_{-0.16}$ & $-19.419^{+0.022}_{-0.021}$ \\
        CC+BAO &$64.7\pm 1.8$ & $0.292^{+0.038}_{-0.041}$ & $-1.25\pm 0.38$ & - \\
  \hline
    \end{tabular}
\end{table}
\renewcommand{\arraystretch}{1} 
\begin{table}[H]
    \centering
    \caption{The statistical comparison between the chosen model and the standard \(\Lambda\)CDM model. Details regarding the \(\Lambda\)CDM model are given in {\color{blue}Appendix}. The first column enumerates the data sets including the \(H_0\) priors. The second column displays the values of \(\chi^{2}_{\text{min}}\). The third and fourth column respectively provides the value of AIC and BIC. The fifth and sixth column respectively illustrates the values of \(\Delta \text{AIC}\) and \(\Delta \text{BIC}\). {\bf(Without SH0ES data points)}}
    \label{tab:model_outputAICBICplus}
      \begin{tabular}{cccccc}
        \hline
		data set& $\chi^{2}_{min}$  &AIC &BIC&$\Delta$AIC &$\Delta$BIC \\ 
		\hline
		CC+PN$^{+}$&1787.76 &1795.76 &  1800.71& -2.54&-1.31 \\ 
		CC+PN$^{+}$+R21& 1780.52 &1798.52 &  1803.48&-3.12 &-1.87\\ 
		CC+PN$^{+}$+TRGB&1788.41 &1796.41 &  1801.37&-2.79 &-1.55\\ 
		\cline{1-6}
       CC+PN$^{+}$+BAO&1798.26 &  1806.26 &  1811.08& -17.49&-16.28 \\ 
		CC+PN$^{+}$+BAO+R21& 1524.96 & 1832.96 &  1837.78&-0.2 &1\\ 
		CC+PN$^{+}$+BAO+TRGB&1801.01 &  1809.01 &  1813.83&-14.99 &-13.78\\
        	CC+BAO&15.98 &  23.98 &    28.80&-2.16 &-0.96\\
  \hline
    \end{tabular}
\end{table}
\begin{table}[H]
    \centering
    \caption{The best-fit values of the parameters explored by MCMC analysis. The first column enumerates a combination of data sets with the \(H_0\) priors. The second column presents the constrained \(H_0\) values. The third column contains the constrained \(\Omega_{m0}\) values. The fourth and fifth columns represent the optimal \(n\) and \(M\) values respectively. {\bf(With SH0ES data points)}}
    \label{tab:model_outputsmodel}
      \begin{tabular}{cccccc}
        \hline
		Data set  & $H_0$ & $\Omega_{m0}$  & n & M \\     
		\hline
		CC+$PN^{+}\& SH0ES$ & $72.5\pm 1.0$ & $0.379^{+0.053}_{-0.069}$ & $-2.30^{+0.36}_{-0.41}$ & $-19.260\pm 0.030$ \\ 
		CC+$PN^{+}\& SH0ES$+R21 & $72.74^{+0.77}_{-0.74}$ & $0.378^{+0.055}_{-0.071}$ & $-2.31^{+0.38}_{-0.42}$ & $-19.254^{+0.023}_{-0.022}$ \\ 
		CC+$PN^{+}\& SH0ES$+TRGB & $71.96^{+0.91}_{-0.90}$ & $0.381^{+0.052}_{-0.071}$ & $-2.27^{+0.36}_{-0.42}$ & $-19.276\pm 0.027$  \\ 
		\cline{1-5}
       CC+$PN^{+}\& SH0ES$+BAO & $69.26^{+0.68}_{-0.66}$ & $0.286^{+0.022}_{-0.021}$ & $-2.05^{+0.17}_{-0.20}$ & $-19.361^{+0.018}_{-0.017}$ \\ 
		CC+$PN^{+}\& SH0ES$+BAO+R21 & $70.28^{+0.59}_{-0.56}$ & $0.280^{+0.020}_{-0.019}$ & $-2.19^{+0.18}_{-0.20}$ & $-19.335\pm 0.015$    \\ 
		CC+$PN^{+}\& SH0ES$+BAO+TRGB &$69.30^{+0.63}_{-0.62}$ & $0.286\pm 0.021$ & $-2.06^{+0.18}_{-0.19}$ & $-19.360\pm 0.017$ \\
  \hline
    \end{tabular}
\end{table}
\renewcommand{\arraystretch}{1} 
\begin{table}[H]
    \centering
    \caption{The statistical comparison between the chosen model and the standard \(\Lambda\)CDM model. Details regarding the \(\Lambda\)CDM model are given in {\color{blue}Appendix}. The first column enumerates the data sets including the \(H_0\) priors. The second column displays the values of \(\chi^{2}_{\text{min}}\). The third and fourth column respectively provides the value of AIC and BIC. The fifth and sixth column respectively illustrates the values of \(\Delta \text{AIC}\) and \(\Delta \text{BIC}\). {\bf(With SH0ES data points)}}
    \label{tab:model_outputAICBIC}
      \begin{tabular}{cccccc}
        \hline
		data set& $\chi^{2}_{min}$  &AIC &BIC&$\Delta$AIC &$\Delta$BIC \\ 
		\hline
		CC+$PN^{+}\& SH0ES$&1538.29 &1546.29 &  1551.25& 1.07&2.32 \\ 
		CC+$PN^{+}\& SH0ES$+R21& 1538.41 &1546.41 &  1551.37&1.16 &2.41\\ 
		CC+$PN^{+}\& SH0ES$+TRGB&1539.99 &1547.99 &  1552.92&0.81 &2.03\\ 
		\cline{1-6}
       CC+$PN^{+}\& SH0ES$+BAO&1571.27 &  1579.27 &  1584.25& 6.1&7.35\\ 
		CC+$PN^{+}\& SH0ES$+BAO+R21& 1580.95 & 1588.95 &  1593.92&12.99 &14.25\\ 
		CC+$PN^{+}\& SH0ES$+BAO+TRGB&1571.35 &  1579.35 &  1584.32&5.85&7.09\\
  \hline
    \end{tabular}
\end{table}
 \section{Linear Matter Perturbations and Large-Scale Structure Evolution}\label{linearperturbationection}
Cosmological models that do not account for interactions within the dark sector, the fundamental equation that governs the growth of matter perturbations in the linear regime at sub-horizon scales during the matter era is,
\begin{equation}\label{deltarhom}
 \ddot{\delta}_m+2 \,H\, \dot{\delta}_m=4\pi G_{eff}\,\rho_m\,\delta\,, 
\end{equation}

where the matter overdensity can be defined as $\delta_m \equiv \frac{\delta \rho_m}{\rho_m}$. In the context of Eq.~\eqref{deltarhom}, the effective Newton's constant is introduced as $G_{eff}(a) = G \,\,P(a)$, where $G$ is the gravitational constant defined in the action of the theory. This formulation accounts for modifications to gravity. The underlying gravitational theory determines the specific form of $P(a)$. In case of GR, we find that $G_{eff}(a) = G$, which corresponds to $P(a) = 1$. Consequently, Eq.~\eqref{deltarhom} simplifies to yield the standard evolution equation for matter density perturbations.

So, it is clear that we can use this general perturbation approach in the context of $f(T, \mathcal{T})$ cosmology, provided the form of $G_{eff}(a)$ is known, or equivalently $P(a)$, in $f(T, \mathcal{T})$ gravity. It can be demonstrated with relative ease that for $f(T, \mathcal{T})$ gravity \cite{Harko_2014a, Junior_2016}.
\begin{equation}\label{Geffective}
P(a)=\frac{G_{eff}}{G}= \frac{1}{1+f_T}\,\bigg(1+\frac{f_{\mathcal{T}}}{2}\bigg)\,.    
\end{equation}
Eq.~\eqref{deltarhom} can also be rewritten as
\begin{equation}\label{deltaloga}
   \delta''_{m}+\bigg(2+ \frac{H'}{H}\bigg)\, \delta'_{N}-\frac{3}{2G}\,G_{eff}\,\Omega_{m}\,\delta_{m} = 0.
\end{equation}

In Eq.~\eqref{deltaloga} prime $(')$ denotes the derivative for $log(a)$. By incorporating the growth rate of matter fluctuations defined as $f_{\delta}\equiv \frac{\delta'_{m}}{\delta_{m}}$, we can express Eq. \eqref{deltaloga} in a different form as,
\begin{equation}
 f_{\delta}'+f_{\delta}^2+\left(2+\frac{H'}{H}\right)f_{\delta}-\frac{3}{2G}\,G_{eff}\,\Omega_{m}=0\,. 
\label{GrowthRate}   
\end{equation}
Where $\Omega_m=\frac{\Omega_{m0}\,a^{-3}}{E^2(a)}$. The measurable $f\sigma_8(z)$ contrasts theoretical predictions with observations data and is specified as
\begin{equation}
f\sigma_{8}(z)\equiv f_{\delta}(a)\cdot \sigma(a)=\frac{\sigma_{8}}{\delta_m{(1)}}\, a\, \delta_m'(a)\,,     
\end{equation}

where $\sigma(a)=\sigma_{8} \delta_m(a)/\delta_m(1)$ represents the amplitude of fluctuations in the matter density within spheres measuring $8~h^{-1}~Mpc$ ($k\sim k_{\sigma_{8}}=0.125$~$h~Mpc^{-1}$), with $\sigma_{8}=\sigma(1)$ being significantly impacted by late-time cosmic expansion and dark energy models. Discrepancies in the $\Lambda$CDM framework, especially concerning $H_{0}$ and $\sigma_8$, arise because large-scale structure (LSS) observations indicate $f\sigma_{8}$ values are roughly $8\%$ lower than anticipated. This implies that the $\Lambda$CDM model might overestimate $\sigma_8$ given the same current growth rate $f_{\delta}(1)$. Nevertheless, since $f_{\delta}(a)$ depends on the specific model \cite{DeFelice:2010aj}, a reduced growth rate might also enhance compatibility with the LSS observations.

\subsection{Numerical results}
\begin{figure}[H]
     \centering
     \begin{subfigure}[b]{0.49\textwidth}
         \centering
         \includegraphics[width=\linewidth]{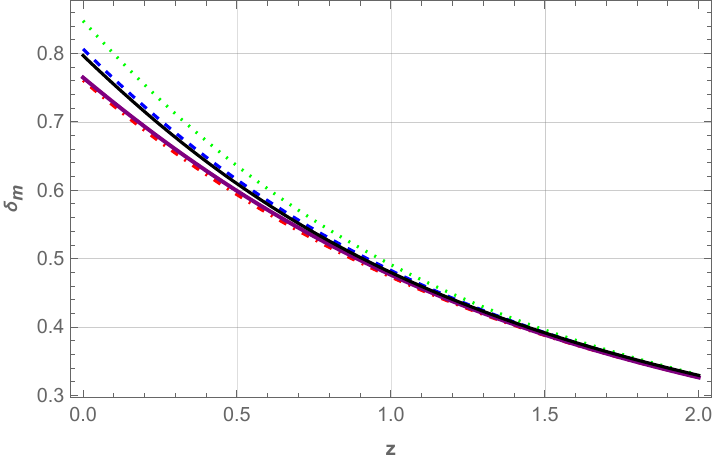}
         \caption{The graph displays the evolution of the matter density fluctuation, referred to as $\delta_m$, to the redshift $z$ across various dataset values, along with the corresponding evolution for the $\Lambda$CDM model.}
         \label{growthrate:CCOm}
     \end{subfigure}
     \hfill
     \begin{subfigure}[b]{0.49\textwidth}
         \centering
         \includegraphics[width=\linewidth]{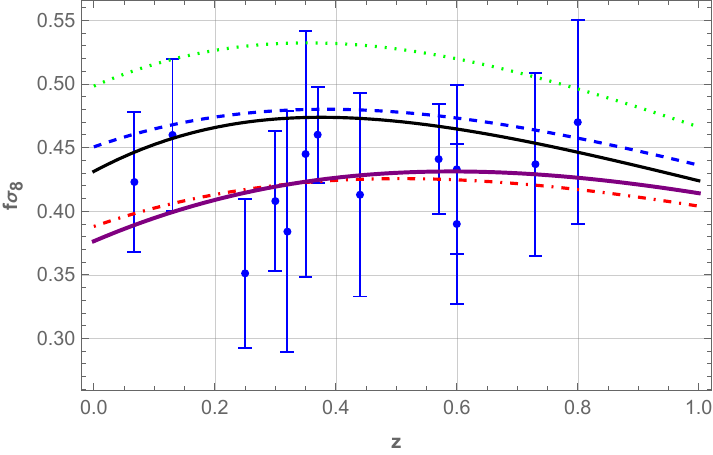}
         \caption{The graph displays the evolution of the weighted growth rate $f\sigma_8$ and the related progression for the $\Lambda$CDM model. In this figure, we have used the complete redshift-space distortion (RSD) $f\sigma_8$ dataset.}
         \label{sigma8:BAOom}
     \end{subfigure}
\caption{Evolution of the matter density perturbation $\delta_m$ and weighted growth rate $f\sigma_8$ with respect to redshift $z$ using the relation $a=\frac{1}{1+z}$. In this graph, the blue-dashed line represents dataset CC+PN$^+$, the red-dot dashed line corresponds to dataset CC+PN$^+$+BAO, the green-dotted line represents dataset CC+PN$^+$\&SH0ES, the purple-thick line corresponds to dataset CC+PN$^+$\&SH0ES+BAO and black-thinning line corresponds to $\Lambda$CDM model. } 
\label{growthratesigma8}
\end{figure}

The theoretical curves for matter density perturbations are shown in Fig.~\ref{growthrate:CCOm}, based on various selections of model parameters. For our analysis, we observe that $\sigma(a) \sim \delta_m(a)$ while $\delta_m(a) \approx a= \frac{1}{1+z}$. For the chosen model \eqref{modelconsidered}, from Fig.~\ref{growthrate:CCOm}, we determine the subsequent values: $\sigma_8 = 0.81$ (blue-dashed line), $\sigma_8 = 0.76$ (red-dot dashed line), $\sigma_8 = 0.85$ (green-dotted line), $\sigma_8 = 0.76$ (purple-thick line). For the $\Lambda$CDM model, we determine that $\sigma_8 = 0.8$ (black-thinning). These values serve as a good approximation that aligns with the findings shown in Ref.\cite{Aghanim:2018eyx}.  

In Fig.~\ref{sigma8:BAOom}, we illustrate the theoretical curves for the weighted linear growth rate \( f \sigma_8(z) \) corresponding to various selections of model parameters. We notice that for the red-dot dashed line and the purple-thick line, the values of \( f \sigma_8(z) \) are lower than the related standard from the \(\Lambda\)CDM model (black-thinning), suggesting that our model might reduce the \(\sigma_8\)-tension. For the selected model \eqref{modelconsidered}, we extract the following values from Fig.~\ref{sigma8:BAOom}: $f\sigma_8(0) \approx 0.45$ (blue-dashed line), $f\sigma_8(0) \approx 0.39$ (red-dot dashed line), $f\sigma_8(0) \approx 0.50$ (green-dotted line), and $f\sigma_8(0) \approx 0.38$ (purple-thick line). In the case of the $\Lambda$CDM model, we find that $f\sigma_8(0) \approx 0.43$ (black-thinning line). To measure this impact, we establish the precise relative difference as,
\begin{equation}
\Delta f\sigma_8 (z) \equiv 100 \times \frac{f\sigma_8(z)_{\text{model}} - f\sigma_8(z)_{\Lambda\text{CDM}}}{f\sigma_8(z)_{\Lambda\text{CDM}}},
\end{equation}

In Fig.~\ref{sigma8:BAOom}, the red-dot dashed line represents $f\sigma_8(0) \approx 0.39$, whereas for the $\Lambda$CDM model, we calculate $f\sigma_8(0) \approx 0.43$, resulting in a relative difference of $\Delta f\sigma_8(0) \approx 9\%$. The purple-thick line indicates that the relative difference with the $\Lambda$CDM model is $\Delta f\sigma_8(0) \approx 11\%$. Consequently, the projection indicated by the red-dot dashed line is approximately 9\% below the $\Lambda$CDM model prediction. The purple-thick line shows a deviation of roughly 11\% below the $\Lambda$CDM model.

\section{Dynamics of Cosmological Parameters} \label{cosmologicalparameters} 
\begin{figure}[ht]
     \centering
         \includegraphics[width=70mm]{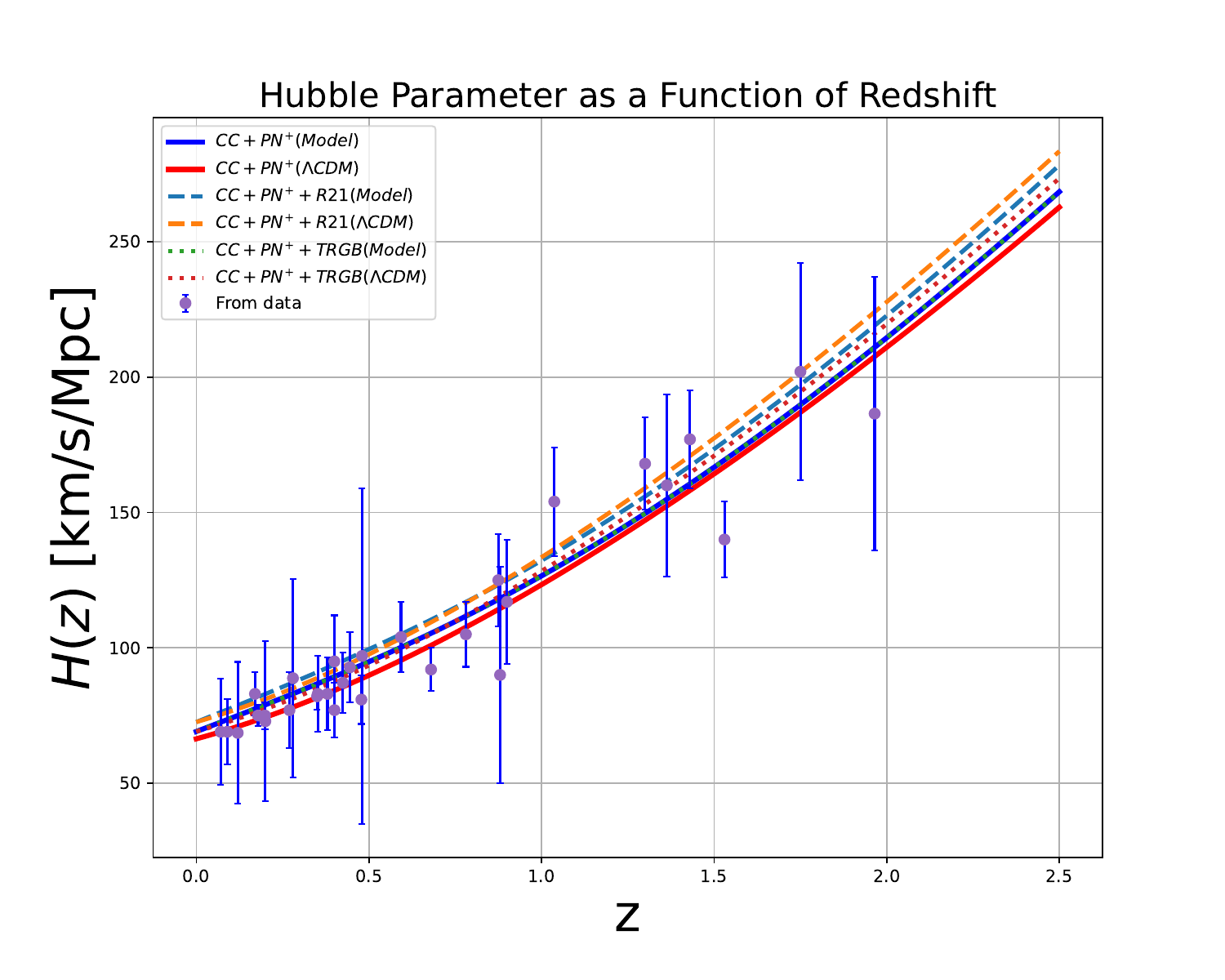}
          \includegraphics[width=70mm]{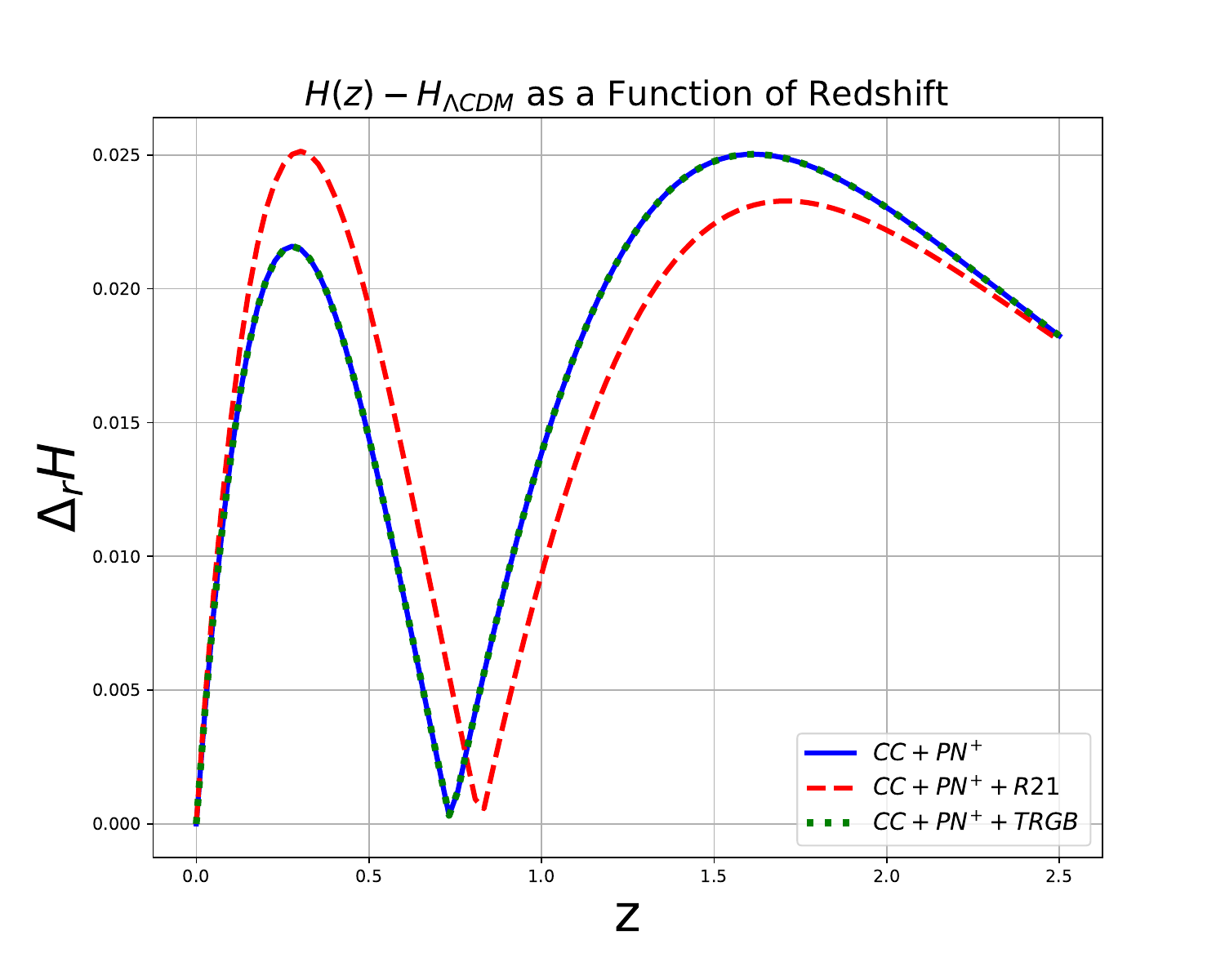}
         \includegraphics[width=70mm]{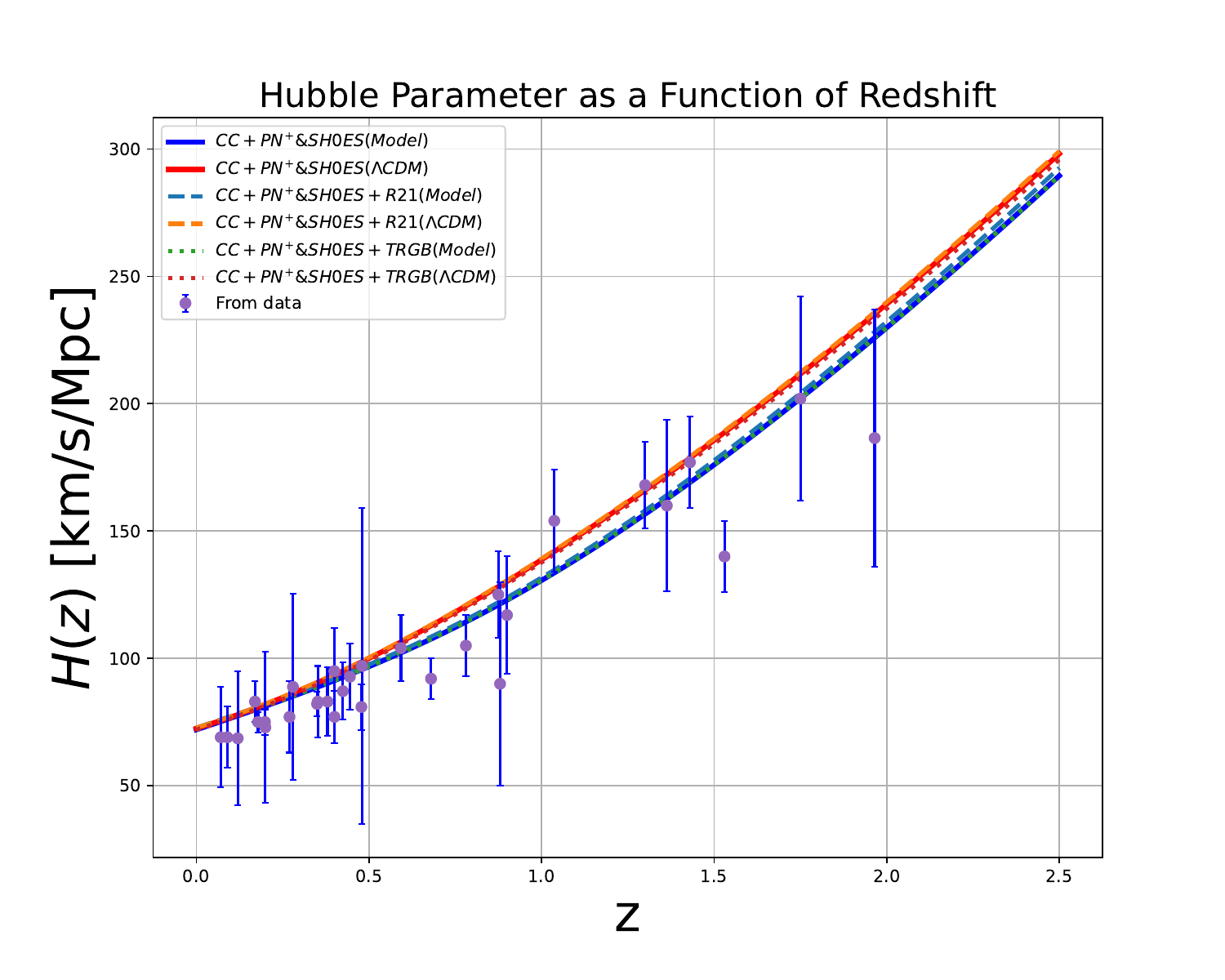}
          \includegraphics[width=70mm]{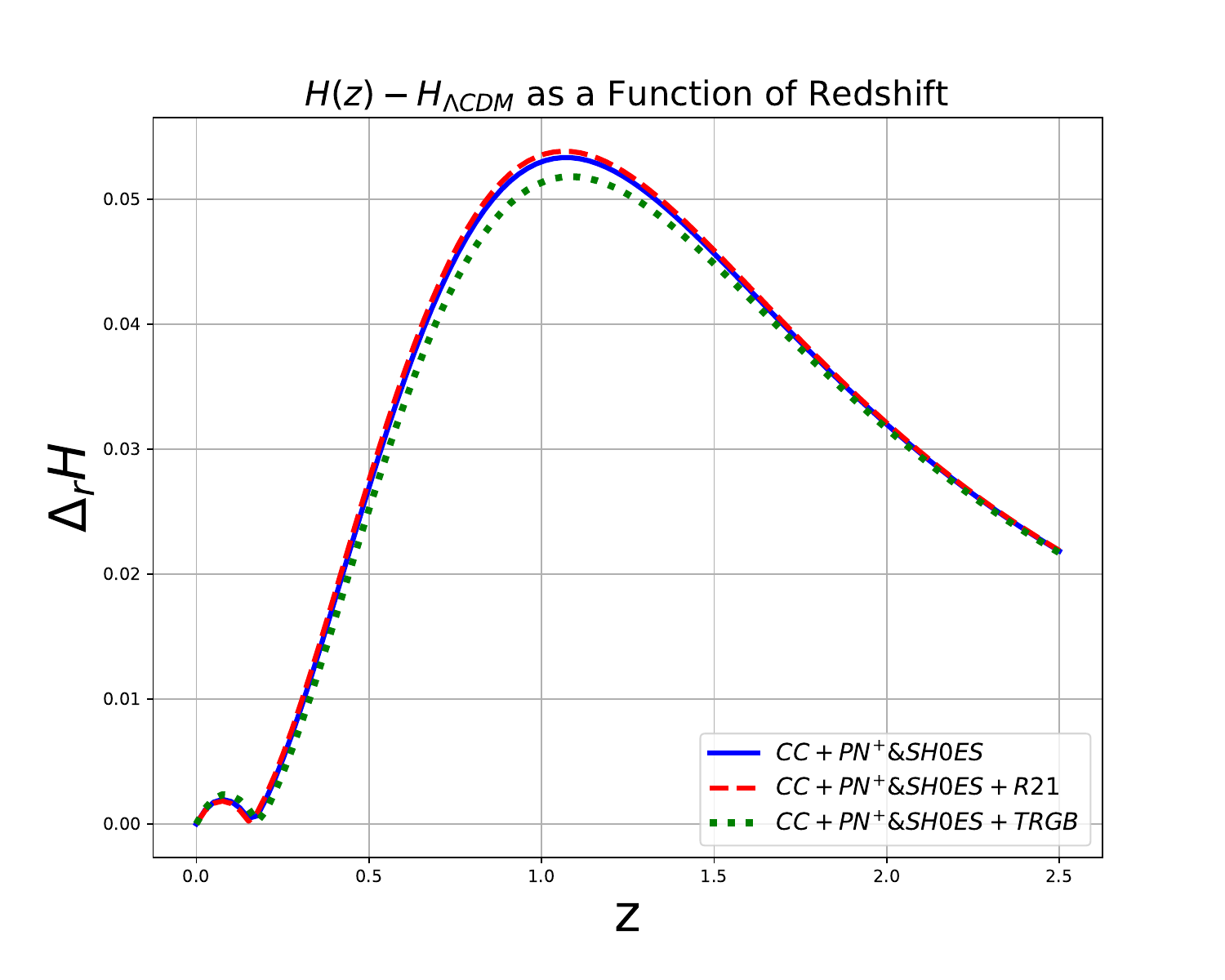}
\caption{Evolutionary behavior of Hubble parameter and comparative analysis of the evolution of the Hubble parameter between the selected model and the $\Lambda$CDM model in redshift for the data set combination: CC, PN$^{+}$ (without SH0ES) and  PN$^{+}$\&SH0ES (with SH0ES). The $H_0$ priors are: R21 and TRGB.} 
\label{plusFighubblediffer}
\end{figure}

In this section, we will examine the background cosmological parameters to investigate the behavior at late time for the $f(T, \mathcal{T})$ model. Additionally, we will juxtapose these results with those from the standard $\Lambda$CDM model. Fig.~\ref{plusFighubblediffer}, displays the behavior of the Hubble parameter and comparative analysis of the evolution of the Hubble parameter between the selected model and the $\Lambda$CDM model for the PN$^+$ and the PN$^+$\& SH0ES data set combinations, respectively, with prior data compared to the $\Lambda$CDM model. It has been noted that the curves follow a similar pattern to that of the $\Lambda$CDM and remain well within the error margins. The analysis of these figures indicates that the model exhibits behavior consistent with that of the $\Lambda$CDM paradigm across the specified combination of data set. We present the relative difference to demonstrate the differences between the selected and standard $\Lambda$CDM models.
\begin{equation}\label{relativedifferenceCC}
\Delta_{r} H(z) = \frac{\left| H_{\text{model}} - H_{\Lambda \text{CDM}} \right|}{H_{\Lambda \text{CDM}}} \,.
\end{equation}

\begin{figure}[ht]
     \centering
         \includegraphics[width=70mm]{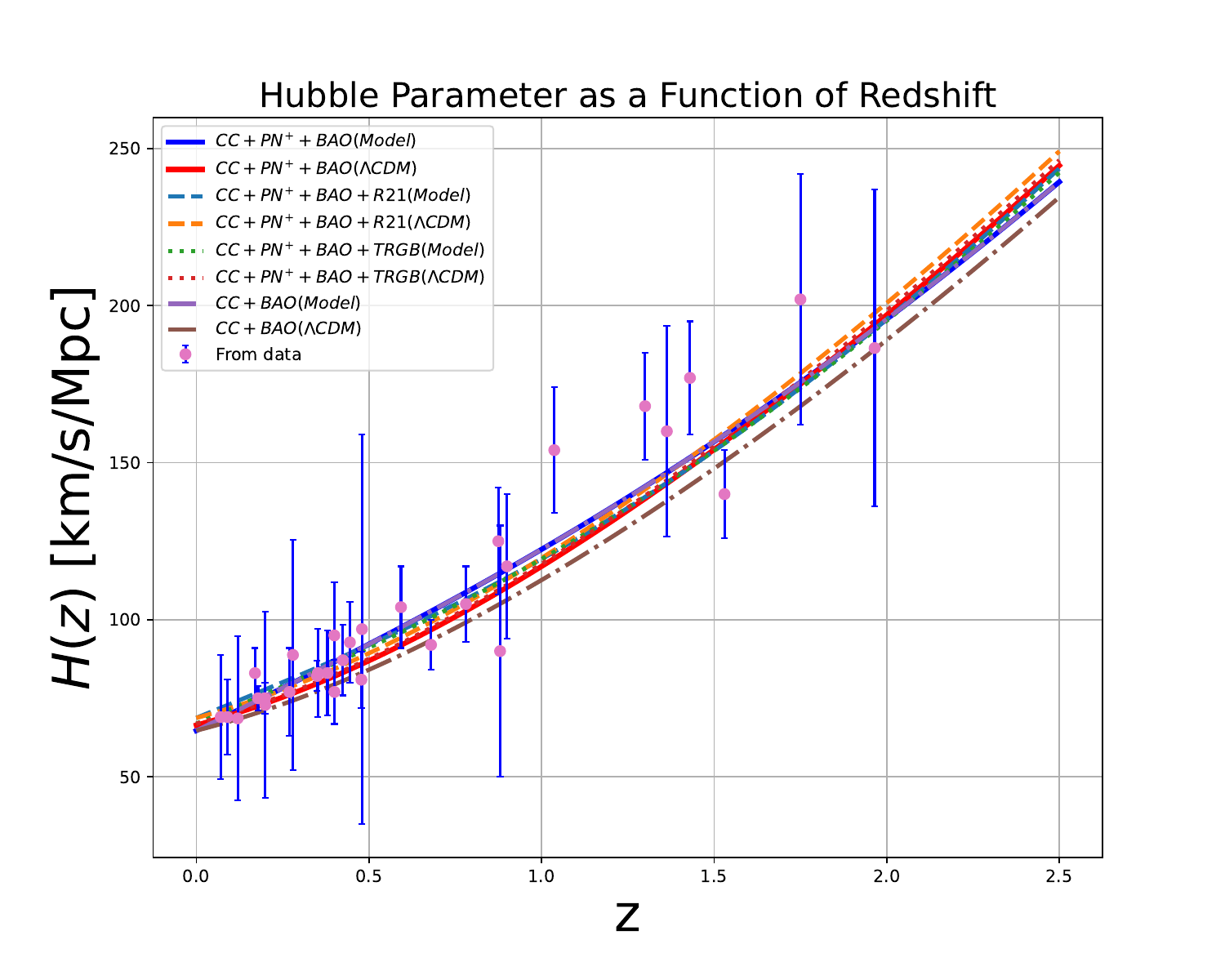}
         \includegraphics[width=70mm]{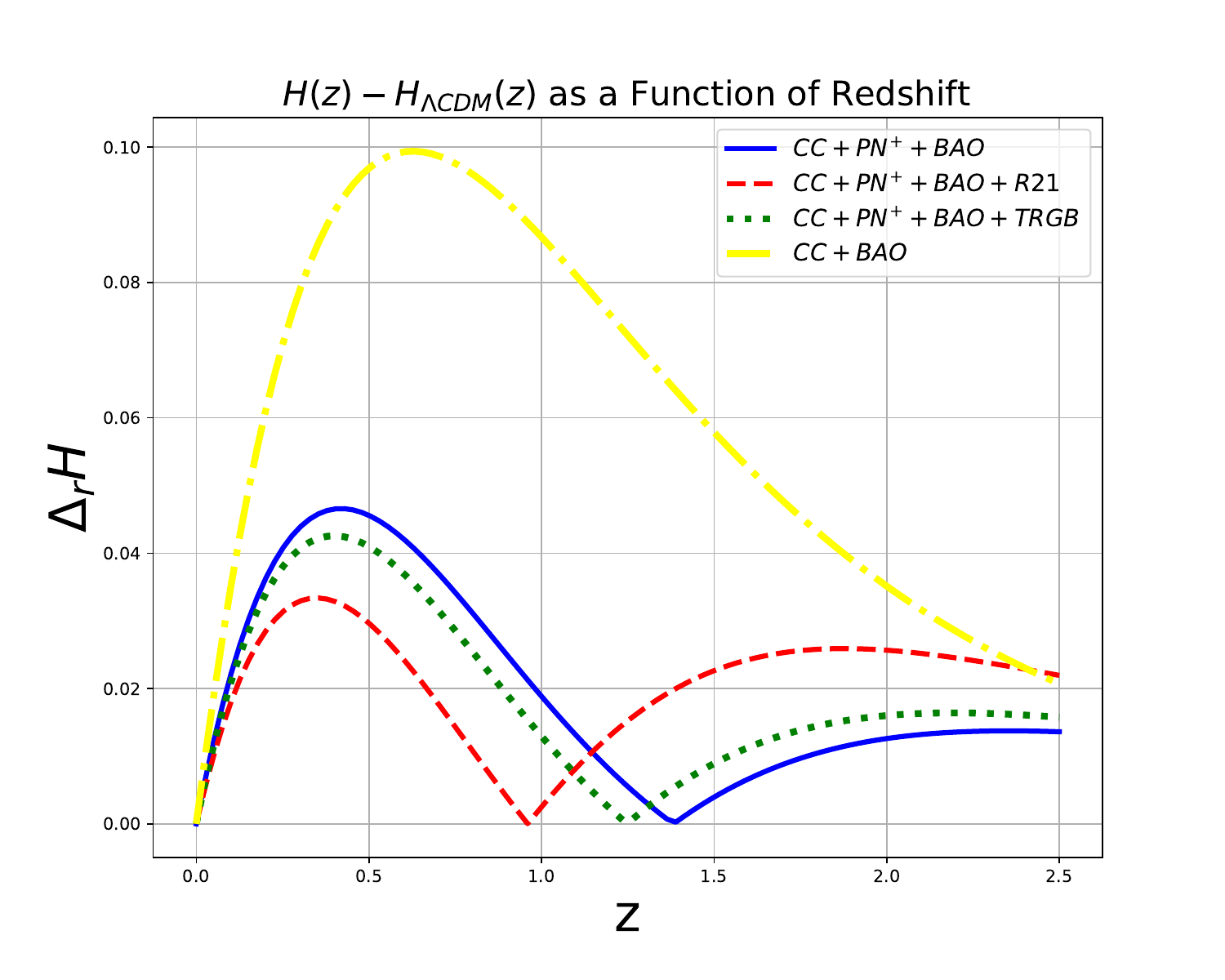}
       \includegraphics[width=70mm]{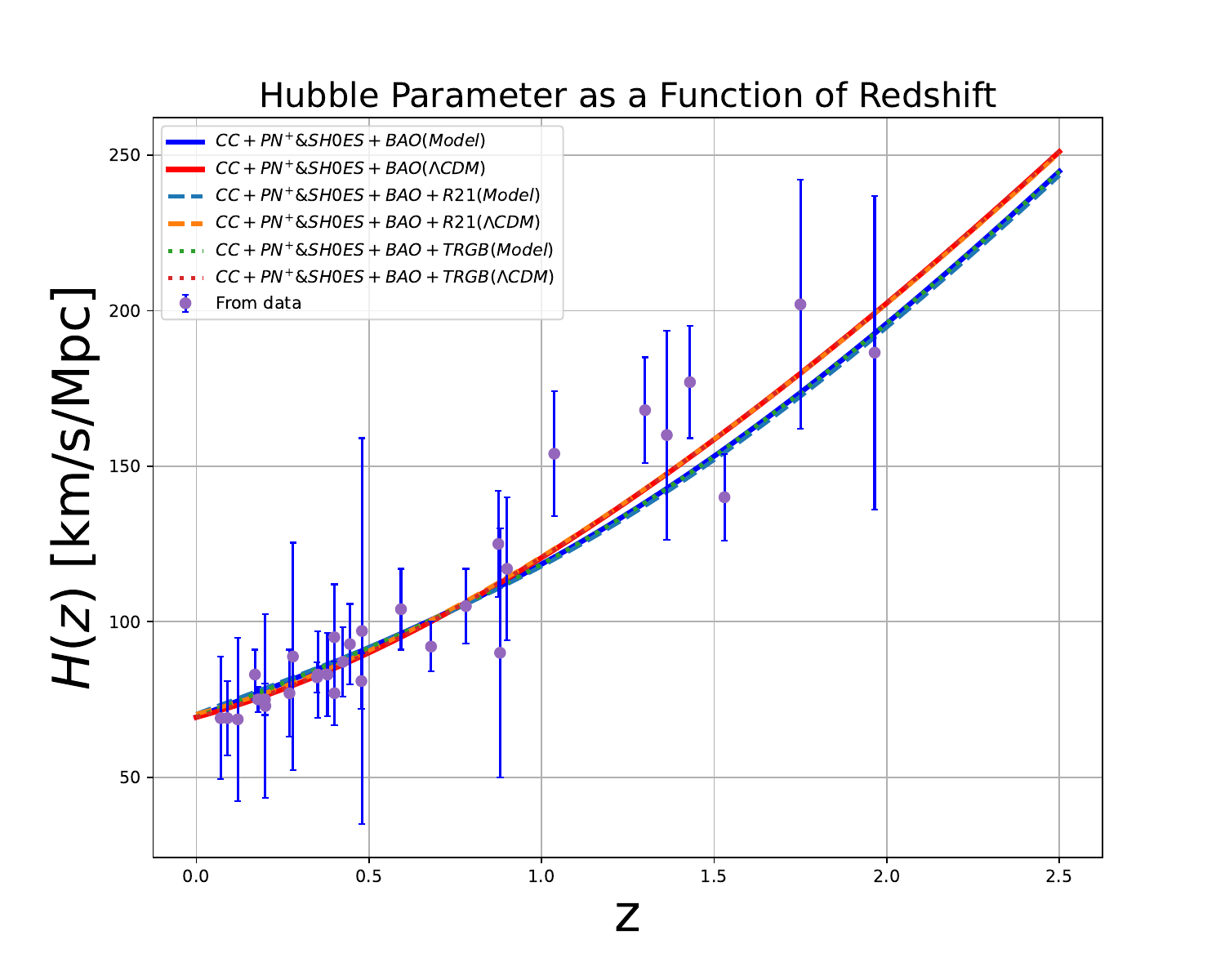}
        \includegraphics[width=70mm]{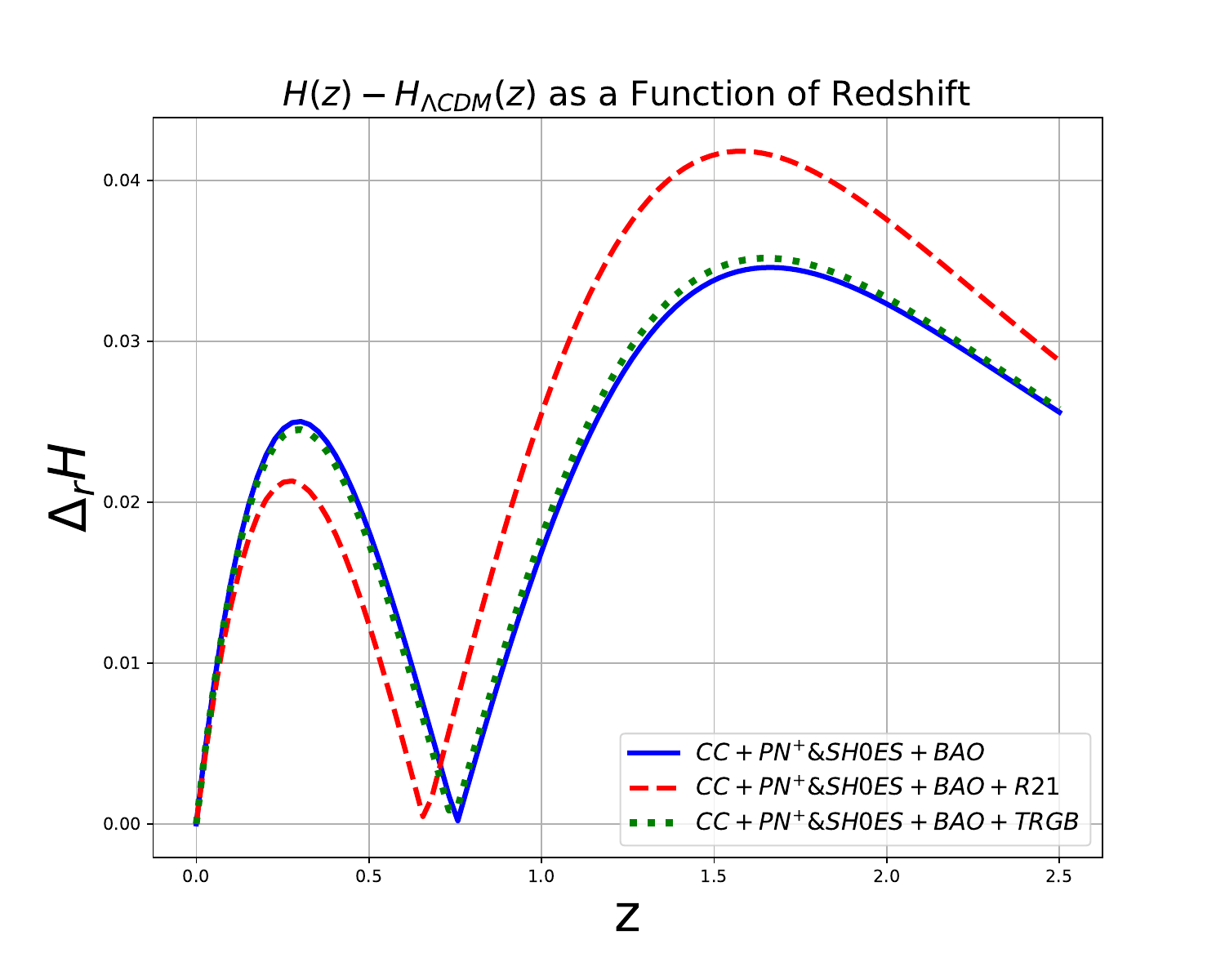}
\caption{Evolutionary behavior of Hubble parameter and comparative analysis of the evolution of the Hubble parameter between the selected model and the $\Lambda$CDM model in redshift for the data sets combination: CC, PN$^{+}$ (without SH0ES), PN$^{+}$\&SH0ES (with SH0ES) and BAO. The $H_0$ priors are: R21 and TRGB.} 
\label{plusFigbaohdiff}
\end{figure}

Additionally, Fig.~\ref{plusFigbaohdiff} showcase the evolution of the Hubble parameter and comparative analysis of the evolution of the Hubble parameter between the selected model and the $\Lambda$CDM model for various data set, including the CC, PN$^{+}$, PN$^{+}$\& SH0ES and BAO, with the incorporation of $H_{0}$ priors.  

\begin{figure}[ht]
     \centering
         \includegraphics[width=70mm]{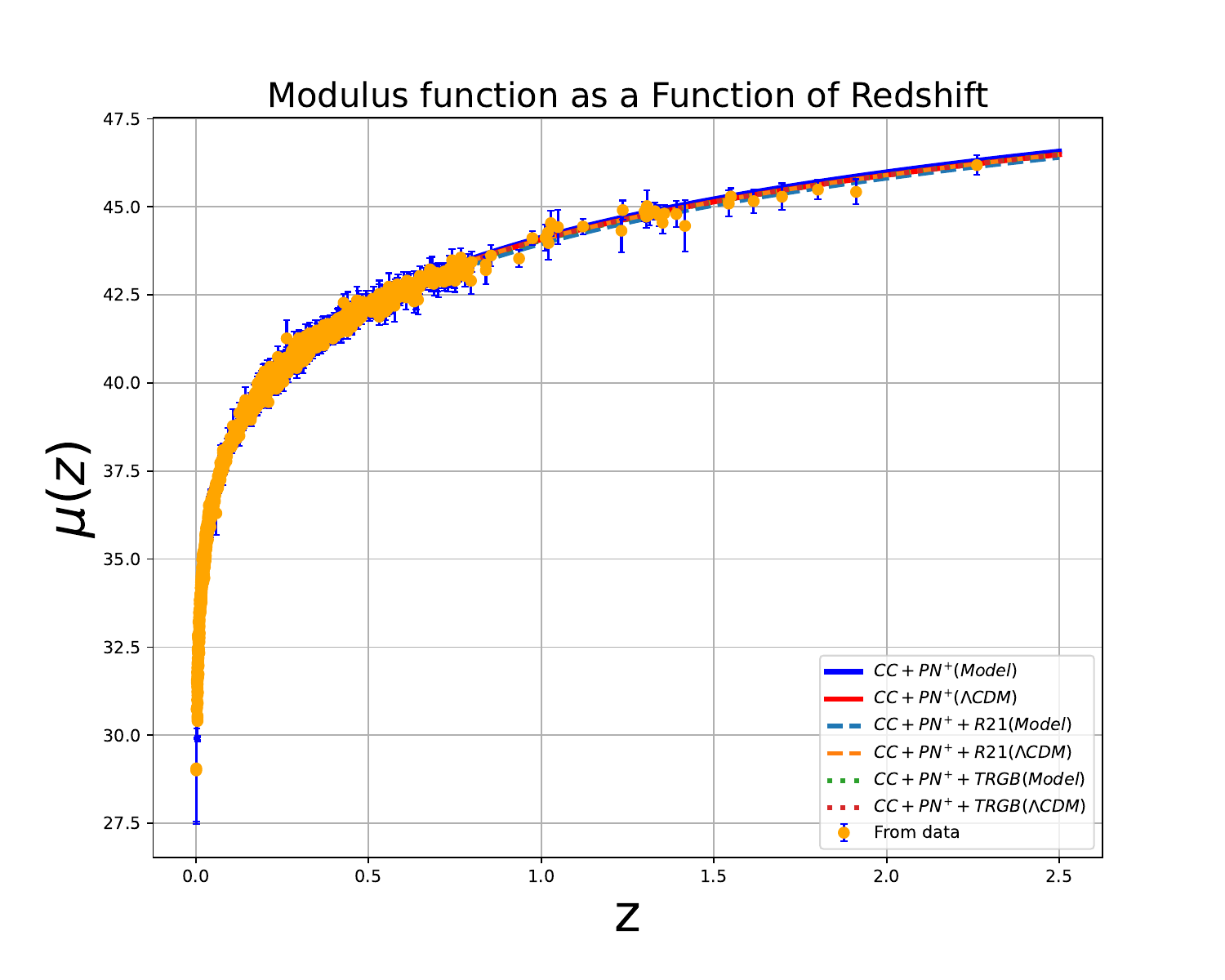}
         \includegraphics[width=70mm]{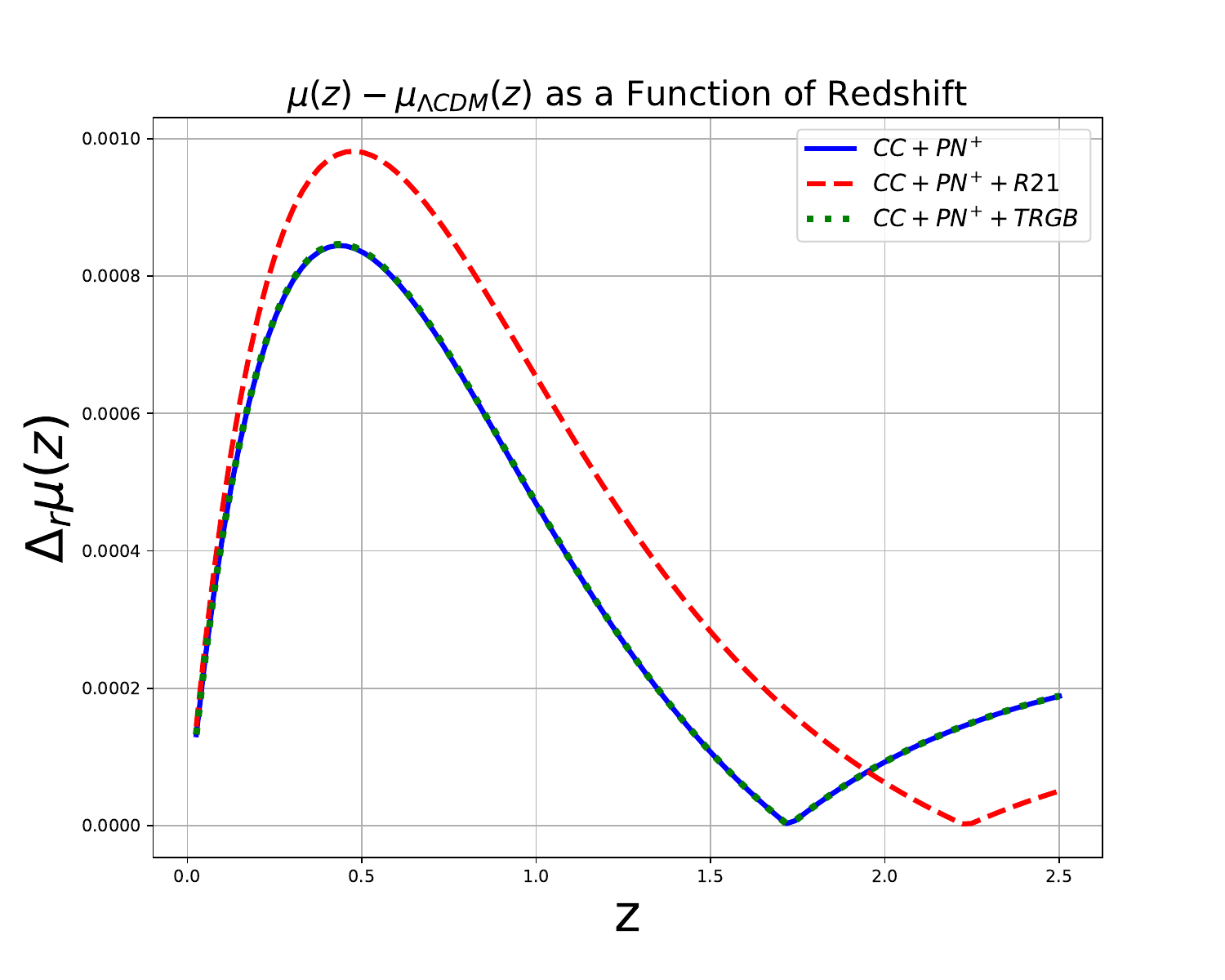}
        \includegraphics[width=70mm]{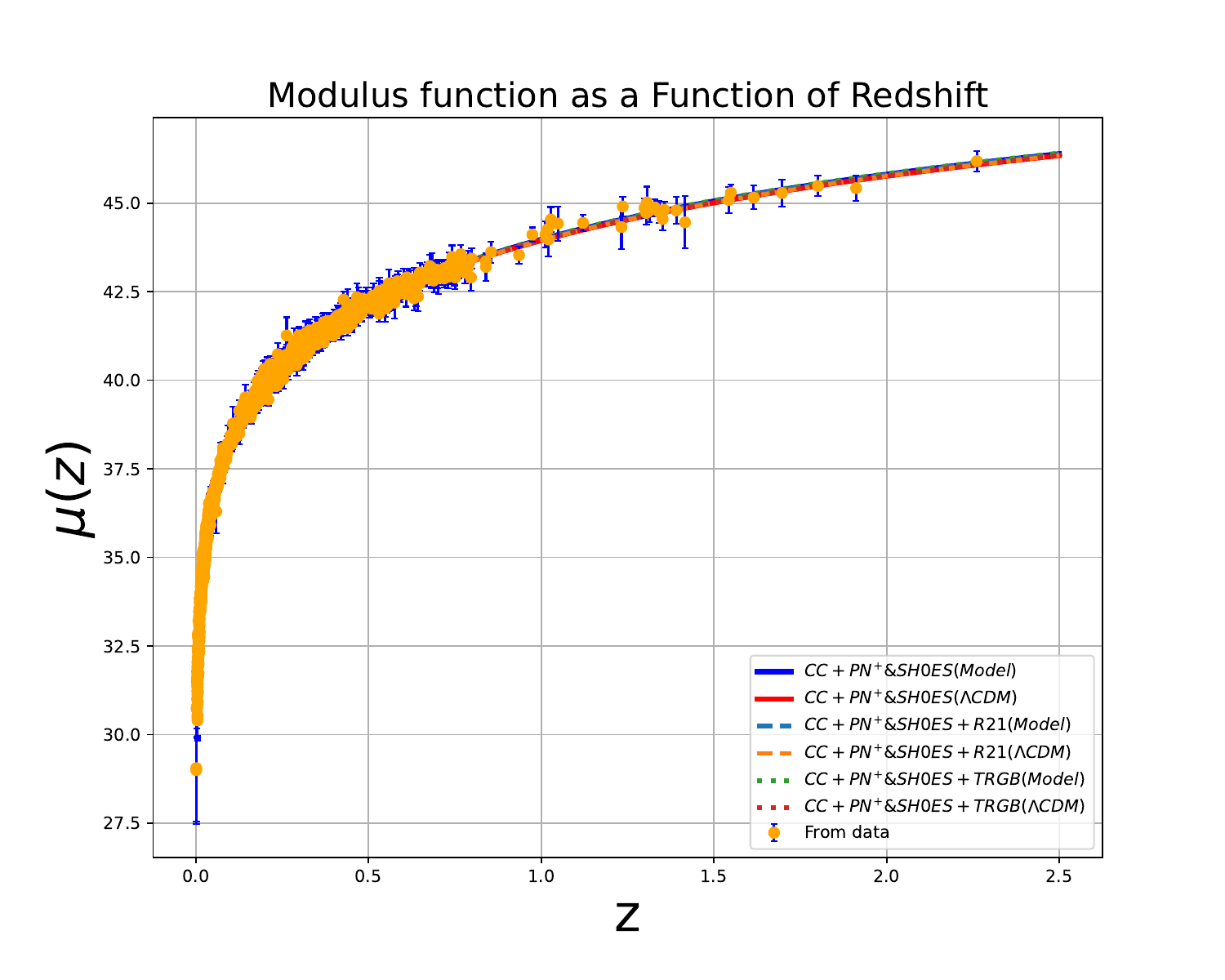}
        \includegraphics[width=70mm]{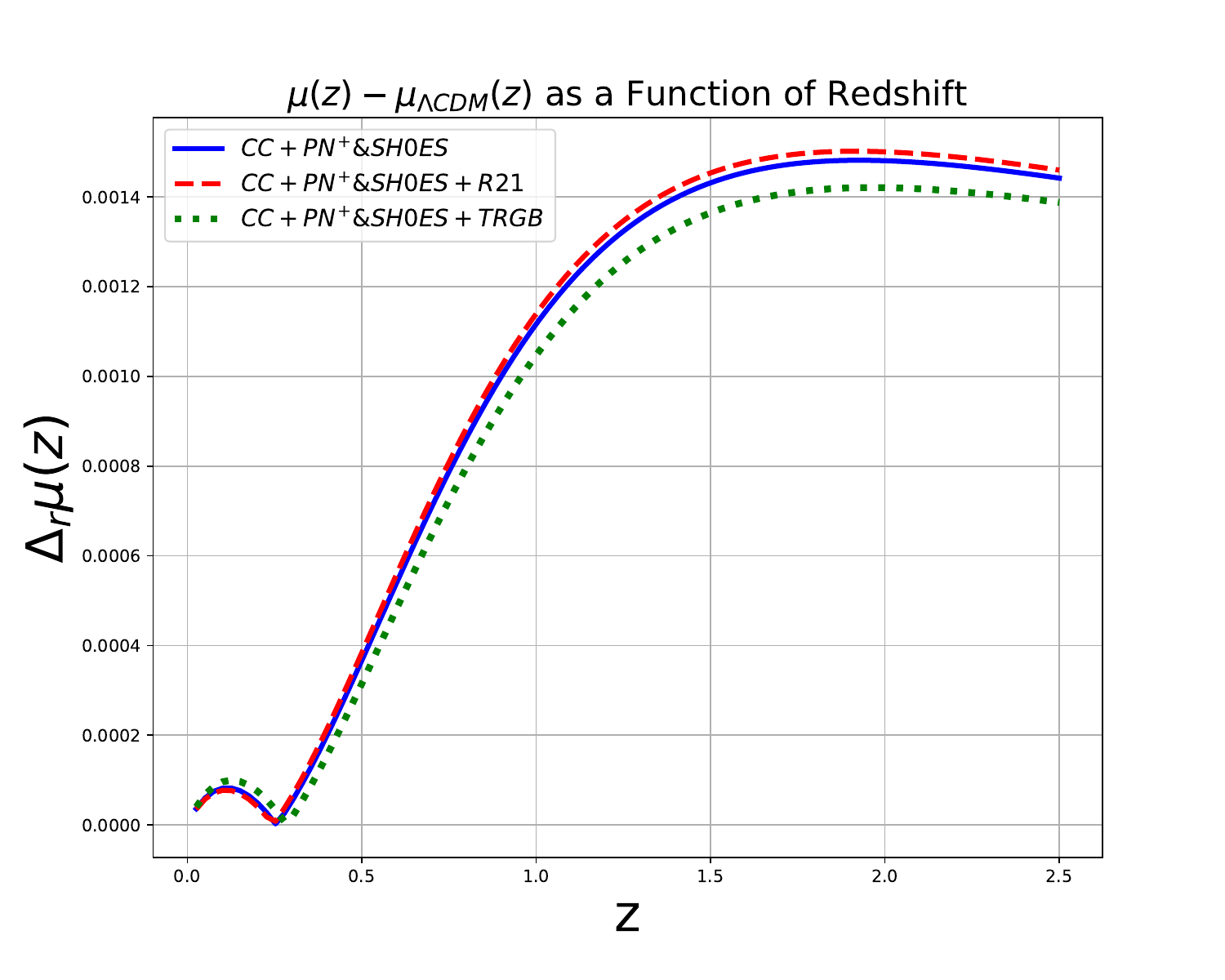}
\caption{Evolutionary behavior of distance modulus and comparative analysis of the evolution of the distance modulus function between the selected model and the $\Lambda$CDM model in redshift for the datasets combination: CC, PN$^{+}$ (without SH0ES) and  PN$^{+}$\&SH0ES (with SH0ES). The $H_0$ priors are: R21 and TRGB. } 
\label{plusFigmudulasdifference}
\end{figure}

 In Figs.~\ref{plusFigmudulasdifference}, we illustrate the evolution of the distance modulus and comparative analysis of the evolution of the distance modulus function between the selected model and the $\Lambda$CDM model for our selected cosmological model in comparison with the $\Lambda$CDM framework, utilizing a data set comprising 1701 data points from the PN$^+$ and PN$^+$ \& SH0ES observations. The analysis reveals a noteworthy concordance between the selected and $\Lambda$CDM models. The calculated relative difference can be defined as  
\begin{equation}\label{relativedifferencemodulus}
\Delta_{r}\mu(z) = \frac{\left| \mu_{\text{model}} - \mu_{\Lambda \text{CDM}} \right|}{\mu_{\Lambda \text{CDM}}} \,.
\end{equation}

\begin{figure}[ht]
     \centering
         \includegraphics[width=70mm]{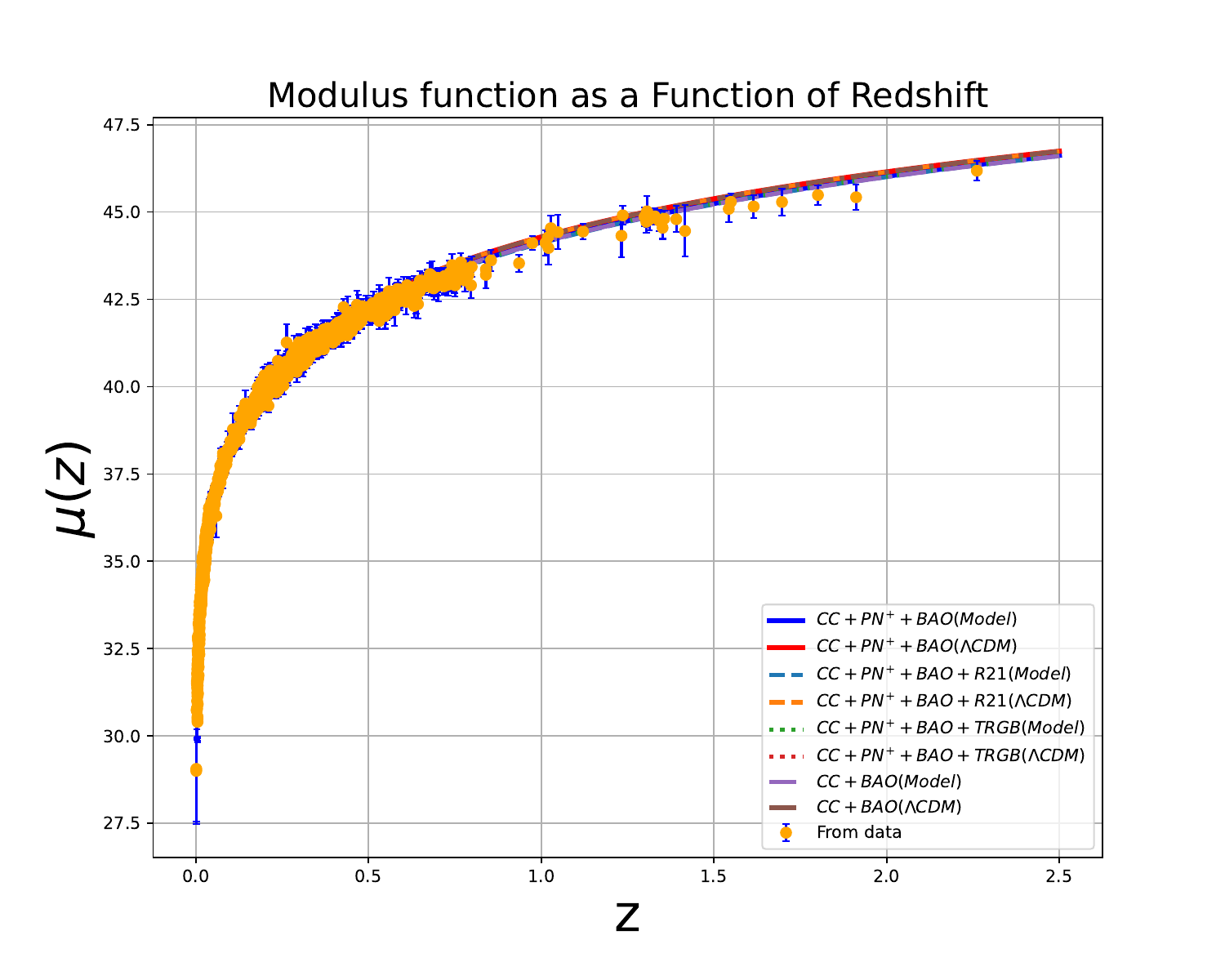}
          \includegraphics[width=70mm]{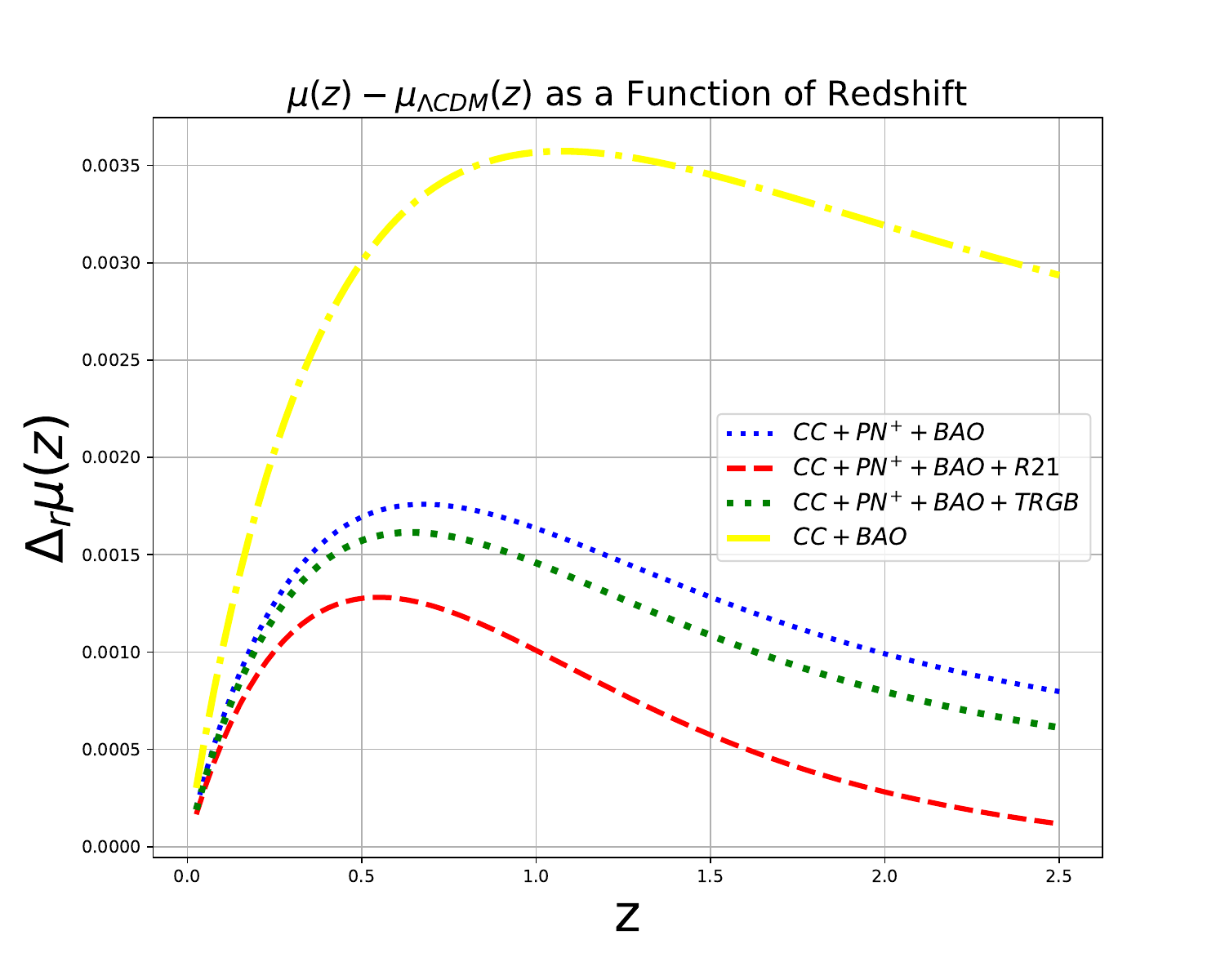}
       \includegraphics[width=70mm]{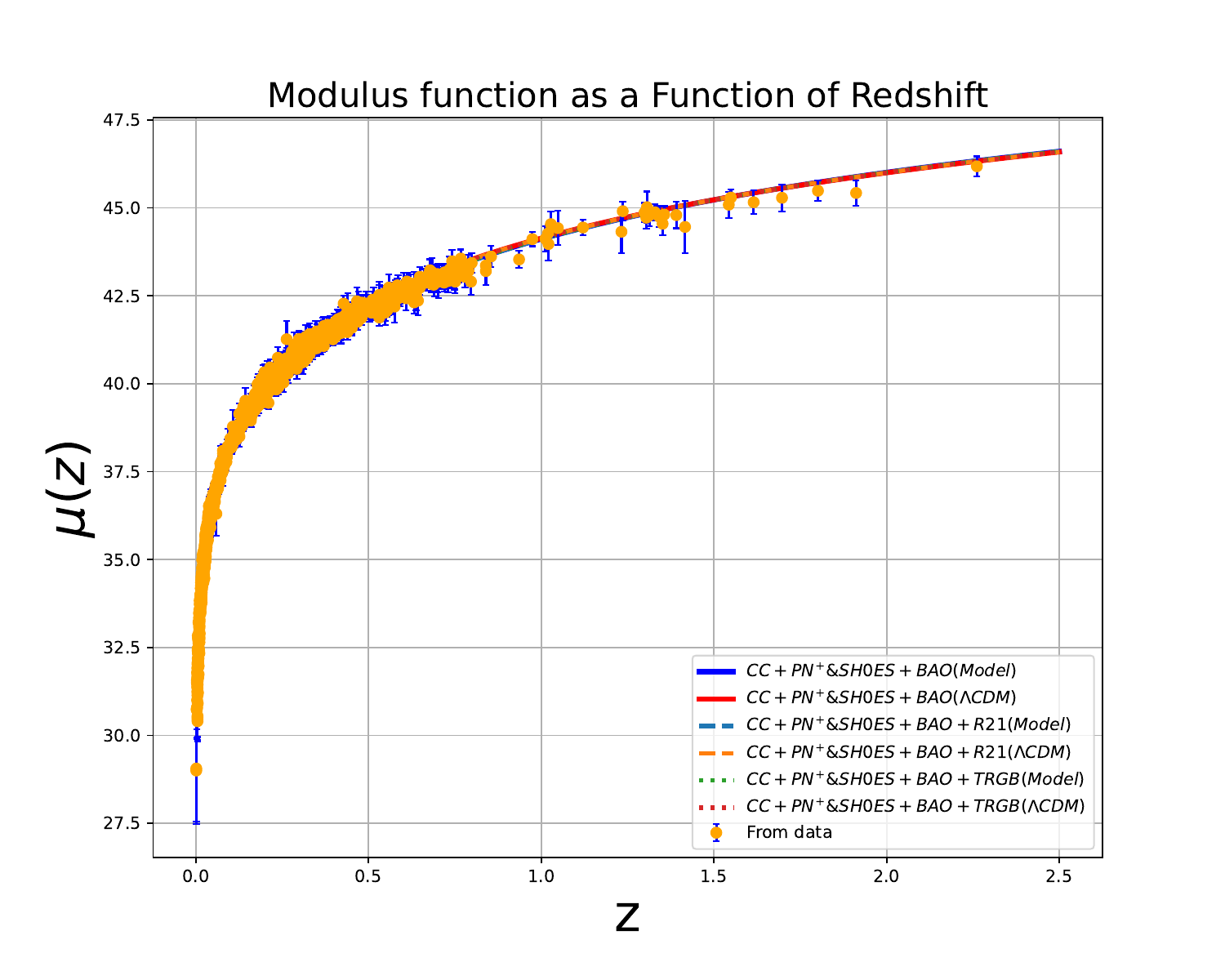}
        \includegraphics[width=70mm]{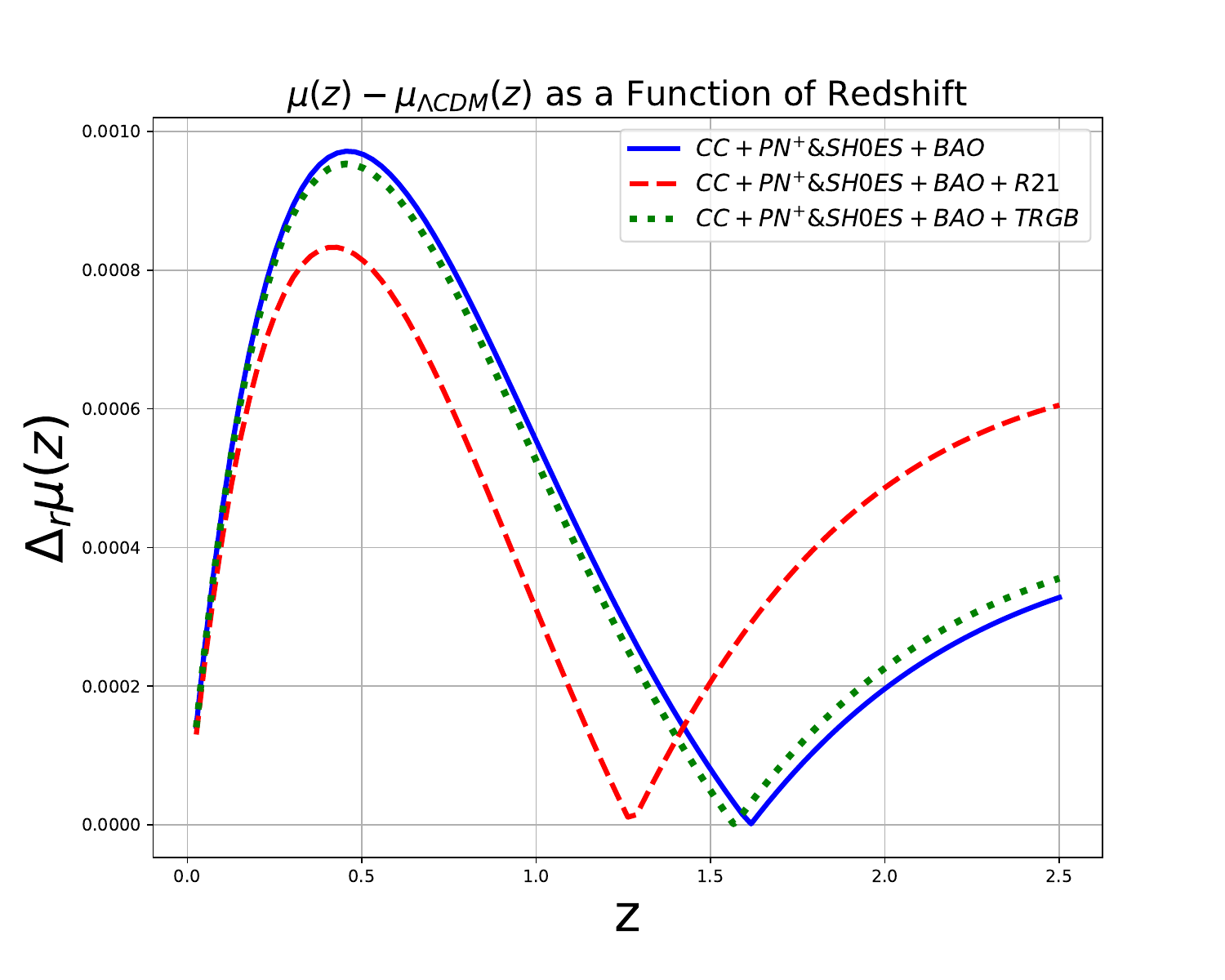}
\caption{Evolutionary behavior of distance modulus and comparative analysis of the evolution of the distance modulus function between the selected model and the $\Lambda$CDM model in redshift for the data sets combination: CC, PN$^{+}$ (without SH0ES), PN$^{+}$\&SH0ES (with SH0ES) and BAO. The $H_0$ priors are: R21 and TRGB. } 
\label{plusFigBAOmudulasdifference}
\end{figure}

 Fig.~\ref{plusFigBAOmudulasdifference} illustrates the progression of the distance modulus function alongside the relative difference in distance modulus for the CC, PN$^{+}$, PN$^{+}$\& SH0ES and BAO data set, taking into account the $H_{0}$ priors. This observed trend is consistent with the patterns seen in Figs.~\ref{plusFigmudulasdifference}, highlighting similar behaviors across the different datasets.
The deceleration parameter, the total EoS and the matter-energy density as a function of redshift can be expressed as,
\begin{eqnarray}
q = -1 + \frac{(1+z) H^{'}(z)}{H(z)} \,, \\ 
\omega_{tot} = -1 + \frac{2 (1+z) H^{'}(z)}{3 H(z)} \,, \\ 
\Omega_{m} = \frac{\Omega{m0} (1+z)^3 H^2_{0}}{H^{2}(z)} \,.  
\end{eqnarray} 

\begin{figure}[ht]
     \centering
         \includegraphics[width=70mm]{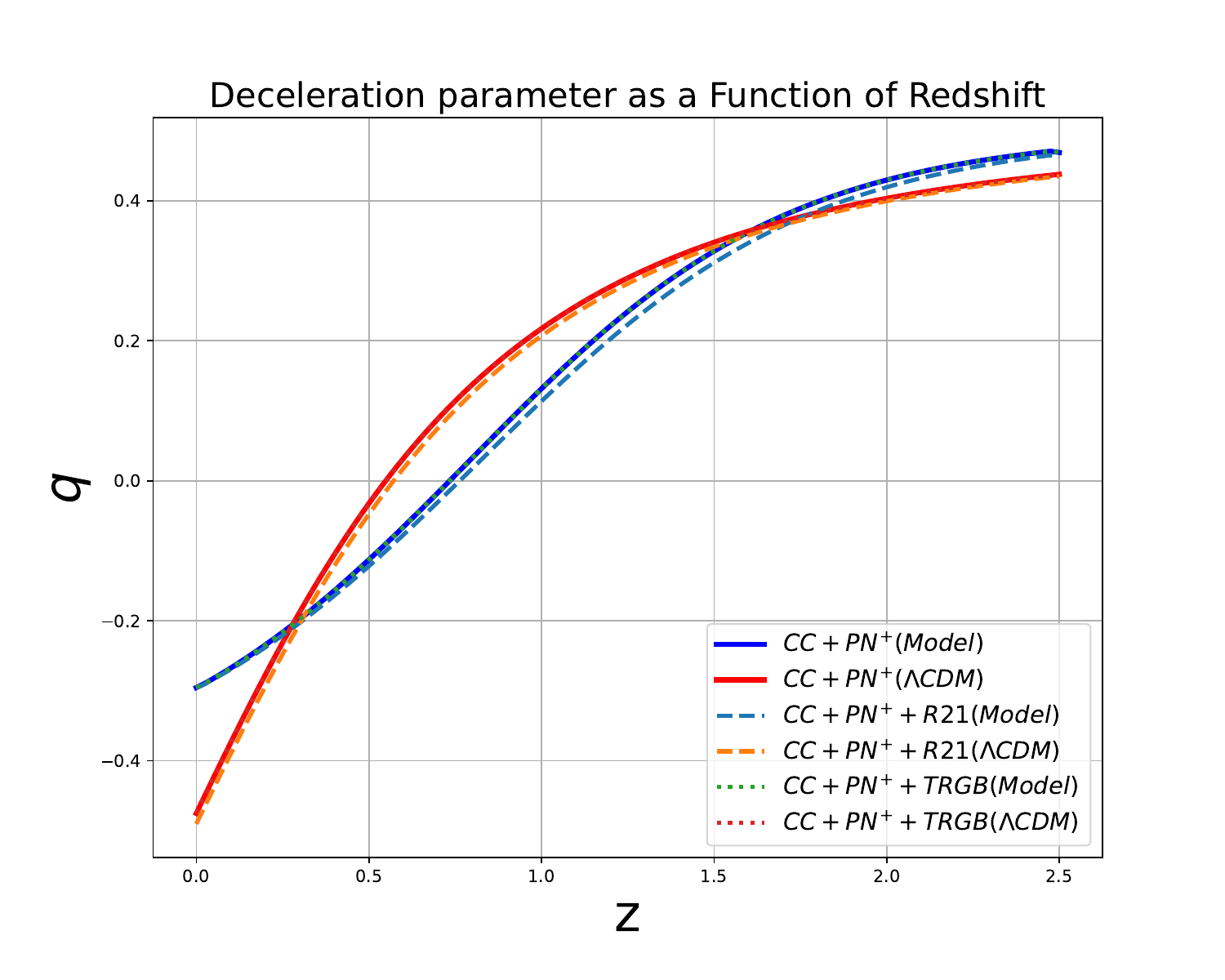}
         \includegraphics[width=70mm]{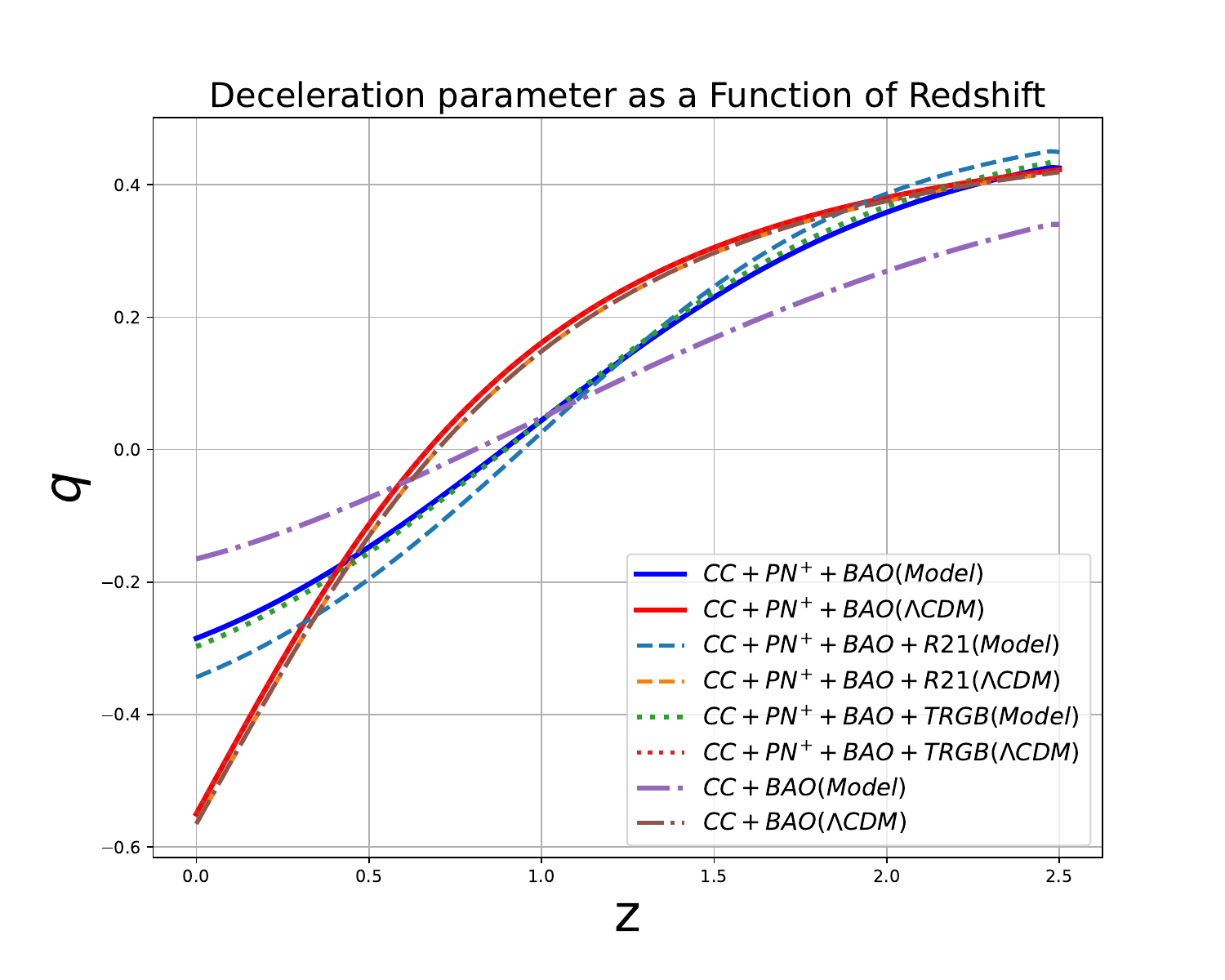}
        \includegraphics[width=70mm]{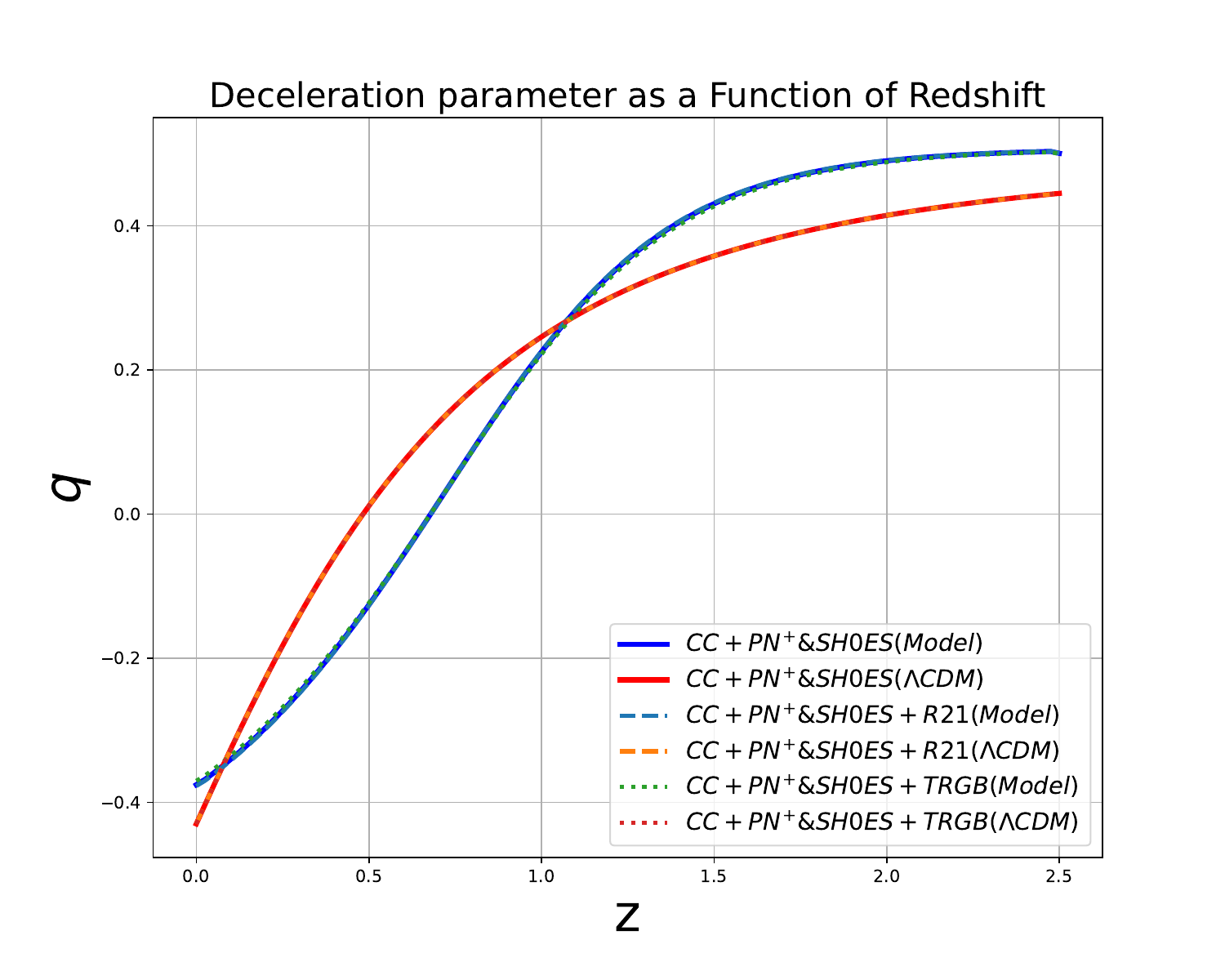}
        \includegraphics[width=70mm]{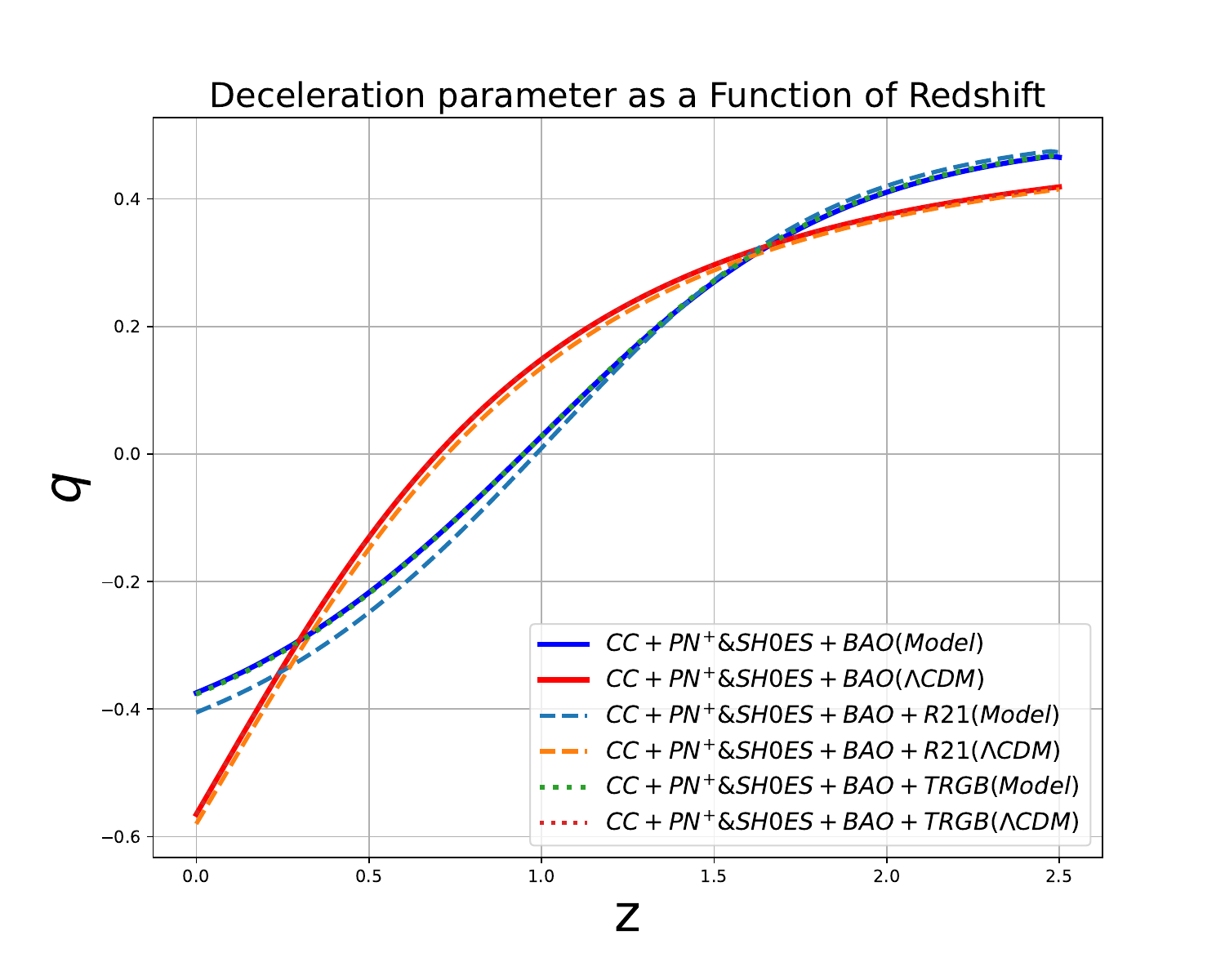}
\caption{Evolutionary behavior of the deceleration parameter and  $\Lambda$CDM model in redshift for the data sets combination: CC, PN$^{+}$ (without SH0ES), PN$^{+}$\&SH0ES (with SH0ES) and BAO. The $H_0$ priors are: R21 and TRGB. } 
\label{plusCCBAOdeceleration}
\end{figure}

In Figs.-\ref{plusCCBAOdeceleration}, we illustrate the evolution of the deceleration parameter for both the selected model and the $\Lambda$CDM model. The selected model indicates a transition from a decelerating period to an accelerating phase of the Universe, implying its potential to represent the accelerated expansion of the Universe. Additionally, we determine the current value of the deceleration parameter. Details regarding the current value of the deceleration parameter and the transition point for various data set combinations can be found in Table~\ref{results}. The outcomes from the selected model concerning the current deceleration parameter value and the transition point are consistent with cosmological observations \cite{PhysRevD.90.044016a, Camarena:2020prr}.
\begin{figure}[ht]
     \centering
         \includegraphics[width=70mm]{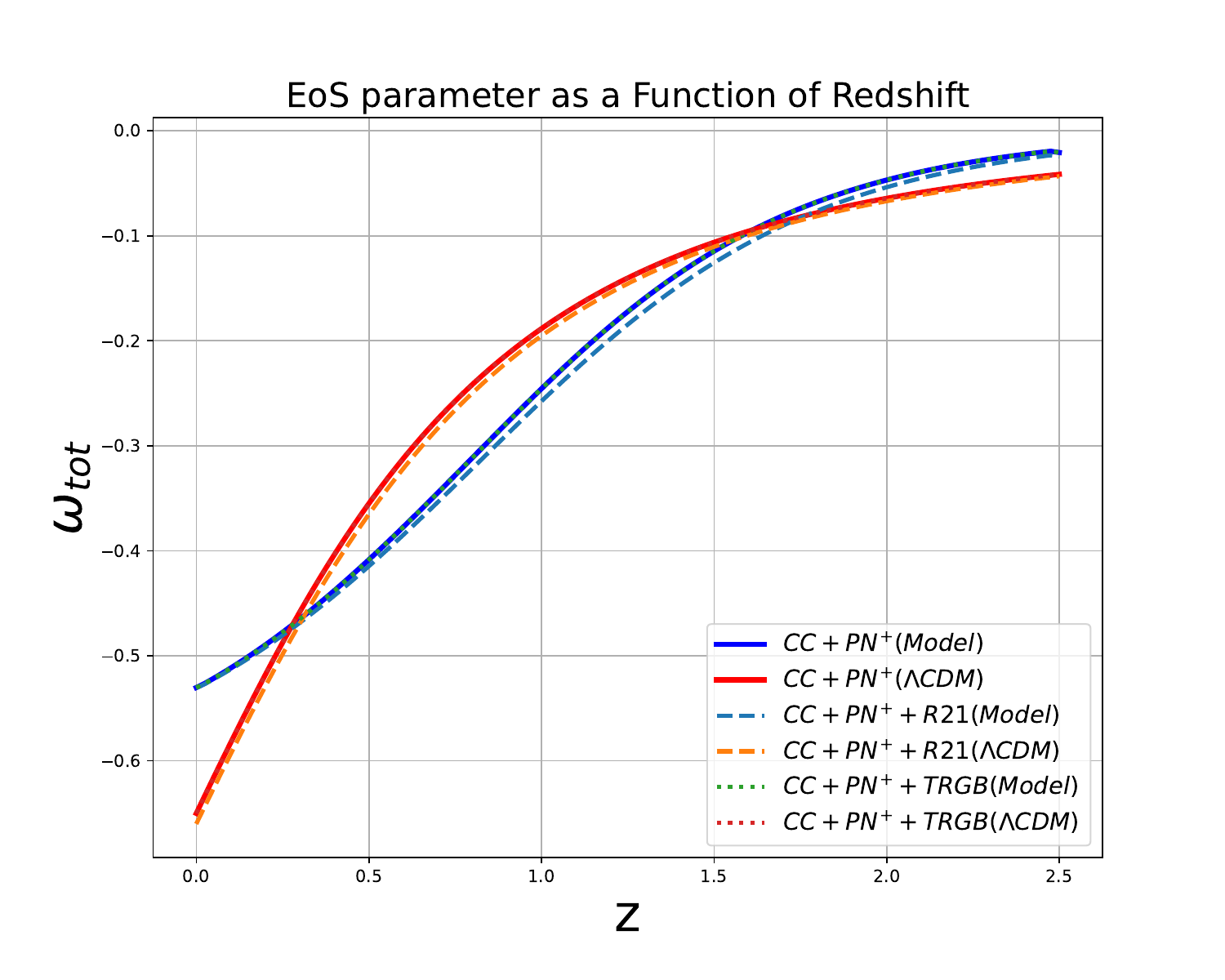}
         \includegraphics[width=70mm]{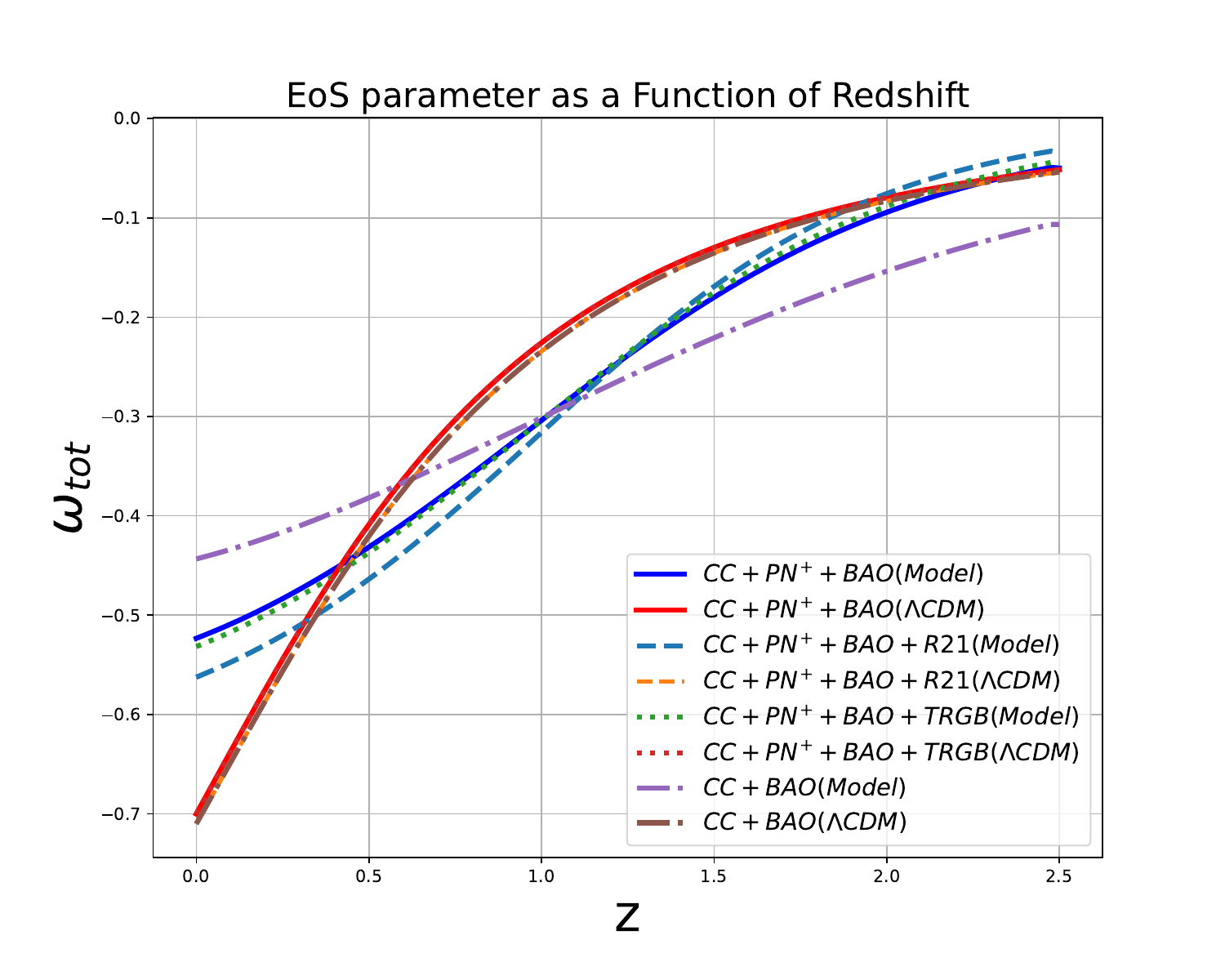}
      \includegraphics[width=70mm]{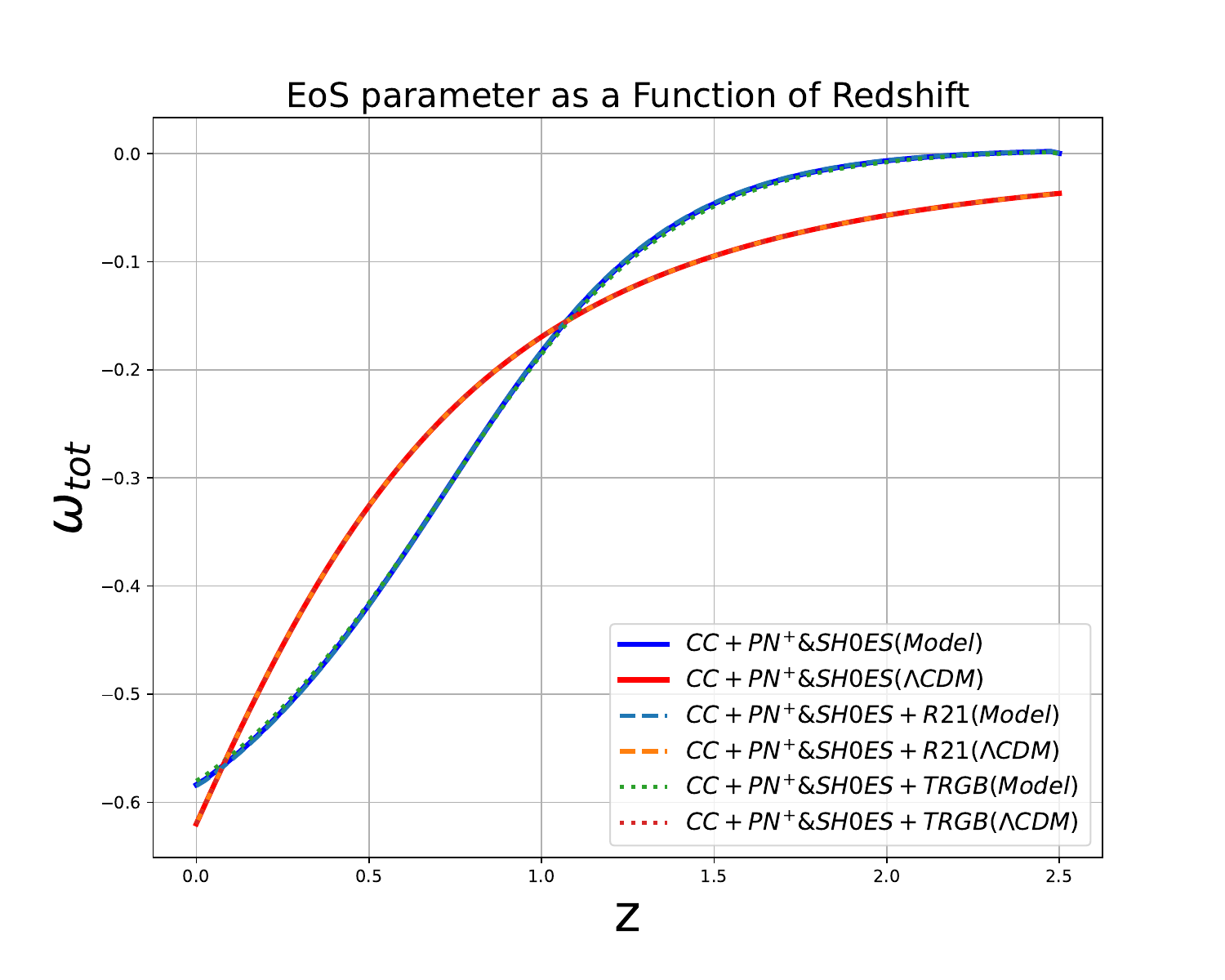}
       \includegraphics[width=70mm]{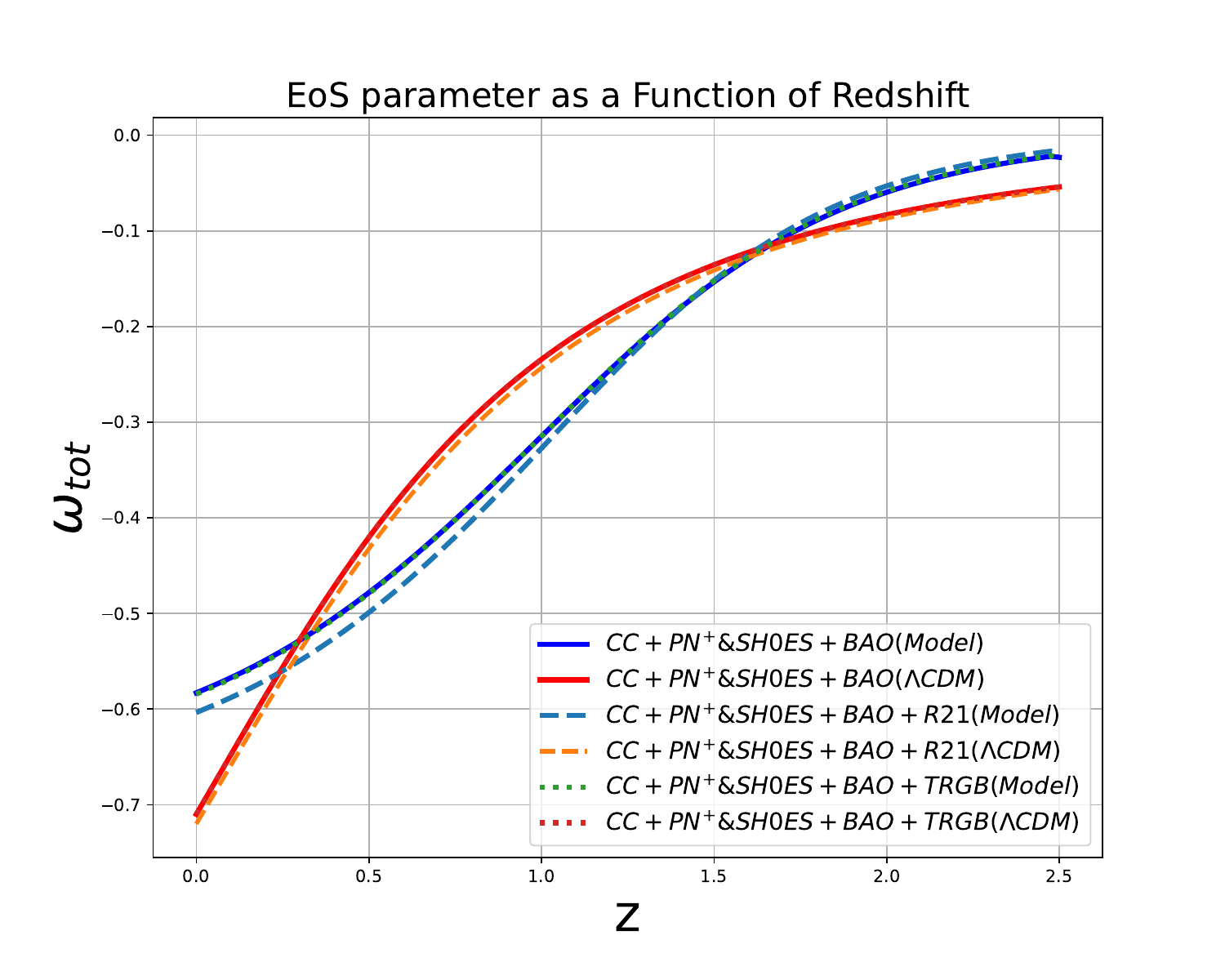}
\caption{Evolutionary behavior of the EoS parameter and  $\Lambda$CDM model in redshift for the data sets combination: CC, PN$^{+}$ (without SH0ES), PN$^{+}$\&SH0ES (with SH0ES) and BAO. The $H_0$ priors are: R21 and TRGB.} 
\label{plusCCBAOEos}
\end{figure}
In Figs.~\ref{plusCCBAOEos}, we depict the evolution of the total EoS parameter for the selected model in comparison with the $\Lambda$CDM model. The EoS parameter indicates that our chosen model shows the quintessence phase of the Universe, as it satisfies the quintessence criterion where \( -1 < \omega_{tot} < -\frac{1}{3} \). A comprehensive summary of the current values of the total EoS parameter is provided in Table \ref{results}. 
\begin{figure}[ht]
     \centering
         \includegraphics[width=70mm]{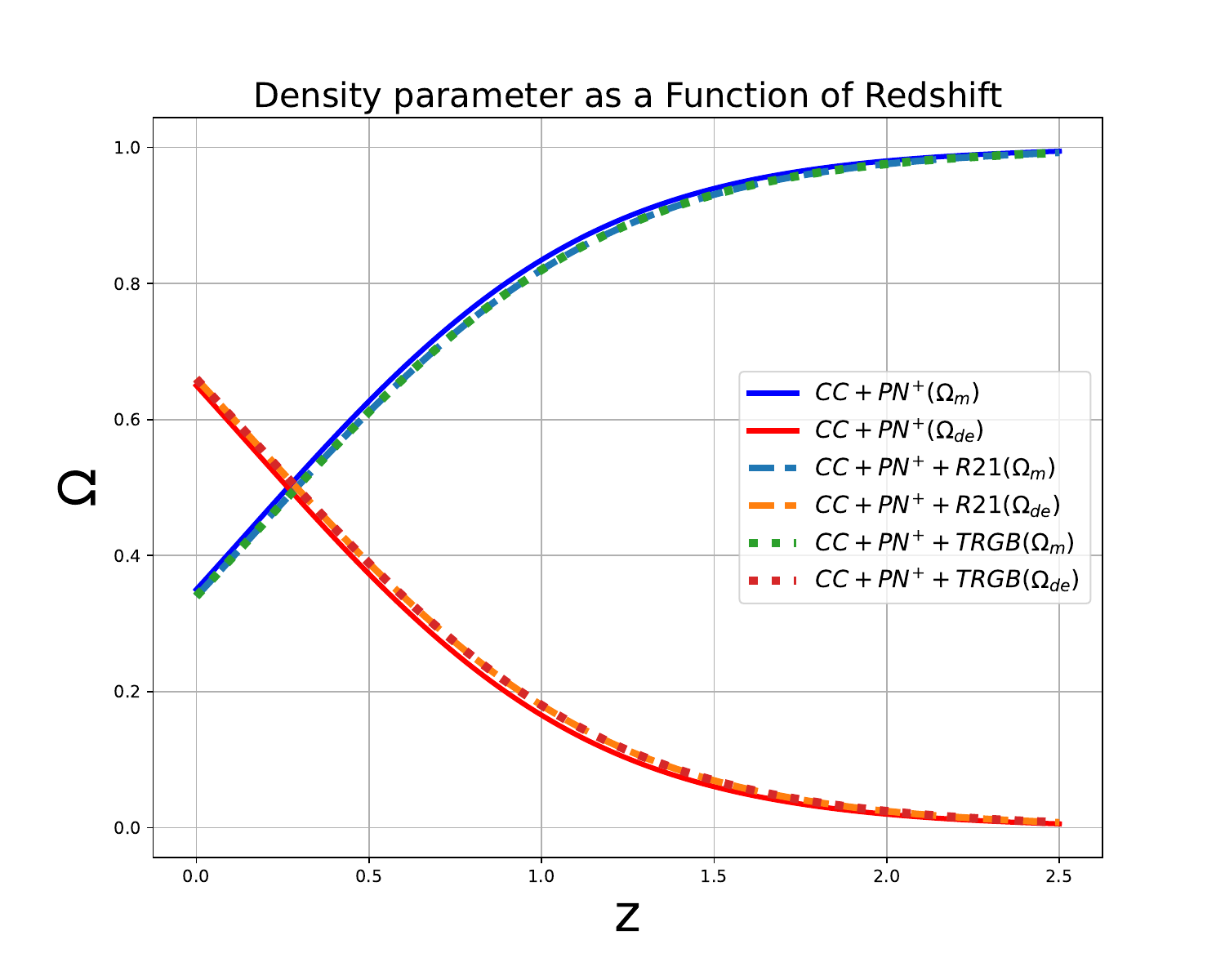}
         \includegraphics[width=70mm]{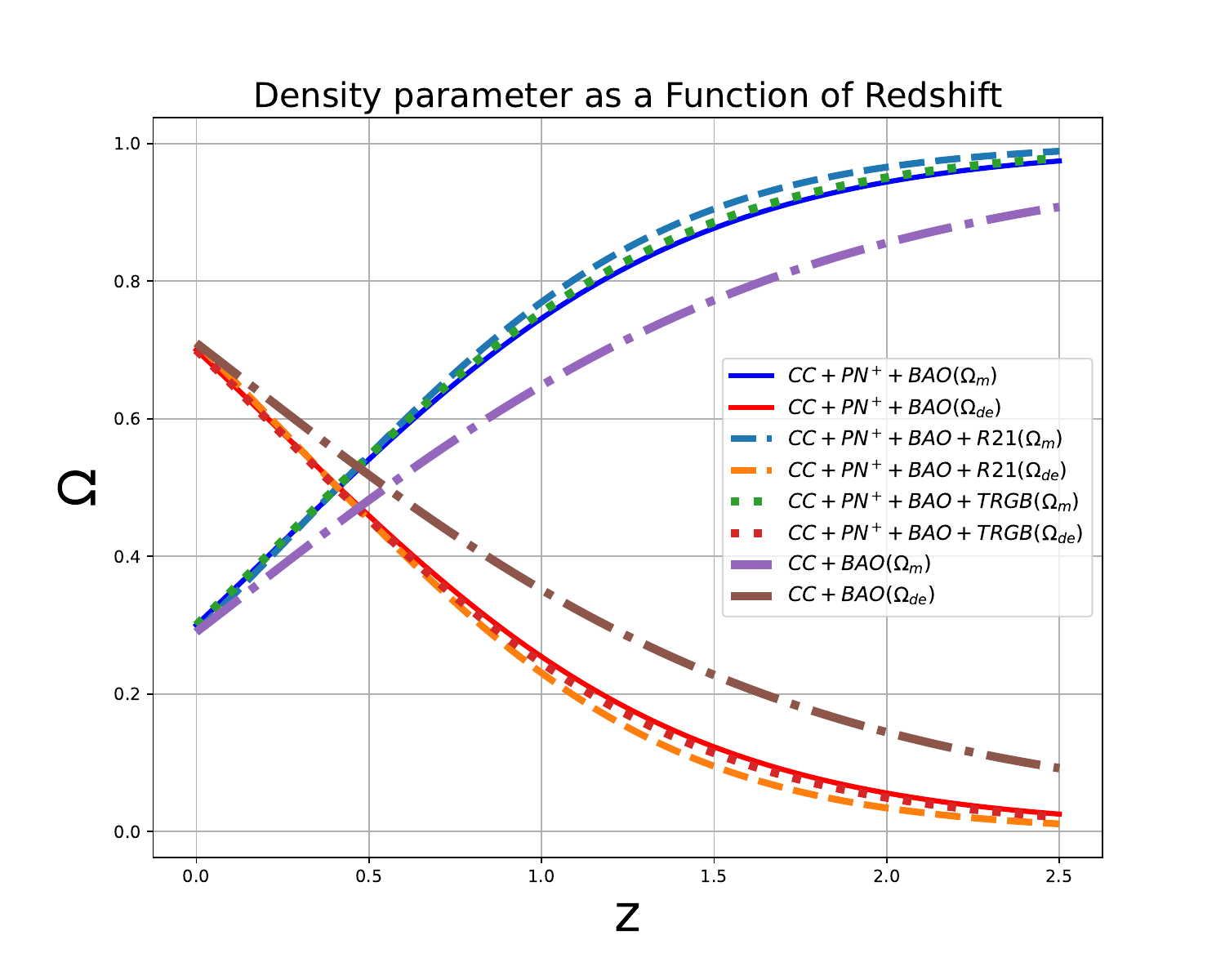}
      \includegraphics[width=70mm]{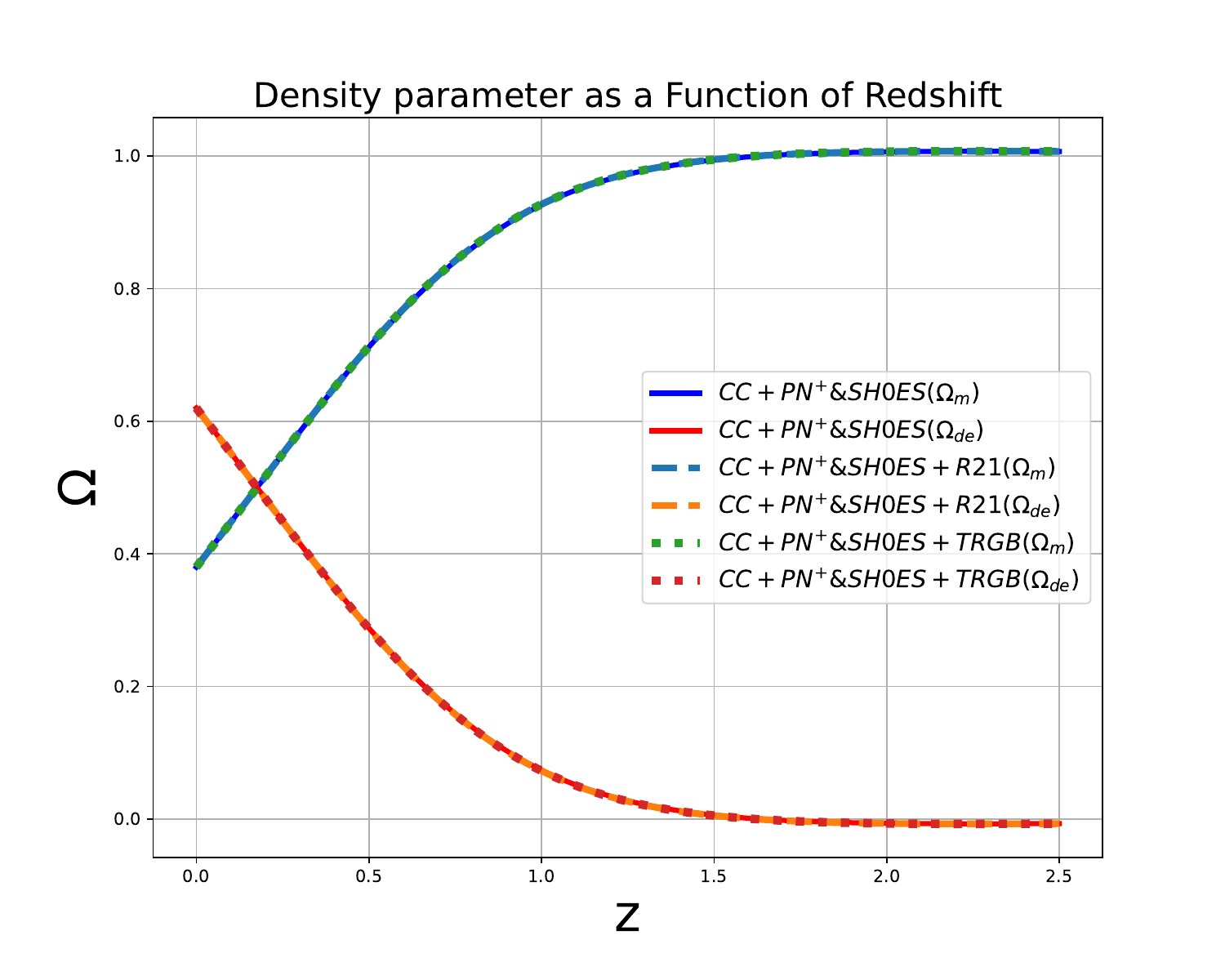}
        \includegraphics[width=70mm]{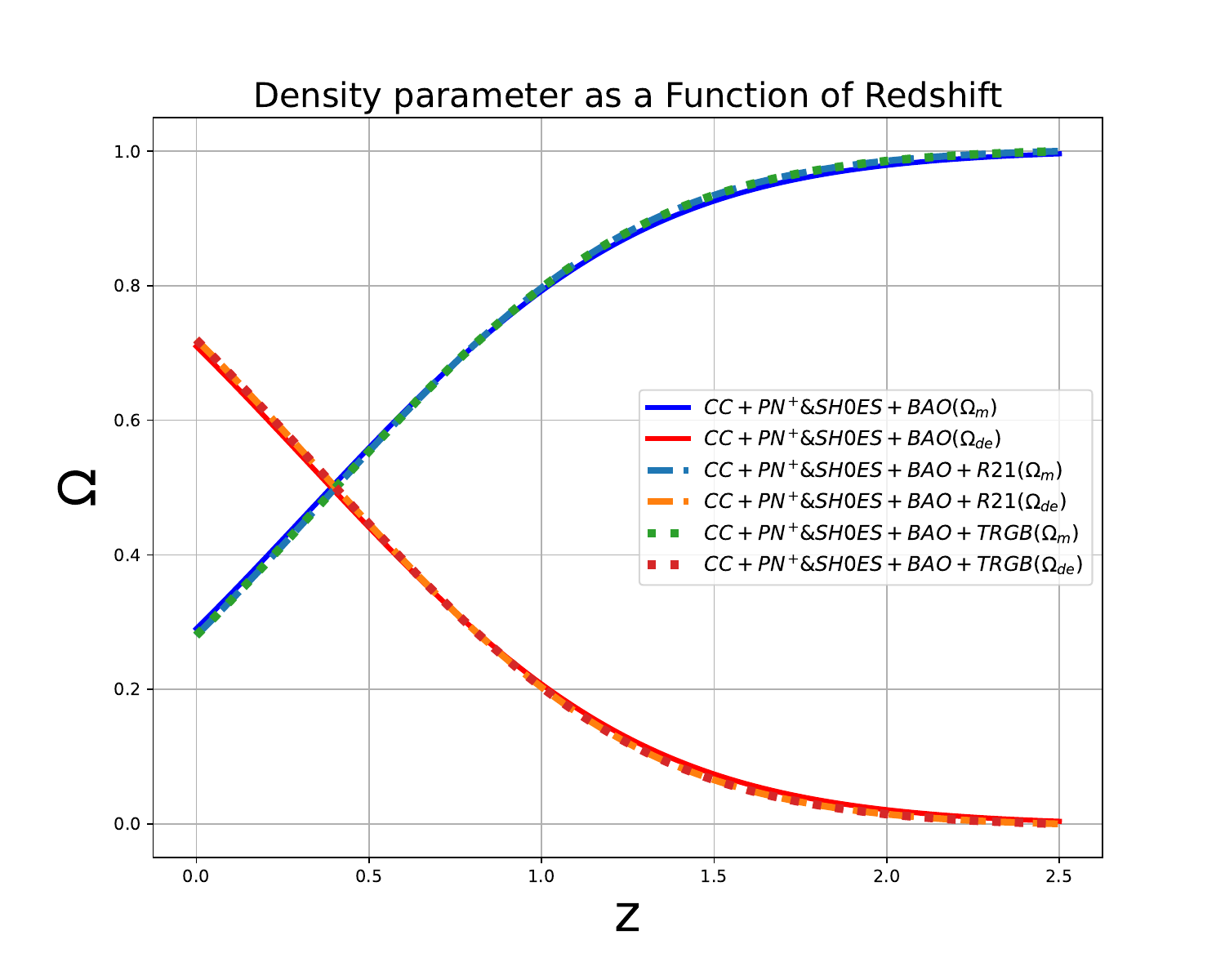}
\caption{Evolutionary behavior of the density parameters and  $\Lambda$CDM model in redshift for the data sets combination: CC, PN$^{+}$ (without SH0ES), PN$^{+}$\&SH0ES (with SH0ES) and BAO. The $H_0$ priors are: R21 and TRGB. }  
\label{plusCCBAOdensity}
\end{figure} 
In Figs.~\ref{plusCCBAOdensity}, we present the evolution of the density parameters for both matter and DE in the Universe as redshift varies. These plots demonstrate the shifting interaction between matter and DE, showing that the density of DE increasingly dominates in later epochs, showing the accelerated expansion. The current values of these density parameters are listed in Table~\ref{results}. During the early stages of the Universe, DM constitutes the main part of the energy density at higher redshifts, far outpacing DE. As redshift decreases, the proportion of DM drops while the influence of DE steadily rises. In the later stages of the Universe, DE becomes the leading component and at lower redshifts, it overtakes DM, driving the accelerated expansion.  

The \( Om(z) \) diagnostic serves as an alternative method to distinguish between various DE cosmological models. This can be represented as \cite{sahniPRD_om, Sahni_2003}.
\begin{equation}\label{Omzequation}
Om(z)= \frac{E^{2}(z)-1}{(1+z)^3-1} \,.   
\end{equation}
Assessing the \( Om(z) \) values across different redshifts can yield insights into the characteristics and dynamics of DE. The procedure for two-point difference diagnostics can be described as follows:
\begin{equation}\label{Omztworedshiftdifference}
Om(z_1-z_2) = Om(z_1)- Om(z_2)\,.   
\end{equation}

\begin{figure}[ht]
     \centering
         \includegraphics[width=70mm]{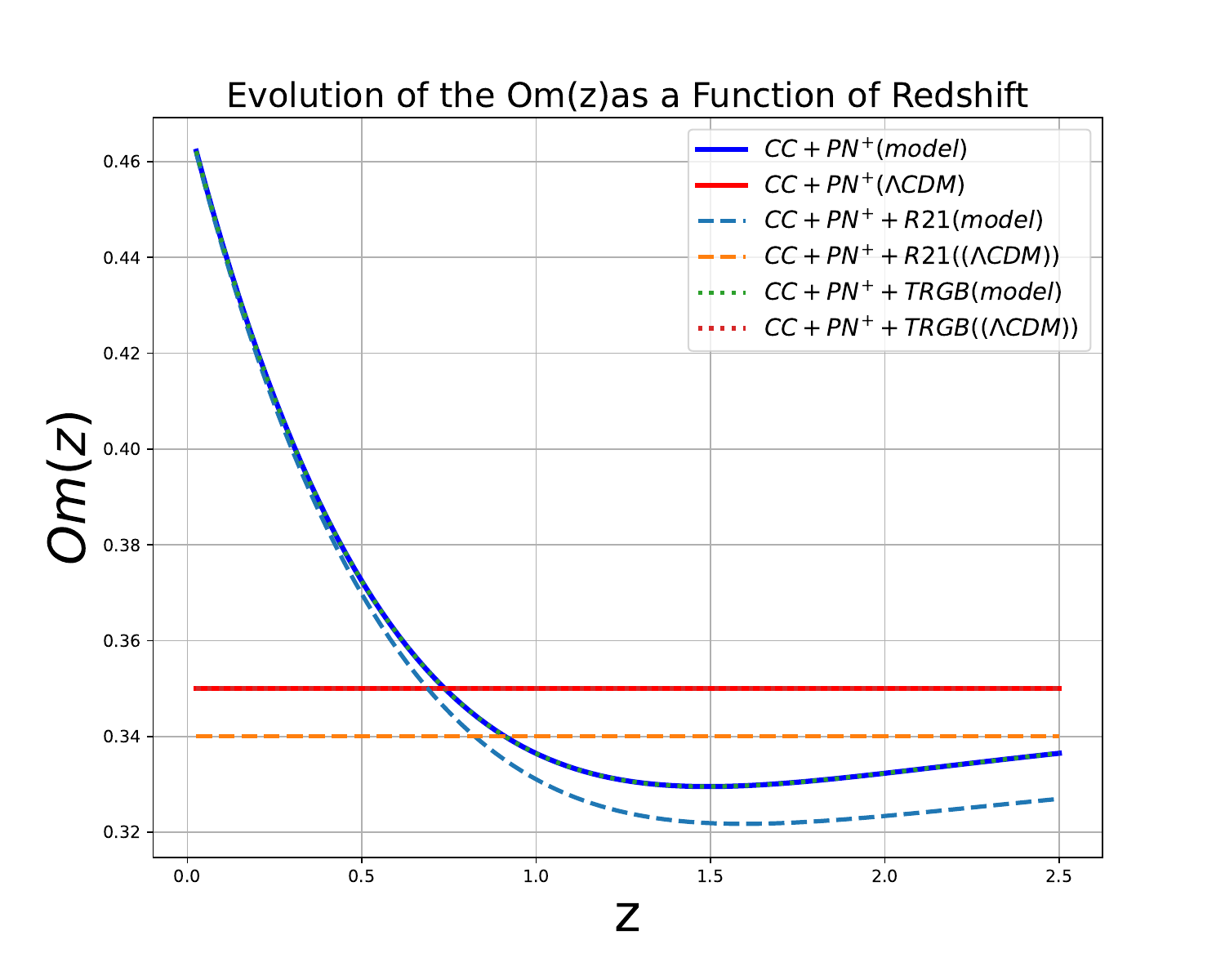}
          \includegraphics[width=70mm]{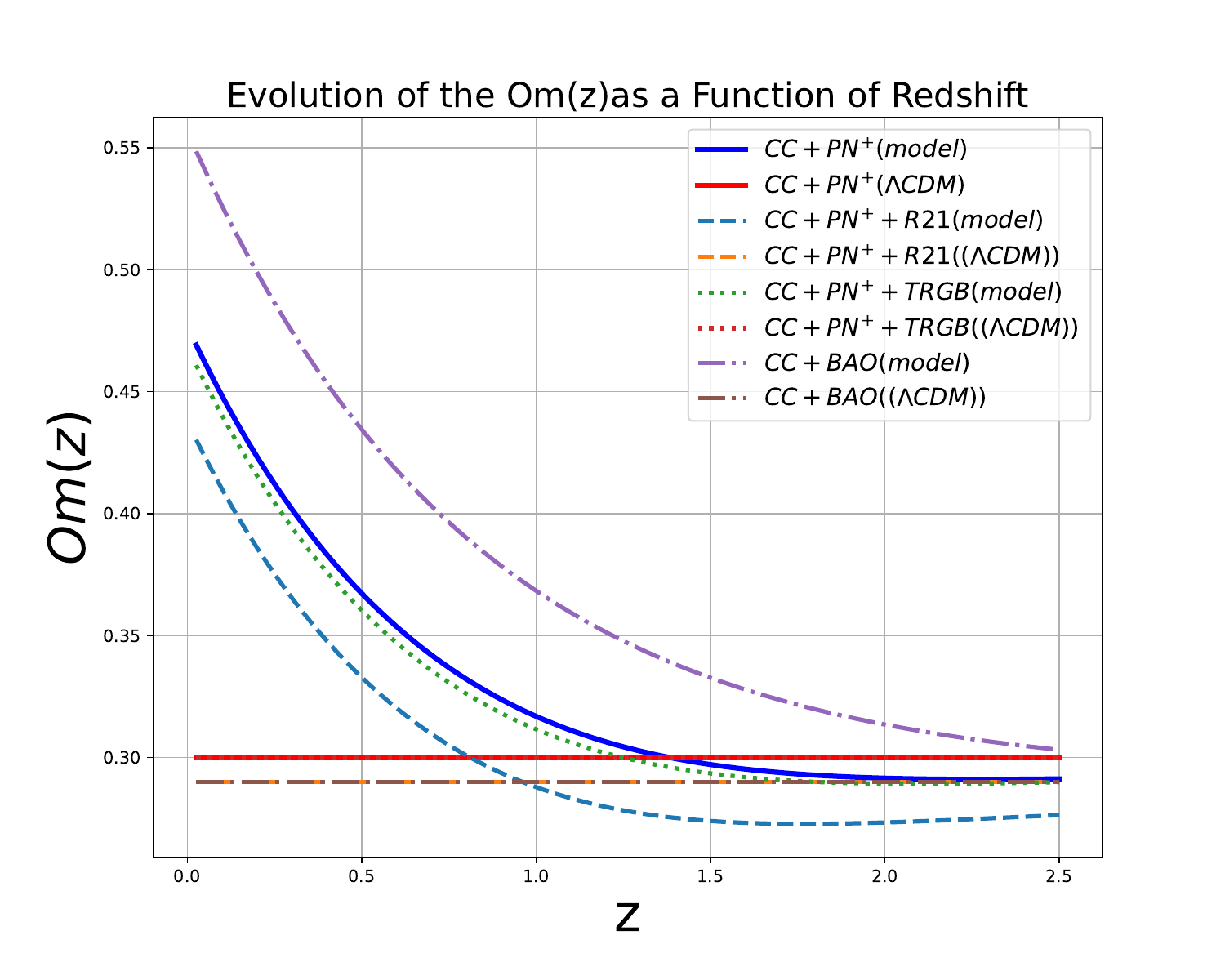}
       \includegraphics[width=70mm]{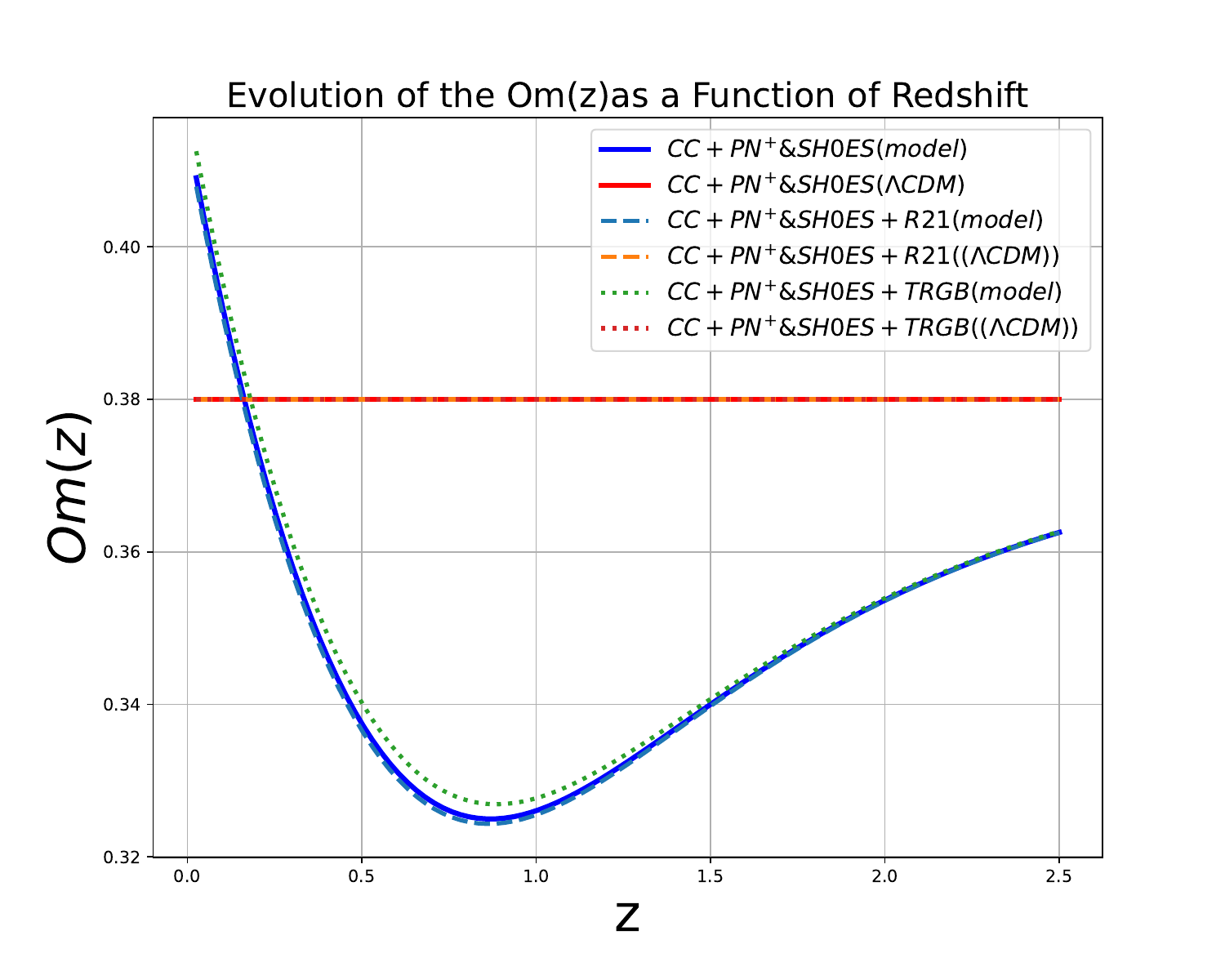}
         \includegraphics[width=70mm]{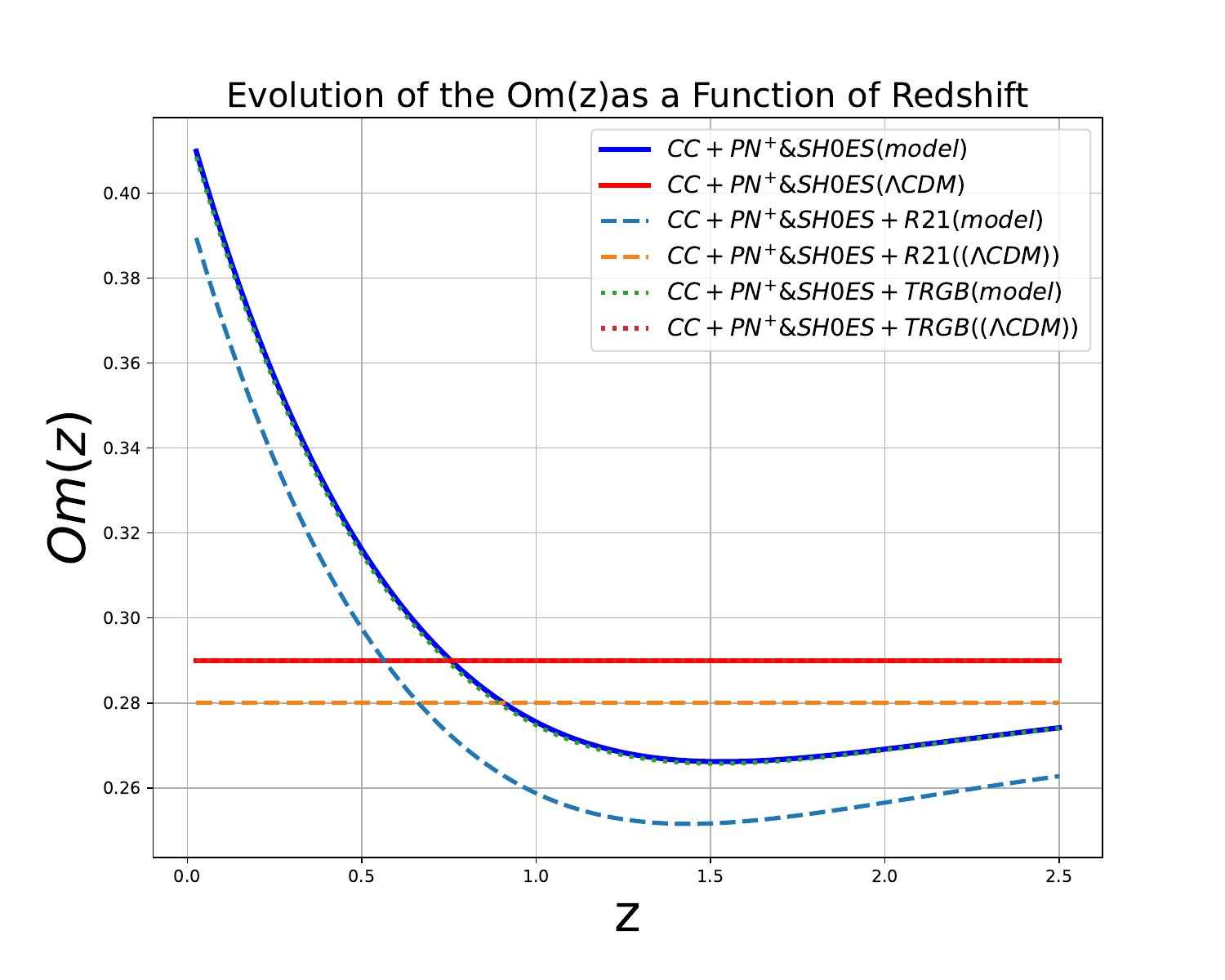}
\caption{Evolutionary behavior of the $Om(z)$ parameter and  $\Lambda$CDM model in redshift for the data sets combination: CC, PN$^{+}$ (without SH0ES), PN$^{+}$\&SH0ES (with SH0ES) and BAO. The $H_0$ priors are: R21 and TRGB. } 
\label{plusFigCCBAOOm}
\end{figure}

If \( Om(z_1, z_2) > 0 \), this suggests the model represents a quintessence scenario, while \( Om(z_1, z_2) < 0 \) implies it exhibits phantom behavior, given that \( z_1 < z_2 \). Additionally, if \( Om(z) \) remains consistent across various redshifts, it indicates a potential connection between DE and the cosmological constant \cite{sahniPRD_om}. We have compared the model discussed here with the $\Lambda$CDM model, as shown in Fig.~\ref{plusFigCCBAOOm}. The \( Om(z) \) parameter displays stability within the redshift range \( 0 < z < 2.5 \). This stability is crucial for understanding the dynamics of the Universe and its accelerated expansion. The slope of \( Om(z) \) is an important indicator for DE models. A positive slope suggests the presence of phantom behavior, which is characterized by an EoS \( \omega < -1 \). Conversely, a negative slope is linked to the quintessence region, where \( \omega > -1 \). In Fig.~\ref {plusFigCCBAOOm}, the slope of the \( Om(z) \) parameter shows a downward trend as redshift increases, suggesting that the impact of DE becomes more significant over time. This reduction indicates a transformation in the dynamics of the Universe, illustrating the evolving roles of matter and DE as the Universe progresses. This decline is consistent with the quintessence phase of the Universe, where the EoS \( \omega \) is greater than \(-1\).

\begin{figure}[H]
 \centering
 \includegraphics[width=100mm]{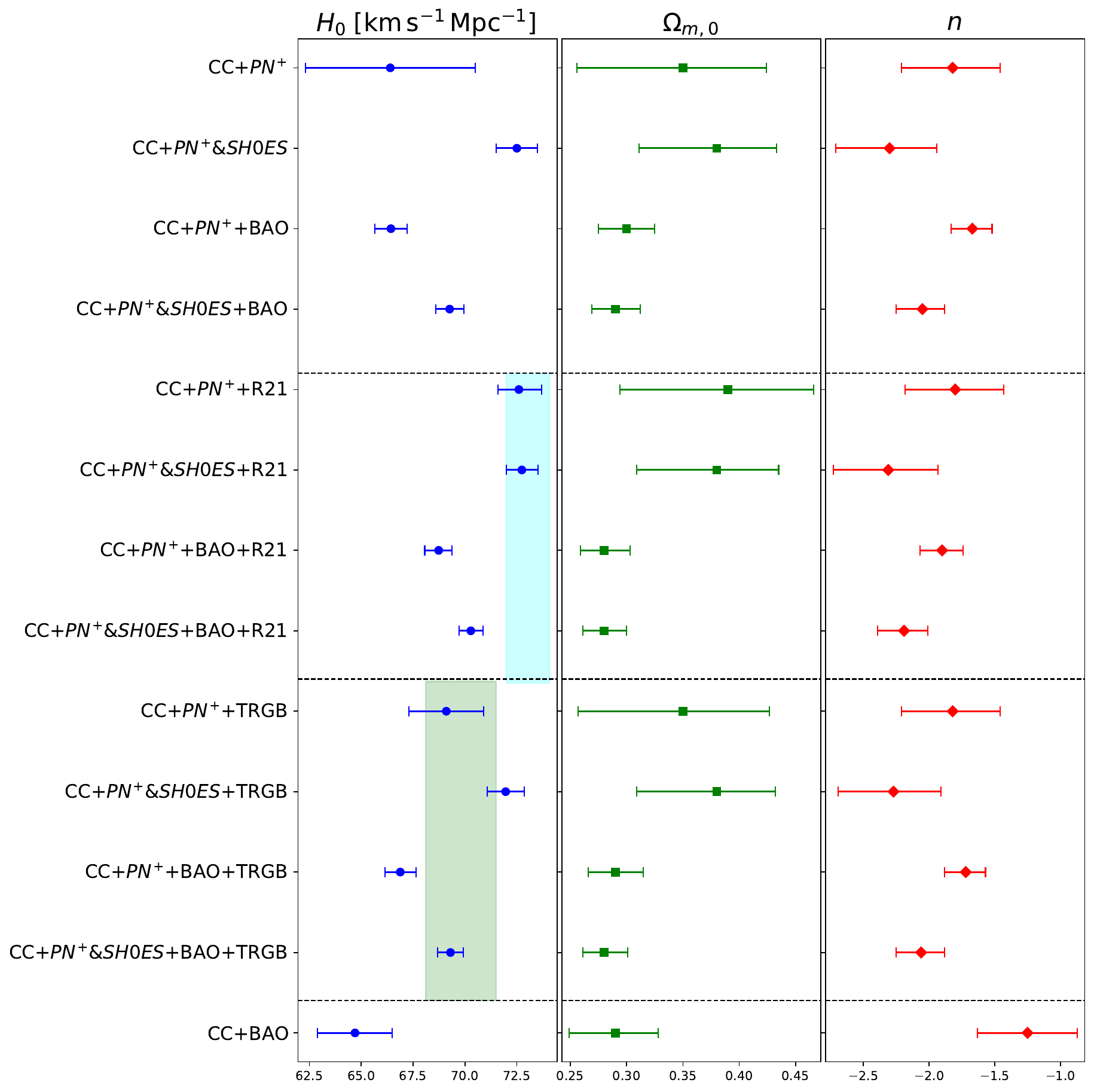} 
 \caption{Whisker plot for the chosen \( f(T, \mathcal{T}) \) model. This plot provides a visual representation of the distributions of several key parameters: the Hubble constant \( H_{0} \), the matter-energy density \( \Omega_{m0} \) and the model parameters \( n \). In the first column, the cyan-shaded region represents the R21 prior, while the green-shaded region indicates the TRGB prior.
} \label{whiskerplot}
 \end{figure}

The whisker plot represents the ranges and central tendencies of the model parameters. In Fig.~\ref{whiskerplot}, the whisker plot effectively distinguishes parameter values derived from different combinations of data sets. Through Fig.~\ref{whiskerplot}, we observe a notable contrast in the values of the parameters ($H_0$, $\Omega_{m0}$, $n$) between the PN$^+$ (without SH0ES) and PN$^+$\& SH0ES data set combinations. Across all combinations of data sets, we find a lower value of $H_0$ in the CC+BAO data set combination where both PN$^+$ and PN$^+$\& SH0ES are absent. Additionally, we can observe how the inclusion of $H_0$ priors influences the values of $H_0$ within the PN$^+$ data set combination, as incorporating the $H_0$ priors increases the $H_0$ values.
\begin{table}[H]
    \centering
    \renewcommand{\arraystretch}{0.65} 
    \begin{tabular}{|p{4.7cm}|p{1cm}|p{1cm}|p{1cm}|p{1cm}|p{1cm}|p{1cm}|p{1cm}|p{1cm}|}
        \hline 
        \textbf{Data set} & \multicolumn{5}{c|}{\textbf{For $f(T, \mathcal{T})$ model}} & \multicolumn{3}{c|}{\textbf{For the $\Lambda$CDM model}} \\
        \hline
        & $q_0$ & $\omega_{0}$ & $z_{tr} $ & $\Omega_{m}^0$ & $\Omega_{de}^0$ & $q_0$ & $\omega_{0}$ & $z_{tr}$\\ [0.5ex]
        \hline
        CC+PN$^{+}$ & $-0.29$ & $-0.53$ &$0.73$  &$0.35$  & $0.65$ & $-0.48$ & $-0.65$ &$0.53$ \\
        \hline
        CC+PN$^{+}$+R21 & $-0.30$ & $-0.53$ & $0.76$ &$0.34$  & $0.66$ & $-0.49$ & $-0.65$ & $0.56$\\
        \hline
        CC+PN$^{+}$+TRGB & $-0.30$ & $-0.53$ &$0.73$  &$0.35$  & $0.65$ & $-0.48$ & $-0.65$ &$0.53$ \\
        \hline
        CC+PN$^{+}$+BAO & $-0.28$ & $-0.52$ &$0.88$  &$0.3$  & $0.7$ & $-0.55$ & $-0.70$ &$0.66$ \\
        \hline
        \begin{tabular}{@{}c@{}}CC+PN$^{+}$+ \\BAO+R21\end{tabular} & $-0.34$ & $-0.56$ &$0.93$  &$0.29$  & $0.71$ & $-0.57$ & $-0.71$ &$0.68$\\
        \hline
        \begin{tabular}{@{}c@{}}CC+PN$^{+}$+ \\BAO+TRGB\end{tabular} & $-0.30$ & $-0.53$ & $0.88$ &$0.3$  & $0.7$ & $-0.55$ & $-0.7$ &$0.66$ \\
        \hline
        CC+BAO & $-0.17$ & $-0.44$ &$0.78$  &$0.29$  & $0.71$ & $-0.565$ & $-0.71$ &$0.68$ \\
        \hline
        CC+$PN^{+}\& SH0ES$ & $-0.37$ & $-0.58$ &$0.66$  &$0.38$  & $0.62$ & $-0.43$ & $-0.62$ &$0.48$ \\
        \hline
        CC+$PN^{+}\& SH0ES$+R21 & $-0.38$ & $-0.58$ & $0.66$ &$0.38$  & $0.62$ & $-0.43$ & $-0.62$ & $0.48$\\
        \hline
        CC+$PN^{+}\& SH0ES$+TRGB & $-0.37$ & $-0.58$ &$0.66$  &$0.38$  & $0.62$ & $-0.43$ & $-0.62$ &$0.48$ \\
        \hline
        CC+$PN^{+}\& SH0ES$+BAO & $-0.37$ & $-0.59$ &$0.93$  &$0.29$  & $0.71$ & $-0.565$ & $-0.71$ &$0.68$ \\
        \hline
        \begin{tabular}{@{}c@{}}CC+$PN^{+}\& SH0ES$+ \\BAO+R21\end{tabular} & $-0.40$ & $-0.60$ &$0.96$  &$0.28$  & $0.72$ & $-0.58$ & $-0.72$ &$0.71$\\
        \hline
        \begin{tabular}{@{}c@{}}CC+$PN^{+}\& SH0ES$+ \\BAO+TRGB\end{tabular} & $-0.37$ & $-0.58$ & $0.93$ &$0.29$  & $0.71$ & $-0.57$ & $-0.71$ &$0.68$ \\
        \hline
    \end{tabular}
    \caption{Present value of the parameters and the transition point. The upper or lower indices $0$ represent the current time at $z=0$.}
    \label{results}
\end{table}
\section{Conclusion} \label{ch3_SEC-IV}
We have analyzed the dynamical system with dynamical variables in the context of modified teleparallel gravity with two forms of $f(T, \mathcal{T})$ has revealed several critical points and their cosmological implications.

In the first model $f(T,\mathcal{T})=\alpha T^n\mathcal{T}+\beta$, three critical points are obtained, out of which two critical points ($A_2, A_3$) are stable and the remaining critical point are unstable. The stable critical points appear in the de-Sitter phase, whereas the unstable behavior is noted in the matter-dominated phase of the Universe. Trajectories illustrate the behavior of critical points, showing stable or unstable behavior based on the values of $n$. The results align with observational constraints for the present values of $\omega_{DE}$ and $q$. For the critical points in the de-Sitter phase, both the values of the DE EoS parameter and deceleration parameter are $-1$, which confirms the accelerating model with the $\Lambda$CDM-like behavior.

The analysis of the dynamical system with an alternative form of $f(T,\mathcal{T})$ represented by $f(T,\mathcal{T})=\gamma T^2+\delta \mathcal{T}$ has provided important insights into the cosmological behavior of the Universe. In this model also the stable critical points appear in the de-Sitter phase and the value of the DE EoS parameter and deceleration parameter is $-1$. So the model appears to be $\Lambda$CDM like accelerating model. We have defined Phantom and Quintessence regions from the total equation of the state parameter value for different choices of $n$. The Phantom region is defined for the range $n>1$ and the Quintessence region shows for the range $n<-\frac{1}{2}$.

In both models, the total and DE EoS parameters merge and the dominance of DE EoS is visible. The deceleration parameter shows the early deceleration and late time acceleration with the present value noted at $q_0=-0.59$ and $q_0=-0.56$ respectively for models I and II and the transition is showing at $0.62$ and $0.51$. For both the models, the present value of the matter and DE density parameters are $\Omega_{m} \approx 0.3$ and $\Omega_{DE}\approx0.7$ and it fits the recent suggestions from cosmological observations. This analysis offers a comprehensive understanding of how modified gravity with the given form of $f(T, \mathcal{T})$ can describe the evolution of the Universe and its various phases based on the chosen model parameters.

In the cosmological observation section, we have presented a cosmological model that supports the late-time cosmic acceleration of the Universe in the framework of $f(T, \mathcal{T})$ gravity using various cosmological data set combinations. Our analysis incorporates data set including the CC, PN$^+$ (without SH0ES), PN$^+$\& SH0ES (with SH0ES) and BAO observations alongside the $H_0$ priors from R21 and TRGB.

In examining the PN$^+$ (without SH0ES) and PN$^+$\& SH0ES (with SH0ES) data set, we have observed that the $H_0$ value is lower for the PN$^+$ (without SH0ES) data set when compared to the PN$^+$\& SH0ES (with SH0ES) data set. The increase in the $H_0$ value for the PN$^+$\& SH0ES (with SH0ES) data set can be attributed to including the SH0ES data points. In Tables~ \ref{tab:model_outputsmodelplus} and \ref{tab:model_outputsmodel}, the results for the PN$^+$ and PN$^+$\& SH0ES data set are presented, respectively. Table~\ref{tab:model_outputsmodelplus} shows a higher $H_0$ value for the data set combination CC+PN$^+$+R21 and a lower value for the CC+BAO data set combination. Consequently, the model yields a lower $H_0$ value without SNIa, SH0ES points and $H_0$ priors. Thus, we can conclude that incorporating the BAO data set results in a decrease in $H_0$ values. In Table~ \ref{tab:model_outputsmodel}, the combination CC+PN$^+$\& SH0ES+R21 yields a higher $H_0$ value. The results from Tables~\ref{tab:model_outputsmodelplus} and \ref{tab:model_outputsmodel} suggest that the CC+PN$^+$\& SH0ES+R21 data set combination leads to a higher $H_0$ value, whereas the BAO data set combination results in a lower value. Our result indicates that the highest $H_0$ value for the CC+PN$^{+}$+R21 aligns with the elevated $H_0$ value reported by the SH0ES team (R22), which is stated as $H_0 = 73.30 \pm 1.04 \,\text{km s}^{-1}\,\text{Mpc}^{-1}$
 \cite{Riess:2021jrx}. Meanwhile, the $H_0$ value from the BAO data set inclusion is consistent with the Planck Collaboration, which reports a Hubble constant of $H_0 =  67.4 \pm 0.5 \,\text{km s}^{-1}\,\text{Mpc}^{-1}$\cite{Aghanim:2018eyx}. In contrast, Aboot et al. \cite{Abbott_2018mnras} present a value of $H_0 = 67.2^{+1.2}_{-1.0} \,\text{km s}^{-1}\,\text{Mpc}^{-1}$. The $H_0$ findings for PN$^{+}$ and PN$^{+}$\& SH0ES in our study resemble those obtained by Brout et al \cite{Brout_2022pan}. 

Alternatively, we delve deeper into our chosen \( f(T, \mathcal{T}) \) model, which exhibits fascinating characteristics due to its absence of a corresponding \( \Lambda \)CDM limit. In particular, this model has no combination of parameter values that can replicate the precise behavior of the \( \Lambda \)CDM model. From a statistical viewpoint, the AIC and BIC values for the data set combination \text{CC+\( \text{PN}^{+}\)} and  \text{CC+\( \text{PN}^{+}\&\text{SH0ES} \)} are shown to be quite close to those of the conventional \( \Lambda \)CDM model, implying that this combination of data supports the \( \Lambda \)CDM model effectively. However, when the BAO data set is incorporated with \text{CC+\( \text{PN}^{+}\)} and \text{CC+\( \text{PN}^{+}\&\text{SH0ES} \)}, both the $\Delta$AIC and $\Delta$BIC values rise. It suggests that the \text{CC+\( \text{PN}^{+}\)+BAO} and \text{CC+\( \text{PN}^{+}\&\text{SH0ES} \)+BAO} data set combination does not provide compelling evidence in favor of the \( \Lambda \)CDM model.

Our analysis of linear matter perturbations in \( f(T, \mathcal{T}) \) gravity demonstrates that the growth rate of matter overdensity is influenced by the effective Newton's constant \( G_{\text{eff}} \), modifying the standard evolution of density fluctuations. The results indicate that the modified \( f(T, \mathcal{T}) \) model can help address the \( \sigma_8 \) tension by predicting a lower growth rate \( f\sigma_8(0) \) compared to \( \Lambda \)CDM. Consequently, in Fig.~\ref{sigma8:BAOom}, the results indicated by the red-dot dashed line are approximately 9\% below the \( \Lambda \)CDM model prediction. In contrast, the purple-thick line shows a deviation of roughly 11\% below \( \Lambda \)CDM. These results indicate that the  \( f(T, \mathcal{T}) \) model could provide a better fit to large-scale structure observations, potentially offering an alternative explanation for the discrepancies in cosmic growth measurements as compared to \( \Lambda \)CDM.

To explore late-time cosmology, we demonstrate the evolution of background cosmological parameters, including the deceleration parameter, the total EoS parameter, the energy density parameters for both matter and DE and the \( Om(z) \) diagnostic parameter for our chosen \( f(T, \mathcal{T}) \) model in comparison to the standard \( \Lambda \)CDM model. The current values for the deceleration and EoS parameters, energy density and the transition redshift from deceleration to acceleration are summarized in Table~\ref{results}. The observed behavior of the deceleration parameter suggests that the selected model successfully captures the shift from deceleration in the early Universe to acceleration in later periods. The dynamics of the EoS parameter indicate a quintessence-like characteristic during the late-time phase of the evolution. The energy density parameters reveal a transition from an early Universe dominated by matter to a late-time phase predominantly governed by DE. Thus, the model can explain the late-time cosmic phenomena of the Universe.

\chapter{Dynamical systems analysis in \texorpdfstring{$f(T,\phi)$}{} gravity} 

\label{Chapter4} 

\lhead{Chapter 4. \emph{Dynamical systems analysis in \texorpdfstring{$f(T,\phi)$}{} gravity}} 
\vspace{10 cm}
*The work in this chapter is covered by the following publications: \\

\textbf{L K Duchaniya}, S A Kadam, B Mishra and Jackson Levi Said ``Dynamical system analysis in $f(T, \phi)$ gravity", \textit{European Physical Journal C}, \textbf{83}, 27 (2023).\\

\clearpage
 \section{Introduction} \label{ch4_SEC I}
 Teleparallel-based cosmological models describe gravity in which torsion is the mediator of gravitation. Several extensions have been made within the so-called Teleparallel equivalent of general relativity, which is equivalent to general relativity at the level of the equations of motion where attempts are made to study the extensions of this form of gravity and to describe more general functions of the torsion scalar $T$. One of the extensions of $f(T)$ gravity is the generalized scalar-torsion $f(T,\phi)$ gravity, where $\phi$ is the canonical scalar. In the gravitational action, the scalar field is non-minimally connected with the torsion scalar \cite{Xu:2012jf}. Further, in the covariant teleparallel framework, a new class of theories has been given where the action depends on the scalar field and arbitrary function of torsion scalar \cite{Hohmann:2018rwf}. Motivated by this non-minimal coupling of torsion scalar and scalar field, in this chapter, we will study the cosmological aspects of the models through the dynamical system analysis. 
\section{Field Equations of the scalar-torsion \texorpdfstring{$f(T,\phi)$}{} gravity}\label{ch4_SECII}
In this study, we examine the $f(T, \phi)$ gravity framework, paying particular attention to its action as described by Eq.~\eqref{ActionEqf(Tphi)}. We employ the flat FLRW metric, detailed in Eq.~\eqref{FLATFLRW}, along with the associated tetrad field presented in Eq.~\eqref{FLRWTETRAD}. By varying the action in Eq.~\eqref{ActionEqf(Tphi)} with respect to the tetrad field and the scalar field $\phi$, we can derive the equations of motion for $f(T,\phi)$ cosmology as follows,
\begin{eqnarray}
    f(T,\phi)-P(\phi)X-2Tf,_{T}&=&\rho_{m}+\rho_{r} \label{ch4_8}\\
    f(T,\phi)+P(\phi)X-2Tf,_{T}-4\dot{H}f,_{T}-4H\dot{f},_{T} &=& -p_{r}\label{ch4_9}\\
    -P,_{\phi}X-3P(\phi)H\dot{\phi}-P(\phi)\ddot{\phi}+f,_{\phi}&=&0\,. \label{ch4_10}
\end{eqnarray}
For brevity $f\equiv f(T,\phi)$ and $f_{,T}=\frac{\partial f}{\partial T}$. In Eqs.~\eqref{ch4_8}--\eqref{ch4_10}, we consider the non-minimal coupling function $f(T,\phi)$ in the form \cite{Hohmann:2018rwf}
\begin{equation}\label{ch4_11}
    f(T,\phi)=-\frac{T}{2\kappa^{2}}-G(T)-V(\phi)\,,
\end{equation}
where $V(\phi)$ is the scalar potential and $G(T)$ is the arbitrary function of torsion scalar. For this $f(T, \phi)$, Eqs.~\eqref{ch4_8}--\eqref{ch4_10} reduce to
\begin{eqnarray}
    \frac{3}{\kappa^{2}}H^{2}=P(\phi)X+V(\phi)-2TG_{,T}+G(T)+\rho_{m}+\rho_{r}\,,\label{ch4_12}\\
    -\frac{2}{\kappa^{2}}\dot{H}=2P(\phi)X+4\dot{H}(G_{T}+2TG_{,TT})+\rho_{m}+\frac{4}{3}\rho_{r}\,,\label{ch4_13}\\
    P(\phi)\ddot{\phi}+P_{,\phi}(\phi)X+3P(\phi)H\dot{\phi}+V_{,\phi}(\phi)=0\,.\label{ch4_14}
\end{eqnarray}
The Friedmann Eqs.~\eqref{ch4_12}--\eqref{ch4_13} are then modified to give
\begin{eqnarray}
    \frac{3}{\kappa^{2}}H^{2}=\rho_{m}+\rho_{r}+\rho_{DE}\,, \label{ch4_15}\\
    -\frac{2}{\kappa^{2}}\dot{H}=\rho_{m}+\frac{4}{3}\rho_{r}+\rho_{DE}+p_{DE}\,. \label{ch4_16}
\end{eqnarray}
Comparing Eq.~\eqref{ch4_12} with Eq.~\eqref{ch4_15} and Eq.~\eqref{ch4_13} with Eq.~\eqref{ch4_16}, the energy density ($\rho_{DE}$) and pressure ($p_{DE}$) for the DE sector can be retrieved as,
\begin{align}
    \rho_{DE} &= P(\phi)X+V(\phi)-2TG_{,T}+G(T)\,, \label{ch4_17}\\
    p_{DE} &= P(\phi)X-V(\phi)+2TG_{,T}-G(T)+4\dot{H}(G_{,T}+2TG_{,TT})\,. \label{ch4_18}
\end{align}
For the sake of brevity, we take $P(\phi)$ = 1. The potential energy, $V(\phi)=V_{0}e^{-\lambda\phi}$, where $\lambda$ is a constant. Our motivation is to construct the cosmological models of the Universe in the DE sector along with its dynamical system analysis. To develop the system, the form of $G(T)$ would be needed and therefore, in the subsequent section, we shall consider two forms of $G(T)$.
\section{Dynamical System Analysis of the Models}\label{ch4_SECIII}
The motivation of this work is to study the cosmological dynamics of some models within the general class of scalar-tensor theories with nontrivial torsion scalar contributions. The dynamical system is a concept that specifies some rules for the development of the system and the possible future behavior of the cosmological models.
\subsection{Model I}\label{ch4_sec:model_1}
For the first model, we consider $G(T)$ as \cite{Zhang2011_jacp}
\begin{equation}\label{ch4_19}
    G(T)= \beta T \ln \left(\frac{T}{T_{0}}\right)\,,
\end{equation}
where $\beta$ be the constant and $T_0$ be the value of $T$ at the initial epoch. This model has been shown to produce \cite{Zhang2011_jacp} physically advantageous critical points and may be interesting to model the evolution of the Universe. Here, the  DE density and the DE pressure terms in Eqs.~\eqref{ch4_17}--\eqref{ch4_18} reduce to
\begin{eqnarray}
    \rho_{DE}&=&\frac{\dot{\phi}^{2}}{2}+V(\phi)-6\beta H^{2} \ln \left(\frac{6 H^{2}}{T_{0}}\right)- 12 H^{2} \beta\,, \label{ch4_20} \\ 
    p_{DE}&=&\frac{\dot{\phi}^{2}}{2}-V(\phi)+6\beta H^{2} \ln \left(\frac{6H^{2}}{T_{0}}\right)+ 12 H^{2} \beta 
    + 4 \dot{H} \left(\beta \ln \left(\frac{6H^{2}}{T_{0}} \right)+3 \beta \right)\,, \label{ch4_21}
\end{eqnarray}
and the scalar field Klein-Gordon equation \eqref{ch4_14} becomes
\begin{equation}\label{ch4_22}
    \ddot{\phi}+3H\dot{\phi}+V,_{\phi}(\phi)=0\,,
\end{equation}
From Eqs.~\eqref{ch4_20}--\eqref{ch4_21}, the EoS can be obtained as 
\begin{equation}\label{ch4_26}
    \omega_{DE}= \frac{\dot{\phi^{2}}-2 V(\phi)+12 H^{2} \beta \ln\left(6\frac{H^{2}}{T_{0}}\right)+24 H^{2} \beta +8 \dot{H} \left(\beta \ln (6\frac{H^{2}}{T_{0}})+3 \beta \right)}{\dot{\phi^{2}}+2 V(\phi)- 12 H^{2} \beta \ln\left (6\frac{H^{2}}{T_{0}}\right)-24 H^{2} \beta }\,.
\end{equation}
To study the dynamics of the model in scalar-torsion $f(T,\phi)$ gravity, we introduce the following phase space variables to frame the autonomous dynamical system,
\begin{align}
    x=\frac{\kappa\dot{\phi}}{\sqrt{6}H}\,, \hspace{1cm} y=\frac{\kappa\sqrt{V}}{\sqrt{3}H}\,, \hspace{1cm}
    z=-4 \beta  \kappa^{2}\,, \hspace{1cm}
    u=-2 \beta \ln \left(\frac{T}{T_{0}}\right)\kappa^{2} \,, \label{ch4_27}\\ 
    \rho=\frac{\kappa\sqrt{\rho_{r}}}{\sqrt{3}H}\,, \hspace{1cm} 
    \lambda= -\frac{V_{,\phi}(\phi)}{\kappa V(\phi)}\,, \hspace{1cm} 
    \Theta= \frac{V(\phi)\,, V_{,\phi \phi}}{V_{,\phi}(\phi)^{2}}\,. \label{ch4_28}
\end{align}
The density parameters for different phases of the evolution of the Universe in terms of dynamical system variable are as follows,
\begin{align}
    \Omega_{DE}&=x^{2}+y^{2}+z+u\,, \label{ch4_29} \\ 
    \Omega_{r}&=\rho^{2}\,, \label{ch4_30} \\ 
    \Omega_{m}&=1-x^{2}-y^{2}-z-u-\rho^{2}\,, \label{ch4_31}
\end{align}
The Friedmann Eqs.~\eqref{ch4_12}--\eqref{ch4_13} and the variables in Eqs.~\eqref{ch4_27}--\eqref{ch4_28} would reproduce
\begin{eqnarray}\label{ch4_32}
    \frac{\dot{H}}{H^{2}}&=\frac{\rho ^2-3 \left(u-x^2+y^2+z-1\right)}{2 u+3 z-2}\,,
\end{eqnarray}
so that the deceleration parameter and EoS parameter can also be expressed in terms of dynamical variables as,
\begin{eqnarray}
    q &=&\frac{\rho ^2-u+3 x^2-3 y^2+1}{-2 u-3 z+2}\,,\\ \label{ch4_33}
    \omega_{tot}&=& \frac{2 \rho ^2+6 x^2-6 y^2+3 z}{-6 u-9 z+6}\,,\\ \label{ch4_34}
    \omega_{DE}&=& -\frac{\rho ^2 (2 u+3 z)+6 x^2-6 y^2+3 z}{3 (2 u+3 z-2) \left(u+x^2+y^2+z\right)}\,. \label{ch4_35}
\end{eqnarray}
The system of autonomous equations that governs the cosmological dynamical system is
\begin{scriptsize}
\begin{align}
    \frac{dx}{dN}&=-\frac{x\rho ^2-3 x\left(u-x^2+y^2+z-1\right)}{2 u+3 z-2}-3 x+\sqrt{\frac{3}{2}} \lambda  y^2\,, \label{ch4_36} \\ 
    \frac{dy}{dN}&=\frac{-y \rho ^2+3y \left(u-x^2+y^2+z-1\right)}{2 u+3 z-2}-\sqrt{\frac{3}{2}} \lambda y x\,, \label{ch4_37} \\
    \frac{du}{dN}&=\frac{z \rho ^2-3 z \left(u-x^2+y^2+z-1\right)}{2 u+3 z-2}\,, \label{ch4_38} \\ 
    \frac{d\rho}{dN}&=-\frac{\rho \left(\rho ^2+u+3 x^2-3 y^2+3 z-1\right)}{2 u+3 z-2}\,, \label{ch4_39} \\
    \frac{dz}{dN}&=0\,,  \label{ch4_40}\\
    \frac{d\lambda}{dN}&= -\sqrt{6}(\Theta-1)x \lambda^{2}\,. \label{ch4_41}
\end{align}    
\end{scriptsize}
To derive the dynamical features of the autonomous system, the coupled equations $x^{\prime}=0$, $y^{\prime}=0$, $z^{\prime}=0$, $u^{\prime}=0$ and $\rho^{\prime}=0$ are to be solved. The special choice of the potential energy function, $V(\phi)=V_{0}e^{-\lambda\phi}$, leads to the value of $\Theta=1$. The corresponding critical points of the above system and its description are given in Table~\ref{ch4_TABLE-I}. The stability condition and the cosmology about the value of deceleration and EoS parameter are given in Table~\ref{ch4_TABLE-II}. The cosmological solution and the corresponding scale factor are also given in Table~\ref{ch4_TABLE-III}. In this work, the eigenvalues are denoted by $\lambda_1$, $\lambda_2$, $\lambda_3$ and $\lambda_4$. 

\begin{table}[H]
\renewcommand{\arraystretch}{0.7} 
    \caption{Critical points for the dynamical system. } 
    \centering 
    \begin{tabular}{|c|c|c|c|c|c|c|} 
    \hline\hline 
    C.P. & $x_{c}$ & $y_{c}$ & $u_{c}$ & $\rho_{c}$ & $z_{c}$ & Exists for \\ [0.5ex] 
    \hline\hline 
    $A$  & $0$ & $0$ & $\alpha$ & 0 & $\beta_{3}$ & $\begin{tabular}{@{}c@{}} $\alpha=1-\beta_{3}$,\\ $ \beta_{3} \neq 0$ \end{tabular}$ \\
    \hline
    $B$ &$0$ & $0$ & $\gamma$ & $0$&0 & $ \gamma \neq 1$ \\
    \hline
    $C$ & 0 & 0 & $\sigma$ & $\tau$ & 0 & $\tau =\sqrt{1-\sigma}$, $\sigma<1$ \\
    \hline
    $D$ & $\delta$ & 0 & $\epsilon$ & 0 & 0 & $\delta \neq 0,$ $\epsilon =1-\delta ^2 $ \\
    \hline
    $E$ & 0 & $\eta$ & $\iota$ & 0 & $\xi$ & $\begin{tabular}{@{}c@{}}$\iota =-\eta ^2-\xi +1,$\\ $2 \eta ^2-\xi \neq 0$, $\lambda=0$\end{tabular}$\\
    \hline
    $F_{+}$  & $\frac{\sqrt{\frac{3}{2}}}{\lambda }$ & $\frac{\sqrt{\frac{3}{2}}}{\lambda }$ & $\mu$ & 0 & $0$ & $\mu -1\neq 0,$ $\lambda \neq 0$\\
    \hline
    $F_{-}$  & $\frac{\sqrt{\frac{3}{2}}}{\lambda }$ & $-\frac{\sqrt{\frac{3}{2}}}{\lambda }$ & $\nu$ & 0 & $0$ & $\nu -1\neq 0,$ $\lambda \neq 0$\\
    \hline
    $G$ & 0 & $\mathbf{f}$ & $\mathbf{e}$ & 0 & 0 & $\begin{tabular}{@{}c@{}} $\mathbf{e}=1-\mathbf{f}^2,$ $\mathbf{f}\neq 0,$\\ $\lambda=0$\end{tabular}$\\
    \hline
    $\mathcal{H}$ & 0 & $\mathbf{i}$& $\mathbf{h}$ & 0 & $\mathbf{j}$& $\begin{tabular}{@{}c@{}}$\lambda \neq 0,$ $\mathbf{h}=-\mathbf{i}^2-\mathbf{j}+1,$\\ $-\mathbf{j} \lambda \neq 0$\end{tabular}$\\
    [1ex] 
    \hline 
    \end{tabular}
    \label{ch4_TABLE-I}
\end{table}

\begin{table}[H]
\renewcommand{\arraystretch}{0.8} 
  \setlength{\tabcolsep}{0.5pt} 
    \caption{Stability conditions, EoS parameter and deceleration parameter } 
    \centering 
    \begin{tabular}{|c|c|c|c|c|} 
    \hline\hline 
    C. P. & Stability Conditions & $q$ & $\omega_{tot}$ & $\omega_{DE}$ \\ [0.5ex] 
    \hline\hline 
    $A$  & Stable & $-1$ & $-1$ & $-1$ \\
    \hline
    $B$  & Unstable & $\frac{1}{2}$ & 0 & 0 \\
    \hline
    $C$  &  Unstable & $1$ & $\frac{1}{3}$ & $\frac{1}{3}$ \\
    \hline
    $D$  &Unstable& $2$ & $1$ & $1$ \\
    \hline
    $E$  & Stable & $-1$ & $-1$ & $-1$ \\
    \hline
    $F_{+}$  & \begin{tabular}{@{}c@{}}Stable for \\ $\mu <1\land \bigg(-2 \sqrt{\frac{6}{7}} \sqrt{-\frac{1}{\mu -1}}\leq \lambda <-\sqrt{3} \sqrt{-\frac{1}{\mu -1}}\lor$ \\ $\sqrt{3} \sqrt{-\frac{1}{\mu -1}}<\lambda \leq 2 \sqrt{\frac{6}{7}} \sqrt{-\frac{1}{\mu -1}}\bigg)$ \end{tabular} & $\frac{1}{2}$ & $0$ & $0$ \\
    \hline
    $F_{-}$  & \begin{tabular}{@{}c@{}}Stable for \\ $\nu <1\land \bigg(-2 \sqrt{\frac{6}{7}} \sqrt{-\frac{1}{\nu -1}}\leq \lambda <-\sqrt{3} \sqrt{-\frac{1}{\nu -1}}\lor$ \\ $\sqrt{3} \sqrt{-\frac{1}{\nu -1}}<\lambda \leq 2 \sqrt{\frac{6}{7}} \sqrt{-\frac{1}{\nu -1}}\bigg)$ \end{tabular} & $\frac{1}{2}$ & $0$ & $0$ \\
    \hline
    $G$  & Stable & $-1$ & $-1$ & $-1$\\
    \hline
    $\mathcal{H}$ & \begin{tabular}{@{}c@{}}Stable for\\ $\left(\mathbf{i}<1\land \mathbf{j}>2 \mathbf{i}^2\right)\lor \left(\mathbf{i}>1\land \mathbf{j}>2 \mathbf{i}^2\right)$\end{tabular} & $\frac{\mathbf{i}^2+\mathbf{j}}{2 \mathbf{i}^2-\mathbf{j}}$ & $\frac{\mathbf{j}}{2 \mathbf{i}^2-\mathbf{j}}$ & $\frac{\mathbf{j}}{\left(\mathbf{i}^2-1\right) \left(\mathbf{j}-2 \mathbf{i}^2\right)}$ \\
    [1ex] 
    \hline 
    \end{tabular}
    \label{ch4_TABLE-II}
\end{table}

\begin{table}[H]
\renewcommand{\arraystretch}{0.4}
    \caption{Cosmological solutions of critical points } 
    \centering 
    \begin{tabular}{|c|c|c|c|} 
    \hline\hline 
    C. P. & Acceleration equation & Scale factor ( Power law solution) & Universe phase \\ [0.5ex] 
    \hline\hline 
    $A$ & $\dot{H}=0$ & $a(t)=t_{0} e^{c_{1}t}$ & de-sitter phase \\
    \hline
    $B$ & $\dot{H}=-\frac{3}{2}H^{2}$ & $a(t)= t_{0} (\frac{3}{2}t+c_{2})^\frac{2}{3}$ & matter-dominated \\
    \hline
    $C$ & $\dot{H}=-2 H^{2}$ & $a(t)= t_{0} (2 t+c_{2})^\frac{1}{2}$ & radiation-dominated \\
    \hline
    $D$ &$\dot{H}=-3 H^{2}$& $a(t)= t_{0} (3 t+c_{2})^\frac{1}{3}$ & stiff-matter \\
    \hline
    $E$ & $\dot{H}=0$ & $a(t)=t_{0} e^{c_{1}t}$ & de-sitter phase \\
    \hline
    $F_{+}$ & $\dot{H}=-\frac{3}{2}H^{2}$ & $a(t)= t_{0} (\frac{3}{2}t+c_{2})^\frac{2}{3}$ & matter-dominated \\
    \hline
    $F_{-}$ & $\dot{H}=-\frac{3}{2}H^{2}$ & $a(t)= t_{0} (\frac{3}{2}t+c_{2})^\frac{2}{3}$ & matter-dominated \\
    \hline
    $G$ &$\dot{H}=0$ & $a(t)=t_{0} e^{c_{1}t}$ & de-sitter phase \\
    \hline
    $\mathcal{H}$ & $\dot{H}=0$ & $a(t)=t_{0} e^{c_{1}t}$ & de-sitter phase \\
    [1ex] 
    \hline 
    \end{tabular}
    \label{ch4_TABLE-III}
\end{table}

\begin{figure}[H]
    \centering
    \includegraphics[width=50mm]{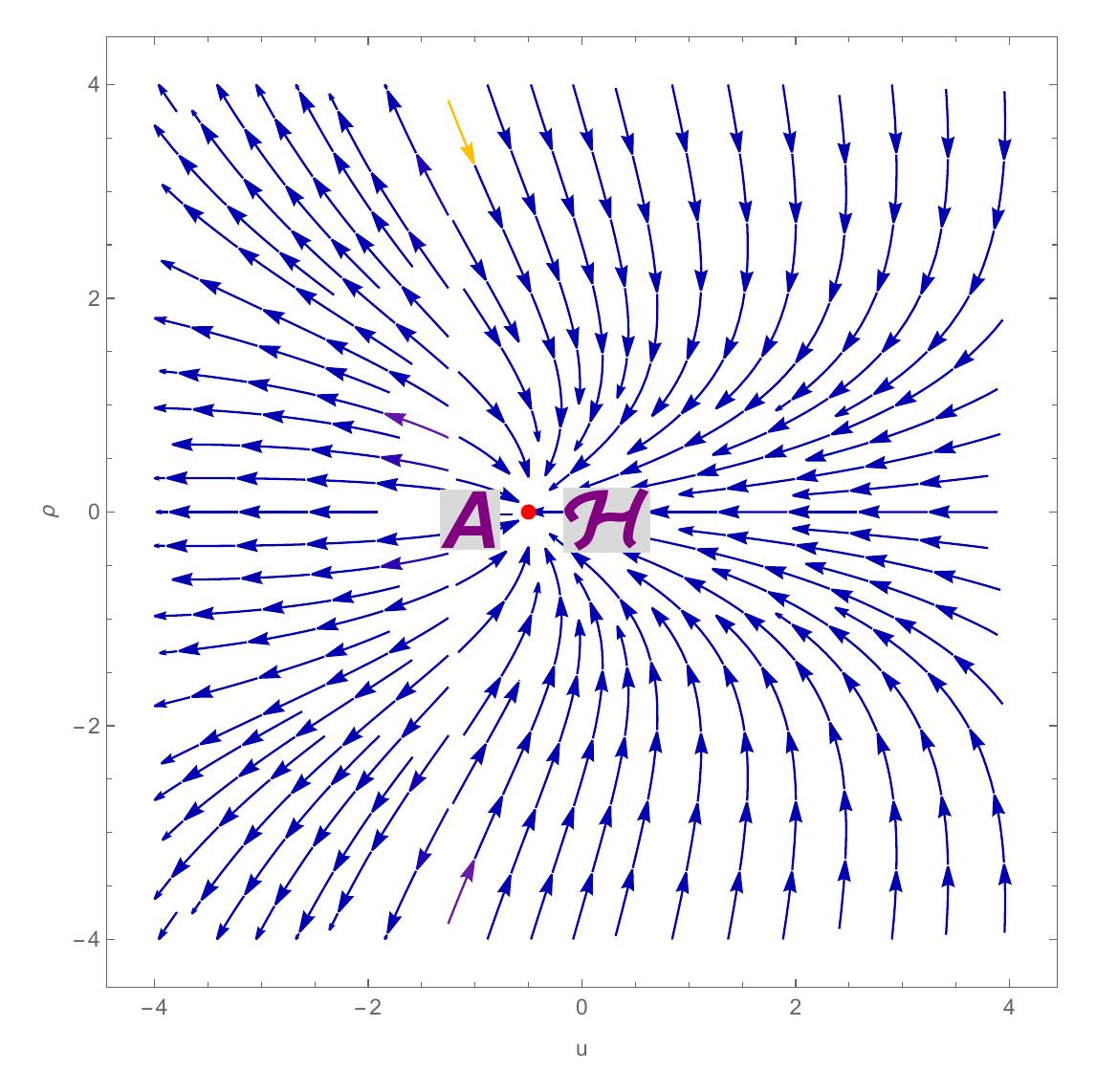}
    \includegraphics[width=50mm]{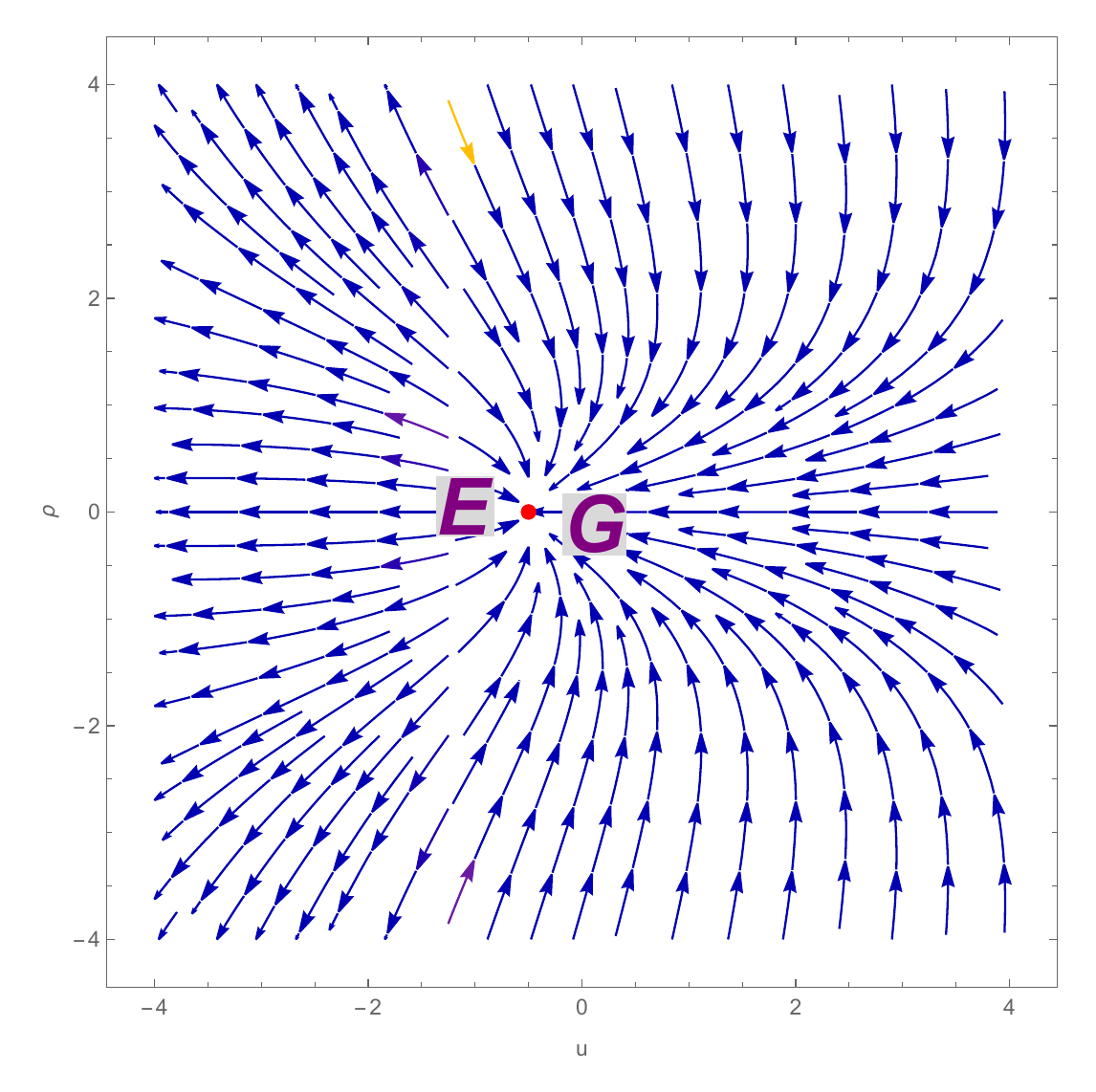}
    \includegraphics[width=50mm]{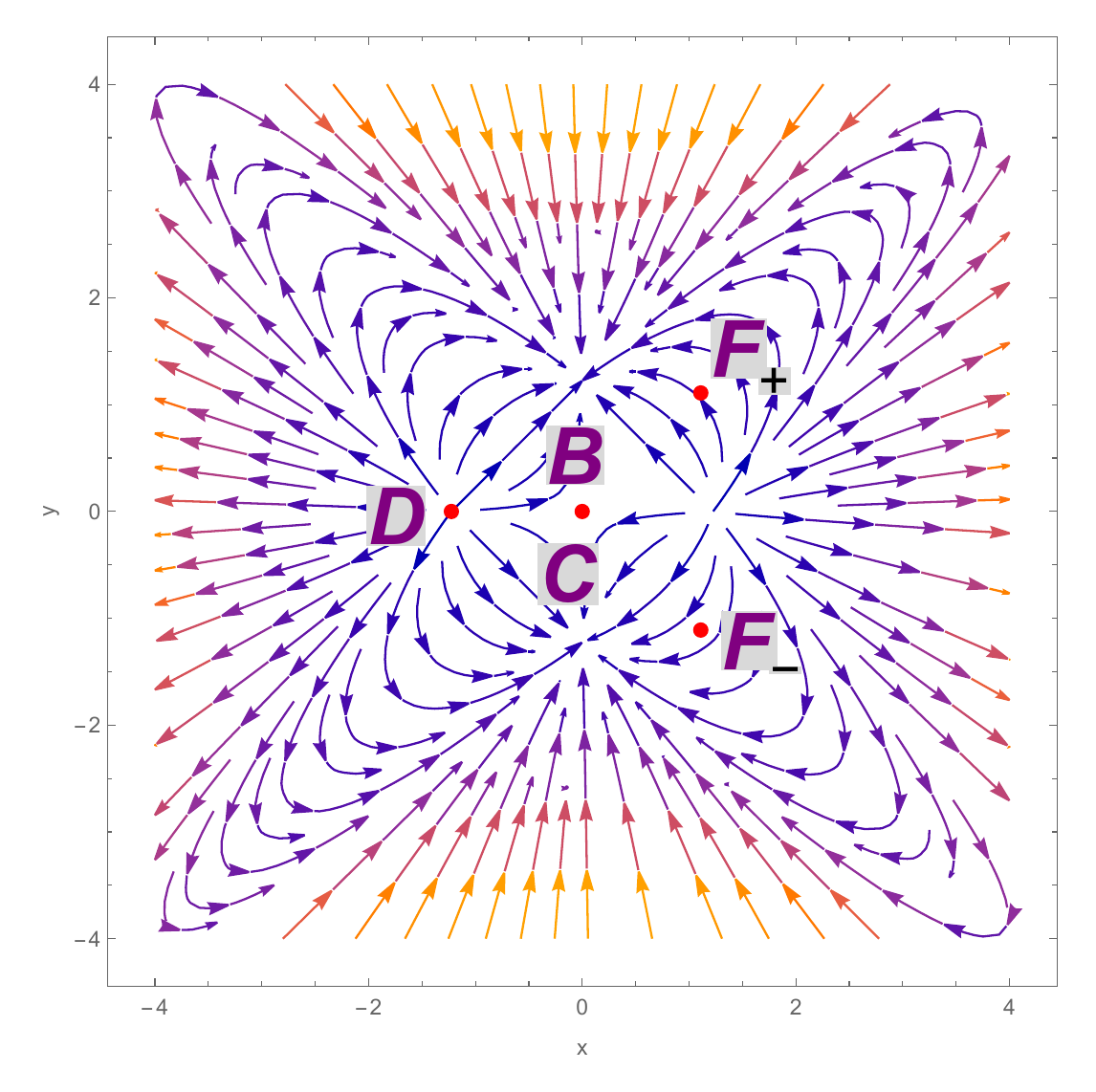}
    \caption{Phase portrait for the dynamical system of Model-I, (i) Left panel($x=0$, $y=0$, $z=1.5$, $\lambda=0.001$); (ii) Middle panel ($x=0$, $y=0$, $\lambda=0.001$) (iii) Right panel ($\rho=0$, $z=1.5$, $\lambda=0.001$.)} \label{ch4_Fig1}
\end{figure}

In the study of dynamical systems, the phase portrait is an important tool, that consists of a plot of typical trajectories in the state space. The stability of the models can be indicated through the phase portrait. Fig.~\ref{ch4_Fig1} shows the phase space portrait diagram for the dynamical system Eqs.~\eqref{ch4_36}--\eqref{ch4_41}. The Left panel shows that the trajectories of critical points $A$ and $\mathcal{H}$ move towards the fixed point, so we conclude that the point $A$ and $\mathcal{H}$ are stable nodes. Similarly, phase portrait in the Middle panel indicates that the trajectories of the critical points $E$ and $G$ move towards the fixed point, showing stable behaviors. The trajectories for the critical points $B$, $C$, $D$ and $F_{+}$,$F_{-}$ move away from the fixed points as in the Right panel. Hence, these points are unstable (saddle). Further, we have described in detail the corresponding cosmology for each critical point as below:

\begin{itemize}
    \item{\bf Critical Point $A$ :} At this point, $\Omega_{DE}=1$, $\Omega_{m}=0$ and $\Omega_{r}=0$, i.e the Universe shows DE dominated phase. The corresponding EoS parameter $\omega_{tot}=-1$ and deceleration parameter $q=-1$ confirms the accelerated DE dominated Universe. The eigenvalues of this critical point are negative real part and zero. Coley and  Aulbach \cite{Coley:1999, aulbach1984} have investigated that the dimension of the set of eigenvalues for non-hyperbolic critical points is one equal to the number of vanishing eigenvalues. As a result, the set of eigenvalues is normally hyperbolic and the critical point associated with it is stable but cannot be a global attractor. In this case, the dimension of the set of eigenvalues is one and only one eigenvalue vanishes. That means the dimension of a set of eigenvalues is equal to the number of vanishing eigenvalues. This critical point is consistent with recent observations and can explain current acceleration of the Universe. The behaviour of this critical point is a stable node.
    \begin{align}
        \{\lambda_1=-3,\quad\lambda_2=-3,\quad\lambda_3=-2,\quad\lambda_4=0\}\,. \nonumber
    \end{align}
    
    \item{\bf Critical Point $B$:} 
    This point exists for $\gamma \neq 1$ and the corresponding deceleration parameter $q=\frac{1}{2}$ and EoS parameter $\omega_{tot}=0$. This behaviour of the critical point leads to the decelerating phase of the Universe. Also, density parameters $\Omega_{DE}=\gamma$, $\Omega_r=0 $ and $\Omega_{m}=1-\gamma$. If we consider $\gamma=0$, the Universe shows the matter-dominated phase. The eigenvalues of the Jacobian matrix for this critical point are given below. The signature of the eigenvalues is both positive and negative, which means it shows unstable saddle behaviour. 
    \begin{align}
        \left\{\lambda_1=-\frac{3}{2},\quad \lambda_2=\frac{3}{2}, \quad \lambda_3=-\frac{1}{2}, \quad \lambda_4=0\right\}\,.\nonumber
    \end{align}
    
    \item {\bf Critical Point $C$:} At this point, the deceleration parameter and EoS parameter are obtained to be $q=1$ and $\omega=\frac{1}{3}$, which demonstrates the decelerating phase of the Universe. The density parameters are: $\Omega_{DE}=\sigma $, $\Omega_{r}=1-\sigma $ and $\Omega_{m}=0$. For the value of $\sigma=0$, the Universe exhibits radiation-dominated phase i.e. $\Omega_{r}=1 $. The eigenvalues of the Jacobian matrix for this critical point are given below and since it contains both negative and positive eigenvalues, this critical point is an unstable saddle.
    \begin{align}
        \{\lambda_1=-1,\quad \lambda_2=1, \quad \lambda_3=2, \quad \lambda_4=0\}\,. \nonumber
    \end{align}
    
    \item{\bf Critical Point $D$:} The value of density parameters for this point are, $\Omega_{m}=0 $, $\Omega_{r}=0 $ and $\Omega_{DE}=1 $. The EoS and deceleration parameter are respectively shown the value $q=2$ and $\omega_{tot}=1$ and so the point behaves as stiff matter and shows the decelerating behaviour. The eigenvalues are obtained to be a positive real part and zero. Due to the presence of a positive eigenvalue, this critical point is showing unstable behaviour.
    \begin{align}
        \left\{\lambda_1=0,\quad \lambda_2=1,\quad \lambda_3=3,\quad \lambda_4=\frac{6-\sqrt{6}\delta \lambda}{2}\right\}\,.\nonumber
    \end{align}

    \item{\bf Critical Point $E$:} The density parameters are $\Omega_{m}=0 $,  $\Omega_{r}=0 $ and $\Omega_{DE}=1 $, which indicates the the DE sector of the Universe. The deceleration parameter value $q=-1$ and the EoS parameter value $\omega_{tot}=-1 $ shows the accelerating behaviour of the Universe at this point. The negative and zero eigenvalues demonstrate the stable behaviour. At this point, the Universe shows the stability behaviour at the accelerating DE phase.
    \begin{align}
        \{\lambda_1=-3,\quad \lambda_2=-3,\quad \lambda_3=-2,\quad \lambda_4=0\}\,.\nonumber
    \end{align}
    
    \item {\bf Critical Point $F_{+}$:} This critical point exists for $\mu \neq 1$ and $\lambda \neq 0$. The decelerating behaviour has been observed since the value of deceleration parameter $q=\frac{1}{2}$ and the EoS parameter vanishes. The density parameters exhibit the value, $\Omega_{m}=1-\frac{3}{\lambda^{2}}-\mu $, $\Omega_{r}=0 $ and $\Omega_{DE}=\frac{3}{ \lambda^{2}}+\mu $. For $\lambda=1$ and $\mu=-3$, the critical point shows the matter-dominated era, else described as a non-standard cold DM-dominated era with $\omega_{tot}=0$. This critical point is stable if it satisfies the stability condition of  Table~\ref{ch4_TABLE-II}, otherwise, unstable saddle behaviour is due to the presence of both positive and negative eigenvalues.
    \begin{align}
        \bigg\{\lambda_1=0, \quad\lambda_2=-\frac{1}{2},\quad \lambda_3=\frac{3}{4} \bigg(-\frac{\sqrt{\lambda ^2 (\mu -1) \bigg(-7 \lambda ^2 (\mu -1)-24\bigg)}}{\lambda ^2 (\mu -1)}-1\bigg),\quad \nonumber\\ \lambda_4=\frac{3}{4} \bigg(\frac{\sqrt{\lambda ^2 (\mu -1) \bigg(-7 \lambda ^2 (\mu -1)-24\bigg)}}{\lambda ^2 (\mu -1)}-1\bigg)\bigg\}\,.\nonumber
    \end{align}
\item { \textbf{Critical Point $F_{-}$}: Similar to the critical point $F_{+}$ this critical point  exists for $\nu \neq 1$ and $\lambda \neq 0$. The value of deceleration parameter and the Eos parameter $\omega_{tot}$ are mentioned in Table~\ref{ch4_TABLE-II}. The density parameters values are, $\Omega_{m}=1-\frac{3}{ \lambda^{2}}-\nu $, $\Omega_{r}=0 $ and $\Omega_{DE}=\frac{3}{\lambda^{2}}+\nu $. For $\lambda=1$ and $\nu=-3$, the critical point indicates the matter-dominated period, else described a non-standard cold DM-dominated era with $\omega_{tot}=0$. From the stability criteria, it is clear that this critical point represents stable behaviour if it satisfies the stability condition which is mentioned in Table~\ref{ch4_TABLE-II}. Otherwise, it exhibits unstable saddle behavior because both positive and negative eigenvalues are present.
    \begin{align}
        \bigg\{\lambda_1=0, \quad \lambda_2=-\frac{1}{2},\quad \lambda_3=\frac{3}{4} \bigg(-\frac{\sqrt{\lambda ^2 (\nu -1) \bigg(-7 \lambda ^2 (\nu -1)-24\bigg)}}{\lambda ^2 (\nu -1)}-1\bigg),\quad \nonumber\\ \lambda_4=\frac{3}{4} \bigg(\frac{\sqrt{\lambda ^2 (\nu -1) \bigg(-7 \lambda ^2 (\nu -1)-24\bigg)}}{\lambda ^2 (\nu -1)}-1\bigg)\bigg\}\,.\nonumber
    \end{align}}\\
    The definition of dimensionless variable $y$ as described in Eq.\eqref{ch4_27} allows us to study the different phases of the Universe evolution. The critical points with the condition on $y$ as if $y>0$ it corresponds to the positive Hubble parameter and can explain the expanding Universe. While the critical points with $y<0$ correspond to the $H<0$ describe the contracting phase of the Universe \cite{dutta2018jcap}. We denote the subscripts $+$ or $-$ corresponding to the
critical point $F$ with $y>0$ or $y<0$.
    \item{ \bf Critical Point $G$:} Here, we obtained $\Omega_{m}=0 $, $\Omega_{r}=0 $ and $\Omega_{DE}=1 $, which shows the DE era of the evolution. The deceleration parameter value $q=-1$ confirms the accelerating behaviour whereas the EoS parameter value $\omega_{tot}=-1$ shows the $\Lambda$CDM like behaviour. The stability of the critical point has been confirmed from the eigenvalues.
    \begin{align}
        \{\lambda_1=0,\quad \lambda_2=-3,\quad \lambda_3=-3,\quad \lambda_4=-2\}\,.\nonumber
    \end{align}
    
    \item{\bf Critical Point $\mathcal{H}$:} It describes the DE dominated phase as, $\Omega_{m}=0 $, $\Omega_{r}=0 $ and $\Omega_{DE}=1 $. The accelerating behaviour and the EoS parameter depend on the relation of $\mathbf{i}$ and $\mathbf{j}$ as described in Table~\ref{ch4_TABLE-II}. For, $\mathbf{j}>2\mathbf{i}^{2}$, the deceleration parameter and EoS parameter exhibit the accelerating phase of the Universe. The eigenvalues, as given below, indicate that there is a region in the parameter space where this point are stable nodes and attractor. Since this is a de-Sitter solution, the values of the parameter listed in Table~\ref{ch4_TABLE-II} will experience an accelerated expansion. The stability behaviour can be observed for $\mathbf{j}>2\mathbf{i}^{2}$.
    \begin{align}
        \left\{\lambda_1=\frac{3 \mathbf{i}^2}{2 \mathbf{i}^2-\mathbf{j}},\quad\lambda_2=-\frac{\mathbf{i}^2-2 \mathbf{j}}{2 \mathbf{i}^2-\mathbf{j}},\quad\lambda_3=-\frac{3 \left(\mathbf{i}^2-\mathbf{j}\right)}{2 \mathbf{i}^2-\mathbf{j}},\quad \lambda_4=-\frac{3 \mathbf{j}^2}{\left(2 \mathbf{i}^2-\mathbf{j}\right)^2}\right\}\,. \nonumber
    \end{align}
\end{itemize}

\begin{figure}[H]
    \centering
    \includegraphics[width=80mm]{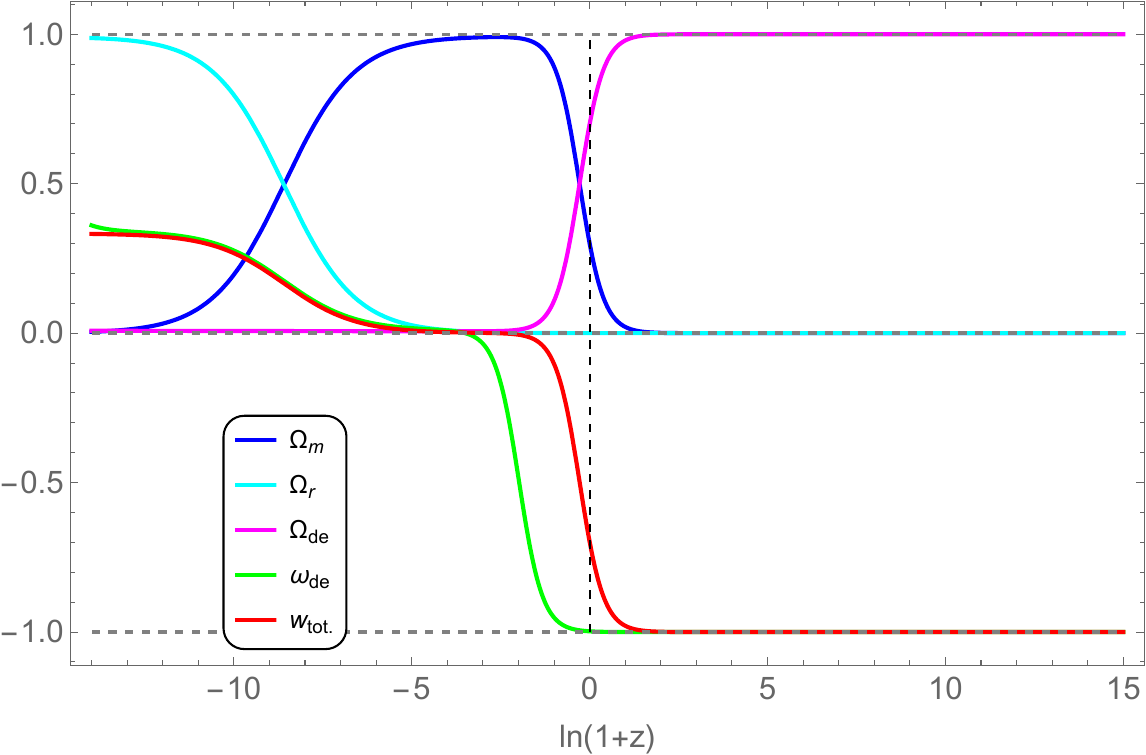}
    \caption{The evolution of the density parameters as well as of the equation-of-state parameter, as functions of the redshift, for the case, $\lambda=0.001$ with the initial condition of dynamical system variables: $x=10^{-4}$, $y=10^{-6}$, $u=0.7 \times 10^{-2}$, $\rho=0.933254$, $z=10^{-4}$, which are representative for their definitions in Eqs.~(\ref{ch4_27},\ref{ch4_28}). The vertical
dashed line denotes the present cosmological time ($z=0$). } \label{ch4_Fig2}
\end{figure}

 The critical points $A$, $E$, $G$ and $\mathcal{H}$ are the last four attractors we found when DE was in charge and the Universe is accelerating. In addition, we have found that the critical points  $B$, $F_{+}$ and $F_{-}$ show a matter-dominated phase and point $C$ represents a radiation-dominated phase of the Universe and observed that the radiation and matter dominated critical points show unstable behavior. In Fig.~\ref{ch4_Fig2} we plot the behavior of the energy densities of DE, DM and radiation, as well as the total EoS ($\omega_{tot.}$) and the EoS of DE ( $\omega_{DE}$) as functions of the redshift. Conveniently, we employ the redshift $z=\frac{a_{0}}{a} -1$ (with $a_{0} = 1$ as the current scale factor) as an independent variable. As is standard, $z = 0$ represents the present time of the Universe. The vertical dashed line in Fig.~\ref{ch4_Fig2} denotes the present cosmological time \cite{Bahamonde:2017ize}. In  Fig.~\ref{ch4_Fig2} we can observe that the cosmos is initially dominated by radiation, then transitions to DM dominance and eventually ends up being dominated by DE. As mentioned above, the Universe provides a scaling-accelerating solution, where the DM and DE density parameters remain around $0.3$ and $0.7$ respectively. Also, It is observed that the $\omega_{tot.} \approx -0.75$ and $\omega_{DE} \approx -1$ at the current time $z = 0$, which is consistent with the
observational constraint from Planck data \cite{Planck:2018vyg}. In Fig.~\ref{ch4_Fig2} we can observe that the Universe first dominated by the radiation era (Cyan curve), followed by a brief phase of matter dominance (Blue curve) and finally the cosmological constant (Pink curve). This behaviour of the density parameter indicates that the present Universe is dominated by DE. The EoS parameter (Red curve) begins with radiation at $\frac{1}{3}$, falls to $0$ during the matter-dominated period and finally rises to $-1$ leading to the $\Lambda$CDM model, which is a candidate for DE models.

\subsection{Model II}\label{ch4_sec:model_2}
In this case, we consider the form of $G(T)$ as, $G(T)= T+\alpha T^{2} $, where $\alpha$ is a constant \cite{Fortes:2021ibz}, which is a small generalization beyond TEGR.  For $\alpha=0$ the model reduces to the TEGR model. The Klein-Gordon equation in this case is the same as in Eq.~(\ref{ch4_22}) and for this $G(T)$, Eqs.~\eqref{ch4_17}-\eqref{ch4_18} become
\begin{eqnarray}
    \rho_{DE}&=&\frac{\dot{\phi^{2}}}{2}+V(\phi)-T(1+3T \alpha)\,, \label{ch4_42}\\
    p_{DE}&=&\frac{\dot{\phi^{2}}}{2}-V(\phi)+T(1+3 T \alpha)+4 \dot{H}(1+6 T \alpha)\,. \label{ch4_43}
\end{eqnarray}
To create the dynamical system, dynamical variables can be specified through the following:
\begin{align}
    x=\frac{\kappa\dot{\phi}}{\sqrt{6}H}\,, \hspace{1cm} y=\frac{\kappa\sqrt{V}}{\sqrt{3}H}\,, \hspace{1cm}
    z=-2 \kappa^{2}\,,  \hspace{1cm} 
    u=-36 H^{2} \alpha \kappa^{2}\,, \label{ch4_44} \\ 
    \rho=\frac{\kappa\sqrt{\rho_{r}}}{\sqrt{3}H}\,,  \hspace{1cm} 
    \lambda= -\frac{V_{,\phi}(\phi)}{\kappa V(\phi)}\,,  \hspace{1cm} 
    \Theta= \frac{V(\phi) V_{,\phi \phi}}{V_{,\phi}(\phi)^{2}}\,. \label{ch4_45}
\end{align}
The dimensionless variables defined in Eqs.~\eqref{ch4_44}-\eqref{ch4_45} also satisfy the constraint equation of density parameters. The EoS parameter and deceleration parameter can be expressed in the form of dimensionless variables as,
\begin{align}
    q&= -1- \frac{3 x^{2}-3 y^{2}-3z-3u+3+\rho^{2}}{-2+2z+4u}\,, \label{ch4_46} \\
    \omega_{tot}&= -1- \frac{2(3 x^{2}-3 y^{2}-3z-3u+3+\rho^{2})}{3(-2+2z+4u)}\,, \label{ch4_47} \\
    \omega_{DE} &= -\frac{3 \left(u+x^2-y^2\right)+\rho ^2 (2 u+z)}{3 (2 u+z-1) \left(u+x^2+y^2+z\right)}\,. \label{ch4_48}
\end{align}
Subsequently, the corresponding dynamical system can be obtained as,
\begin{align}
    \frac{dx}{dN}&=-\frac{x \left(\rho ^2-3 \left(u-x^2+y^2+z-1\right)\right)}{2 (2 u+z-1)}-3 x+\sqrt{\frac{3}{2}} \lambda y^2\,, \label{ch4_49} \\ 
    \frac{dy}{dN}&=-\frac{1}{2} y \left(\frac{\rho ^2-3 \left(u-x^2+y^2+z-1\right)}{2 u+z-1}+\sqrt{6} \lambda x\right)\,, \label{ch4_50} \\ 
    \frac{du}{dN}&=\frac{u \left(\rho ^2-3 \left(u-x^2+y^2+z-1\right)\right)}{2 (2 u+z-1)}\,, \label{ch4_51} \\
    \frac{d\rho}{dN}&=-\frac{\rho \left(\rho ^2+5 u+3 x^2-3 y^2+z-1\right)}{2 (2 u+z-1)}\,, \label{ch4_52} \\ 
    \frac{dz}{dN}&=0\,, \label{ch4_53} \\
    \frac{d\lambda}{dN}&= -\sqrt{6}(\Theta-1)x \lambda^{2}\,. \label{ch4_54}
\end{align}

\begin{figure}[H]
    \centering
    \includegraphics[width=50mm]{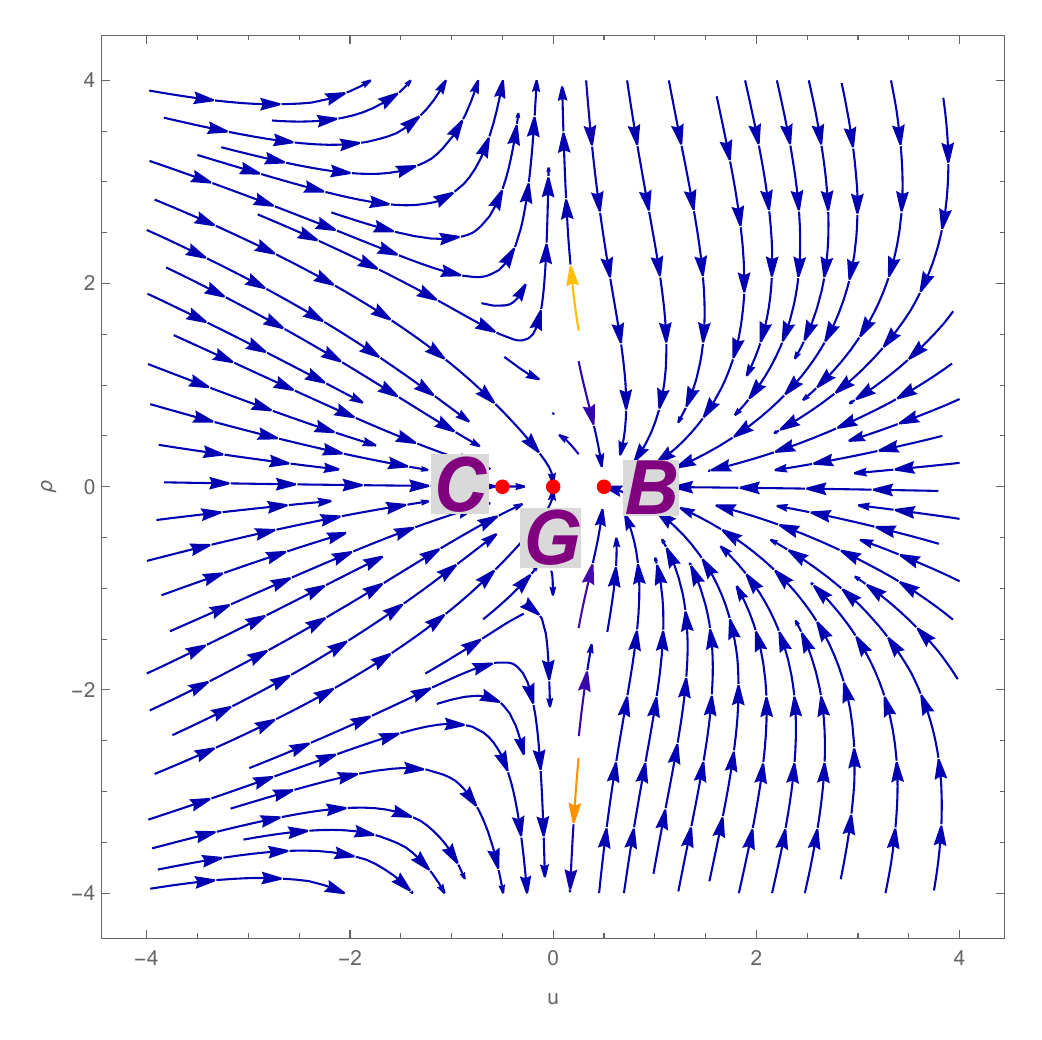}
    \includegraphics[width=50mm]{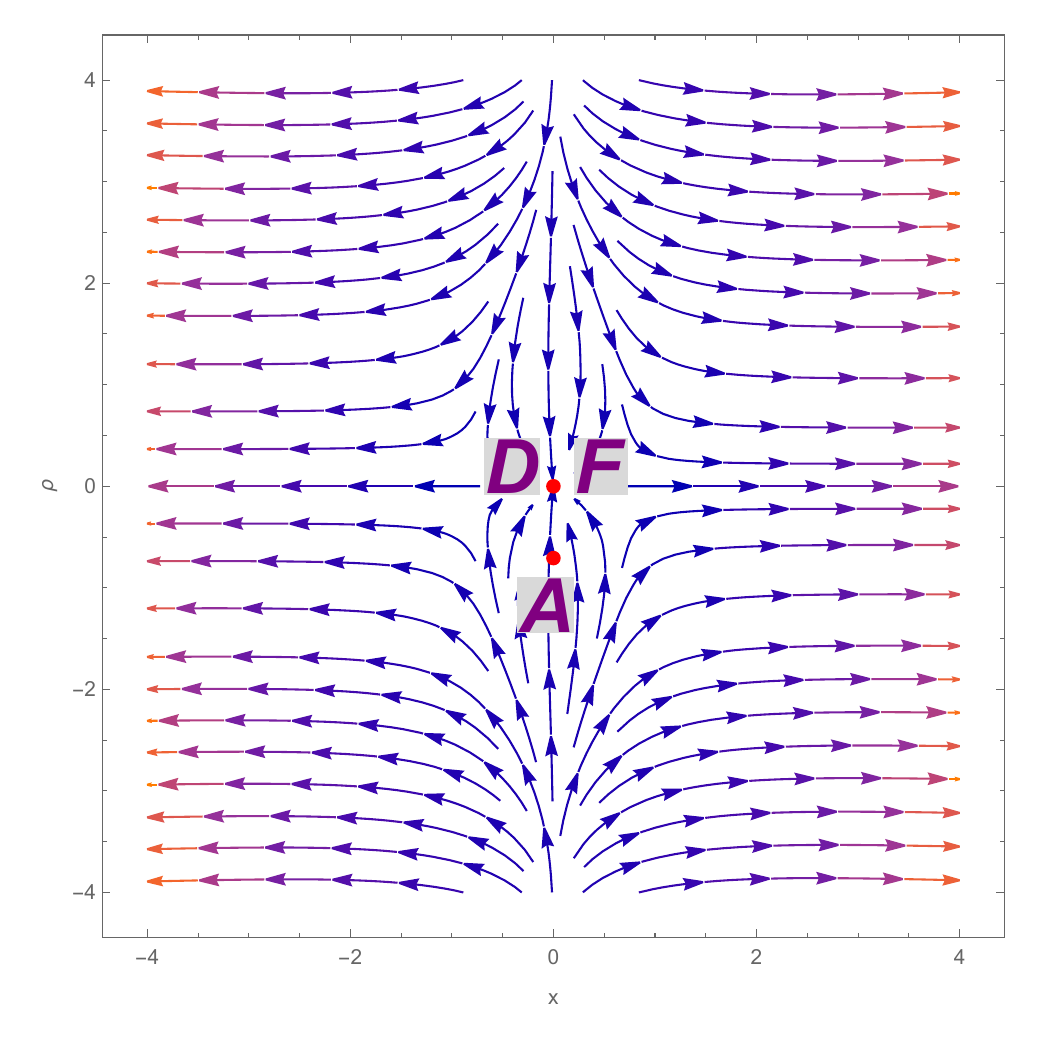}
    \includegraphics[width=50mm]{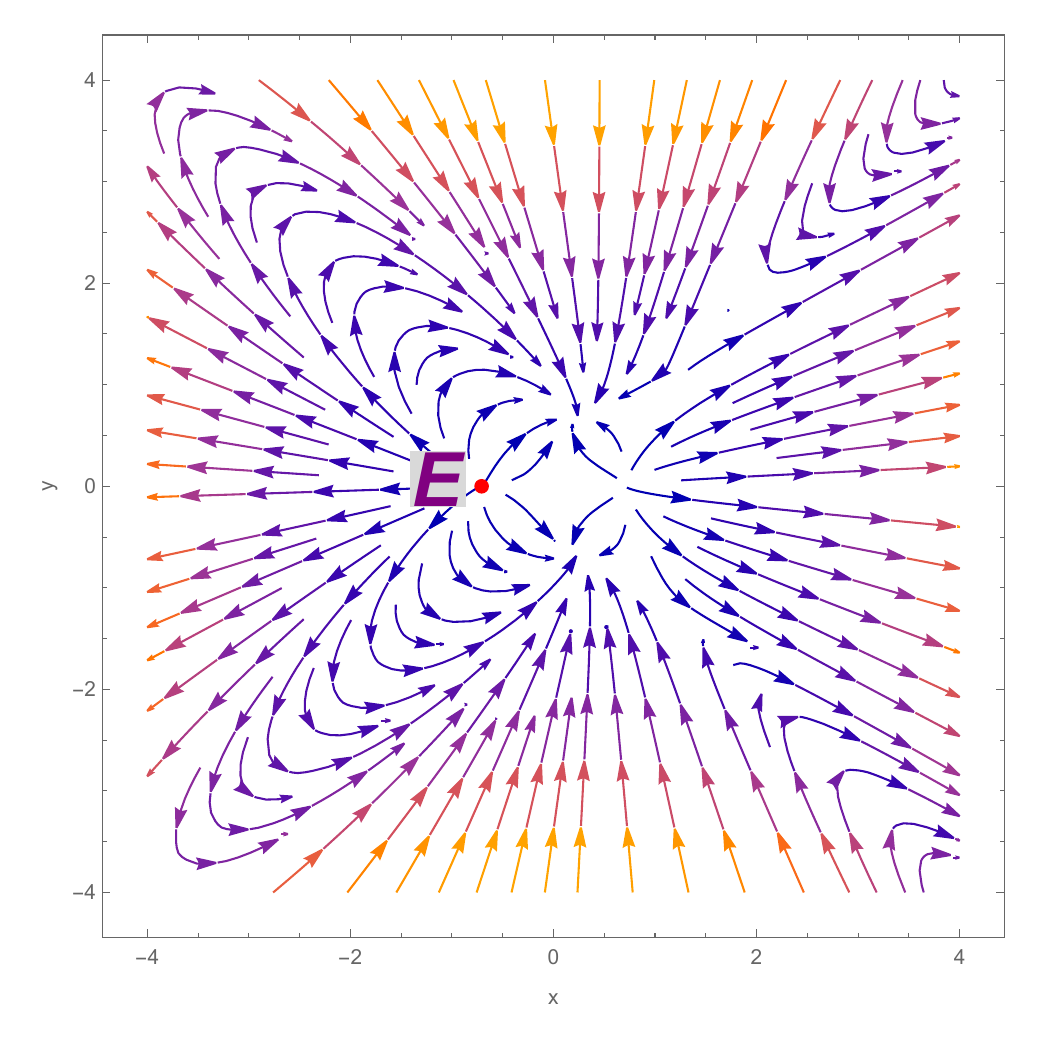}
    \caption{Phase portrait for the dynamical system of Model-II (i) Left panel ($x=0$, $y=0$, $z=0.5$, $\lambda=0.001$) ; (ii) Middle panel ($y=0$, $u=0$, $z=0.5$, $\lambda=0.001$) (ii) Right panel ($u=0$, $\rho=0$, $z=0.5$, $\lambda=0.001$).} \label{ch4_Fig3}
\end{figure}

Using the same approach as in Model I, the critical points of the autonomous dynamical system Eqs.~(\ref{ch4_49}--\ref{ch4_54}) are listed in Table~\ref{ch4_TABLE-IV}. 
\begin{table}[H]
 \renewcommand{\arraystretch}{0.5} 
    \caption{Critical Points for Dynamical System. } 
    \centering 
    \begin{tabular}{|c|c|c|c|c|c|c|} 
    \hline\hline 
    Critical Points & $x_{c}$ & $y_{c}$ & $u_{c}$ & $\rho_{c}$ & $z_{c}$ & Exists for \\ [0.5ex] 
    \hline\hline 
    $A$ & 0 & 0 & 0 & $\gamma_{1}$ & $\beta_{2}$ & $\begin{tabular}{@{}c@{}} $\gamma_{1}=-\sqrt{1-\beta_{2}}$,\\ $ \beta_{2}<1$ \end{tabular}$ \\
    \hline
    $B$ &0 & 0 & $\gamma_{2}$ & 0 &$\gamma $& $\begin{tabular}{@{}c@{}} $\gamma_{2}=1-\gamma$,\\ $-1+ \gamma \neq 0$ \end{tabular}$ \\
    \hline
    $C$ & 0 & $\xi$ & $\tau$ & 0 & $\sigma$ & $\begin{tabular}{@{}c@{}} $-1+2 \xi^{2}+\sigma \neq 0$,\\ $ \lambda=0$ \end{tabular}$ \\
    \hline
    $D$ & 0 & 0 & 0 & 0 & $\epsilon$ & $\epsilon \neq 1 $ \\
    \hline
    $E$ & $\gamma_{3}$ & 0 & 0 & 0 & $\alpha_{1}$ &  $\begin{tabular}{@{}c@{}}$\gamma_{3} =-\sqrt{1-\alpha_{1}},$\\ $ \alpha_{1}<1$ \end{tabular}$\\
    \hline
    $F$ & 0 & $\alpha_{2}$ & $\gamma_{4}$ & 0 & $\alpha_{3}$ & $\begin{tabular}{@{}c@{}}$\gamma_{4} =1-\alpha_{2}^{2}-\alpha_{3},$\\ $ -1+\alpha_{3} \neq 0,$ $\lambda \neq 0$ \end{tabular}$\\
    \hline
    $G$ & 0 & $\gamma_{5}$ & 0 & 0 & $\beta_{1}$ & $\begin{tabular}{@{}c@{}} $-1+\beta_{1} \neq 0,$\\ $\lambda=0$\end{tabular}$\\
    [0.5ex] 
    \hline 
    \end{tabular}
    \label{ch4_TABLE-IV}
\end{table} 
\begin{table}[H]
 \renewcommand{\arraystretch}{0.6}
    \caption{Stability conditions, EoS Parameter and deceleration parameter } 
    \centering 
    \begin{tabular}{|c|c|c|c|c|} 
    \hline\hline 
    C. P. & Stability Conditions & $q$ & $\omega_{tot}$ & $\omega_{DE}$ \\ [0.5ex] 
    \hline\hline 
    $A$ & \begin{tabular}{@{}c@{}}Stable for \\ $\frac{2}{5} < \beta_{2} < 1$ \end{tabular} & $1$ & $\frac{1}{3}$ & $\frac{1}{3}$ \\
    \hline
    $B$ & Stable &$-1$ &$ -1 $ & $-1$ \\
    \hline
    $C$ & Stable & $-1$ & $-1$ & $-1$ \\
    \hline
    $D$ & \begin{tabular}{@{}c@{}}Stable for \\ $\frac{2}{3} < \epsilon < 1$ \end{tabular}& $\frac{1}{2}$ & $0$ & $0$ \\
    \hline
    $E$ & Unstable & $2$ & $1$ & $1$ \\
    \hline
    $F$ & \begin{tabular}{@{}c@{}}Stable for \\ $\alpha _3>2 \alpha _2^2$ \end{tabular} & $\frac{-\alpha _2^2-2 \alpha _3+2}{-4 \left(-\alpha _2^2-\alpha _3+1\right)-2 \alpha _3+2}$ & $\frac{3 \left(-\alpha _2^2-\alpha _3+1\right)}{-6 \left(-\alpha _2^2-\alpha _3+1\right)-3 \alpha _3+3}$ & $\frac{\alpha _2^2+\alpha _3-1}{\left(\alpha _2^2-1\right) \left(2 \alpha _2^2+\alpha _3-1\right)}$ \\
    \hline
    $G$ & Stable & $-1$ & $-1$ & $-1$ \\
    [1ex] 
    \hline 
    \end{tabular}
    \label{ch4_TABLE-V}
\end{table}
For each critical point, the stability condition and to understand the corresponding cosmology, the deceleration and EoS parameter values are listed in Table~\ref{ch4_TABLE-V}. In Table~\ref{ch4_TABLE-VI}, the scale factor and the evolutionary phase of each critical point have been listed. Further to observe the stability behavior of the critical points, the phase portrait is given in Fig.~\ref{ch4_Fig3}.

\begin{table}[H]
 \renewcommand{\arraystretch}{0.5}
    \caption{Cosmological solutions of critical points } 
    \centering 
    \begin{tabular}{|c|c|c|c|} 
    \hline\hline 
    C. P. & Acceleration equation & Scale factor(Power law solution) & Universe phase \\ [0.5ex] 
    \hline\hline 
    $A$ & $\dot{H}=-2 H^{2}$ & $a(t)= t_{0} (2 t+c_{2})^\frac{1}{2}$ & radiation-dominated \\
    \hline
    $B$ & $\dot{H}=0$ & $a(t)=t_{0} e^{c_{1}t}$ & de-sitter phase \\
    \hline
    $C$ & $\dot{H}=0$ & $a(t)=t_{0} e^{c_{1}t}$ & de-sitter phase \\
    \hline
    $D$ &$\dot{H}=-\frac{3}{2}H^{2}$ & $a(t)= t_{0} (\frac{3}{2}t+c_{2})^\frac{2}{3}$ & matter-dominated \\
    \hline
    $E$ & $\dot{H}=-2 H^{2}$ & $a(t)= t_{0} (3 t+c_{2})^\frac{1}{3}$ & stiff-matter   \\
    \hline
    $F$ & $\dot{H}=0$ & $a(t)=t_{0} e^{c_{1}t}$ & de-sitter phase \\
    \hline
    $G$ &$\dot{H}=0$ & $a(t)=t_{0} e^{c_{1}t}$ & de-sitter phase \\
    \hline 
    \end{tabular}
    \label{ch4_TABLE-VI}
\end{table}

\begin{figure}[H]
    \centering
    \includegraphics[width=80mm]{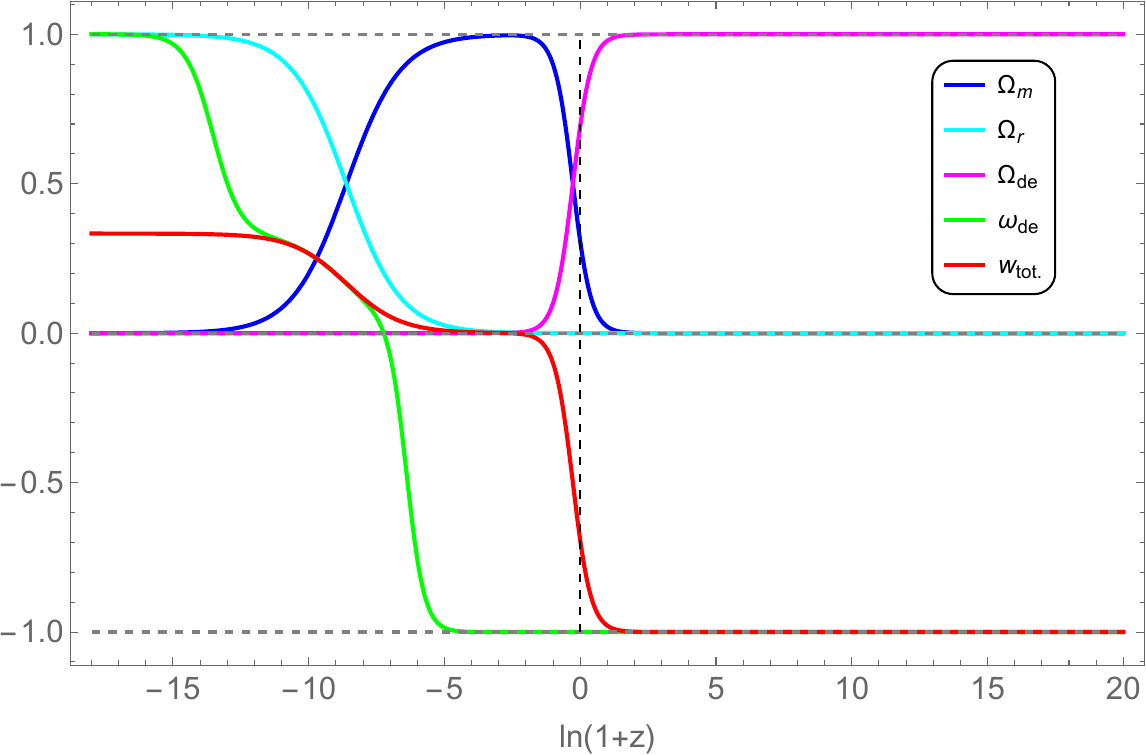}
    \caption{The evolution of the density parameters ($\Omega_{DE}$), ($\Omega_{m}$) and, ($\Omega_{r}$)  as well as of the equation-of-state parameter, as functions of the redshift, $\lambda=0.001$ with the initial conditions of dynamical system variables: $x=10^{-6}$, $y=10^{-6}$, $u=10^{-15}$, $\rho=0.933234$, $z=10^{-8}$. The vertical
dashed line denotes the present cosmological time ($z=0$).} \label{ch4_Fig4}
\end{figure}

The phase portrait diagram Fig.~\ref{ch4_Fig3} displays the critical points. Plots of these phase space trajectories are shown for the dynamical system indicated in Eqs.~\eqref{ch4_49}--\eqref{ch4_54}. The Left panel plot shows that the phase space trajectories are moving towards the critical points $B$, $C$ and $G$ hence these points represent stability with stable node point behaviour. If the critical points $A$, $D$ and $F$ satisfy the stability condition given in Table~\ref{ch4_TABLE-IV}, then phase space trajectories are moving towards the critical points $A$, $D$ and $F$. Otherwise, phase portraits are moving away from these critical points Middle panel, we can observe that the critical point $A$, $D$ and $F$ are showing unstable behaviour. The Right panel phase portrait shows that the critical point $E$ trajectories deviate from the fixed point, indicating unstable behaviour. Additionally, we have included detailed descriptions of the associated cosmology at each critical point, below:
\begin{itemize}
    \item{\bf Critical Point A :} The density parameters for this point are $\Omega_{m}=0$, $\Omega_{r}= 1-\beta_{2}$ and $\Omega_{DE}=\beta_{2}$. The behaviour depends on the value of the parameter $\beta_{2}$. For $\beta_{2}=0$, the critical point satisfies the radiation dominated phase. The positive deceleration parameter shows the decelerating phase of the Universe and the EoS parameter yields the value, $\omega_{tot}=\frac{1}{3}$. The eigenvalues for this critical point are given below, which can be interpreted as if the parameter $\beta$ satisfies the stability condition mentioned in Table~\ref{ch4_TABLE-IV}, then this critical point is stable, otherwise unstable.
    \begin{align}
        \left\{\lambda_1=-\frac{\beta_{2} (\beta_{2} +2)}{(3 \beta_{2} -2)^2},\quad \lambda_2=-\frac{2}{3 \beta_{2} -2},\quad \lambda_3=\frac{4 (\beta_{2} -1)}{3 \beta_{2} -2},\quad \lambda_4=-\frac{5 \beta_{2} -2}{3 \beta_{2} -2}\right\}\,. \nonumber
    \end{align}
    
    \item{\bf Critical Point B :} Both the deceleration parameter and EoS parameter are showing the accelerating $\Lambda$CDM like behaviour. The DE phase has been confirmed from the density parameters, which are $\Omega_{m} = 0$, $\Omega_{r} = 0$ and $\Omega_{DE} = 1$. The eigenvalues are either negative or zero, hence it confirms the stability behaviour.
    \begin{align}
        \{\lambda_1=-3,\quad \lambda_2=-3,\quad \lambda_3=-2,\quad \lambda_4=0\}\,.\nonumber
    \end{align}
    
    \item{\bf Critical Point C :} Similar behaviour has been obtained for this point as in the critical point $B$, i.e. the accelerating $\Lambda$CDM like behaviour. The nature of the eigenvalues confirms the stability.
    \begin{align}
        \{\lambda_1=-3,\quad \lambda_2=-3,\quad \lambda_3=-2,\quad \lambda_4=0\}\,.\nonumber
    \end{align}
    
    \item{\bf Critical Point D :} This point exists for $\epsilon \neq 1$. For this condition, the vanishing EoS parameter shows the matter-dominated Universe with the deceleration parameter $q=\frac{1}{2}$. Hence, the density parameters $\Omega_{m}=1-\epsilon$ and $\Omega_{DE}=\epsilon$. From the eigenvalues of the critical point, we can conclude that for $\frac{2}{3}<\epsilon<1$, it shows the stability, otherwise the unstable behavior.
    \begin{align}
        \left\{\lambda_1=-\frac{3 \epsilon ^2}{(3 \epsilon -2)^2},\quad \lambda_2=\frac{3 (\epsilon -1)}{3 \epsilon -2},\quad \lambda_3=-\frac{3 (2 \epsilon -1)}{3 \epsilon -2},\quad \lambda_4=-\frac{3 \epsilon -1}{3 \epsilon -2}\right\}\,.\nonumber
    \end{align}
    
    \item{\bf Critical Point E :} At this point, $\Omega_{m}=0$, $\Omega_{r}=0$ and $\Omega_{DE}=1$ with $\omega_{tot}=1$ and $q=2$. The behavior of this critical point is always unstable due to the presence of positive and negative eigenvalues. At the point when DE dominates the Universe, the EoS parameter reduces to a stiff fluid and there is no sign of acceleration.

\begin{eqnarray}
&\Bigg\{ 
\lambda_1=-\frac{2}{3 \alpha _1-2}, \quad \lambda_2=\frac{-2 \sqrt{6} \sqrt{1-\alpha _1} \lambda +3 \sqrt{6} \alpha _1 \sqrt{1-\alpha _1} \lambda + 12 \alpha _1-12} 
  {2 (3 \alpha _1-2)},\nonumber \\& \lambda_3= \frac{3 \bigg(2+2 \alpha _1^2-5 \alpha _1-\sqrt{7 \alpha _1^4- 28 \alpha _1^3+37 \alpha _1^2-20 \alpha _1+4}\bigg)}
  {(3 \alpha _1-2)^{2}}, 
 \lambda_4=\frac{3 \bigg(2+2 \alpha _1^2-5 \alpha _1+ \sqrt{7 \alpha _1^4-28 \alpha _1^3+37 \alpha _1^2-20 \alpha _1+4}\bigg)}
  {(3 \alpha _1-2)^2} 
\Bigg\} \,. \nonumber
\end{eqnarray}
\item{\bf Critical Point $F$:} The solution to this critical point is $\Omega_{r}=0$, $\Omega_{DE}=1- \alpha_{2}^{2}$ and $\Omega_{m}=\alpha_{2}^{2}$ with the EoS and and deceleration parameter are as in Table~\ref{ch4_TABLE-V}. The EoS parameter satisfying this condition $\alpha _3<1-2 \alpha _2^2$. It is interesting to note that in this case, the final value of $\omega_{tot}$ ranges between -$\frac{1}{3}$ to $-1$. For this condition, the EoS parameter and deceleration parameters indicate the accelerated phase of the Universe. For $\alpha_{2}=0$, the critical point indicates a period where the Universe is dominated by a DE era ($\Omega_{DE}=1$). Also, the behaviour of the EoS and deceleration parameters for $\alpha_{2}= 0$ shows an accelerating phase of the Universe. The critical point is stable for $\alpha _3>2 \alpha _2^2$ and for this condition, all the eigenvalues are negative, which confirms the stability behaviour.
    \begin{align}
        \left\{\lambda_1=\frac{3 \alpha _2^2}{2 \alpha _2^2-\alpha _3},\quad \lambda_2=-\frac{\alpha _2^2-2 \alpha _3}{2 \alpha _2^2-\alpha _3},\quad \lambda_3=-\frac{3 \left(\alpha _2^2-\alpha _3\right)}{2 \alpha _2^2-\alpha _3},\quad \lambda_4=-\frac{3 \alpha _3^2}{\left(2 \alpha _2^2-\alpha _3\right){}^2}\right\}\,. \nonumber
    \end{align}
    
    \item{\bf Critical Point $G$:} As the values of the density parameters, deceleration parameter and EoS parameter become same as that of the critical point $B$ and $C$ and also the eigenvalues, therefore the behaviour of this critical point $G$ remains same as that of $B$ and $C$.
    \begin{align}
        \{\lambda_1=-3,\quad \lambda_2=-3,\quad \lambda_3=-2,\quad \lambda_4=0\}\,.\nonumber
    \end{align}
\end{itemize}
The critical points $B$, $C$, $F$ and $G$ are representing the DE sector and showing late-time cosmic acceleration behaviour of the Universe. These critical points show the attractor phase (stable). The critical points $A$ and $D$ indicate the matter and radiation phase respectively and show unstable behaviour of the Universe. In Fig.~\ref{ch4_Fig4}, the evolution of the energy densities as well as EoS parameter as a function of redshift has been shown. The EoS parameter ($\omega_{tot}$) (Red curve) of the cosmos together with the relative energy densities of DM ($\Omega_{m}$), radiation ($\Omega_{r}$) and DE ($\Omega_{DE}$) are shown. The evolution shows the radiation phase (Cyan curve), followed by a brief period of domination by the matter (Blue curve) and after that the domination of the DE sector (Pink curve). We observe that the Universe transitions from a matter dominated phase to an acceleration era at late times. The present value of the DM and DE density parameters remain respectively, around $0.3$ and $0.7$ at $z=0$. Also, we have found $\omega_{tot.}\approx -0.76$ and $\omega_{DE}\approx -1$ at the present cosmic time. The EoS parameter approaches $-1$ leading to the $\Lambda$CDM behaviour of the model. 

\section{Conclusion} \label{ch4_SECIV}
In this Chapter, we explored the dynamical systems analysis of two particular models within the general class of scalar-tensor theories coupled with the torsion scalar, as prescribed in Eq.~\eqref{ch4_11}. The effective Friedmann and Klein-Gordon equations provided in Eqs.~(\ref{ch4_12}-\ref{ch4_14}) describe fully the background dynamics of the system but are beyond analytic techniques and so we explore their dynamics using dynamical systems analysis. Models in this class of theories may offer some advantages such as the scalar field and torsion scalar freedoms being associated with different epochs of the evolution of the Universe, or with different mechanisms within the Universe.

The scalar field is ultimately described canonically with an exponential potential. On the other hand, building on the proposals in Ref.~\cite{Zhang2011_jacp}, we use logarithmic and power-law models to describe the form of the torsion scalar term beyond TEGR. These were first probed in an $f(T)$ gravity context in Ref.~\cite{Zhang2011_jacp} where they were found to have some advantageous properties that were correlated with the evolution of the Universe. Adding a scalar field may produce more realistic cosmology since scalar fields have been suggested to be responsible for a variety of different mechanisms in the Universe such as inflation and late-time accelerated expansion. In this analysis, we find that the logarithmic model developed in Sec.~\ref{ch4_sec:model_1} produces a rich cosmology as shown through the critical points in Table~\ref{ch4_TABLE-I} which are then further studied for the nature of their critical points in Table~\ref{ch4_TABLE-II}. To show these properties in fuller detail, we also include phase portraits in Fig.~\ref{ch4_Fig1} where the behavior at those points is more clearly represented. The behavior of the scale factor at each critical point is shown in Table~\ref{ch4_TABLE-III}.  If we compare the analysis made in  Ref.~\cite{Zhang2011_jacp} for the logarithmic model, we can describe that there are eight more critical points. The study made in  Ref.\cite{Zhang2011_jacp}, successfully explains the de-Sitter solution through the dynamical system analysis of the logarithmic model and concludes that this study will not explain the radiation and matter-dominated era of the Universe evolution. The cosmology based on this study of the logarithmic model along with the addition of a scalar field successfully explains the de-Sitter solution in the matter and radiation-dominated phases of the evolution of the Universe. In this study, we have added the scalar field to explain both the radiation and matter-dominated era. We close the discussion with the figure that describes the evolutionary behavior of various density parameters and EoS parameters.

For the second model explored in Sec.~\ref{ch4_sec:model_2}, we take a square torsion scalar extension to the TEGR term. This would represent many other extensions as a leading order term in most circumstances such as background cosmology. Again, here we define suitable dynamical variables and provide the autonomous dynamical system in Eqs.~(\ref{ch4_49}-\ref{ch4_54}). This leads to the critical points presented in Table~\ref{ch4_TABLE-IV} together with their properties as described in Table~\ref{ch4_TABLE-V}. Similarly, we describe the behavior of each scale factor at each critical point in Table~\ref{ch4_TABLE-VI}. Finally, the phase portraits of Fig.~\ref{ch4_Fig3} are shown where the nature of each critical point is shown more fully through the evolutionary contours. Finally, we conclude with a diagram showing the evolution of each density parameter in Fig.~\ref{ch4_Fig4}.

\chapter{Quintessence dark energy models} 

\label{Chapter5} 

\lhead{Chapter 5. \emph{Quintessence dark energy models}} 
\vspace{10 cm}
*The work in this chapter is covered by the following article: \\

\textbf{L K Duchaniya}, Jackson Levi Said and B. Mishra, ``Quintessence dark energy models", \textbf{(Communicated)}. 


\clearpage
\section{Introduction} \label{ch5_SEC_intro}
In this chapter, we reconsider the action of a single scalar field in the local Universe through the realization of different potentials for a canonical scalar field. It is increasingly becoming less likely that the series of cosmological tensions is the result of a single systematic issue with the statistical treatment, there have been a diversity of possibilities of directions beyond $\Lambda$CDM in terms of additional or new physical mechanisms. In the latter case, there have been several promising proposals including the modification of physics beyond recombination such as in early dark energy, the additional of extra relativistic degrees of freedom, as well as the modification of gravitational physics at various scales. In these scenarios, scalar-tensor theories have been prominent as providing a possible avenue for confronting the problem of cosmic tensions. The simplest of these models involves a canonical scalar field as an additional ingredient to $\Lambda$CDM. In these scenarios, the cosmological constant is supplanted by the scalar field. This is performed by considering the Friedmann and Klein-Gordon equations in Sec.~\ref{ch5_SEC_matheforma} together with a series of late-time data sets in Sec.~\ref{ObservationalCosmology}. MCMC approach is taken in Sec.~\ref{ch5_cosmologicalmodels} where three potentials are considered for the scalar field. These are firstly a regular power-law form, then a hyperbolic form inspired by its tracker-like solution behavior in some circumstances and finally an axion-like potential. Through these potentials, we hope to generally reassess the potential of these scalar field cosmologies to meet some elements of the observational challenges posed in recent years. In Sec.~\ref{ch5_modelcomparison} we compare these models in the context of $\Lambda$CDM and finally close with a summary in Sec.~\ref{ch5_conclusion}.

\section{Scalar-tensor cosmology}\label{ch5_SEC_matheforma}
The scalar-tensor mathematical formalism is defined in Sec.~\ref{quintessencemodel}. In this Chapter, we will study various cosmological observation datasets, including $H_0$ prior, in the context of the quintessence model. Formalisms for the datasets are described in the Sec.~\ref{ObservationalCosmology}. To do so, we need to calculate the Hubble parameter $H(z)$ in terms of redshift with the help of Eqs.~ (\ref{quinfriedmann1},\ref{kleingordanequation}). Additionally, we need to specify the particular form of the potential function $V(\phi)$. Therefore, we will use three different potential function forms and constrain their parameters using observation data. As a function of redshift, the Klein-Gordon Eq.~\eqref{kleingordanequation} can be written as 
\begin{equation}\label{ch5_redshift_Klein_gordan_equation}
(1+z)^2 H^2(z) \frac{d^2\phi}{dz^2}+ (1+z)^2 H(z) \frac{dH}{dz}\frac{d\phi}{dz}-2(1+z)H^2(z)\frac{d\phi}{dz}+V_{,\phi}(\phi) \frac{d\phi}{dz}=0 \,, 
\end{equation}
where $V_{,\phi}(\phi)$ represents the derivative with respect to scalar field $\phi$. The Eq.~\eqref{quinfriedmann1} can be written as a function of redshift as,
\begin{equation}\label{ch5_model_Hz_first_friedmann_equation}
H^2(z)=\frac{3H_0^2(\Omega_{m0} (1+z)^3+\Omega_{r0} (1+z)^4)+8\pi G V(\phi)}{3-4 \pi G (1+z)^2 \left(\frac{d\phi}{dz}\right)^2}\,.
\end{equation}
In the above equation, $H_0$ refers to the Hubble parameter value, $\Omega_{m0}$ and $\Omega_{r0}$ refer to the matter density parameter and the radiation density parameter, respectively, at present. The next section explores cosmological observation datasets within the framework of three well-known potential functions 
$V(\phi)$.

\section{Cosmological models} \label{ch5_cosmologicalmodels}
In this section, we present and examine the results based on the approach described in Sec.~\ref{ObservationalCosmology} and utilizing the observational data mentioned earlier. We will examine how the choice of an $H_{0}$ prior value affects the parameter constraints of the potential function $V(\phi)$ and the datasets mentioned earlier. We will consider the recent local measurement from SH0ES which gives $H_0 = 73.04 \pm 1.04 \, \text{km} \, \text{s}^{-1} \text{Mpc}^{-1}$ \textbf{(R21)} \cite{Riess:2021jrx} and the estimate of $H_0 = 69.8 \pm 1.7 \, \text{km} \, \text{s}^{-1} \text{Mpc}^{-1}$ \textbf{(F21)} \cite{Freedman_2021apjh0tension, Di_Valentino_2022apj} derived from supernovae in the Hubble flow. Each subsection emphasizes the most promising models of potential functions $V(\phi)$, featuring contour plots of the constrained parameters with uncertainties $1\sigma$ and $2\sigma$ and tables displaying the final results. These models have become significant in literature and are often examined due to their ability to reflect cosmological history. In all the tables and posterior plots, we present results for the Hubble constant $H_0$, the current matter density parameter $\Omega_{m,0}$ and the model parameters. This setup will enable us to evaluate how various independent data sets and cosmological models influence the Hubble tension.
\subsection{Model-I: Power Law Potential}
Let us consider the power law potential \cite{Copeland:2006wr}, where the potential function is defined as,
\begin{equation}\label{ch5_model_I}
 V(\phi)=V_{0} \phi^{n}\,.   
\end{equation}
This specific form of the potential is defined by its reliance on the scalar field $\phi$ raised to the power of $n$, with $V_0$ serving as the constant coefficient. This potential is frequently utilized in inflation and DE models, as it accommodates various dynamical behaviors based on the selection of $n$. Many authors have studied cosmological evolution history \cite{Hossain:2024prd} by considering power law potential function.  In our investigation, we will examine the consequences of this potential on cosmological development and the associated observational data set. For this potential, the Hubble parameter $H(z)$ is expressed as a function of redshift,
\begin{equation}\label{ch5_modelI_Hz}
H^2(z)=\frac{3H_0^2(\Omega_{m0} (1+z)^3+\Omega_{r0} (1+z)^4)+8\pi G V_0 \phi^n}{3-4 \pi G (1+z)^2 \left(\frac{d\phi}{dz}\right)^2}\,,    
\end{equation}
where $\Omega_{m0}$ and $\Omega_{r0}$ represent the density parameters for matter and radiation at present, respectively. Using this potential, we can find the value of $V_0$ to be,
\begin{eqnarray}\label{ch5_modelI_LCDMlimit}
V_0=\frac{3 H_0^2 (1-\Omega_{m0}-\Omega_{r0})}{8\pi G}\,.  \end{eqnarray}
The Hubble parameter function described in Eq.~\eqref{ch5_modelI_Hz} reduced to the standard $\Lambda$CDM model when we set the model parameter $n=0$ and utilize the corresponding $V_{0}$ value as indicated in Eq.~\eqref{ch5_modelI_LCDMlimit}. In this analysis, we have fixed the parameter $V_0$ due to the difficulties associated with adequately constraining it across the available datasets. Subsequently, we employed the MCMC approach to constrain the model parameters, specifically $H_0$, $\Omega_{m0}$ and $n$. In this scenario, the Hubble parameter $ H(z)$ \eqref{ch5_modelI_Hz} is influenced by the scalar field function $\phi(z)$ and its derivative $ \phi'(z) $. Consequently, we have derived the solutions for  $\phi(z) $ and $ \phi'(z)$ from the Klein-Gordon  Eq.~\eqref{ch5_redshift_Klein_gordan_equation}. To find the solution to the second-order non-linear differential equation, we utilized numerical methods to determine $\phi(z) $ and $ \phi'(z)$ based on Eq.~\eqref{ch5_redshift_Klein_gordan_equation}. We simultaneously solved the Hubble parameter $ H(z)$ and the Klein-Gordon equation for each redshift value $z_i$ within the range $0<z<2.4$ to obtain the posterior distribution and the best-fit values of the model parameters for various combinations of datasets using the MCMC technique. 

\begin{figure}[H]
 \centering
 \includegraphics[width=68mm]{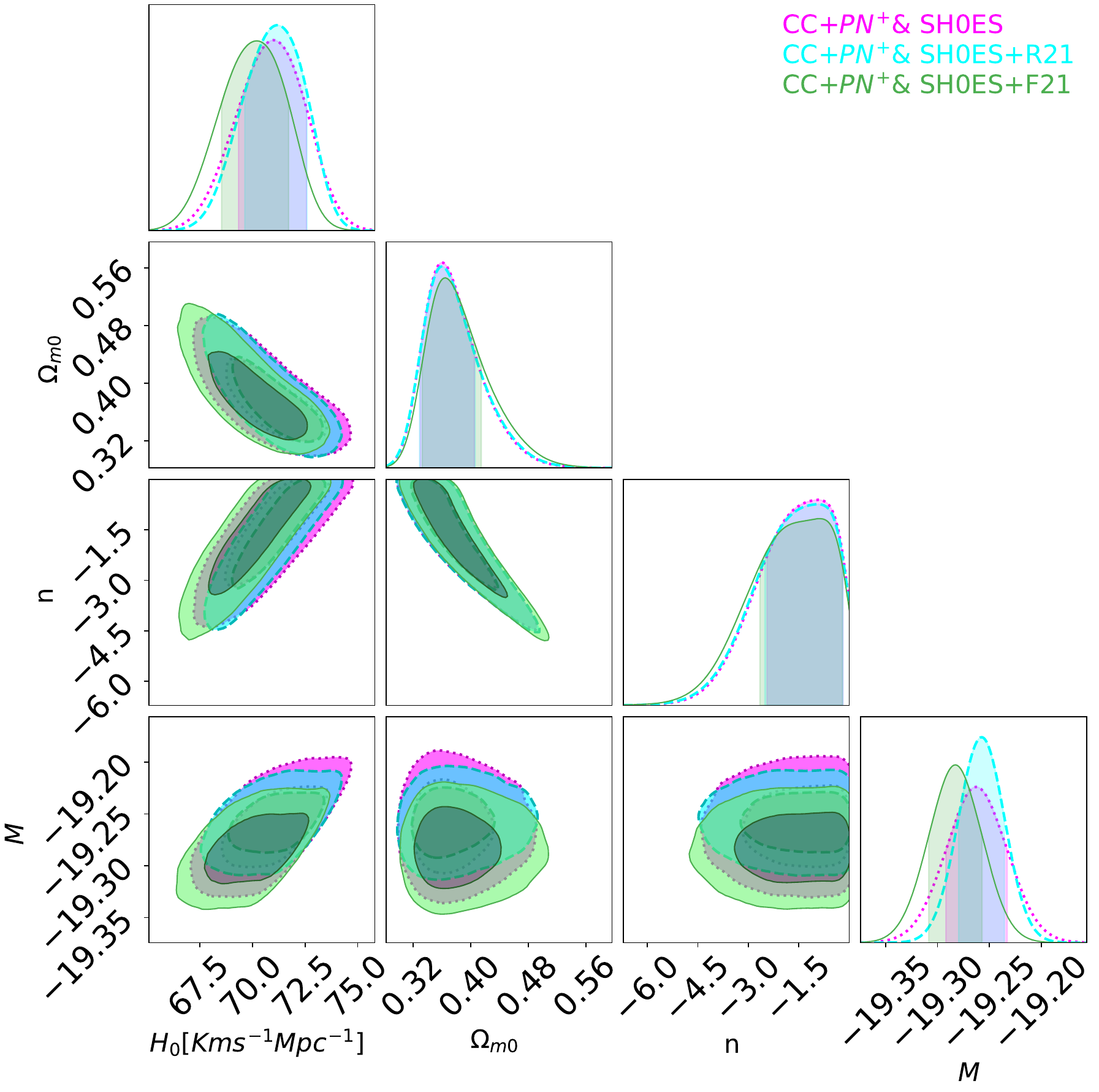}
  \includegraphics[width=68mm]{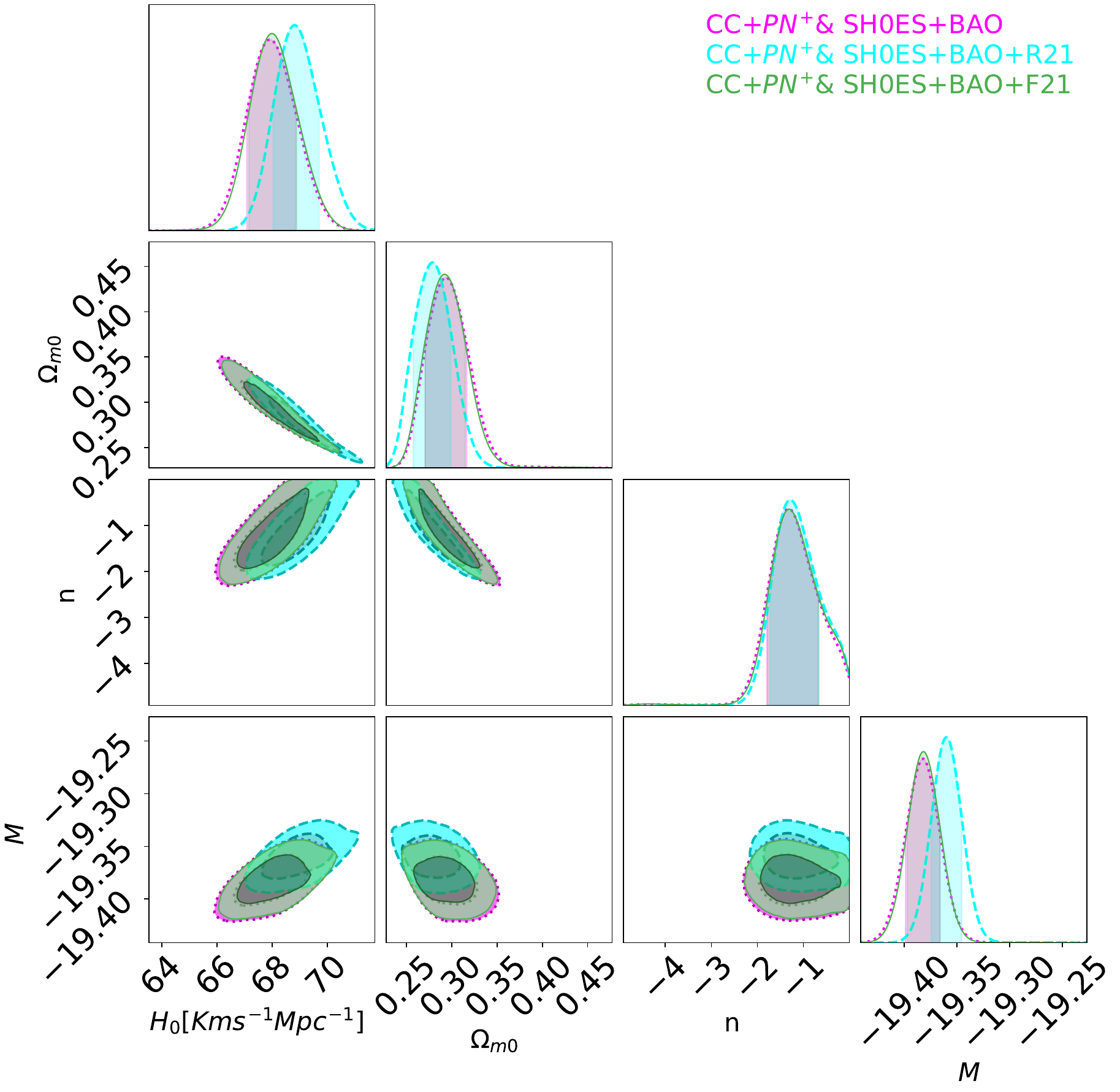}
 \caption{Left panel: Confidence intervals and posterior distributions for the power law model derived from the combined datasets CC and PN$^{+}$\&SH0ES, incorporating the $H_0$ priors R21 and F21. Right panel: Confidence intervals and posterior distributions for the power law model utilizing CC + PN$^{+}$\&SH0ES + BAO, again under the same prior assumptions.} \label{ch5_powerlawCCSNBAO}
 \end{figure}

The constraints on the defined parameters for the power law model are illustrated in Fig.~\eqref{ch5_powerlawCCSNBAO}. This figure displays the confidence regions along with the posteriors for various combinations of observational data sets. In these figures, we additionally display the outcomes for each prior on $H_0$ discussed earlier. In these illustrations, we observed the effect of the $H_0$ prior when combining different data sets. We also noted that when the BAO data set was included, the $H_0$ and $\Omega_{m0}$ values were lower than the CC + $PN^{+} \&$ SH0ES results due to the impact of the early Universe measurement data set.

 The exact values for the model parameters, along with the nuisance parameter $M$, for the power law model are presented in Table~\ref{ch5_powerlaw_outputs}. The results indicate that the $H_0$ values for data set combinations that involve CC+PN$^{+}$\&SH0ES are notably higher than those for their respective $H_0$ values. This observation aligns with the elevated $H_0$ value reported by the SH0ES team (R22), which presents $H_0 = 73.30 \pm 1.04\, \text{km s}^{-1} \, \text{Mpc}^{-1}$ \cite{Riess:2021jrx}. The results suggest that the maximum value of $H_0$ is attained with the combination CC+PN$^{+}$\&SH0ES+R21, yielding $H_0=71.2^{+1.4}_{-1.5}$. This value is somewhat elevated compared to the data set combination CC+PN$^{+}$\&SH0ES due to the addition of the $H_0$ prior $R21$. Meanwhile, for CC+PN$^{+}$\&SH0ES+F21, we find $H_0=70.2^{+1.5}_{-1.6}$, which is slightly lower than the value from the CC+PN$^{+}$\&SH0ES data set combination. From these findings, we deduce that the $R21$ prior increases the $H_0$ value, while the $F21$ prior decreases it.  A comparable pattern was noted with the inclusion of the BAO dataset, further reinforcing these conclusions, although including the BAO dataset shifts the $H_0$ value downward. The upcoming section will present a more in-depth statistical examination of these results and a comparison to the $\Lambda$CDM model. 
 \renewcommand{\arraystretch}{1.4} 
\begin{table}
    \centering
    \caption{The table displays outcomes for the power law model, with the initial column enumerating the combinations of data sets. The second column indicates the restrictions on the Hubble constant $H_0$, while the third and fourth columns show the values for the matter density $\Omega_{m0}$ and the model parameter $n$, respectively. The final column illustrates the Nuisance parameter $M$.}
    \label{ch5_powerlaw_outputs}
    \begin{tabular}{ccccc}
        \hline
		Data sets & $H_0$ & $\Omega_{m0}$ & $n$ & $M$ \\ 
		\hline
		CC+PN$^{+}$\&SH0ES & $71.0^{+1.6}_{-1.7}$ & $0.361^{+0.045}_{-0.031}$ & $-0.91^{+0.72}_{-1.53}$ & $-19.263^{+0.030}_{-0.029}$ \\
		CC+PN$^{+}$\&SH0ES+R21 & $71.2^{+1.4}_{-1.5}$ & $0.360^{+0.046}_{-0.031}$ & $-0.93^{+0.74}_{-1.58}$ & $-19.257\pm 0.022$ \\ 
		CC+PN$^{+}$\&SH0ES+F21 & $70.2^{+1.5}_{-1.6}$ & $0.364^{+0.051}_{-0.031}$ & $-0.90^{+0.70}_{-1.75}$ & $-19.282^{+0.025}_{-0.026}$ \\ 
          \cline{1-5}
		CC+PN$^{+}$\&SH0ES+BAO & $67.93^{+0.92}_{-0.86}$ & $0.294\pm 0.023$ & $-1.31^{+0.63}_{-0.49}$ & $-19.383\pm 0.016$ \\ 
		CC+PN$^{+}$\&SH0ES+BAO+R21 & $68.82^{+0.88}_{-0.79}$ & $0.279^{+0.020}_{-0.021}$ & $-1.28^{+0.61}_{-0.44}$ & $-19.360^{+0.014}_{-0.015}$\\ 
		CC+PN$^{+}$\&SH0ES+BAO+F21 & $67.98^{+0.91}_{-0.83}$ & $0.292\pm 0.022$ & $-1.31^{+0.65}_{-0.47}$ & $-19.382^{+0.016}_{-0.015}$ \\ 
		\hline
    \end{tabular}
\end{table}
 \renewcommand{\arraystretch}{1} 
\begin{table}[H]
    \centering
    \caption{This table presents a statistical comparison between the selected model and the standard $\Lambda$CDM model. Further information about the $\Lambda$CDM model can be found in {\color{blue}Appendix}. The first column lists the data sets, including the $H_0$ priors. The second column shows the values of $\chi^{2}_{\text{min}}$. The third and fourth columns indicate the AIC and BIC values, respectively. Finally, the fifth and sixth columns demonstrate the values for $\Delta \text{AIC}$ and $\Delta \text{BIC}$.}
    \label{ch5_powerlaw_outputAICBIC}
      \begin{tabular}{cccccc}
        \hline
		Data sets& $\chi^{2}_{min}$  &AIC &BIC&$\Delta$AIC &$\Delta$BIC \\ 
		\hline
		CC+PN$^{+}$\&SH0ES&1539.22 &1547.22 &  1552.18& 2&3.25 \\ 
		CC+PN$^{+}$\&SH0ES+R21& 1539.27 &1547.27 &  1552.22&2.02 &3.25\\ 
		CC+PN$^{+}$\&SH0ES+F21&1541.43 &1549.43 &  1554.38&1.88 &3.12\\ 
		\cline{1-6}
       CC+PN$^{+}$\&SH0ES+BAO&1588.78 &  1596.78 &  1601.75& 23.61&24.85 \\ 
		CC+PN$^{+}$\&SH0ES+BAO+R21& 1597.61 &1605.61 &  1610.59&29.65 &30.92\\ 
		CC+PN$^{+}$\&SH0ES+BAO+F21&1588.81 &  1596.81 &  1601.78&23.24 &24.49\\
  \hline
    \end{tabular}
\end{table}
\subsection{Model-II: Hyperbolic Potential function}
The hyperbolic function discussed in Ref. \cite{Urel:2000sinh} is characterized by
\begin{equation}\label{ch5_model_II}
V(\phi)=V_{0}\, sinh^{\alpha}(\phi\, \gamma) \,,    
\end{equation}
where $\alpha$, $\gamma$ and $V_0$ represent the parameters of the model. This potential is applied in the early Universe context, where the scalar field $\phi$ dynamics can trigger cosmic inflation \cite{SAHNI_2000}. According to Ref. \cite{roy2018dynamical}, this model is also suitable for examining the DE domain or explaining the late-time acceleration of the Universe. For this model, we have obtained the following form of Hubble parameter $H(z)$ 
\begin{equation}\label{ch5_modelII_Hz}
H^2(z)=\frac{3H_0^2(\Omega_{m0} (1+z)^3+\Omega_{r0} (1+z)^4)+8\pi G V_{0}\, sinh^{\alpha}(\phi\, \gamma)}{3-4 \pi G (1+z)^2 \left(\frac{d\phi}{dz}\right)^2}\,.    
\end{equation}  

\begin{figure}[H]
 \centering
 \includegraphics[width=78mm]{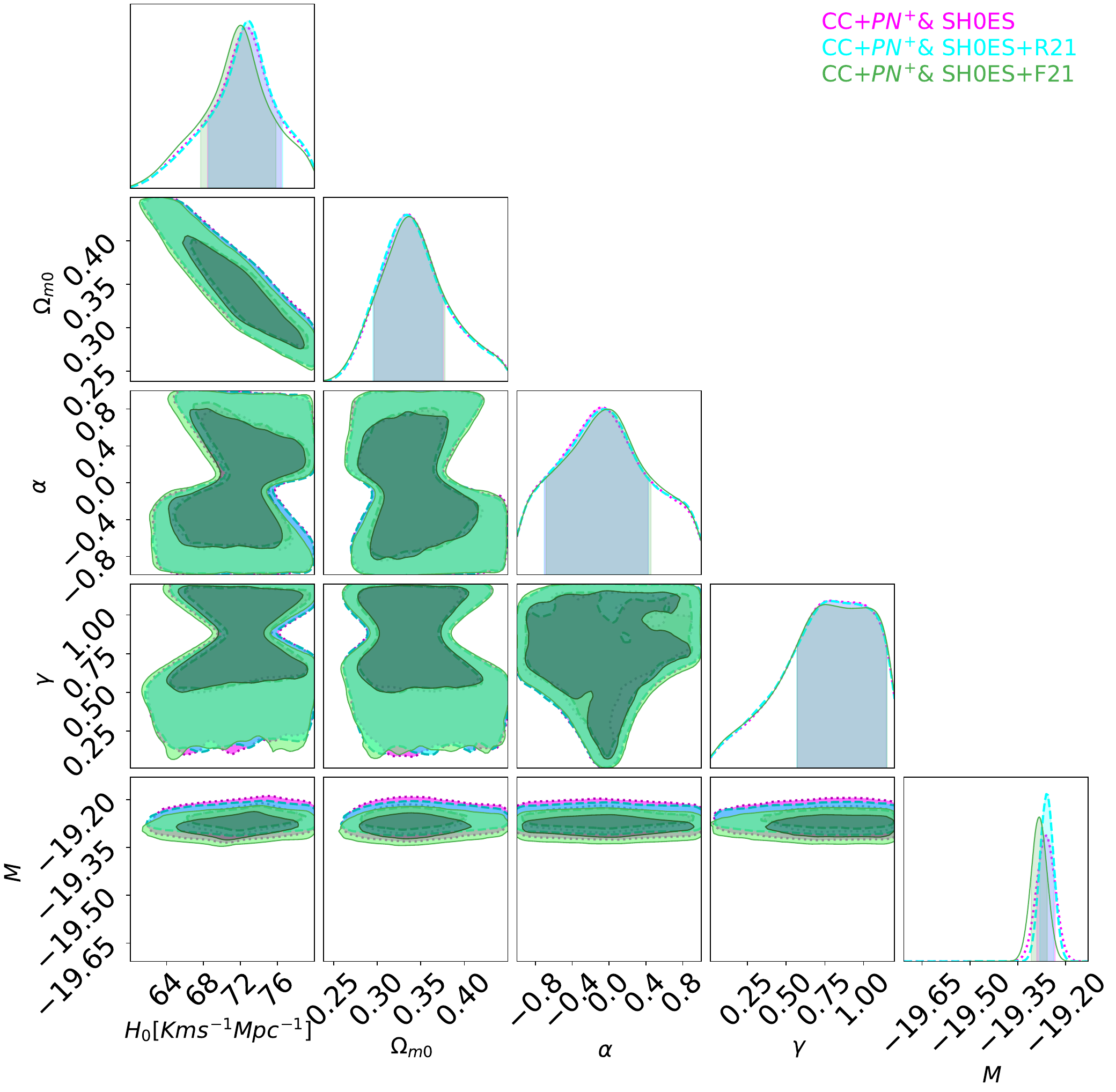}
  \includegraphics[width=78mm]{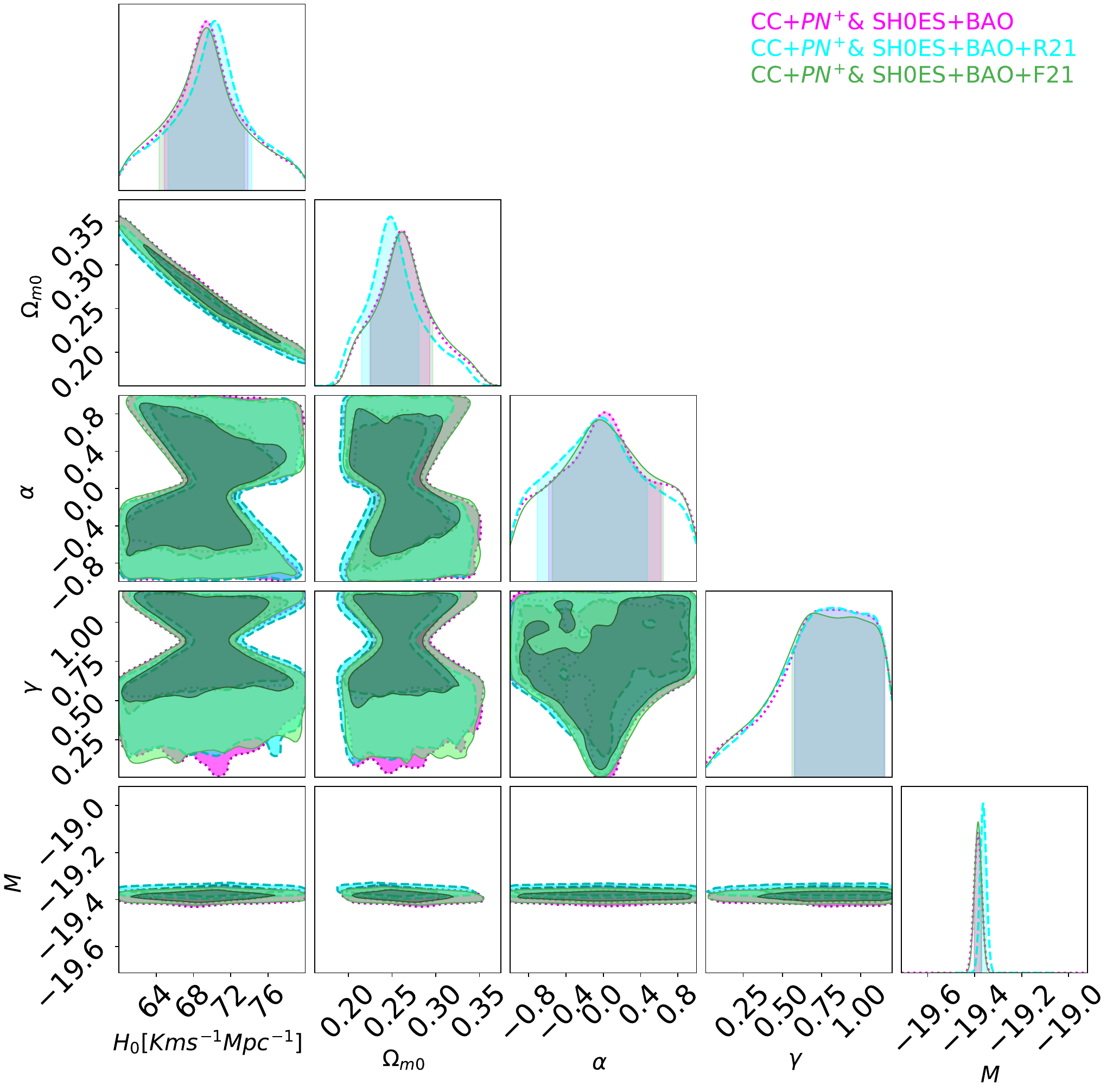}
 \caption{Left panel: Confidence intervals and posterior distributions for the hyperbolic model derived from the combined datasets CC and PN$^{+}$\&SH0ES, incorporating the $H_0$ priors R21 and F21. Right panel: Confidence intervals and posterior distributions for the hyperbolic model utilizing CC + PN$^{+}$\&SH0ES + BAO, again under the same prior assumptions.} \label{ch5_hyperbolicCCSNBAO}
 \end{figure}

This model is reduced to $\Lambda$CDM model when $\alpha = 0$, alongside the specific value of $V_0$ outlined in Eq.~\eqref{ch5_modelI_LCDMlimit}. 
 Fig.~\eqref{ch5_hyperbolicCCSNBAO} illustrates the posterior distributions and confidence intervals of the constrained parameters for model II. In contrast to model-I, model-II exhibits tightly constrained parameters due to the periodic nature of the sinh function. As with model-I, we have also fixed the $V_0$ value, as outlined in Eq.~\eqref{ch5_modelI_LCDMlimit}. 

 Table~\ref{ch5_hyperbolic_outputs} displays the precise numerical values for the parameters illustrated in Fig.~\eqref{ch5_hyperbolicCCSNBAO}, encompassing the nuisance parameter M. The findings indicate that the estimated values of $H_0$ and $\Omega_{m0}$ are similar to those derived from model-I. The Hubble constant $H_0$ in this model is slightly increased, while the matter density parameter $\Omega_{m0}$ is somewhat lower than in Model I. The analysis of the dataset reveals an inverse correlation between $H_0$ and $\Omega_{m0}$: when $H_0$ rises, $\Omega_{m0}$ usually falls and vice versa, a decrease in $H_0$ is linked to an increase in $\Omega_{m0}$. For this model, the combination of the CC+PN$^{+}$\&SH0ES+R21 data set yields a higher estimate for $H_0 $, specifically $H_0 = 72.8^{+3.8}_{-4.2} $. As observed in model I, including the $H_0$ prior significantly influences the results similarly for model II. Additionally, including the BAO data set leads to a reduction in the estimated value of $H_0$. Additional comparisons and statistical evaluations involving the $\Lambda$CDM are presented in Sec.~\ref{ch5_modelcomparison}.
\renewcommand{\arraystretch}{1.4} 
\begin{table}
    \centering
    \caption{The table displays results about the hyperbolic potential function, where the first column lists the combinations of data sets. The second column indicates the Hubble constant $H_0 $, while the third and fourth columns provide the values for the matter density $\Omega_{m0}$  and the model parameter $\alpha$, respectively. The fifth column represents $ \gamma$. Finally, the sixth column shows the nuisance parameter $M$.}
    \label{ch5_hyperbolic_outputs}
    \begin{tabular}{p{5.4cm} p{1.5cm} p{2cm} p{1.5cm} p{1.5cm} p{2.5cm}}
        \hline
		Data sets & $H_0$ & $\Omega_{m0}$ & $\alpha$ & $\gamma$ &$M$ \\ 
		\hline
		CC+PN$^{+}$\&SH0ES & $72.7^{+3.7}_{-4.3}$ & $0.334^{+0.042}_{-0.038}$ & $-0.07^{+0.50}_{-0.62}$ & $0.78^{+0.36}_{-0.21}$ & $-19.260^{+0.027}_{-0.030}$   \\
		CC+PN$^{+}$\&SH0ES+R21 & $72.8^{+3.8}_{-4.2}$ & $0.334^{+0.041}_{-0.039}$ & $-0.05^{+0.48}_{-0.65}$ & $0.78^{+0.37}_{-0.21}$ & $-19.258^{+0.020}_{-0.023}$  \\ 
		CC+PN$^{+}$\&SH0ES+F21 & $72.0^{+3.8}_{-4.3}$ & $0.337\pm 0.04$ & $-0.01^{+0.47}_{-0.67}$ & $0.75^{+0.40}_{-0.18}$ & $-19.283\pm 0.02$\\ 
          \cline{1-6}
		CC+PN$^{+}$\&SH0ES+BAO & $69.4\pm 4.5$ & $0.261^{+0.032}_{-0.037}$ & $0.02^{+0.61}_{-0.60}$ & $0.78^{+0.38}_{-0.20}$ & $-19.386^{+0.016}_{-0.017}$ \\ 
		CC+PN$^{+}$\&SH0ES+BAO+R21 & $70.3^{+4.0}_{-5.0}$ & $0.248\pm 0.03$ & $-0.02^{+0.50}_{-0.68}$ & $0.83^{+0.32}_{-0.26}$ & $-19.364^{+0.014}_{-0.013}$  \\ 
		CC+PN$^{+}$\&SH0ES+BAO+F21 & $69.4^{+4.1}_{-5.1}$ & $0.261^{+0.07}_{-0.035}$ & $-0.04^{+0.69}_{-0.50}$ & $0.70^{+0.45}_{-0.14}$ & $-19.386^{+0.016}_{-0.014}$   \\ 
		\hline
    \end{tabular}
\end{table}
 \renewcommand{\arraystretch}{1} 
\begin{table}[H]
    \centering
    \caption{This table provides a statistical comparison between the chosen model and the standard $\Lambda$CDM model. More details about the $\Lambda$CDM model can be found in {\color{blue}Appendix}. The first column displays the data sets, which include the $H_0$ priors. The second column presents the values of $\chi^{2}_{\text{min}}$. The third and fourth columns represent the AIC and BIC values, respectively. Lastly, the fifth and sixth columns show the values for $\Delta \text{AIC}$ and $\Delta \text{BIC}$.}
    \label{ch5_hyperbolic_outputAICBIC}
      \begin{tabular}{cccccc}
        \hline
		Data sets& $\chi^{2}_{min}$  &AIC &BIC&$\Delta$AIC &$\Delta$BIC \\ 
		\hline
		CC+PN$^{+}$\&SH0ES& 1539.22&1549.22 &  1555.41&4 &6.48 \\ 		
        CC+PN$^{+}$\&SH0ES+R21&1539.25  &1549.25 &  1555.44&4 &6.47\\ 
		CC+PN$^{+}$\&SH0ES+F21&1541.55 &1551.55 &  1557.74& 4&6.48\\ 
		\cline{1-6}
       CC+PN$^{+}$\&SH0ES+BAO&1591.95 &  1601.45 &  1607.67& 28.28& 30.77\\ 
		CC+PN$^{+}$\&SH0ES+BAO+R21& 1600.35 &1610.35 &  1616.57& 34.39&36.9\\ 
		CC+PN$^{+}$\&SH0ES+BAO+F21& 1591.48&  1601.48 &  1607.70&27.91 &30.29\\
  \hline
    \end{tabular}
\end{table}
\subsection{Model-III: Axionlike potential}
The axionlike potential function addressed in Ref. \cite{Marsh_2016axion} is formulated as follows,
\begin{equation}\label{ch5_model_III}
V(\phi)=V_{0} \bigg(1-\cos\bigg(\frac{\phi}{F_{EDE}}\bigg)\bigg)^{\beta}\,,   
\end{equation}
where $\beta$, $F_{EDE}$ and $V_0$ represent the parameters of the model. Numerous researchers have examined the cosmological timeline, including cosmic inflation, late-time cosmology and the Hubble constant tension issue in the context of the axionlike potential function. Poulin et al. \cite{Poulin:2018prd} analyzed how the axionlike field influences cosmological observations as it becomes dynamic at various times in the axionlike potential function for specific values of $\beta=(1, 2, 3)$. In Ref.\cite{Poulin:2019prl}, they investigated how the early dark model can address the Hubble tension. They selected the axionlike potential function with $\beta$ values $(2,3, \infty)$. Herold and Ferreira \cite{Herold:2023prd} studied how the axionlike potential function offers a solution to the Hubble tension for the particular value of $\beta=3$. Inspired by these studies, we have focused on the axionlike potential function. However, in this investigation, we explore without fixing the value of $\beta$. We have determined the best-fit value for all these model parameters through various combinations of datasets. For the axionlike potential function, we have derived the following $H(z)$
\begin{align}\label{ch5_modelIII_Hz}
H^2(z)=\frac{3H_0^2(\Omega_{m0} (1+z)^3+\Omega_{r0} (1+z)^4)+8\pi G V_0 \bigg(1-\cos\bigg(\frac{\phi}{F_{EDE}}\bigg)\bigg)^{\beta}}{3-4 \pi G (1+z)^2 \left(\frac{d\phi}{dz}\right)^2}\,.    
\end{align}  

\begin{figure}[H]
 \centering
 \includegraphics[width=75mm]{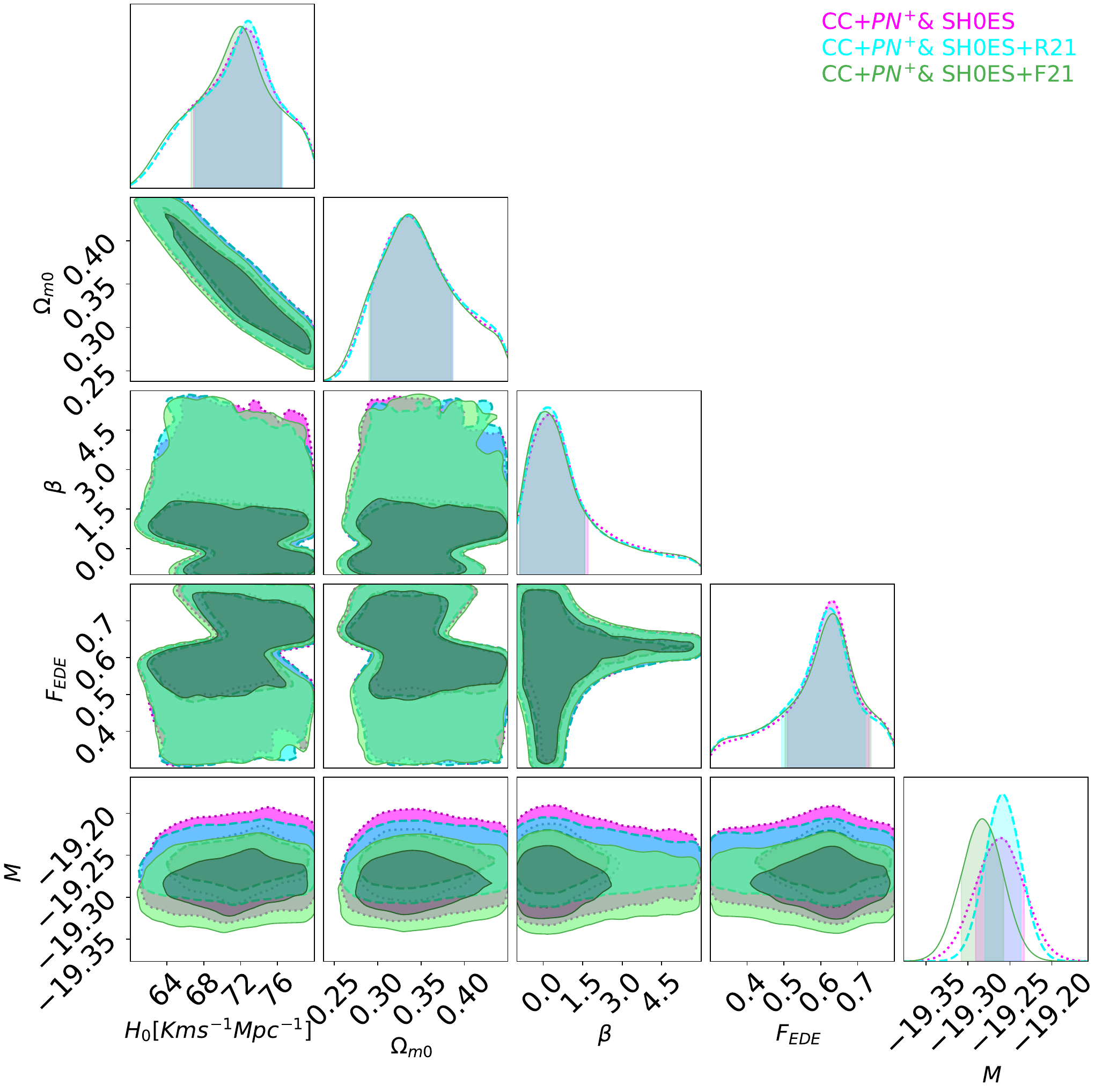}
 \includegraphics[width=75mm]{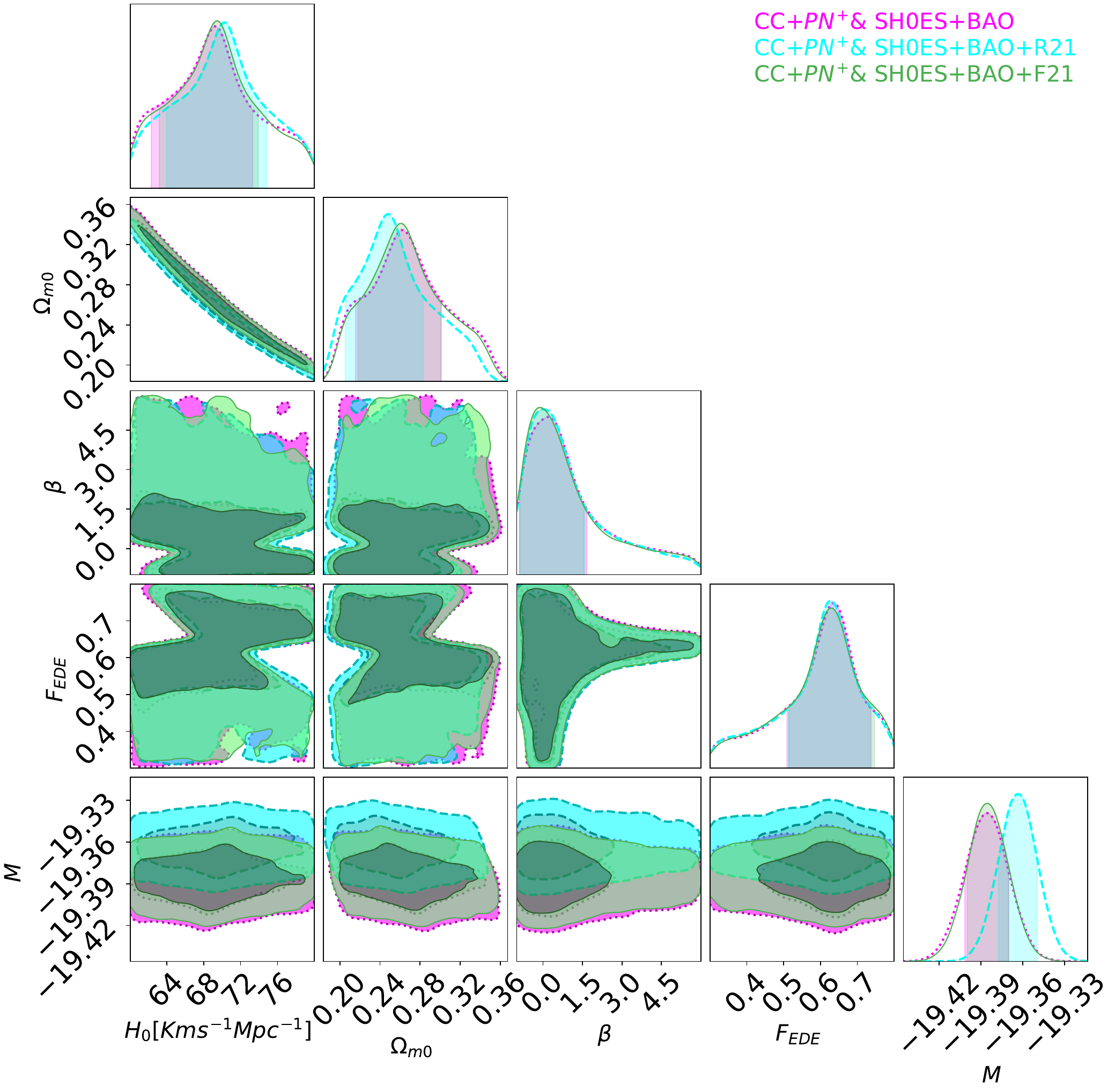}
 \caption{Left: Confidence intervals and posterior distributions for the axion-like model derived from the combined datasets CC and PN$^{+}$\& SH0ES, incorporating the $H_0$ priors R21 and F21. Right: Confidence intervals and posterior distributions for the axion-like model utilizing CC + PN$^{+}$\& SH0ES + BAO, again under the same prior assumptions.} \label{ch5_axionlikeCCSNBAO}
 \end{figure}

This model is reduced to the $\Lambda$CDM model when $\beta = 0$, along with the specific value of $V_0$ specified in Eq.~\eqref{ch5_modelI_LCDMlimit}. Fig.~\eqref{ch5_axionlikeCCSNBAO} displays the posterior distributions and confidence intervals for the constrained parameters of model III. The behavior shown in Fig.~\eqref{ch5_axionlikeCCSNBAO} is similar to model-II. Consistent with models I and II, we have also fixed the $V_0$ value, as described in Eq.~\eqref{ch5_modelI_LCDMlimit}.

Table~\ref{ch5_hyperbolic_outputs} provides the precise numerical values for the parameters illustrated in Fig.~\eqref{ch5_hyperbolicCCSNBAO}, including the nuisance parameter $M$. The analysis reveals that the model parameters $\beta$ and $F_{EDE}$ are tightly constrained, as shown by the contour plots. Notably, the inclusion of the R21 prior results in a higher $H_0$ value for the data set combination CC+PN$^{+}$\&SH0ES+R21 compared to CC+PN$^{+}$\&SH0ES. Similarly, the addition of F21 to CC+PN$^{+}$\&SH0ES slightly lowers the $H_0$ value relative to CC+PN$^{+}$\&SH0ES. The findings suggest that R21 increases the estimate of $H_0$, whereas F21 restricts it to a lower value. This highlights the substantial impact that the chosen $H_0$ priors have on the selected model.

The $H_0$ estimate derived from the CC+PN$^{+}$\&SH0ES combination, augmented with $H_0$ priors, aligns closely with the elevated $H_0$ reported by the SH0ES team (R22), specifically $H_0 = 73.30 \pm 1.04 \text{km s}^{-1} \, \text{Mpc}^{-1}$ \cite{Riess:2021jrx}. Consistently with models I and II, the incorporation of the BAO data with the CC+PN$^{+}$\&SH0ES combination yields a reduction in the $H_0$ values when compared to the CC+PN$^{+}$\&SH0ES data set. The $H_0$ values derived from the combination of the CC+PN$^{+}$\&SH0ES+BAO datasets, along with the relevant priors, align with the higher estimates of $H_0$ reported in \cite{Planck:2018vyg}. 
 \renewcommand{\arraystretch}{1.2} 
\begin{table}
    \centering
    \caption{The table displays results related to the axion-like model, where the first column identifies the combinations of data sets. The second column indicates the Hubble constant $H_0$, while the third and fourth columns provide the values for the matter density $\Omega_{m0}$ and the model parameter $\beta$, respectively. The fifth column represents $F_{EDE}$. The sixth column presents the Nuisance parameter $M$.}
    \label{ch5_Axion_outputs}
    \begin{tabular}{p{5.4cm} p{1.5cm} p{1.8cm} p{1.5cm} p{1.5cm} p{2.5cm}}
        \hline
		Data sets & $H_0$ & $\Omega_{m0}$ & $\beta$ & $F_{EDE}$ &$M$ \\ 
		\hline
		CC+PN$^{+}$\&SH0ES & $72.6^{+3.9}_{-5.7}$ & $0.335^{+0.051}_{-0.043}$ & $0.2^{+1.5}_{-1.1}$ & $0.63^{+0.10}_{-0.12}$ & $-19.261^{+0.028}_{-0.030}$\\
		CC+PN$^{+}$\&SH0ES+R21 & $72.8^{+3.8}_{-5.7}$ & $0.334^{+0.053}_{-0.042}$ & $0.2^{+1.4}_{-1.1}$ & $0.624^{+0.096}_{-0.131}$ & $-19.259^{+0.022}_{-0.021}$ \\ 
		CC+PN$^{+}$\&SH0ES+F21 & $72.0^{+4.4}_{-5.4}$ & $0.337^{+0.048}_{-0.047}$ & $0.06^{+1.52}_{-0.96}$ & $0.63^{+0.10}_{-0.13}$ & $-19.283^{+0.026}_{-0.025}$ \\ 
          \cline{1-6}
		CC+PN$^{+}$\&SH0ES+BAO & $69.3^{+4.0}_{-7.0}$ & $0.263^{+0.038}_{-0.047}$ & $0.2^{+1.4}_{-1.1}$ & $0.637^{+0.100}_{-0.128}$ & $-19.386\pm 0.01$  \\ 
		CC+PN$^{+}$\&SH0ES+BAO+R21 & $70.2^{+4.6}_{-6.3}$ & $0.248^{+0.035}_{-0.043}$ & $0.03^{+1.58}_{-0.92}$ & $0.63\pm 0.1$ & $-19.363^{+0.013}_{-0.014}$\\ 
		CC+PN$^{+}$\&SH0ES+BAO+F21 & $69.5^{+4.4}_{-6.3}$ & $0.261^{+0.040}_{-0.043}$ & $-0.11^{+1.67}_{-0.78}$ & $0.63\pm 0.1$ & $-19.386^{+0.016}_{-0.014}$\\ 
		\hline
    \end{tabular}
\end{table}
 \renewcommand{\arraystretch}{1} 
\begin{table}[H]
    \centering
    \caption{This table provides a statistical comparison between the chosen model and the standard $\Lambda$CDM model. More details about the $\Lambda$CDM model can be found in {\color{blue}Appendix}. The first column displays the data sets, which include the $H_0$ priors. The second column presents the values of $\chi^{2}_{\text{min}}$. The third and fourth columns represent the AIC and BIC values, respectively. Lastly, the fifth and sixth columns show the values for $\Delta \text{AIC}$ and $\Delta \text{BIC}$.}
    \label{ch5_Axion_outputAICBIC}
      \begin{tabular}{cccccc}
        \hline
		Data sets& $\chi^{2}_{min}$  &AIC &BIC&$\Delta$AIC &$\Delta$BIC \\ 
		\hline
		CC+PN$^{+}$\&SH0ES&1539.22 &1549.22 &  1555.41&4 &6.48 \\ 		CC+PN$^{+}$\&SH0ES+R21& 1539.25 &1549.25 &  1555.45& 4&6.48\\ 
		CC+PN$^{+}$\&SH0ES+F21&1541.51 &1551.51 &  1557.74& 3.96&6.48\\ 
		\cline{1-6}
       CC+PN$^{+}$\&SH0ES+BAO& 1591.44&  1601.44 &  1607.66&28.27 &30.76 \\ 
		CC+PN$^{+}$\&SH0ES+BAO+R21& 1600.35 &1610.35&  1616.58& 34.39&36.91\\ 
		CC+PN$^{+}$\&SH0ES+BAO+F21& 1591.49&  1601.49 &  1607.71& 27.92&30.42\\
  \hline
    \end{tabular}
\end{table}
\section{Model Comparison} \label{ch5_modelcomparison}
We assess the effectiveness of every potential function and dataset by calculating their corresponding minimum $\chi^{2}_{\text{min}}$ values derived from the maximum likelihood $L_{\text{max}}$, as $\chi^{2}_{\text{min}} = -2 \ln L_{\text{max}}$. We also evaluate the models in comparison to the standard $\Lambda$CDM by utilizing the AIC, which considers both the fit quality (assessed through $\chi^{2}_{min}$) and the complexity of models. The mathematical formalism of the AIC and BIC is defined in Section~\ref{AICBICIntro}.

To evaluate the performance of different models with various combinations of data sets, we determine the differences in AIC and BIC between each model and the reference model $\Lambda$CDM. The constrained parameters for the $\Lambda$CDM model corresponding to each combination of data sets are detailed in {\color{blue}Appendix}. Lower values of $\Delta$AIC and $\Delta$BIC indicate that the model using the selected data set aligns more closely with the $\Lambda$CDM model, reflecting superior performance. Table~\ref{ch5_powerlaw_outputAICBIC}, Table~\ref{ch5_hyperbolic_outputAICBIC} and Table~\ref{ch5_Axion_outputAICBIC} present the values for different statistical measures including $\chi^{2}_{\text{min}}$, $\Delta$AIC and $\Delta$BIC for each model. 

In Table~\ref{ch5_powerlaw_outputAICBIC}, we observe that the combination of CC+PN$^{+}$\&SH0ES with $H0$ priors yields lower values for both $\Delta\text{AIC}$ and $\Delta\text{BIC}$. This suggests that this data set configuration aligns more closely with the standard $\Lambda$CDM model. Conversely, when incorporating the BAO data set alongside CC+PN$^{+}$\&SH0ES, we see an increase in $\Delta\text{AIC}$ and $\Delta\text{BIC}$, which implies that this particular observational combination provides weaker support for the model compared to the $\Lambda$CDM model. 

In Table~\ref{ch5_hyperbolic_outputAICBIC} and Table~\ref{ch5_Axion_outputAICBIC}, we present the statistical results for model-II and model-III, respectively. Both models exhibit comparable values for $\chi^{2}_{\text{min}}$, $\text{AIC}$, $\text{BIC}$, $\Delta\text{AIC}$ and $\Delta\text{BIC}$, indicating that they demonstrate similar performance relative to the $\Lambda$CDM model. When contrasting the values of $\Delta\text{AIC}$ and $\Delta\text{BIC}$ for model-II and model-III with those of model-I, it becomes evident that model I aligns more closely with the $\Lambda$CDM model than either model-II or model-III. 

In model-II and model-III, similar to the model-I, we find that the combination of CC+PN$^{+}$\&SH0ES with $H_0$ priors results in lower values for both $\Delta\text{AIC}$ and $\Delta\text{BIC}$. This indicates that this data set aligns more closely with the standard $\Lambda$CDM model. On the other hand, when we include the BAO data set together with CC+PN$^{+}$\&SH0ES, there is an increase in $\Delta\text{AIC}$ and $\Delta\text{BIC}$, suggesting that this particular combination of observations offers less support for the model in comparison to the $\Lambda$CDM model.  

\begin{figure}[H]
 \centering
 \includegraphics[width=83mm]{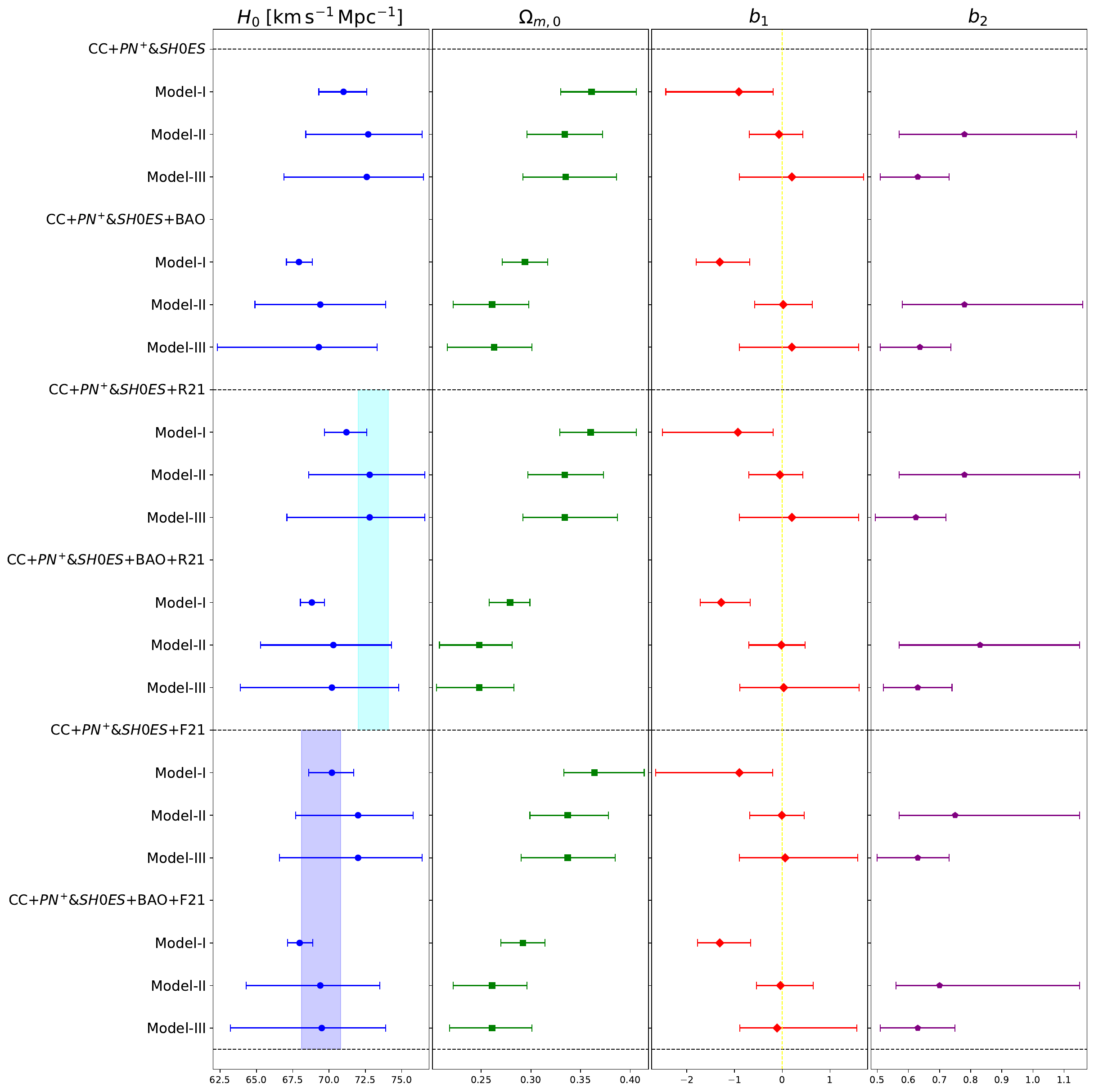} 
 \caption{The whisker plot illustrates the distribution of model parameters $H_0$, $\Omega_{m0}$, $ b_1 $ and $ b_2 $. The parameters $b_1 $ and $b_2 $ correspond to different functional forms of the potential function across various models. Specifically, for Model I, $ b_1 = n $; for Model II, $ b_1 = \alpha $; and for Model III, $ b_1 = \beta $. In terms of the $ b_2 $ parameter, Model II utilizes $ b_2 = \gamma $ and Model III employs $ b_2 = F_{EDE} $. The first column features a cyan-shaded region representing the R21 prior, while the blue-shaded region delineates the F21 prior.} \label{ch5_whisker}
 \end{figure}

To simplify the cross-analysis of the different models, data sets and prior selections, we present a whisker plot in Fig.~\ref{ch5_whisker} that displays each cosmological parameter against one another. Additionally, we illustrate the value of each prior in shaded areas, which helps clarify their direct influence on the cosmological parameters for each model. In the whisker plot, the yellow vertical dashed line in the third column indicates the $\Lambda$CDM limit at $b_1=0$, as well as the value of $V_0=\frac{3 H_0^2 (1-\Omega_{m0}-\Omega_{r0})}{8\pi G}$ applicable to all three models. The analysis reveals that the $b_1$ values for model-II and model-III fall within the $1\sigma$ confidence region of the $\Lambda$CDM predictions. Conversely, $b_1$ values of model-I do not consistently fall within the same $1\sigma$ interval as those predicted by the $\Lambda$CDM model. In all three models, we observed that the value of $H_0$ is higher for the dataset combination CC +  PN$^+\&$SH0ES and $H_0$ priors, as opposed to when the BAO dataset is included. Specifically, integrating the BAO dataset with CC +  PN$^+\&$SH0ES yields a lower $H_0$ value relative to the CC +  PN$^+\&$SH0ES combination. This reduction can be attributed to the influence of the early Universe measurements provided by the BAO dataset. Additionally, we found a correlation where an increase in $H_0$ is associated with a rise in $ \Omega_{m0} $. Conversely, a decrease in $H_0$ decreases $ \Omega_{m0} $. This relationship implies that by accelerating the expansion of the Universe, the contribution of DE diminishes. In contrast, a deceleration in the expansion rate is associated with an increased influence of the DE.
\section{Conclusion} \label{ch5_conclusion}
In recent years, scalar-tensor theories have shown great promise in meeting the growing challenges in the observational predictions of $\Lambda$CDM. This has taken various forms in terms of coupled and uncoupled scalar fields, as well as both early and late time centered fields. In this work, we explore the possible observations and impacts of a late time acting scalar field by adopting 3 physically motivated potentials and using data sets located in the late Universe.

The models explored in this work are driven by a power-law potential, a hyperbolic sinh function, and an axion-like function, which encompass many of the different possible general behaviors that are of interest in the literature. The power-law model incorporates the general trend of models that have a healthy $\Lambda$CDM limit but which may express deviations when the scalar field takes on large values such as in the late Universe. The hyperbolic model produces a smooth transition in the sign of the potential for different values of the scalar field, rescaled by the $\gamma$ model parameter. As for the final axion-like function, this incorporates possible oscillatory behavior in this regime of the Universe. 

These potentials are constrained by the consideration of CC, $PN^{+}\& SH0ES$, and BAO data sets together with priors imposed using R21 and F21 literature estimates. These priors consistently raise the value of the Hubble constant while the data sets alone produce lower values of this parameter and a correspondingly lower value of the matter density parameter. For the power-law model, the models are largely consistent with each other to a high statistical confidence level. This is also true for the other two models. On the other, the hyperbolic and axion-like potentials produce a wider fit for these parameters while also giving good fits for the model-specific parameters.

The three models show good performance in comparison with $\Lambda$ as evidenced by the statistical metrics under consideration. These show promising possibilities for the models. The next phase of this analysis would be to consider the effect of the perturbative sector in comparison with observational data and how large-scale structure formation would be impacted. We intend to do this in future work, which will provide a more robust analysis of the models.



\chapter{Dynamical system analysis in modified Galileon cosmology} 

\label{Chapter6} 

\lhead{Chapter 6. \emph{Dynamical system analysis in modified Galileon cosmology}} 

\vspace{10 cm}
*The work in this chapter is covered by the following publication:\\

\textbf{L K Duchaniya}, B Mishra, IV Fomin and  SV Chervon, "Dynamical system analysis in modified Galileon cosmology", \textit{Classical and Quantum Gravity}, \textbf{41}, 253016 (2024).

\clearpage
\section{Introduction}\label{ch6_Introchapt_1}
In this chapter, we have investigated the phase space analysis in teleparallel Galileon cosmology, where the Galileon term is considered a coupled scalar field $F(\phi)$. We focus on the exponential type function of $F(\phi)$ and the three well-motivated potential functions $V(\phi)$. We obtain the critical points of the autonomous system, along with their stability conditions and cosmological properties. The critical points of the autonomous system describe different phases of the Universe. The scaling solution for critical points was found in our analysis to determine the matter-DE and radiation-DE-dominated eras of the Universe. These solutions were originally obtained by Wetterich (1988) \cite{Wetterich_1988} and Ferreira and Joyce(1998) \cite{Ferreira_1998}. These scaling solutions typically involve the proportional evolution of DE alongside another component, such as radiation or matter and help explain the transition between different cosmological eras. The DE-dominated critical points show stable behavior and indicate the late-time cosmic acceleration phase of the Universe. We employ CMT to address cases where some fixed points display eigenvalues with zero and negative real parts. Further, the results are examined with the Hubble rate $H(z)$ and the SNIa cosmological data sets. In the literature, the Galileon cosmology has been thoroughly examined within the cosmological framework. The late-time cosmic acceleration of the Universe studied in Ref. \cite{Silva_2009,  Burrage_2010, Germani_2012, Appleby_2012, Iorio_2012, Easson_2021}. Refs. \cite{Burrage_2011, Renaux_Petel_2013, Gonzalez_Espinoza_2019,  Choudhury_2024, Chaadaeva_2024STF, Nicolis:2008in, Deffayet:2009wt, Leon_2013, Baker:2017hug} have explained inflation phase of the Universe.
\section{Mathematical formalism}\label{ch6_SECII} 
This chapter will study the phase space analysis in torsion-Galileon theory. In this process, the mathematical formalism of the teleparallel Horndeski theory is defined in Sec.~\ref{Teleparallel FrameworkofHorndeskitheory}. Varying the action (\ref{Horndeski_action_Lagran}) with respect to lapse function $N$, the first Friedmann equation can be obtained as \cite{Bahamonde:2019shr},
\begin{eqnarray}\label{ch6_first_friedmann_equation}     
&2X G_{2,X}-G_{2}+6X\dot{\phi}HG_{3,X}-2XG_{3,\phi}-6H^{2}G_{4}+24H^{2}X(G_{4,X}+XG_{4,XX})-12HX\dot{\phi}G_{4,\phi X}\nonumber\\&-6H\dot{\phi}G_{4,\phi} +2H^{3}X \dot{\phi}(5G_{5,X}+2XG_{5,XX})-6H^{2}X(3G_{5,\phi}+2XG_{5,\phi X})=\rho_{m}+\rho_{r}\,,    
\end{eqnarray}  
and varying the action (\ref{Horndeski_action_Lagran}) with respect to scale factor $a(t)$, one can obtain
\begin{eqnarray}\label{ch6_second_friedmann_equation}
&G_{2}-2X(G_{3, \phi}+\ddot{\phi}G_{3, X})+2G_{4}(3H^{2}+2\dot{H})-12H^{2}XG_{4, X}-4H\dot{X}G_{4, X}-8\dot{H}XG_{4,X}-8HX\dot{X}G_{4, XX}\nonumber\\&+2G_{4, \phi}(\ddot{\phi}+2H\dot{\phi})+4XG_{4, \phi \phi}+4X(\ddot{\phi}-2H\dot{\phi})G_{4, \phi X}-2X(2H^{3}\dot{\phi}+2H\dot{H}\dot{\phi}+3H^{2}\ddot{\phi})G_{5, X}\nonumber\\&-4H^{2}X^{2} \ddot{\phi} G_{5, XX}+4HX(\dot{X}-HX)G_{5,\phi X} + 2\left[2\frac{d}{dt}(HX)+3H^{2}X\right]+4HX\dot{\phi}G_{5,\phi \phi}=-p_{m}-p_{r}\,.    
\end{eqnarray}
Now taking variation of (\ref{Horndeski_action_Lagran}) with respect to scalar field $\phi$, we can get the Klein-Gordan equation as, 
\begin{equation}\label{ch6_Klein-Gordan-equation}
\frac{1}{a^{3}}\frac{d}{dt}(a^{3}J)=P_{\phi}\,,    
\end{equation}
where
\begin{eqnarray}
J&= \dot{\phi}G_{2, X}+6HXG_{3, X}-2\dot{\phi}G_{3, \phi}+6H^{2}\dot{\phi}(G_{4, X}+2XG_{4, XX})-12HXG_{4,\phi X}  \nonumber \\ &+2H^{3}X(3G_{5,X}+2XG_{5, XX})-6H^{2}\dot{\phi}(G_{5,\phi}+XG_{5, \phi X})\,, \nonumber \\
P_{\phi}&= G_{2, \phi}-2X(G_{3, \phi \phi}+\ddot{\phi}G_{3,\phi X})+6G_{4, \phi}(2H^{2}+\dot{H})+6HG_{4,\phi X}(\dot{X}+2HX)\nonumber \\ &-6H^{2}XG_{5,\phi \phi}+2H^{3}X \dot{\phi} G_{5, \phi X}\,.
\end{eqnarray}
To study the phase space analysis in the torsion Galileon theory, we need to define the function $G_{i}$, $i = 2, 3, 4, 5$ as,
\begin{eqnarray} \label{ch6_G_i_functions}
G_{2}(\phi, X)=X-V(\phi)\,,\quad
G_{3}(\phi, X)=-F(\phi)X\,,\quad
G_{4}(\phi, X)=\frac{1}{2 \kappa^{2}}\,,\quad
G_{5}(\phi, X)=0\,.
\end{eqnarray}
Eq.~(\ref{ch6_first_friedmann_equation}), Eq.~(\ref{ch6_second_friedmann_equation}) and Eq.~ (\ref{ch6_Klein-Gordan-equation}) can be rewritten by using Eq.~(\ref{ch6_G_i_functions})  as,
\begin{eqnarray}
\frac{3H^{2}}{\kappa^{2}}&=& \frac{\dot{\phi}^{2}}{2}+V(\phi)-3\dot{\phi}^{3}HF(\phi)+\frac{\dot{\phi}^{4}}{2}F_{,\phi}+\rho_{m}+\rho_{r} \label{ch6_model_first_friedmann}\,,\\
-\frac{\dot{2H}}{\kappa^{2}}&= &\dot{\phi}^{2}+\dot{\phi}^{4}F_{,\phi}+\ddot{\phi}\dot{\phi}^{2}F(\phi)-3\dot{\phi}^{3} H F(\phi)+\rho_m+\rho_{r}+p_{r} \label{ch6_model_second_friedmann}\,,
\end{eqnarray}
\begin{eqnarray}\label{ch6_model_Klein_gordan}
\ddot{\phi}+3H\dot{\phi}+2\ddot{\phi}\dot{\phi}^{2}F_{,\phi}+\frac{1}{2}\dot{\phi}^{4} F_{,\phi \phi}-3\dot{H}\dot{\phi}^{2}F(\phi)-6H\ddot{\phi}\dot{\phi}F(\phi)-9H^{2}\dot{\phi}^{2}F(\phi)+V_{,\phi}=0\,.    
\end{eqnarray}
The Friedmann [Eq.~\ref{ch6_model_first_friedmann}-Eq.~\ref{ch6_model_second_friedmann}] can also be respectively represented as,
\begin{eqnarray}
    \frac{3}{\kappa^{2}}H^{2}&=&\rho_{m}+\rho_{r}+\rho_{DE}\,, \label{ch6_Einstein_first_friedmann}\\
    -\frac{2}{\kappa^{2}}\dot{H}&=&\rho_{m}+\frac{4}{3}\rho_{r}+\rho_{DE}+p_{DE}\,. \label{ch6_Einstein_second_friedmann}
\end{eqnarray}
Now, comparing Eq.~(\ref{ch6_model_first_friedmann}) with Eq.~(\ref{ch6_Einstein_first_friedmann}) and Eq.~  (\ref{ch6_model_second_friedmann}) with Eq.~ (\ref{ch6_Einstein_second_friedmann}), the energy density ($\rho_{DE}$) and pressure ($p_{DE}$) for the DE sector can be determined as, 
\begin{eqnarray}
\rho_{DE}&=& \frac{\dot{\phi}^{2}}{2}+V(\phi)-3\dot{\phi}^{3}HF(\phi)+\frac{\dot{\phi}^{4}}{2}F_{,\phi} \label{ch6_rho_de}\,, \\ 
p_{DE}&=& \frac{\dot{\phi}^{2}}{2}-V(\phi)+\ddot{\phi}\dot{\phi}^{2}F(\phi)+\frac{\dot{\phi}^{4}}{2}F_{,\phi}  \label{ch6_p_de}\,.
\end{eqnarray}
Furthermore, we can define the DE EoS parameter as
\begin{eqnarray}\label{ch6_w_de}
\omega_{DE}=\frac{p_{DE}}{\rho_{DE}}= \frac{\frac{\dot{\phi}^{2}}{2}-V(\phi)+\ddot{\phi}\dot{\phi}^{2}F(\phi)+\frac{\dot{\phi}^{4}}{2}F_{,\phi}}{\frac{\dot{\phi}^{2}}{2}+V(\phi)-3\dot{\phi}^{3}HF(\phi)+\frac{\dot{\phi}^{4}}{2}F_{,\phi}}\,.    
\end{eqnarray}
One advantage of Galileon cosmology is that $\omega_{DE}$ can be quintessence-like or phantom-like or undergo the phantom divide crossing throughout the evolution according to $F(\phi)$.  We shall now define the dynamical variables and autonomous systems to study the different phases and behaviors of the Universe through dynamical system analysis in Galileon cosmology. 
\section{Dynamical System Framework}\label{ch6_SECIII}
Here, we propose the following dynamical variables to generate the relevant autonomous system associated with the set of cosmological equations,
\begin{eqnarray}\label{ch6_dynamical_variable}
x=\frac{\kappa\dot{\phi}}{\sqrt{6}H}\,, \hspace{1cm} y=\frac{\kappa\sqrt{V}}{\sqrt{3}H}\,, \hspace{1cm} u=H\dot{\phi} F(\phi)\,, \hspace{1cm} \rho=\frac{\kappa \sqrt{\rho_{r}}}{\sqrt{3}H} \,,
\end{eqnarray}
\begin{eqnarray}\label{ch6_dynamical_variable1}
\alpha=-\frac{F^{'}(\phi)}{\kappa F(\phi)}\,, \hspace{0.2cm} \lambda=-\frac{V^{'}(\phi)}{\kappa V(\phi)}\,, \hspace{0.2cm} \Theta=\frac{F^{''}(\phi)F(\phi)}{F^{'}(\phi)^{2}}\,, \hspace{0.2cm} \Gamma=\frac{V^{''}(\phi)V(\phi)}{V^{'}(\phi)^{2}}\,.
\end{eqnarray}
In terms of dynamical system variables, the density parameters can be expressed as,
\begin{eqnarray}
\Omega_{DE}&=&(1-6u)x^{2}+y^{2}-\sqrt{6}x^{3}u\alpha\,, \label{ch6_density-parameter-dynamical variable1} \\ 
\Omega_{r}&=&\rho^{2}\,, \label{ch6_density-parameter-dynamical variable2} \\ 
\Omega_{m}&=&1-(1-6u)x^{2}-y^{2}+\sqrt{6}x^{3}u\alpha-\rho^{2}\,. \label{ch6_density-parameter-dynamical variable3}     
\end{eqnarray}
The constraint equation can be written as,
\begin{equation} \label{ch6_Constraint equation}
\Omega_{DE}+\Omega_{m}+\Omega_{r}=1 .   
\end{equation}
To obtain an autonomous system, we need to take the derivative of Eq.~(\ref{ch6_dynamical_variable}) and Eq.~ (\ref{ch6_dynamical_variable1}) for $N$,
\begin{eqnarray}
\frac{dx}{dN}&=&\frac{\sqrt{6} \lambda  y^2-6 \left(\alpha ^2 \Theta  u x^3-3 u x+x\right)}{4 u \left(\sqrt{6} \alpha  x+3\right)+2}+\frac{x \left(\frac{3 u}{2 \sqrt{6} \alpha  u x+6 u+1}-1\right)}{2 u \left(9 u x^2+2 \sqrt{6} \alpha  x+6\right)+2} \nonumber \\ &&\bigg[18 \alpha ^2 (\Theta +2) u^2 x^4-3 \left(18 u^2+1\right) x^2 +3 \sqrt{6} \alpha  u (6 u-1) x^3 +\sqrt{6} u x \nonumber \\ &&\left(-2 \alpha  \left(\rho ^2+3\right)+6 \alpha  y^2-3 \lambda  y^2\right)+(6 u+1) \left(-\rho ^2+3 y^2-3\right)\bigg]\,, \label{ch6_autonomous-system1}\\
\frac{dy}{dN}&=& -\sqrt{\frac{3}{2}} \lambda  x y+\frac{1}{2 u \left(9 u x^2+2 \sqrt{6} \alpha  x+6\right)+2}\bigg[y \bigg(\rho ^2+u x \bigg(\sqrt{6} \bigg(\alpha  \bigg(2 \rho ^2 \nonumber \\&&+3 x^2-6 y^2+6\bigg) +3 \lambda  y^2\bigg) -18 u x \bigg(\alpha  x \left(\alpha  (\Theta +2)x+\sqrt{6}\right)-3\bigg)\bigg)\nonumber \\&&+6 u \left(\rho ^2-3 y^2+3\right)+3 x^2-3 y^2+3\bigg)\bigg]\,, \label{ch6_autonomous-system2} 
\end{eqnarray}
\begin{eqnarray}
\frac{du}{dN}&=& -\frac{3 \alpha ^2 \Theta  u^2 x^2}{2 \sqrt{6} \alpha  u x+6 u+1}+\frac{9 u^2}{2 \sqrt{6} \alpha  u x+6 u+1}-\sqrt{6} \alpha  u x -\frac{3 u}{2 \sqrt{6} \alpha  u x+6 u+1}\nonumber \\&&+\frac{\sqrt{\frac{3}{2}} \lambda  u y^2}{2 u x \left(\sqrt{6} \alpha  x+3\right)+x}+\frac{\frac{3 u^2}{2 \sqrt{6} \alpha  u x+6 u+1}+u}{2 u \left(9 u x^2+2 \sqrt{6} \alpha  x+6\right)+2} \bigg[18 \alpha ^2 (\Theta +2) u^2 x^4\nonumber \\&&-3 \left(18 u^2+1\right) x^2+3 \sqrt{6} \alpha  u (6 u-1) x^3 +\sqrt{6} u x \bigg(-2 \alpha  \left(\rho ^2+3\right) \nonumber \\&&+6 \alpha  y^2-3 \lambda  y^2\bigg)+(6 u+1) \left(-\rho ^2+3 y^2-3\right)\bigg]\,, \label{ch6_autonomous-system3} \\
\frac{d\rho}{dN}&=&\frac{1}{2 u \left(9 u x^2+2 \sqrt{6} \alpha  x+6\right)+2} \bigg[\rho  \bigg(\rho ^2+u x \bigg(\sqrt{6} \bigg(\alpha  \left(2 \rho ^2+3 x^2-6 y^2-2\right)\nonumber \\&&+3 \lambda  y^2\bigg)-18 u x \left(\alpha  x \left(\alpha  (\Theta +2) x+\sqrt{6}\right)-1\right)\bigg)\nonumber \\&&+6 u \left(\rho ^2-3 y^2-1\right)+3 x^2-3 y^2-1\bigg)\bigg]\,,\label{ch6_autonomous-system4} \\ 
\frac{d\alpha}{dN}&=& -\sqrt{6} \alpha ^2 (\Theta -1) x \,,\label{ch6_autonomous-system5} \\  
\frac{d\lambda}{dN}&=& -\sqrt{6} \lambda ^2 (\Gamma -1) x \,.\label{ch6_autonomous-system6}
\end{eqnarray}
Dynamical systems [Eq.~\ref{ch6_autonomous-system1}-Eq.~\ref{ch6_autonomous-system6}] are not autonomous systems without knowing $\Gamma$ and $\Theta$ parameters. The parameters depend on the coupling scalar field $F(\phi)$ with the Galileon term and the potential scalar field $V(\phi)$. So, we need to define the functions $F(\phi)$ and $V(\phi)$ to get the autonomous systems. We have considered the exponential form of the coupling scalar field function $F(\phi)=F_{0}e^{-\alpha \kappa \phi(t)}$, where $\alpha$ is a dimensionless parameter. For this exponential form of the coupling scalar field, we obtain $\Theta=1$, which means $\frac{d\alpha}{dN}=0$. 
In addition, we have taken different forms of the potential function $V(\phi)$ to define the value of the parameter $\Gamma$. Other potential functions yield different forms of $\Gamma$, where $\Gamma$ depends on $\lambda$. Remember that the phase space analysis only applies to potentials where $\Gamma$ can be expressed as a function of $\lambda$. To close the system [Eq.~\ref{ch6_autonomous-system1}-Eq.~\ref{ch6_autonomous-system6}], we consider three well-motivated forms of $V(\phi)$ in three different models and will provide a detailed analysis for the torsion Galileon theory. 

In terms of dynamical variables, the background cosmological parameters can be defined as,
\begin{eqnarray}
 q&=&\frac{1}{2 u \left(9 u x^2+2 \sqrt{6} \alpha  x+6\right)+2} \bigg[\rho ^2+u x \bigg(\sqrt{6} \bigg(\alpha  \left(2 \rho ^2+3 x^2-6 y^2+2\right) +3 \lambda  y^2\bigg)  -18 u x \bigg(\nonumber\\ &&\alpha  x \bigg(\alpha  (\Theta +2) x+\sqrt{6}\bigg)-2\bigg)\bigg)+6 u \left(\rho ^2-3 y^2+1\right)+3 x^2-3 y^2+1\bigg] \,,\label{ch6_qmodel} \\   
\omega_{tot}&=&\frac{1}{3 u \left(9 u x^2+2 \sqrt{6} \alpha  x+6\right)+3}\bigg[-18 \alpha ^2 (\Theta +2) u^2 x^4+3 \left(9 u^2+1\right) x^2 +3 \sqrt{6} \alpha  u (1-6 u) x^3 \nonumber\\ && +\sqrt{6} u x \left(2 \alpha  \rho ^2+3 y^2 (\lambda -2 \alpha )\right) -(6 u+1) \left(3 y^2-\rho ^2\right)\bigg] \,,\label{ch6_omegatotmodel}  
\end{eqnarray}
\begin{eqnarray}
\omega_{DE}&=&\frac{1}{-\sqrt{6} \alpha  u x^3+(1-6 u) x^2+y^2} \bigg[\frac{\sqrt{6} \lambda  u^3 y^2}{x \left(2 \sqrt{6} \alpha  u x+6 u+1\right)}-\frac{6 \alpha ^2 \Theta  u^2 x^4}{2 \sqrt{6} \alpha  u x +6 u+1} +\frac{18 u^2 x^2}{2 \sqrt{6} \alpha  u x+6 u+1} \nonumber\\ &&-\sqrt{6} \alpha  u x^3-\frac{6 u x^2}{2 \sqrt{6} \alpha  u x+6 u+1} +x^2-y^2+\frac{1}{\left(2 \sqrt{6} \alpha  u x+6 u+1\right) \left(2 u \left(9 u x^2+2 \sqrt{6} \alpha  x+6\right)+2\right)} \nonumber\\ &&\bigg(6 u^2 x^2 \bigg(18 \alpha ^2 (\Theta +2) u^2 x^4 -3 \left(18 u^2+1\right) x^2 +3 \sqrt{6} \alpha  u (6 u-1) x^3+\sqrt{6} u x \bigg(-2 \alpha  \bigg(\rho ^2+3\bigg)\nonumber\\ &&+6 \alpha  y^2-3 \lambda  y^2\bigg)+(6 u+1) \left(-\rho ^2+3 y^2-3\right)\bigg)\bigg) \bigg] \label{ch6_omegademodel}.
\end{eqnarray}
The background parameters display the evolution of the Universe through the fixed points of the autonomous system. It includes various phases of the Universe, accelerating and decelerating phases, quintessence, phantom and quintom eras of the Universe. 
\subsection{Model-I: \texorpdfstring{$V(\phi)=V_{0}e^{-\lambda \kappa \phi(t)}$}{case1}} \label{ch6_case1exp}
For exponential form of $V(\phi)$ \cite{copelandLiddle:1998, Ng_2001}, $\Gamma=1$ and so $\frac{d\lambda}{dN}=0$. The dynamical system [Eq.~\ref{ch6_autonomous-system1}-Eq.~\ref{ch6_autonomous-system6}] reduces to four dimensions as well as the autonomous system. To obtain the critical points ($x_c, y_c, u_c, \rho_c$) of the autonomous system [Eq.~\ref{ch6_autonomous-system1}-Eq.~\ref{ch6_autonomous-system6}], we need to take the condition $\frac{dx}{dN}=\frac{dy}{dN}=\frac{du}{dN}=\frac{d\rho}{dN}=0$. After imposing the conditions, we have obtained the six critical points with their existence conditions as given in Table~\ref{ch6_TABLE-I}. According to cosmological observations \cite{Riess:1998cb, SupernovaCosmologyProject:1998vns}, the phase of the Universe is accelerating expansion ($H>0$). Thus, the condition on the dynamical variables $x_c$, $y_c$, $u_c$ and $\rho_c$ must be real with $y_c>0$ and $\rho_c>0$. We have taken only positive values of $y_c$ since the negative value of $y_c$ describes the contraction phase ($H<0$) of the Universe. The corresponding value of the background cosmological parameters of each critical point has been displayed in Table~\ref{ch6_TABLE-II}. Finally, it should be noted from the physical viability condition [$0<\Omega_{i}<1$, where i= matter (m), radiation (r), dark energy (DE)] of the energy density parameters that the physical condition $\rho_{r} \geq 0$ implies $\rho \geq 0$ and $(-\sqrt{6} \alpha u x^3 + (1-6u) x^2 + y^2) < 1$ and $y > 0$. 
\begin{table}[H]
     \renewcommand{\arraystretch}{1} 
     \setlength{\tabcolsep}{8pt} 
    \caption{Critical points and existence condition (Model-I).} 
    \centering 
    \begin{tabular}{|c|c|c|c|c|c|} 
    \hline\hline 
    C.P. & $x_{c}$ & $y_{c}$ & $u_{c}$ & $\rho_{c}$ & Exists for \\ [0.5ex] 
    \hline\hline 
    $A_{1}$  & $0$ & $0$ & $0$ & $0$ &$Always$ \\
    \hline
    $A_{2}$  & $0$ & $1$ & $0$ & $0$ &$\lambda=0$ \\
    \hline
    $A_{3 \pm}$  & $\pm1$ & $0$ & $0$ & $0$ &$Always$ \\
    \hline
    $A_{4}$  & $\frac{2 \sqrt{\frac{2}{3}}}{\lambda }$ & $\frac{2}{\sqrt{3} \lambda }$ & $0$ & $\frac{\sqrt{\lambda ^2-4}}{\lambda }$ &$\lambda \neq 0, \hspace{0.2cm} \lambda^2 \geq 4 $ \\
    \hline
    $A_{5}$  & $\frac{\sqrt{\frac{3}{2}}}{\lambda }$ & $\frac{\sqrt{\frac{3}{2}}}{\lambda }$ & $0$ & $0$ &$\lambda \neq 0$ \\
    \hline
    $A_{6}$  & $\frac{\lambda }{\sqrt{6}}$ & $\frac{\sqrt{6-\lambda ^2}}{\sqrt{6}}$ & $0$ & $0$ &$6 \geq \lambda^2 > 0$ \\
     [1ex] 
    \hline 
    \end{tabular}
    \label{ch6_TABLE-I}
\end{table}

\begin{table}[H]
     \renewcommand{\arraystretch}{1} 
     \setlength{\tabcolsep}{8pt} 
    \caption{Density parameters, Deceleration parameter, EoS parameters (Model-I).} 
    \centering 
    \begin{tabular}{|c|c|c|c|c|c|c|} 
    \hline\hline 
    C.P. & $\Omega_{DE}$ & $\Omega_{m}$ & $\Omega_{r}$ & $q$ & $\omega_{DE}$ & $\omega_{tot}$ \\ [0.5ex] 
    \hline\hline 
    $A_{1}$  & $0$ & $1$ & $0$ & $\frac{1}{2}$ &$1$ & $0$\\
    \hline
    $A_{2}$  & $1$ & $0$ & $0$ & $-1$ &$-1$ & $-1$\\
    \hline
    $A_{3 \pm}$  & $1$ & $0$ & $0$ & $2$ &$1$ & $1$\\
    \hline
    $A_{4}$  & $\frac{4}{\lambda ^2}$ & $0$ & $1-\frac{4}{\lambda ^2}$ & $1$ &$\frac{1}{3}$ & $\frac{1}{3}$\\
    \hline
    $A_{5}$  & $\frac{3}{\lambda ^2}$ & $1-\frac{3}{\lambda ^2}$ & $0$ & $\frac{1}{2}$ &$0$ & $0$\\
    \hline
    $A_{6}$  & $1$ & $0$ & $0$ & $\frac{1}{2} \left(\lambda ^2-2\right)$ &$\frac{1}{3} \left(\lambda ^2-3\right)$& $\frac{1}{3} \left(\lambda ^2-3\right)$ \\
     [1ex] 
    \hline 
    \end{tabular}
    \label{ch6_TABLE-II}
\end{table}

{\bf \large {Description of Critical Points:}}
\begin{itemize}
\item Critical point $A_{1}$ provides matter-dominated solution $\Omega_{m}=1$ with total EoS parameter $\omega_{tot}=0$ and DE sector EoS parameter $\omega_{DE}=1$. The positive value of the deceleration parameter ($q=\frac{1}{2}$) shows the decelerating phase of the Universe. It exists always.

\item Critical point $A_{2}$ represents DE-dominated state $\Omega_{DE}=1$ with the value of the total and DE-dominated EoS parameter is $-1$. For this critical point, the negative value of the deceleration parameter indicates that the Universe is having accelerating behavior. It exists for $\lambda=0$.

\item The solution of the density parameter for the critical point $A_{3\pm}$ behaves $\Omega_{DE}=1$. Simultaneously, the solution of the total and DE-dominated EoS parameter is $1$. This critical point shows the stiff-matter phase. It cannot explain the accelerated phase of the Universe since $q>0$. It exists always.

\item The solution of the density parameter for the critical point $A_{4}$ is $\Omega_{DE}=\frac{4}{\lambda^{2}}$, $\Omega_{m}=0$ and $\Omega_{r}=1-\frac{4}{\lambda^{2}}$. The solution depends on the parameter $\lambda$, hence the scaling solution. This scaling solution reflects the radiation-DE-dominated era of the Universe. For this point, we have obtained $\omega_{tot}= \omega_{DE}=\frac{1}{3}$. So, $A_{4}$ depicts a non-standard radiation-dominated epoch of the Universe with a negligible DE contribution. The positive value of the deceleration parameter indicates the decelerating era of the Universe. We have obtained the restriction $ \lambda^2  > 4$ on the model parameter imposed by the physical condition $0<\omega_{DE}^{r}<1$.

\item The scaling solution of the density parameters for the critical point $A_{5}$ is $\Omega_{DE}=\frac{3}{\lambda^{2}}$, $\Omega_{m}=1-\frac{3}{\lambda^{2}}$ and $\Omega_{r}=0$ with the solution of the total and DE dominated EoS parameter is $0$. The point shows the non-standard matter-dominated phase of the Universe. This scaling solution represents the matter-DE-dominated era of the Universe. The positive value of the deceleration parameter indicates the decelerating phase. On imposing the condition $0<\omega_{DE}<1$, we obtain the constraint $\lambda^{2}>3$. It exists for $\lambda \neq 0$.

\item The critical point $A_{6}$ represented the DE-dominated phase of the Universe with $\Omega_{DE}=1$ and $\omega_{tot}=\omega_{DE}=-1+\frac{\lambda^2}{3}$. This critical point can explain the accelerated phase of the Universe for $\lambda^2 <2$. The condition $6\geq\lambda^{2}> 0$ indicates its existence. This critical point shows the quintessence phase ($-1<\omega_{tot}<-\frac{1}{3}$) of the Universe for the conditions $-\sqrt{2}<\lambda <0 $ and $0<\lambda <\sqrt{2}$ and shows the Phantom phase ($\omega_{tot}<-1$) for $\lambda <-\sqrt{6}$ and $\lambda >\sqrt{6}$. 
\end{itemize} 
{\bf \large{Stability Analysis:}}\\
Small perturbations around the critical points have been introduced to assess the stability of critical points and the equations of the system have been linearized. So, we can determine the stability of the matrix $\mathcal{M}$ by determining its eigenvalues $\lambda_1$, $\lambda_2$, $\lambda_3$ and $\lambda_4$. The stability features are classified as (a) stable node; all the eigenvalues are negative; (b) unstable node, all the eigenvalues are positive; (iii) saddle point, in the presence of both positive and negative eigenvalues; and (iv) stable spiral: complex with negative real part eigenvalues. Stable nodes and stable spirals describe the late-time cosmic acceleration phase of the Universe without depending on the initial conditions. The eigenvalues of the Jacobean matrix and stability conditions for each critical point are listed below.

\begin{figure}[H]
 \centering
 \includegraphics[width=65mm]{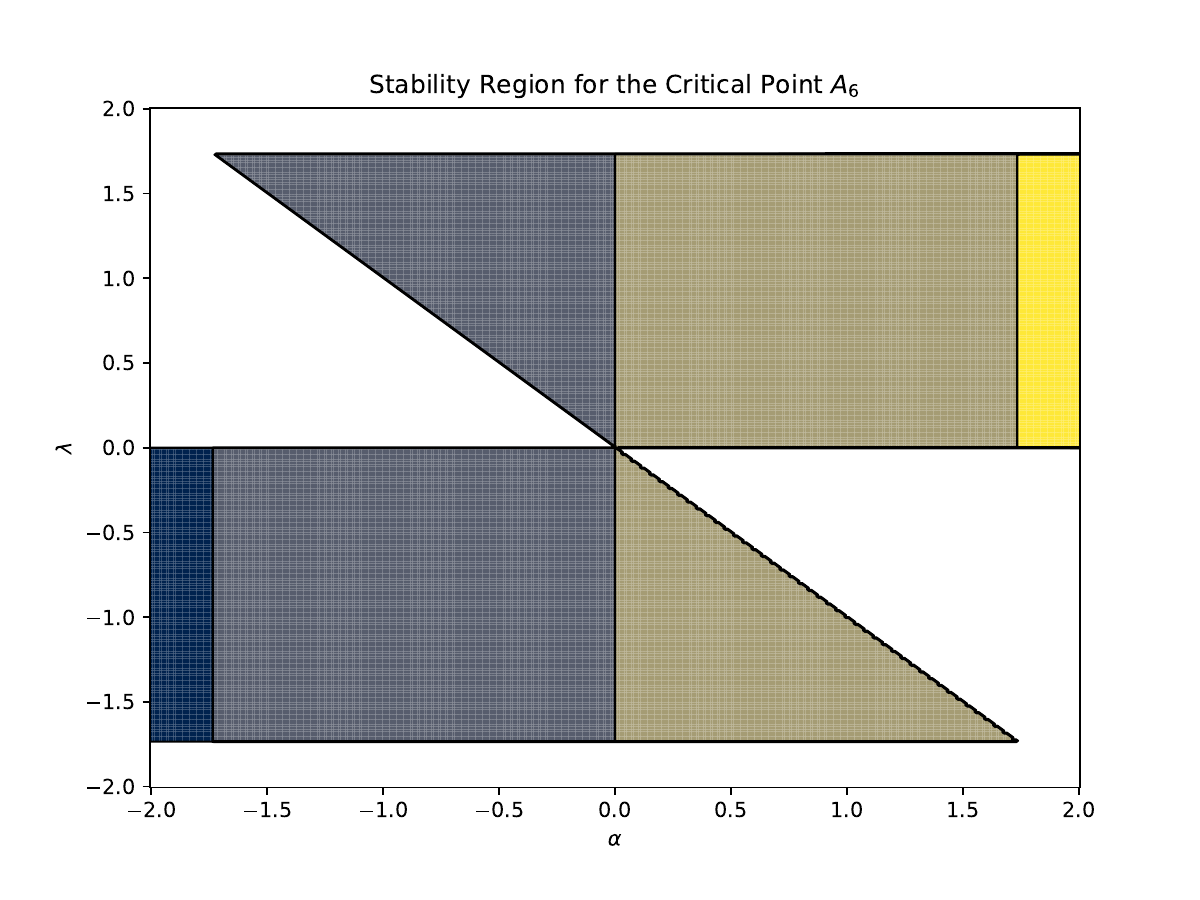}
 \caption{Stability region of the critical point $A_{6}$ between the model parameters $\alpha$ and $\lambda$.} \label{ch6_FigA6}
 \end{figure}

\begin{itemize}
\item  Eigenvalues of critical point $A_{1}$
\begin{eqnarray*}
\lambda_{1} = -\frac{1}{2}, \hspace{0.2cm} \lambda_{2} = -\frac{3}{2}, \hspace{0.2cm} \lambda_{3} = -\frac{9}{2} , \hspace{0.2cm} \lambda_{4} = \frac{3}{2} \,.     
\end{eqnarray*}
This point exhibits saddle behavior according to the linear stability theory as both positive and negative eigenvalues are present. 

\item Eigenvalues of critical point $A_{2}$
\begin{eqnarray*}
\lambda_{1} = -2, \hspace{0.2cm} \lambda_{2} = -3, \hspace{0.2cm} \lambda_{3} = -3 , \hspace{0.2cm} \lambda_{4} = -3 \,.   
\end{eqnarray*}
This point shows stable node behavior since all the eigenvalues are negative. This point can explain the accelerated phase of the Universe.

\item Eigenvalues of critical point $A_{3\pm}$
\begin{eqnarray*}
\lambda_{1} = 3, \hspace{0.2cm} \lambda_{2} = 1, \hspace{0.2cm} \lambda_{3} = -6\pm\sqrt{6} \alpha , \hspace{0.2cm} \lambda_{4} = 3\pm\sqrt{\frac{3}{2}} \lambda \,.   
\end{eqnarray*}

Sign ($-$) in $\lambda_3$ and $\lambda_{4}$ denote the eigenvalues of the critical point $A_{3+}$, whereas sign ($+$) describes the eigenvalues of the critical point $A_{3-}$. For $\alpha > -\sqrt{6}$ and $\lambda >\sqrt{6}$, the saddle behaviour is represented by the point $A_{3+}$. For the conditions $\alpha <\sqrt{6}$ and $\lambda <-\sqrt{6}$, the saddle behaviour is described by the point $A_{3-}$. The behavior of the unstable (node) if one of these two key points fails to satisfy the requirements mentioned above. 
    
\item Eigenvalues of critical point $A_{4}$
\begin{eqnarray*}
\lambda_{1} = 1, \hspace{0.2cm} \lambda_{2} = -\frac{4 (\alpha +\lambda )}{\lambda }, \hspace{0.2cm} \lambda_{3} =-\frac{1}{2} -\frac{\sqrt{64 \lambda ^6-15 \lambda ^8}}{2 \lambda ^4},\hspace{0.2cm} \lambda_{4} = -\frac{1}{2} +\frac{\sqrt{64 \lambda ^6-15 \lambda ^8}}{2 \lambda ^4} \,.   
\end{eqnarray*}

This critical point shows unstable (saddle) behavior for the conditions $\bigg(-\frac{8}{\sqrt{15}}\leq \lambda <-2\land \alpha <-\lambda\bigg)$ and $\bigg(2<\lambda  \leq \frac{8}{\sqrt{15}}\land \alpha >-\lambda\bigg)$. Failing to satisfy the above conditions, the point exhibits unstable (node) behavior.
\item Eigenvalues of critical point $A_{5}$
\begin{eqnarray*}
\lambda_{1} = -\frac{1}{2}, \hspace{0.2cm} \lambda_{2} = -\frac{3 (\alpha +\lambda )}{\lambda }, \hspace{0.2cm} \lambda_{3} = -\frac{3 \left(\lambda ^4+\sqrt{24 \lambda ^6-7 \lambda ^8}\right)}{4 \lambda ^4}, \hspace{0.2cm} \lambda_{4} = -\frac{3}{4}+\frac{3 \sqrt{24 \lambda ^6-7 \lambda ^8}}{4 \lambda ^4} \,. 
\end{eqnarray*}
This critical point describes the stable node behavior for the conditions  
$\bigg(-2 \sqrt{\frac{6}{7}}\leq \lambda <-\sqrt{3}\land \alpha <-\lambda \bigg)$ and $\left(\sqrt{3}<\lambda \leq 2 \sqrt{\frac{6}{7}}\land \alpha >-\lambda \right)$. The point exhibits saddle behavior if it does not fulfill the above-mentioned conditions. 

\item Eigenvalues of critical point $A_{6}$
\begin{eqnarray*}
\lambda_{1} = \frac{1}{2} \left(\lambda ^2-6\right), \hspace{0.2cm} \lambda_{2} = \frac{1}{2} \left(\lambda ^2-4\right), \hspace{0.2cm} \lambda_{3} =\lambda ^2-3 , \hspace{0.2cm}  \lambda_{4} = -\lambda  (\alpha +\lambda ) \,.
\end{eqnarray*}
This critical point indicates stable node behavior for the conditions $\bigg(\alpha \leq -\sqrt{3}\land -\sqrt{3}<\lambda <0\bigg) $,  $\bigg(-\sqrt{3}<\alpha \leq 0\land \bigg(-\sqrt{3}<\lambda <0\lor -\alpha <\lambda <\sqrt{3}\bigg)\bigg) $,$ \bigg(0<\alpha <\sqrt{3}\land \left(-\sqrt{3}<\lambda <-\alpha \lor 0<\lambda <\sqrt{3}\right)\bigg)$ and $ \bigg(\alpha \geq \sqrt{3}\land 0<\lambda <\sqrt{3}\bigg)$.  If the point fails to satisfy the above conditions, it behaves as an unstable node or saddle. For this critical point, we have displayed a stability region [Fig.~\ref{ch6_FigA6}] between the model parameters $\alpha$ and $\lambda$ for the above-mentioned stable node conditions. The shaded region in  Fig.~\ref{ch6_FigA6} shows the stable node behavior. Late-time cosmic acceleration can be studied through this critical point.
\end{itemize}

{\bf { \large Numerical Solutions:}}

According to the solution of background parameters, the critical points represent different phases of the Universe. Among these critical points, $A_{2}$ and $A_{6}$ represent the DE-dominated phase of the Universe. Now, we wish to analyze the autonomous system [Eq.~\ref{ch6_autonomous-system1}-Eq.~\ref{ch6_autonomous-system6}] with the numerical solution. To obtain this, we used the NDSolve command in Mathematica. In addition, we compared our results with Hubble and SNIa observational data sets [Details in section--\ref{ObservationalCosmology}]. 

\begin{figure}
     \centering
     \begin{subfigure}[b]{0.3\textwidth}
         \centering
         \includegraphics[width=\linewidth]{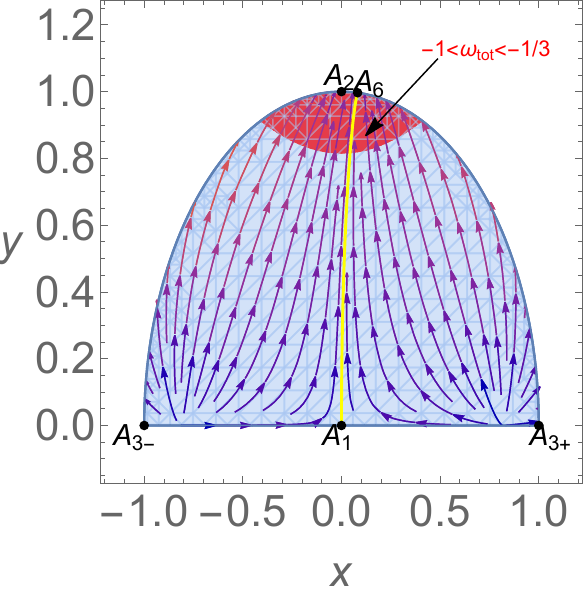}
         \caption{Phase space with $\alpha=-5$ and $\lambda=0.2$}
         \label{ch6_fig:case1_phaseportrait}
     \end{subfigure}
     \hfill
     \begin{subfigure}[b]{0.3\textwidth}
         \centering
         \includegraphics[width=\linewidth]{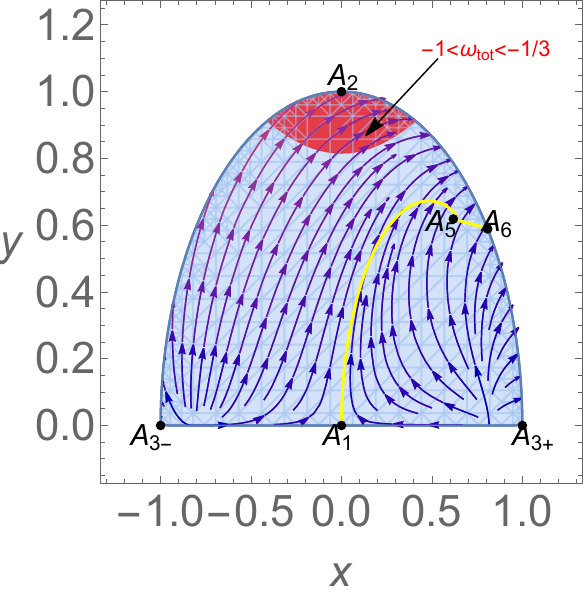}
         \caption{Phase space with $\alpha=-5$ and $\lambda=1.98$}
         \label{ch6_fig:case1_phaseA5}
     \end{subfigure}
     \hfill
     \begin{subfigure}[b]{0.3\textwidth}
         \centering
         \includegraphics[width=\linewidth]{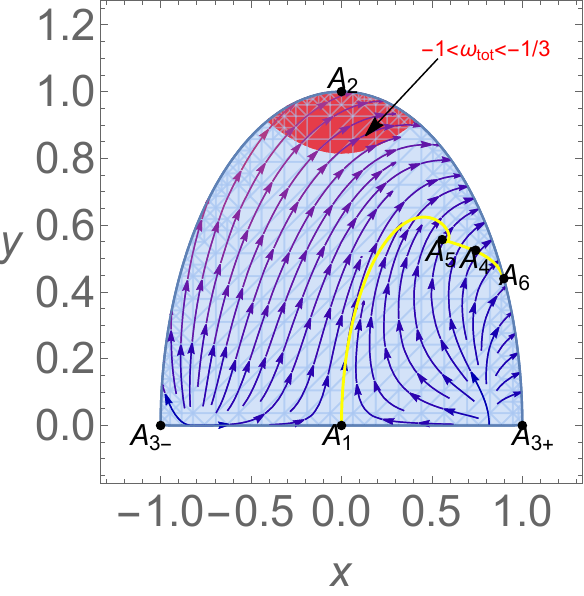}
         \caption{Phase space with $\alpha=-5$ and $\lambda=2.2$}
         \label{ch6_fig:case1_phaseA4}
     \end{subfigure}
\caption{Model-I: 2D phase space portrait of the autonomous system [Eq.~\ref{ch6_autonomous-system1}-Eq.~\ref{ch6_autonomous-system6}]. The red/shaded region indicates where the Universe experiences accelerated expansion ($-1 < \omega_{tot} < -\frac{1}{3}$). The initial conditions are set to: $x = 10^{-5}$, $y = 9 \times 10^{-13}$, $u = 10^{-5}$ and $\rho = \sqrt{0.999661}$.} 
\label{ch6_Fig2phasepor}
\end{figure}

In Fig.~\ref{ch6_Fig2phasepor}, we analyze the phase space portrait for different model parameter values $\lambda$. The behavior within the phase space is categorized into three distinct regions, each corresponding to specific ranges of $\lambda$. This variation in $\lambda$ is crucial as it underscores the existence of critical points and delineates the accelerating region, represented in red/shaded areas, which includes the critical points $A_4$, $A_5$ and $A_6$. The critical points within this accelerating region are significant for understanding the dynamics of the Universe, as they indicate the conditions under which cosmic acceleration occurs. 

In Fig.~\ref{ch6_fig:case1_phaseportrait}, we illustrate the phase portrait for $\lambda = 0.2$. At this value, the critical points $A_4$ and $A_5$ are absent as these points arise only when $\lambda^2 > 4$ and $\lambda^2 > 3$, respectively. These conditions adhere to the physical constraints on the density parameters $(0 < \Omega_r < 1)$ and $(0 < \Omega_m < 1)$. Consequently, in the accelerating region associated with the critical point $A_6$, where $\lambda^2 < 2$, the critical points $A_4$ and $A_5$ do not emerge. For $\lambda = 0.2$, the critical point resides in the accelerating region (red/shaded), with $A_6$ corresponding to the accelerating phase of the Universe under the condition $\lambda^2 < 2$. Thus, we have selected $\lambda = 0.2$ since it meets the stability criteria for the critical point $A_6$. All trajectories in the phase space consist of heteroclinic orbits that start from the points $A_{3\pm}$ and end at $A_6$. Two heteroclinic orbits are observed: $A_{3\pm} \to A_1 \to A_6$. These orbits can serve as physical models for the transition from DM to DE, effectively capturing the late-time evolution of the Universe, with the total EoS parameter, $\omega_{tot} = -1 + \frac{\lambda^2}{3}$. However, at early times, the model predicts stiff fluid domination represented by the points $A_{3\pm}$, which is unfavorable from a phenomenological viewpoint. For values of $\lambda^2 > 2$, the point $A_6$ would fall outside the acceleration region (red/shaded) and would not represent an inflationary solution. At $\lambda=0.2$, the critical point $A_6$ exhibits stable node behavior, indicating late-time cosmic acceleration of the Universe. The heteroclinic orbit solution (yellow line) is derived from the numerical solution of the autonomous system [Eq.~\ref{ch6_autonomous-system1}-Eq.~\ref{ch6_autonomous-system6}] with initial conditions $x = 10^{-5}$, $y = 9 \times 10^{-13}$, $u = 10^{-5}$ and $\rho = \sqrt{0.999661}$.

In Fig.~\ref{ch6_fig:case1_phaseA5}, we can see six critical points in the phase space for the condition $\lambda^2 > 3$. For $\lambda=1.98$, the critical point $A_4$ does not exist and the critical point $A_6$ lies outside the accelerating region (red/shaded). So, the critical points $A_5$ and $A_6$ show saddle behavior (unstable) and indicate the decelerating phase. There are also two heteroclinic orbits $A_{3\pm} \to A_1 \to A_6$.

In Fig.~\ref{ch6_fig:case1_phaseA4}, we observe seven critical points in the phase space for the condition $\lambda^2 > 4$. For $\lambda = 2.2$, the critical point $A_6$ is located outside the accelerating region (red/shaded). At this value, the critical points $A_4$, $A_5$ and $A_6$ exhibit saddle-like (unstable) behavior and show the decelerating phase. Here also, $A_6$ remains outside the acceleration region (red/shaded) and hence never corresponds to an inflationary solution. Two heteroclinic orbits exist: $A_{3\pm} \to A_1 \to A_6$. 

\begin{figure}
     \centering
     \begin{subfigure}[b]{0.46\textwidth}
         \centering
         \includegraphics[width=70mm]{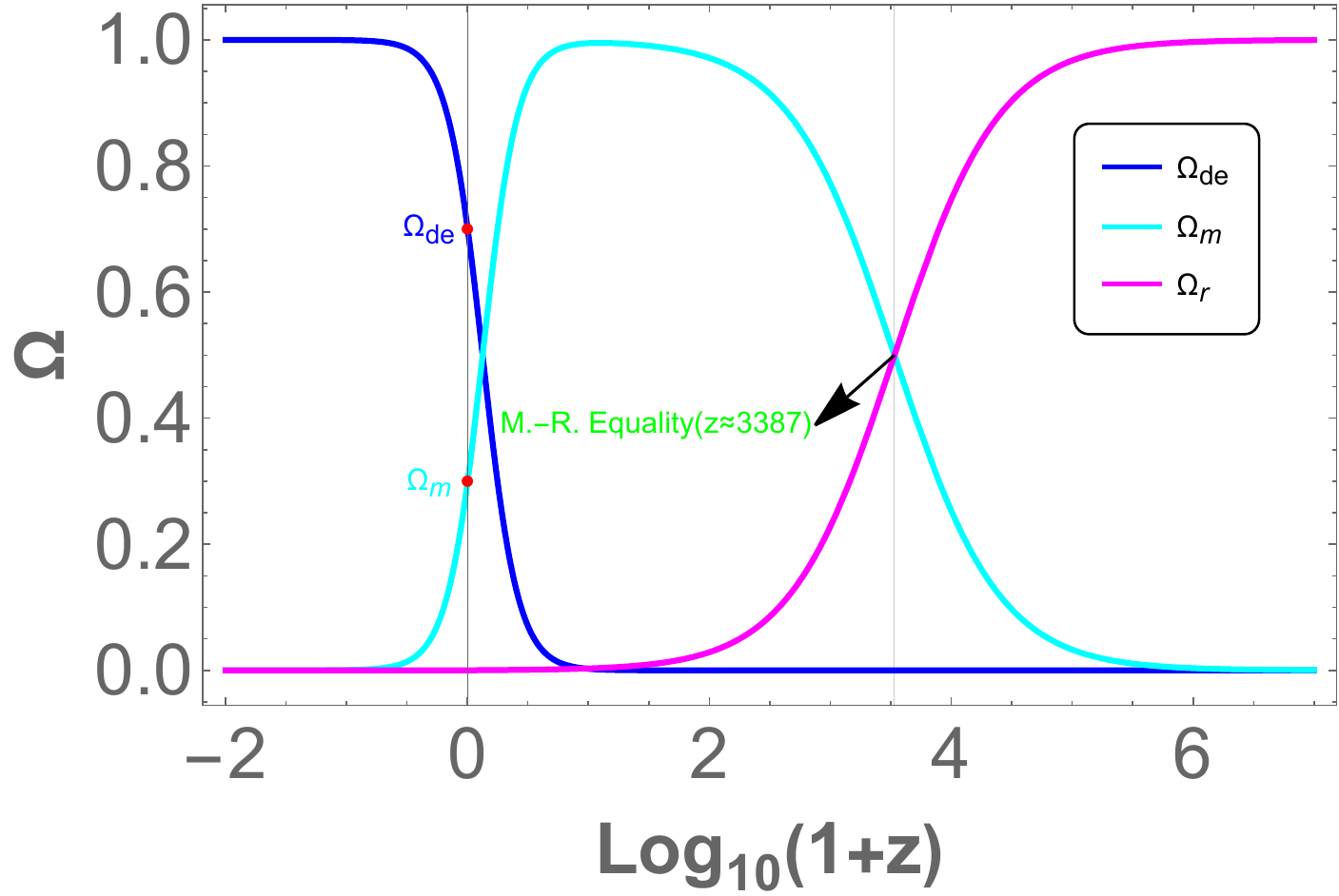}
         \caption{Evolution of density parameters.}
         \label{ch6_fig:case1_density}
     \end{subfigure}
     \begin{subfigure}[b]{0.46\textwidth}
         \centering
         \includegraphics[width=70mm]{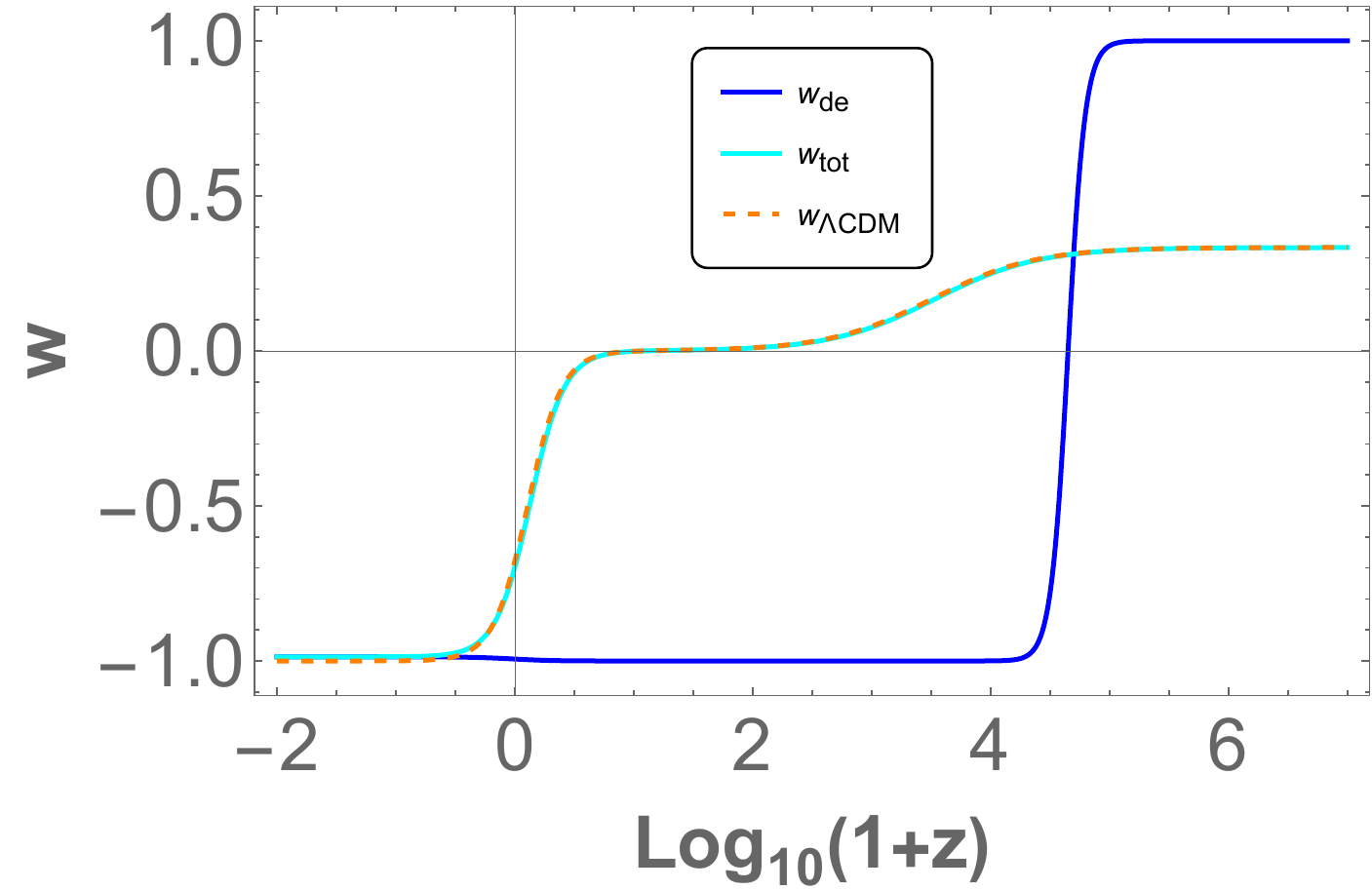}
         \caption{Evolution of EoS parameters.}
         \label{ch6_fig:case1_statepara}
     \end{subfigure}
\caption{In this figure, we set $\alpha=-5$ and $\lambda=0.2$ with the initial conditions are the same as in Fig.~\ref{ch6_Fig2phasepor}. The red dot indicates the present-day value of the density parameter at redshift $z = 0$, while the vertical black line corresponds to the current cosmological time. M.R. indicates the time of matter-radiation equality.} 
\label{ch6_Fig1}
\end{figure}

\begin{figure}
     \centering
     \begin{subfigure}[b]{0.49\textwidth}
         \centering
         \includegraphics[width=70mm]{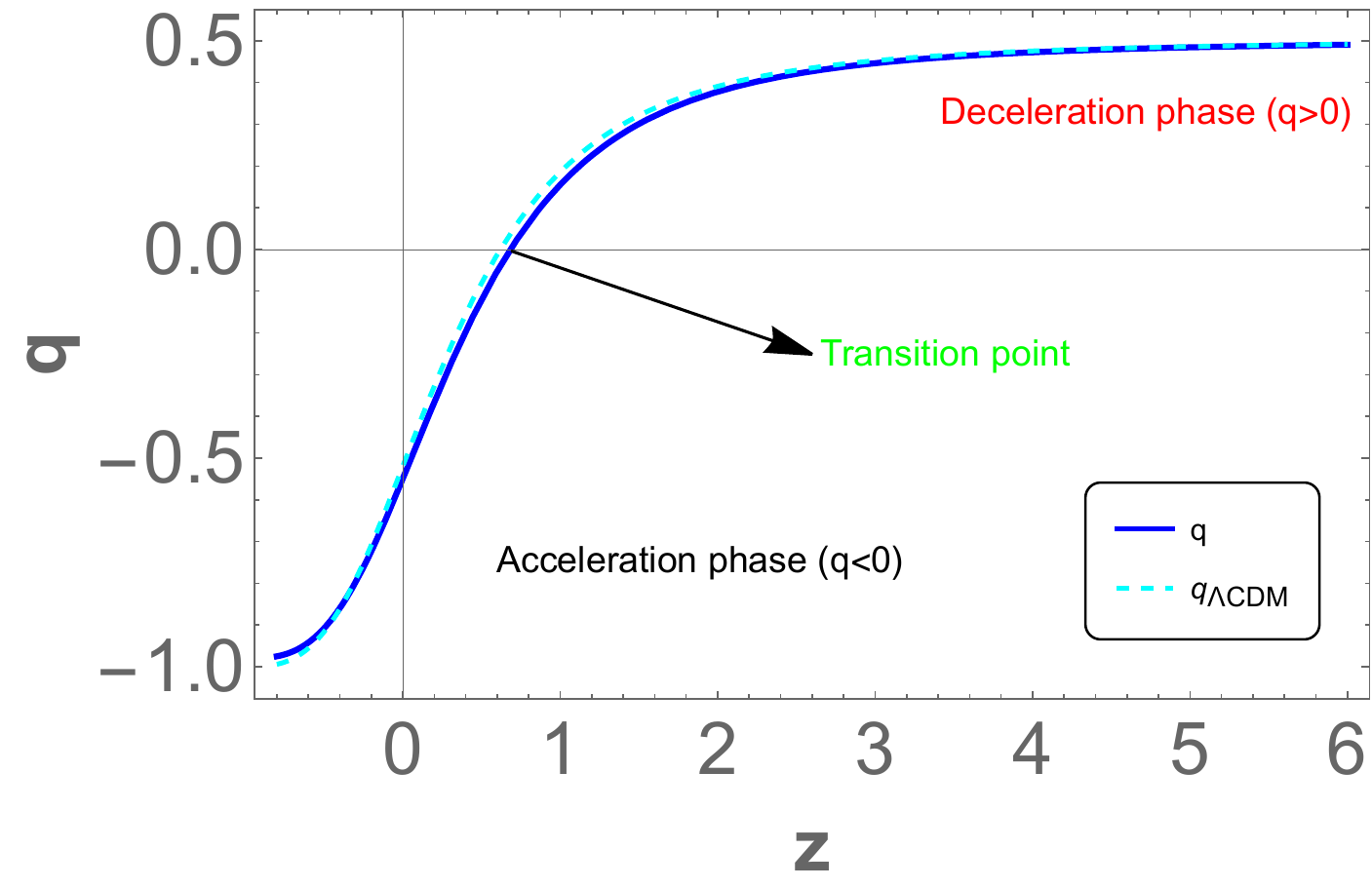}
         \caption{Evolution of deceleration parameter.}
         \label{ch6_fig:case1_qz}
     \end{subfigure}
     \begin{subfigure}[b]{0.49\textwidth}
         \centering
         \includegraphics[width=70mm]{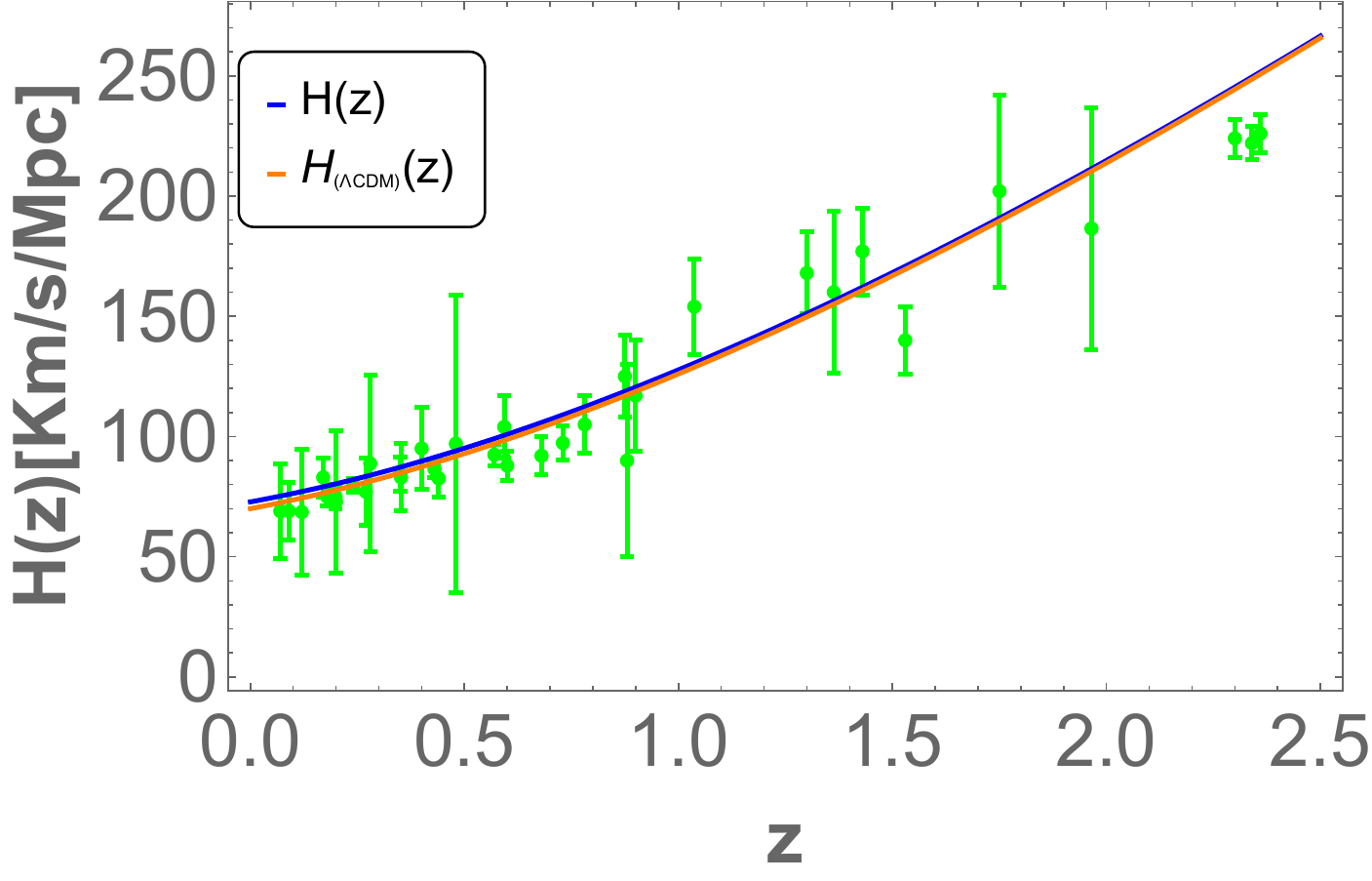}
         \caption{Evolution of Hubble rate $H(z)$.}
         \label{ch6_fig:case1_Hz}
     \end{subfigure}
\caption{In this figure, we set $\alpha=-5$ and $\lambda=0.2$ with the initial conditions are the same as in Fig.~\ref{ch6_Fig2phasepor}.} 
\label{ch6_Fig2}
\end{figure}

\begin{figure}[H]
 \centering
 \includegraphics[width=80mm]{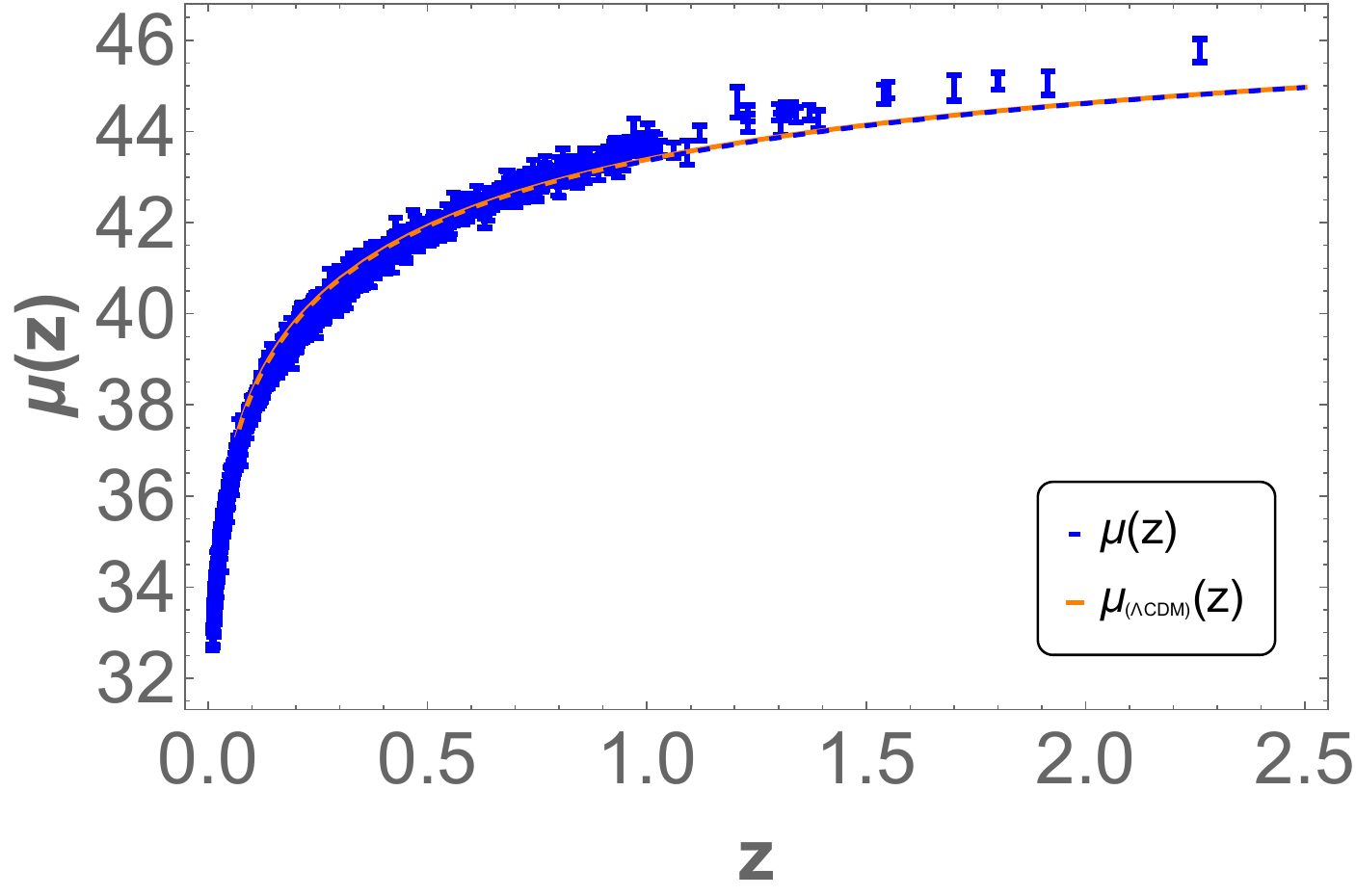} 
 \caption{In this figure, we set $\alpha=-5$ and $\lambda=0.2$ with the initial conditions are the same as in Fig.~\ref{ch6_Fig2phasepor}. The figure shows the evolution of distance modulus function $\mu(z)$ ( dashed blue) and $\Lambda$CDM model distance modulus function $\mu_{\Lambda CDM}(z)$ along with 1048 SNIa data points \cite{Scolnic_2018}. } \label{ch6_Fig2pan}
 \end{figure}

From a physics perspective, the cosmological dynamics of the exponential potential are intriguing due to the emergence of late-time accelerated solutions that can be utilized to model DE. For these solutions to be cosmologically viable, a sufficiently flat potential ($\lambda^2<2$) is necessary, along with a strong fine-tuning of initial conditions to ensure DE domination persists. This solution must resemble the sequence:$A_{3\pm} \to A_1 \to A_6$ (Fig.~\ref{ch6_fig:case1_phaseportrait}). At early times, the only feasible solutions are the non-physical stiff fluid Universe. We have plotted the background cosmological parameters such as the  EoS parameters, energy densities, deceleration parameters, Hubble rate and modulus function for a solution shadowing the heteroclinic sequence $A_{3\pm} \to A_1 \to A_6$ shown in Fig.~\ref{ch6_Fig1}, Fig.~\ref{ch6_Fig2} and Fig.~\ref{ch6_Fig2pan}.

Graphically, the relative energy densities of radiation, DE and DM are shown in Fig.~\ref{ch6_fig:case1_density}. Radiation occurs in the early Universe, followed by a brief period of dominance over DM and, finally, the cosmological constant. The matter and DE sector density parameters are currently $\Omega_{m}\approx 0.3$ and $\Omega_{DE}\approx 0.7$, respectively. The time of matter-radiation equality is around $z\approx 3387$. In Fig.~\ref{ch6_fig:case1_statepara}, the behaviour of EoS parameter shows that  $\omega_{tot}$ (cyan) starts from the radiation value of $\frac{1}{3}$, drops to $0$ during the matter-dominated era and eventually reaches $-1$. Also, both $\omega_{DE}$ (blue) and $\omega_{\Lambda CDM}$(dashed orange) approaches $-1$ at late-time. At present, $\omega_{DE}(z=0)=-0.99$, which is compatible with the present Planck Collaboration result [$\omega_{DE}(z=0)= -1.028 \pm 0.032$ \cite{Aghanim:2018eyx}]. In Fig.~\ref{ch6_fig:case1_qz}, the evolutionary behavior of the deceleration parameter shows transition behavior at $z\approx 0.66$, compatible with the current observational data \cite{PhysRevD.90.044016a}. The present value of the deceleration parameter is $q(z=0) \approx -0.55$, consistent with the visualized cosmological observations \cite{Camarena:2020prr}. In Fig.~\ref{ch6_fig:case1_Hz}, we illustrate the  Hubble rate evolution with the Hubble rate $H_{\Lambda CDM}(z)$ and the 31 Hubble data points \cite{Moresco_2022_H0},$H_{0} = 71.88 \ \text{Km} \, \text{s}^{-1} \text{Mpc}^{-1}$ \cite{Aghanim:2018eyx}. We find our model to be fairly close to the standard $\Lambda$CDM model. In Fig.~\ref{ch6_Fig2pan}, we plot the evolution of the modulus function $\mu(z)$ and observe that the model curve and the $\Lambda$CDM model modulus function $\lambda_{\Lambda CDM}$ are well within error bar. 

\subsection{Model-II: \texorpdfstring{$V(\phi)=V_{0}\phi^{n}$}{caseII}} \label{ch6_case2power}

To find the value of $\Gamma$ in the power-law form of potential function $V(\phi)=V_{0}\phi^{n}$ \cite{Copeland:2006wr}, we consider the function  $f(\lambda)=\lambda^{2}(\Gamma-1)$ [Eq.~ (\ref{ch6_autonomous-system6})]. The general structure of the function $f(\lambda)$ becomes $f(\lambda)=\beta_1 \lambda^2+ \beta_{2} \lambda+ \beta_{3}$. Here $\lambda^\pm=\frac{1}{2\beta_{1}}\left(-\beta_{2}\pm \sqrt{\beta_2^{2}-4\beta_{1} \beta_{3}}\right)$ are two roots of $f(\lambda)$. These roots always satisfy the condition $(\beta_2^{2}-4\beta_{1} \beta_{3}) \geq 0$ since the interest is only the real solutions of $f(\lambda)$. The autonomous system is reduced to five-dimension space for this form of $V(\phi)$. Now, we need to impose the conditions, $\frac{dx}{dN}=\frac{dy}{dN}=\frac{du}{dN}=\frac{d\rho}{dN}=\frac{d\lambda}{dN}=0$ to determine the critical points of the autonomous system [Eq.~\ref{ch6_autonomous-system1}-Eq.~\ref{ch6_autonomous-system6}]. In Table~\ref{ch6_TABLE-III}, the eigenvalues of the critical points are presented along with the existence condition, whereas the values of the cosmological parameters are given in Table~\ref{ch6_TABLE-IV}. Now, the roots of the potential function $V(\phi)=V_{0}\phi^{n}$ are as follows,
\begin{eqnarray}\label{ch6_powerlawpotentialroot}
f(\lambda)=-\frac{\lambda^{2}}{2}\,, \hspace{0.5cm} \beta_{1}=-\frac{1}{n}\,, \hspace{0.5cm} \beta_{2}=\beta_{3}=0\,, \hspace{0.2cm} \lambda^{+}=\lambda^{-}=-\frac{n}{2}
\end{eqnarray}
Since the roots $\lambda^{+}$ and $\lambda^{-}$ are same, therefore the behaviour of critical points $B_{3\pm}$, $B_{5}$, $B_{7}$ and $B_{9}$ respectively same as that of $B_{4\pm}$, $B_{6}$, $B_{8}$ and $B_{10}$.   
\begin{table}[H]
     \renewcommand{\arraystretch}{1.2} 
     \setlength{\tabcolsep}{7.5pt} 
    \caption{Critical points and existence condition.} 
    \centering 
    \begin{tabular}{|c|c|c|c|c|c|c|} 
    \hline\hline 
    C.P. & $x_{c}$ & $y_{c}$ & $u_{c}$ & $\rho_{c}$&$\lambda_c$ & Exists for \\ [0.5ex] 
    \hline\hline 
    $B_{1}$  & $0$ & $0$ & $0$ & $0$ &$\lambda_c$&$Always$ \\
    \hline
    $B_{2}$  & $0$ & $1$ & $0$ & $0$& $0$ &$\lambda=0$ \\
    \hline
    $B_{3 \pm}$  & $\pm1$ & $0$ & $0$ & $0$&$\lambda^{+}$ &$Always$ \\
    \hline
    $B_{4 \pm}$  & $\pm1$ & $0$ & $0$ & $0$&$\lambda^{-}$ &$Always$ \\
    \hline
    $B_{5}$  & $\frac{\sqrt{\frac{3}{2}}}{\lambda^{+} }$ & $\frac{\sqrt{\frac{3}{2}}}{\lambda^{+} }$ & $0$ & $0$ &$\lambda^{+}$&$\lambda^{+} \neq 0$ \\
      \hline
    $B_{6}$  & $\frac{\sqrt{\frac{3}{2}}}{\lambda^{-} }$ & $\frac{\sqrt{\frac{3}{2}}}{\lambda^{-} }$ & $0$ & $0$ &$\lambda^{-}$&$\lambda^{-} \neq 0$ \\
    \hline
    $B_{7}$  & $\frac{2 \sqrt{\frac{2}{3}}}{\lambda^{+} }$ & $\frac{2}{\sqrt{3} \lambda^{+} }$ & $0$ & $\frac{\sqrt{(\lambda^{+}) ^2-4}}{\lambda^{+} }$ &$\lambda^{+}$&$\lambda^{+} \neq 0, \hspace{0.2cm} (\lambda^{+})^2 \geq 4 $ \\
    \hline
    $B_{8}$  & $\frac{2 \sqrt{\frac{2}{3}}}{\lambda^{-} }$ & $\frac{2}{\sqrt{3} \lambda^{-} }$ & $0$ & $\frac{\sqrt{(\lambda^{-}) ^2-4}}{\lambda^{-} }$ &$\lambda^{-}$&$\lambda^{-} \neq 0, \hspace{0.2cm} (\lambda^{-})^2 \geq 4 $ \\
    \hline
    $B_{9}$  & $\frac{\lambda^{+} }{\sqrt{6}}$ & $\frac{\sqrt{6-(\lambda^{+} )^2}}{\sqrt{6}}$ & $0$ & $0$&$\lambda^{+}$ &$6 \geq (\lambda^{+})^2 > 0$ \\
      \hline
    $B_{10}$  & $\frac{\lambda^{-} }{\sqrt{6}}$ & $\frac{\sqrt{6-(\lambda^{-} )^2}}{\sqrt{6}}$ & $0$ & $0$&$\lambda^{-}$ &$6 \geq (\lambda^{-})^2 > 0$ \\
     [1ex] 
    \hline 
    \end{tabular}
    \label{ch6_TABLE-III}
\end{table}

\begin{table}[H]
     \renewcommand{\arraystretch}{1.2} 
     \setlength{\tabcolsep}{7.5pt} 
    \caption{Density parameters, Deceleration parameter, EoS parameters.} 
    \centering 
    \begin{tabular}{|c|c|c|c|c|c|c|} 
    \hline\hline 
    C.P. & $\Omega_{DE}$ & $\Omega_{m}$ & $\Omega_{r}$ & $q$ & $\omega_{DE}$ & $\omega_{tot}$ \\ [0.5ex] 
    \hline\hline 
    $B_{1}$  & $0$ & $1$ & $0$ & $\frac{1}{2}$ &$1$ & $0$\\
    \hline
    $B_{2}$  & $1$ & $0$ & $0$ & $-1$ &$-1$ & $-1$\\
    \hline
    $B_{3 \pm}$  & $1$ & $0$ & $0$ & $2$ &$1$ & $1$\\
    \hline
    $B_{4 \pm}$  & $1$ & $0$ & $0$ & $2$ &$1$ & $1$\\
    \hline
    $B_{5}$  & $\frac{3}{(\lambda^{+}) ^2}$ & $1-\frac{3}{(\lambda^{+}) ^2}$ & $0$ & $\frac{1}{2}$ &$0$ & $0$\\
     \hline
    $B_{6}$  & $\frac{3}{(\lambda^{-}) ^2}$ & $1-\frac{3}{(\lambda^{-}) ^2}$ & $0$ & $\frac{1}{2}$ &$0$ & $0$\\
    \hline
    $B_{7}$  & $\frac{4}{(\lambda^{+}) ^2}$ & $0$ & $1-\frac{4}{(\lambda^{+}) ^2}$ & $1$ &$\frac{1}{3}$ & $\frac{1}{3}$\\
    \hline
    $B_{8}$  & $\frac{4}{(\lambda^{-}) ^2}$ & $0$ & $1-\frac{4}{(\lambda^{-}) ^2}$ & $1$ &$\frac{1}{3}$ & $\frac{1}{3}$\\
    \hline
    $B_{9}$  & $1$ & $0$ & $0$ & $\frac{1}{2} \left((\lambda^{+}) ^2-2\right)$ &$\frac{1}{3} \left((\lambda^{+} )^2-3\right)$& $\frac{1}{3} \left((\lambda^{+}) ^2-3\right)$ \\
     \hline
    $B_{10}$  & $1$ & $0$ & $0$ & $\frac{1}{2} \left((\lambda^{-}) ^2-2\right)$ &$\frac{1}{3} \left((\lambda^{-} )^2-3\right)$& $\frac{1}{3} \left((\lambda^{-}) ^2-3\right)$ \\
     [1ex] 
    \hline 
    \end{tabular}
    \label{ch6_TABLE-IV}
\end{table}
{\bf{\large Description of Critical Points:}}
\begin{itemize}
\item At the background level, the critical points $B_{1}$, $B_{2}$ and $B_{3\pm}$ show similar behavior respectively to that of the critical points $A_{1}$, $A_{2}$ and $A_{3\pm}$ same as in  Model-I.

\item The scaling solutions of density parameters for the critical point $B_{5}$ are $\Omega_{DE}=\frac{12}{n^{2}}$, $\Omega_{m}=1-\frac{12}{n^{2}}$ and $\Omega_{r}=0$, which indicates the non-standard phase of matter-dominated Universe. The scaling solution shows the evolution between the matter-DE-dominated eras of the Universe. It shows the decelerated era of the Universe. The existence condition of $B_{5}$ is $n^2 > 12$. 

\item The density parameter scaling solution at the critical point $B_{7}$ is, $\Omega_{DE}=\frac{16}{n^{2}}$, $\Omega_{m}=0$ and $\Omega_{r}=1-\frac{16}{n^{2}}$. Scaling solutions represent the evolution of the Universe between radiation-DE phases. This indicates the non-standard radiation and decelerating phase of the Universe. It exists for $n^{2}\geq 16$.

\item The critical point $B_{9}$ shows the DE-dominated phase of the Universe. The value of the deceleration parameter for this critical point is $\frac{1}{8} \left(n^2-8\right)$. The condition for accelerated era is, $n^2<8 $ whereas for decelerated era, $n<-2 \sqrt{2}$ and $ n>2 \sqrt{2}$. Both the EoS parameters become $\omega_{DE}=\omega_{tot}=-1+\frac{n^2}{12}$. This point shows quintessence phase for $-2 \sqrt{2}<n<0 $ and $ 0<n<2 \sqrt{2}$. The phantom phase is not visible at this critical point as it does not satisfy $\omega_{tot}<-1$ for any value of $n$.   
\end{itemize}

{\bf{\large Stability Analysis:}}

The eigenvalues of the Jacobean matrix for each critical point are given below:

\begin{itemize}
\item Eigenvalues of critical point $B_{1}$
\begin{eqnarray*}
\lambda_{1} = -\frac{1}{2}, \hspace{0.2cm} \lambda_{2} = -\frac{3}{2}, \hspace{0.2cm} \lambda_{3} = -\frac{9}{2} , \hspace{0.2cm} \lambda_{4} = \frac{3}{2}\,, \hspace{0.2cm} \lambda_{5} = 0  \,.   
\end{eqnarray*}
The presence of the positive eigenvalue shows the unstable behavior.
\item Eigenvalues of critical point $B_{2}$
\begin{eqnarray*}
\lambda_{1} = -2, \hspace{0.2cm} \lambda_{2} = -3, \hspace{0.2cm} \lambda_{3} = -3 , \hspace{0.2cm} \lambda_{4} = -3, \hspace{0.2cm} \lambda_{5} = 0\,.  
\end{eqnarray*}

The eigenvalues of this critical point are zero and the negative real part is known as non-hyperbolic. The stability of this point cannot be explained by linear stability theory and needs to be obtained through CMT [{\color{blue}See appendix}]. With this approach, the point shows stable behavior.

\item Eigenvalues of critical points $B_{3+}$ and $B_{4+}$
\begin{eqnarray*}
\lambda_{1} = 3, \hspace{0.2cm} \lambda_{2} = 1, \hspace{0.2cm} \lambda_{3} = -6-\sqrt{6} \alpha , \hspace{0.2cm} \lambda_{4} = \frac{1}{4} \left(\sqrt{6} n+12\right), \hspace{0.2cm} \lambda_{5} = -\sqrt{6} \,.  
\end{eqnarray*}
These critical points indicate saddle behavior.

\item Eigenvalues of critical point $B_{3-}$ and $B_{4-}$
\begin{eqnarray*}
\lambda_{1} = 3, \hspace{0.2cm} \lambda_{2} = 1, \hspace{0.2cm} \lambda_{3} = -6 +\sqrt{6} \alpha , \hspace{0.2cm} \lambda_{4} = 3-\frac{1}{2} \sqrt{\frac{3}{2}} n, \hspace{0.2cm} \lambda_{5} = \sqrt{6} \,. 
\end{eqnarray*}
The critical points show saddle behavior for the condition $\alpha <\sqrt{6}$ and $ n>2 \sqrt{6}$, else unstable behavior.
    
\item Eigenvalues of critical point $B_{5}$ and $B_{6}$
\begin{eqnarray*}
&\lambda_{1} = -\frac{1}{2}, \hspace{0.2cm} \lambda_{2} =-3+ \frac{6 \alpha }{n}, \hspace{0.2cm} \lambda_{3} =\frac{Root[\mathcal{Q}\&,1]}{4 n^4}, \hspace{0.2cm} \lambda_{4} =\frac{Root[\mathcal{Q}\&,2]}{4 n^4}, \hspace{0.2cm} \lambda_{5} =\frac{Root[\mathcal{Q}\&,3]}{4 n^4} \,, \nonumber \\ 
&\mathcal{Q}=\#1^3+\#1^2 \left(6 n^4-24 n^3\right)+\#1 \left(72 n^8-216 n^7-864 n^6\right)\hspace{0.2cm} -864 n^{11}+10368 n^9
\end{eqnarray*} 

We have obtained the above output in Mathematica, where $Root[\mathcal{Q}\&,1]$, $Root[\mathcal{Q}\&,2]$ and $Root[\mathcal{Q}\&,3]$ represents the first, second and third roots of the function $\mathcal{Q}$ respectively. The $\#1$  represented is an anonymous function with coefficients that depend on $n$. So, these critical point shows unstable behavior.

\item Eigenvalues of critical point $B_{7}$ and $B_{8}$
\begin{eqnarray*}
&\lambda_{1} = 1, \hspace{0.2cm} \lambda_{2} =-4+ \frac{8 \alpha }{n}, \hspace{0.2cm} \lambda_{3} =\frac{Root[\mathcal{P}\&,1]}{12 n^4}, \hspace{0.2cm}  \lambda_{4} =\frac{Root[\mathcal{P}\&,2]}{12 n^4}, \hspace{0.2cm} \lambda_{5} =\frac{Root[\mathcal{P}\&,3]}{12 n^4} \,, \nonumber \\ &
\mathcal{P}=\#1^3+\#1^2 \left(12 n^4-96 n^3\right)+\#1 \left(576 n^8-1728 n^7-9216 n^6\right)-27648 n^{11}+442368 n^9.
\end{eqnarray*}
According to the behavior of the eigenvalues, these critical point indicates unstable behavior.

\item Eigenvalues of critical point $B_{9}$ and $B_{10}$
\begin{eqnarray*}
\lambda_{1} = -2+\frac{n^2}{8}, \hspace{0.4cm} \lambda_{2} = -3+\frac{n^2}{4} , \hspace{0.4cm} \lambda_{3} =-\frac{1}{4} n (n-2 \alpha )\,,\nonumber \\
\lambda_{4} = \frac{1}{16} \left(n (n+4)-\sqrt{((n-12) n-72) ((n-4) n-8)}-24\right) \,, \nonumber \\
\lambda_{5} = \frac{1}{16} \left(n (n+4)+\sqrt{((n-12) n-72) ((n-4) n-8)}-24\right) \,.
\end{eqnarray*}
The critical points $B_{9}$ and $B_{10}$ show stable  behavior for the conditions $2 \left(1-\sqrt{3}\right)\leq n<0$ and $\alpha >\frac{n}{2}$. The stability of this point depends on the parameters $\alpha$ and $n$. The stability region plot is given in Fig.~\ref{ch6_Figb9b10M2}, where the green/shaded region shows the stable phase.   
 \begin{figure}[H]
 \centering
 \includegraphics[width=80mm]{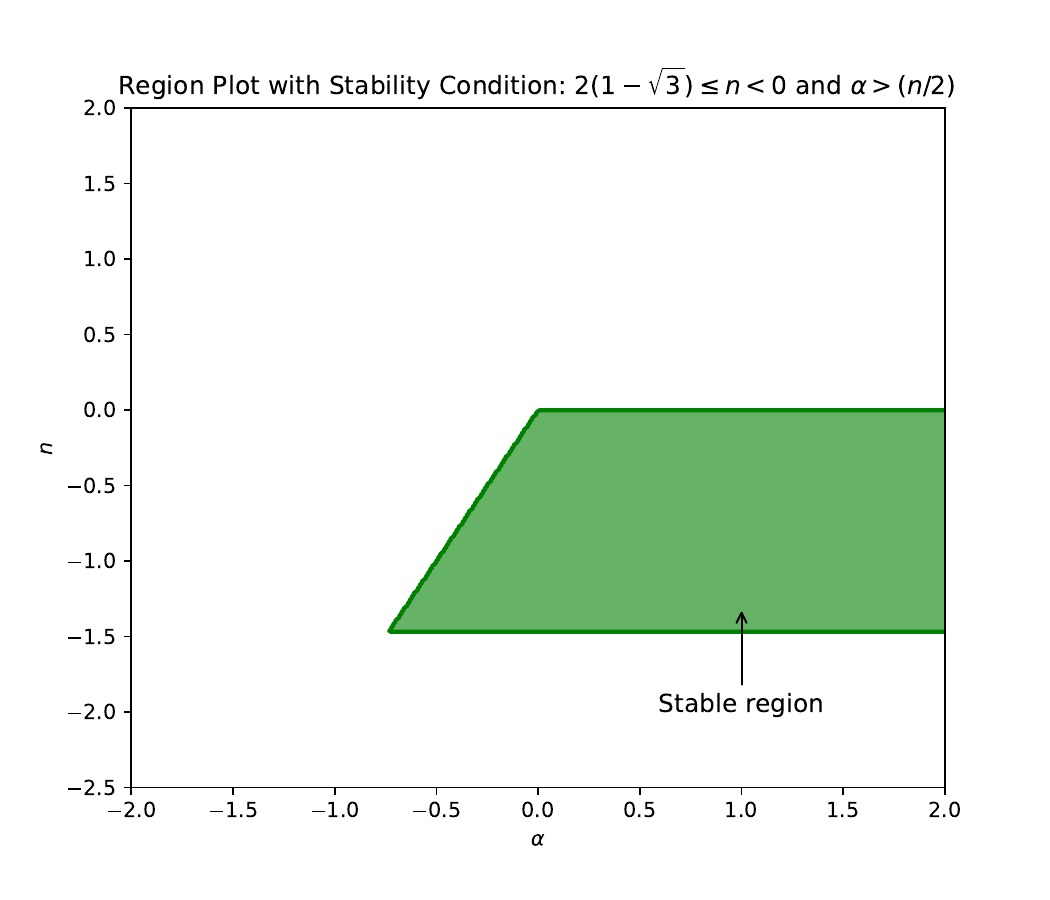}
 \caption{Stability region of the critical points $B_{9}$ and $B_{10}$  between the model parameter $\alpha$ and $n$.} \label{ch6_Figb9b10M2}
 \end{figure}
\end{itemize}

{\bf{\large Numerical Solutions:}}\\

In this case, three attractor points ($B_{2}$, $B_{9}$, $B_{10}$) are obtained which describe the accelerating phase of the Universe. 

\begin{figure}
     \centering
     \begin{subfigure}[b]{0.3\textwidth}
         \centering
         \includegraphics[width=\linewidth]{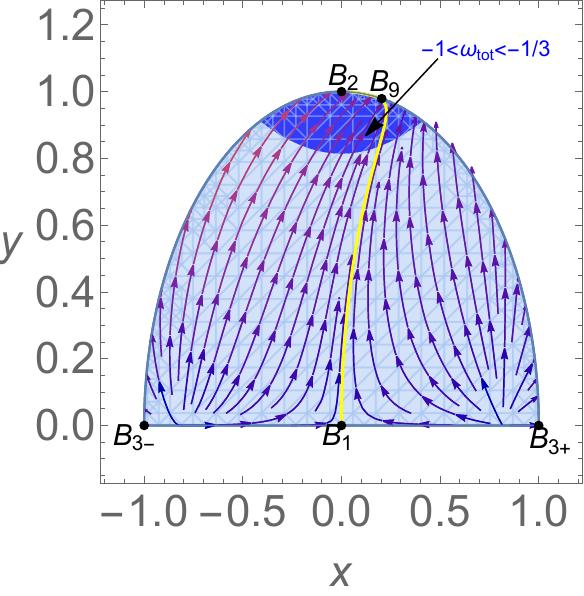}
         \caption{Phase space analysis with $\alpha=-0.6$, $n=-1$}
         \label{ch6_fig:case2_phaseportrait}
     \end{subfigure}
     \hfill
     \begin{subfigure}[b]{0.3\textwidth}
         \centering
         \includegraphics[width=\linewidth]{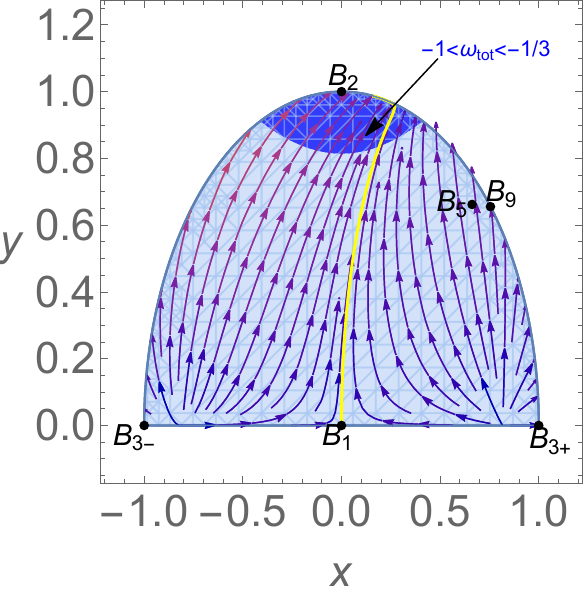}
         \caption{Phase space with $\alpha=-0.6$, $n=-3.7$}
         \label{ch6_fig:case2_phaseB5}
     \end{subfigure}
     \hfill
     \begin{subfigure}[b]{0.3\textwidth}
         \centering
         \includegraphics[width=\linewidth]{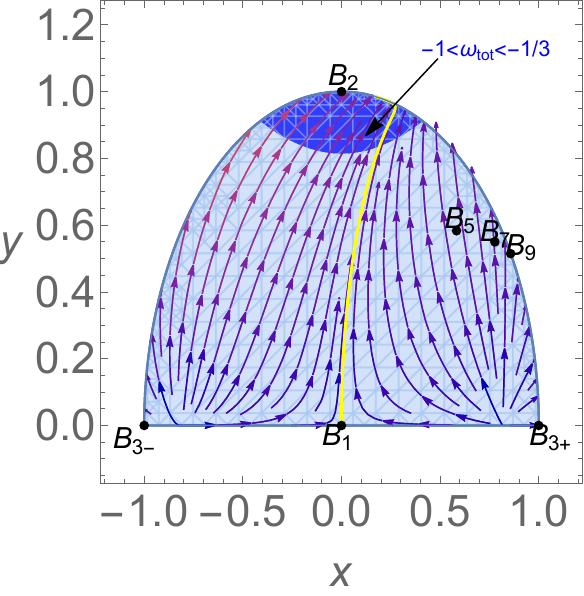}
         \caption{Phase space with $\alpha=-0.6$, $n=-4.2$}
         \label{ch6_fig:case2_phaseB7}
     \end{subfigure}
\caption{Model-II: 2D phase space portrait for the autonomous system [Eq.~\ref{ch6_autonomous-system1}-Eq.~\ref{ch6_autonomous-system6}]. The blue/shaded region represents the portion of the phase space where the Universe undergoes accelerated expansion $(-1<\omega_{tot}<-\frac{1}{3})$. The initial conditions are: $x = 10^{-5}$, $y = 9 \times 10^{-13}$, $u=10^{-5}$, $\rho=\sqrt{0.999661}$, $\lambda=0.8$.} 
 \label{ch6_Fig4phase_portrait}
\end{figure}

In Fig.~\ref{ch6_Fig4phase_portrait}, we examine the phase space portrait for different values of the model parameter $n$. Based on the value of $n$, the phase space has been divided into three regions. The variation in $n$ emphasizes the presence of the critical points $B_5-B_8$ and an accelerating region (blue/shaded) associated with the critical points $B_9$ and $B_{10}$.

In Fig.~\ref{ch6_fig:case2_phaseportrait}, we illustrate the phase portrait for $n = -1$. At this value, the critical points ($B_5$–$B_8$) do not exist because the critical points ($B_5$–$B_6$) and ($B_7$–$B_8$) only emerge when $n^2 > 12$ and $n^2 > 16$ respectively. These conditions are consistent with the physical viability of the density parameters $(0 < \Omega_r < 1)$ and $(0 < \Omega_m < 1)$. In the accelerating region associated with the critical points $B_9$ and $B_{10}$ where $n^2 < 8$, the critical points ($B_5$–$B_6$) and ($B_7$–$B_8$) do not appear. For $n = -1$, the critical point $B_9$ lies within the accelerating region (blue/shaded). Specifically, the critical point $B_9$ corresponds to the accelerating phase when $n^2 < 8$. Thus, we have chosen $n = -1$ since it satisfies the stability conditions for a critical point $B_9$.

All the trajectories in the phase space are heteroclinic orbits that begins at $B_{3\pm}$ and ends at $B_9$. There are two heteroclinic orbits, $B_{3\pm} \to B_1 \to B_9$. These orbits can serve as physical models for the transition from DM to DE, effectively characterizing the late-time evolution of the Universe. The total EoS, $\omega_{tot} = -1 + \frac{n^2}{12}$. However at early times the model always predicts stiff fluid domination, represented by $B_{3\pm}$ and it does not favor from the phenomenological perspective.

For $n^2 > 8$, point $B_9$ lies outside the acceleration region (blue/shaded) and do not represent an inflationary solution. At $n=-1$, the critical point $B_9$ exhibits stable node behavior and indicates the late-time cosmic acceleration of the Universe. The heteroclinic orbit solution (yellow line) is obtained from the numerical solution of the autonomous system [Eq.~\ref{ch6_autonomous-system1}-Eq.~\ref{ch6_autonomous-system6}] with the initial conditions $x = 10^{-5}$, $y = 9 \times 10^{-13}$, $u = 10^{-5}$, $\rho = \sqrt{0.999661}$ and $\lambda = 0.8$.

In Fig.~\ref{ch6_fig:case2_phaseB5}, we observe six critical points in the phase space under the condition $n^2 > 12$. For $n = -3.7$, the critical points $B_7$ and $B_8$ do not appear and the critical point $B_9$ lies outside the accelerating region (blue/shaded). At $n=-3.7$, the critical points $B_5$, $B_6$ and $B_9$ exhibit saddle-like (unstable) behavior, indicating a decelerating phase of the Universe. Point $B_9$ consistently falls outside the acceleration region (blue/shaded) and thus never represents an inflationary solution. There are two heteroclinic orbits: $B_{3\pm} \to B_1 \to B_2$.

In Fig.~\ref{ch6_fig:case2_phaseB7}, we identify seven critical points in the phase space for the condition $n^2 > 16$. For $n = -4.2$, the critical point $B_9$ is positioned outside the accelerating region (blue/shaded). At this value, the critical points $B_5$, $B_7$ and $B_9$ exhibit saddle-like (unstable) behavior, signaling the decelerating phase of the Universe. Importantly, point $B_9$ lies outside the acceleration region (blue/shaded) and, therefore, cannot represent an inflationary solution. There are two heteroclinic orbits: $B_{3\pm} \to B_1 \to B_2$.

\begin{figure}
     \centering
     \begin{subfigure}[b]{0.44\textwidth}
         \centering
         \includegraphics[width=70mm]{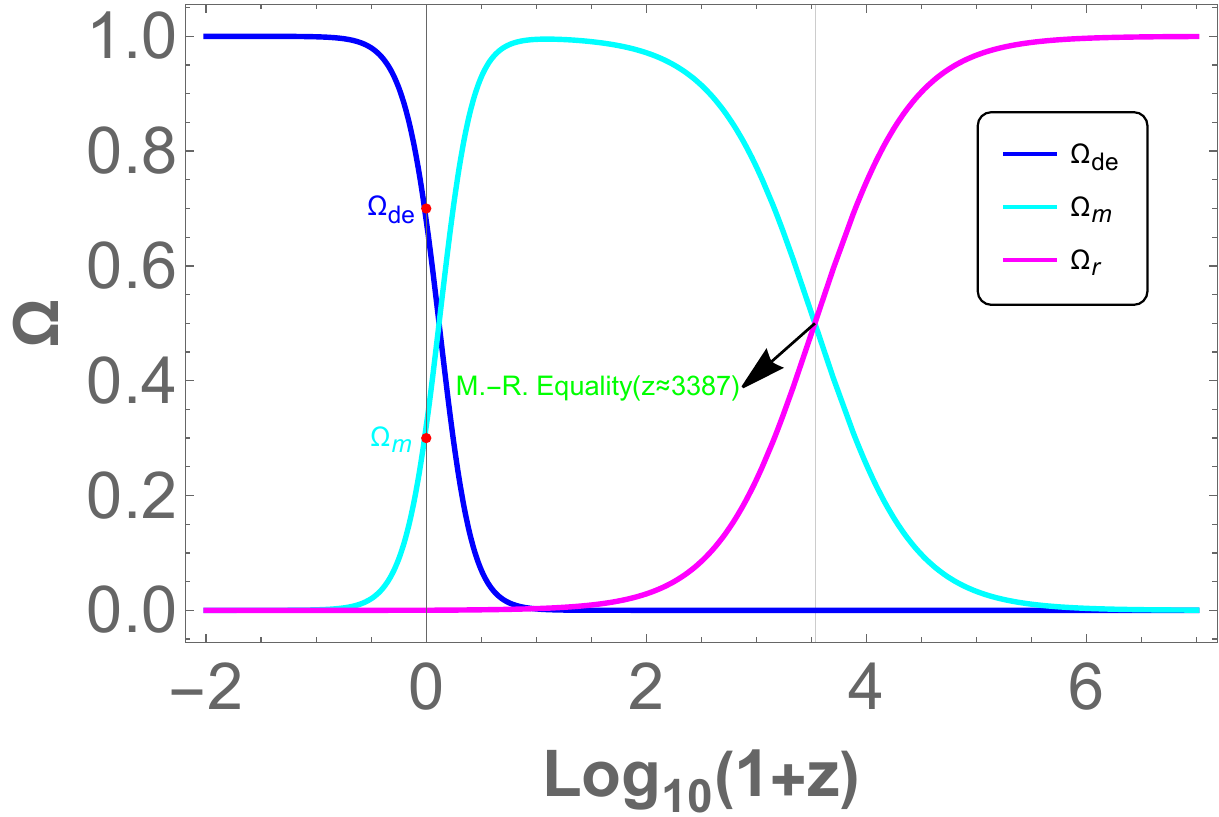}
         \caption{Evolution of density parameters.}
         \label{ch6_fig:case2_density}
     \end{subfigure}
     \begin{subfigure}[b]{0.44\textwidth}
         \centering
         \includegraphics[width=70mm]{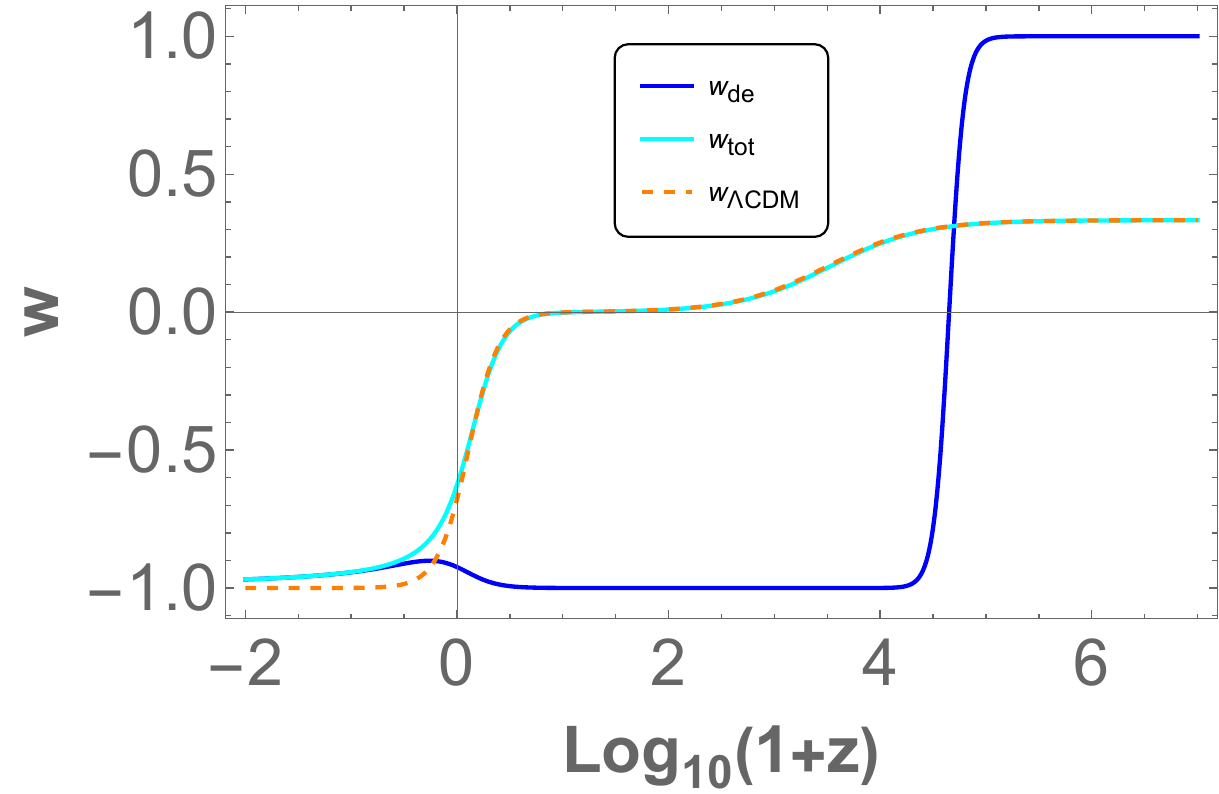}
         \caption{Evolution of EoS parameters.}
         \label{ch6_fig:case2_statepara}
     \end{subfigure}
\caption{In this figure, we set $\alpha=-0.6$ and $n=-1$ with the initial conditions are the same as in Fig.~\ref{ch6_Fig4phase_portrait}.} 
\label{ch6_Fig3}
\end{figure}

\begin{figure}
     \centering
     \begin{subfigure}[b]{0.49\textwidth}
         \centering
         \includegraphics[width=70mm]{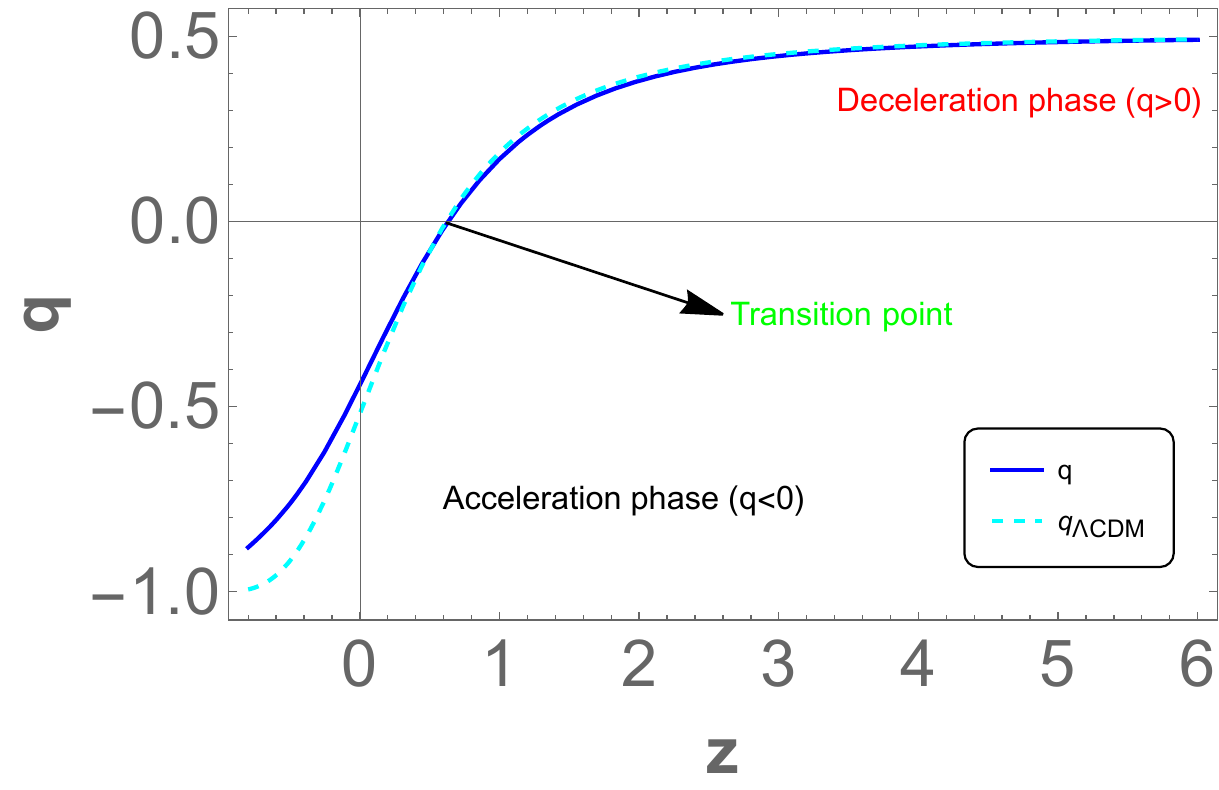}
         \caption{Evolution of deceleration parameter $q$.}
         \label{ch6_fig:case2_qz}
     \end{subfigure}
     \begin{subfigure}[b]{0.49\textwidth}
         \centering
         \includegraphics[width=70mm]{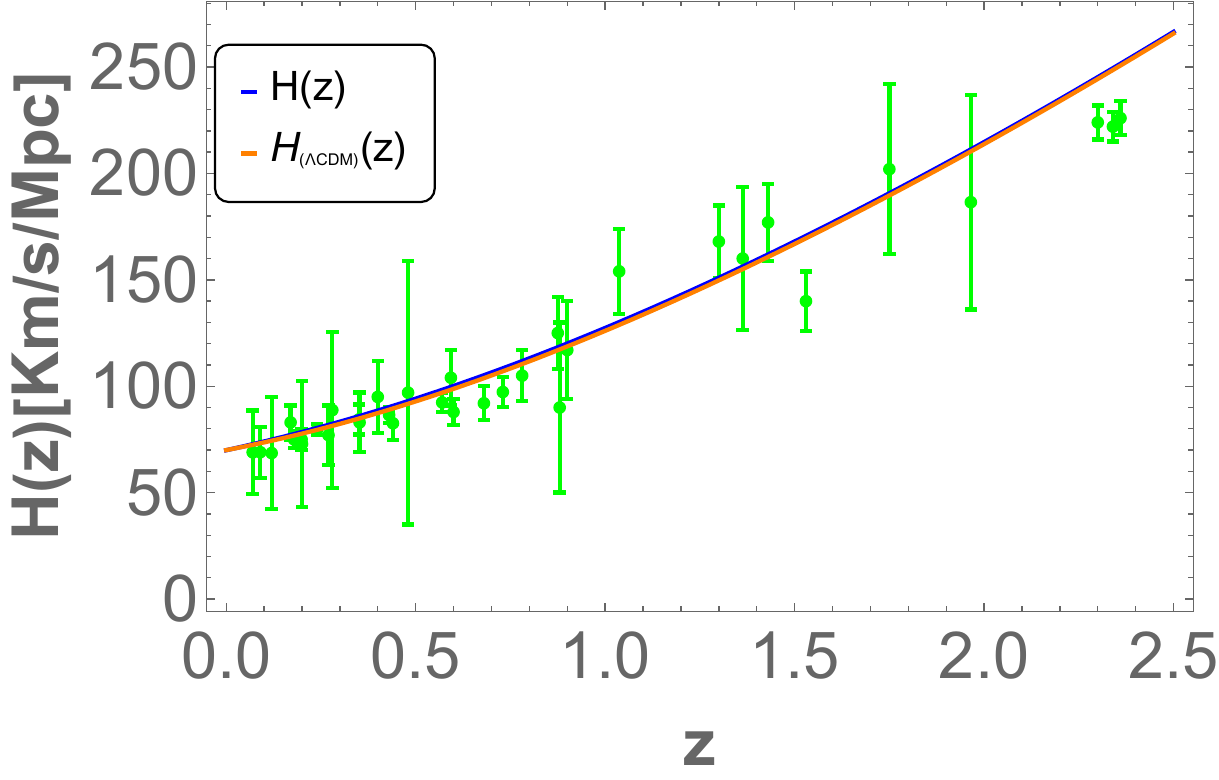}
         \caption{Evolution of the Hubble rate $H(z)$.}
         \label{ch6_fig:case2_Hz}
     \end{subfigure}
\caption{In this figure, we set $\alpha=-0.6$ and $n=-1$ with the initial conditions are the same as in Fig.~\ref{ch6_Fig4phase_portrait}.} 
\label{ch6_Fig4}
\end{figure}

\begin{figure}[H]
 \centering
 \includegraphics[width=70mm]{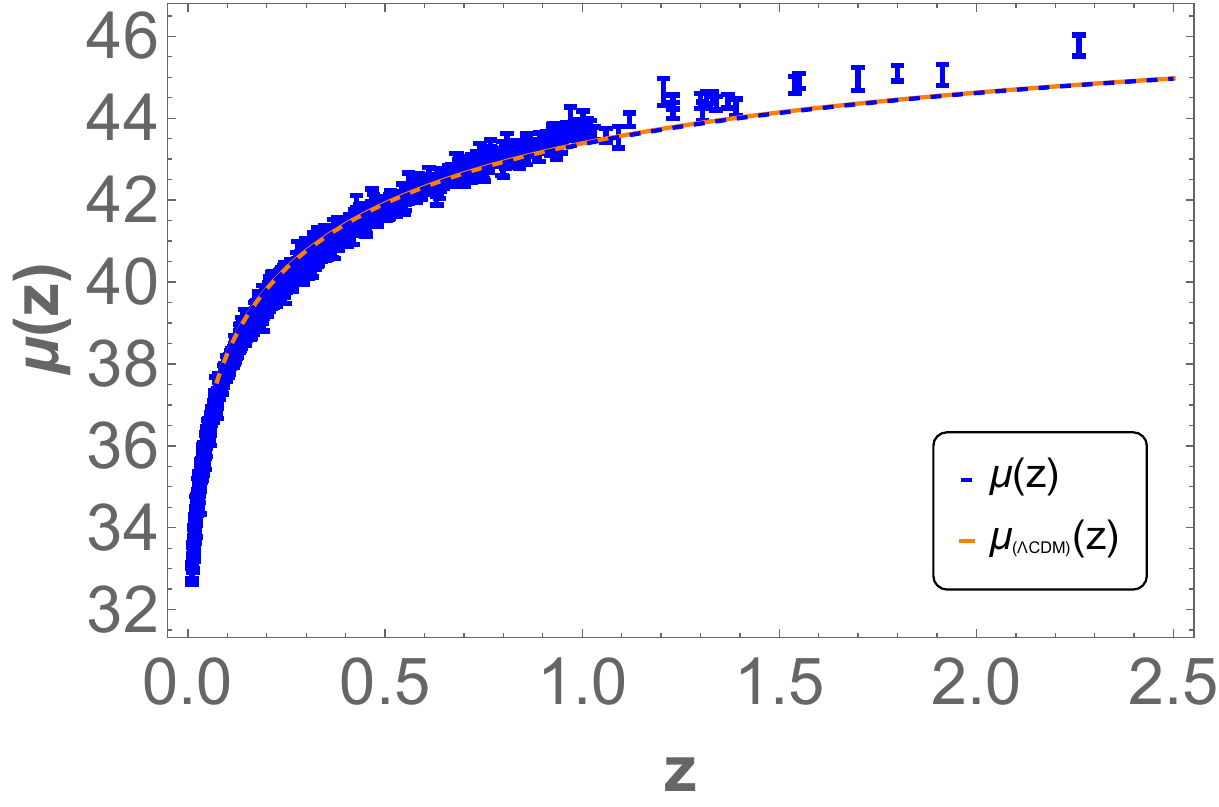} 
 \caption{We show the evolution of the distance modulus function $\mu(z)$ ( dashed blue) and the $\Lambda$CDM model distance modulus function $\mu_{\Lambda CDM}(z)$ along with the 1048 SNIa data points \cite{Scolnic_2018}. In this figure, we set $\alpha=-0.6$ and $n=-1$ with the initial conditions are the same as in Fig.~\ref{ch6_Fig4phase_portrait}.} \label{ch6_Fig4pan}
 \end{figure}

 For these solutions to be cosmologically viable, the potential must be sufficiently flat ($n^2 < 8$) and the initial conditions are fine-tuned to ensure the prolonged domination of DE. The solution should follow the sequence: $B_{3\pm} \to B_1 \to B_9$ (see Fig.~\ref{ch6_fig:case2_phaseportrait}). Further, at early times, the only admissible solutions correspond to non-physical stiff fluid Universe fails to describe accurately. The EoS parameters, energy densities, deceleration parameters, Hubble rate and the modulus function for a solution that shadows the heteroclinic sequence $B_{3\pm} \to B_1 \to B_9$ are shown in Fig.~\ref{ch6_Fig3}, Fig.~\ref{ch6_Fig4}, Fig.~\ref{ch6_Fig4pan}.

Fig.~\ref{ch6_fig:case2_density} shows the evolutionary behavior of radiation, DE and DM. The radiation appears first in the early cosmos then the DM takes over for a short while and eventually the cosmological constant. The present value of density parameters, $\Omega_{m}\approx 0.33$, $\Omega_{DE}\approx 0.67$ and the matter-radiation equality at $z\approx 3387$. In Fig.~\ref{ch6_fig:case2_statepara}, the evolution of EoS parameters are shown. We observe that $\omega_{tot}$ (cyan) begins at $\frac{1}{3}$ for radiation, decreases to 0 during the matter-dominated period and ultimately reaches approximately $-1$. The EoS parameter $\omega_{\Lambda CDM}$ and the DE dominated EoS parameter $\omega_{DE}$ (blue) approach about to $-1$ at late-time. The current value of the EoS parameter for the DE sector, $\omega_{DE}(z=0)=-0.92$, is consistent with the current Planck collaboration [$\omega_{DE}(z=0)= -1.028 \pm 0.032$ \cite{Aghanim:2018eyx}]. The deceleration parameter [Fig.~\ref{ch6_fig:case2_qz}] shows the transition at $z\approx 0.62$ from decelerating to accelerating aligning with the recent findings \cite{PhysRevD.90.044016a}. The present value, $q(z=0) \approx -0.45$, agrees with the visualized cosmic identification \cite{Camarena:2020prr}. The visualization of the Hubble rate evolution, Hubble rate $H_{\Lambda CDM}(z)$  and the Hubble data points \cite{Moresco_2022_H0} are shown in Fig.~\ref{ch6_fig:case2_Hz}. We have used $H_{0} = 70 \ \text{Km} \, \text{s}^{-1} \text{Mpc}^{-1}$ as the current Hubble parameter \cite{Aghanim:2018eyx}. The outcomes of the standard $\Lambda$CDM model are extremely similar to the obtained result. The evolution of the modulus function $\mu(z)$ is presented in Fig.~\ref{ch6_Fig4pan} along with the $\Lambda$CDM model modulus function $\lambda_{\Lambda CDM}$ and 1048 pantheon data points. The mathematical formalism of the Hubble and Pantheon data sets is presented in section--\ref{ObservationalCosmology}. 
\subsection{Model-III: \texorpdfstring{$V(\phi)=V_{0}\sinh^{-\gamma}(\beta \phi)$}{Case-III: V(phi)}}
 \label{ch6_case3sinh}
For this form of $V(\phi)=V_{0}\sinh^{-\gamma}(\beta \phi)$ \cite{sahni2000case, Ure_a_L_pez_2016}, the roots of the function $f(\lambda)$ are, 
\begin{eqnarray}\label{ch6_sinnhpotentialroot}
f(\lambda)=\frac{\lambda^{2}}{\gamma}-\gamma \beta^{2}\,, \hspace{0.5cm} \beta_{1}=\frac{1}{\gamma}\,, \hspace{0.5cm} \beta_{2}=0 \,, \hspace{0.2cm} \beta_{3}=-\gamma \beta^{2}\,, \hspace{0.2cm} 
\lambda^{+}= \gamma\beta \,,\hspace{0.2cm}\lambda^{-}=-\gamma \beta,
\end{eqnarray}

In Table~\ref{ch6_TABLE-III}, the eigenvalues of the critical points and their corresponding existence conditions are given. The values of the cosmological parameters are listed in  Table~\ref{ch6_TABLE-IV}. The eigenvalues and background cosmological parameters for the critical points $B_{1}$ and $B_{2}$ are obtained to be the same as that of the power-law potential function ($V(\phi)=V_{0}\phi^{n}$) [Model-II]. Since the model parameter does not affect the critical points $B_{1}$ and $B_{2}$, only the other critical points will be investigated.

{\bf{\large Description of Critical Points:}}

\begin{itemize}
\item  For the critical points $B_{3\pm}$ and $B_{4\pm}$,  $\omega_{DE}=\omega_{tot}=1$, $\Omega_{DE}=1$ and so these points described the stiff-matter. The positive value of $q$ indicates the deceleration phase. 

\item The density parameters scaling solutions of the critical points $B_{5}$ and $B_{6}$ are $\Omega_{DE}=\frac{3}{\gamma^2 \beta^2}$ and $\Omega_{m}=1-\frac{3}{\gamma^2 \beta^2}$. Also, we have  $\omega_{DE}=\omega_{tot}=0$. The scaling solution shows the evolution between matter-DE-dominated phases of the Universe. These critical points show a deceleration phase, indicated by the positive value of the deceleration parameter. The points $B_5$ and $B_6$ exist if they satisfy the condition $\gamma^2 \beta^2 > 3$.

\item The critical points $B_{7}$ and $B_{8}$ have density parameter scaling solutions of $\Omega_{DE}=\frac{4}{\gamma^2 \beta^2}$ and $\Omega_{r}=1-\frac{4}{\gamma^2 \beta^2}$. The critical points reveal non-standard radiation and the decelerating phase of the Universe. Scaling solutions illustrate the transition between radiation-DE-dominated phases of the Universe. The existence of the points $B_7$ and $B_8$ is contingent upon the condition $\gamma^2 \beta^2 > 4$ being satisfied.

\item The critical points $B_{9}$ and $B_{10}$ reveal the DE dominated era of the Universe. The solution of the DE and total EoS parameters is the same, i.e. $\omega_{DE}=\omega_{tot}=-1+\frac{\beta ^2 \gamma ^2}{3}$ and the solution of the deceleration parameter is $q=-1+\frac{\beta ^2 \gamma ^2}{2}$. The points show the accelerated phase for the conditions $\left(\gamma <0\land -\sqrt{2} \sqrt{\frac{1}{\gamma ^2}}<\beta <\sqrt{2} \sqrt{\frac{1}{\gamma ^2}}\right)$, $\gamma=0$ and $\bigg(\gamma >0\land -\sqrt{2} \sqrt{\frac{1}{\gamma ^2}}<\beta <\sqrt{2} \sqrt{\frac{1}{\gamma ^2}}\bigg)$;  decelerated phase for the conditions $\bigg(\gamma <0\land \left(\beta <-\sqrt{2} \sqrt{\frac{1}{\gamma ^2}}\lor \beta >\sqrt{2} \sqrt{\frac{1}{\gamma ^2}}\right)\bigg)$ and $\bigg(\gamma >0\land \bigg(\beta <-\sqrt{2} \sqrt{\frac{1}{\gamma ^2}}\lor \beta >\sqrt{2} \sqrt{\frac{1}{\gamma ^2}}\bigg)\bigg)$. For the condition on the model parameter $\bigg(\beta <0\land \bigg(-\sqrt{2} \sqrt{\frac{1}{\beta ^2}}<\gamma <0\lor 0<\gamma <\sqrt{2} \sqrt{\frac{1}{\beta ^2}}\bigg)\bigg)$ and $\bigg(\beta >0\land \bigg(-\sqrt{2} \sqrt{\frac{1}{\beta ^2}}<\gamma <0\lor 0<\gamma <\sqrt{2} \sqrt{\frac{1}{\beta ^2}}\bigg)\bigg)$ points $B_{9}$ and $B_{10}$ imply the quintessence phase of the Universe. In Fig.~\ref{ch6_fig:case3_q_region}, we present the region plot of the model parameters $\beta$ and $\gamma$, showing the deceleration parameter for both the acceleration phase ($q < 0$) and the deceleration phase ($q > 0$).  In Fig.~\ref{ch6_fig:case3_omega_region}, we show the region plot of the model parameters $\beta$ and $\gamma$ for the quintessence phase ($-1 < \omega_{tot} < -\frac{1}{3}$).
\begin{figure}
     \centering
     \begin{subfigure}[b]{0.4\textwidth}
         \centering
         \includegraphics[width=60mm]{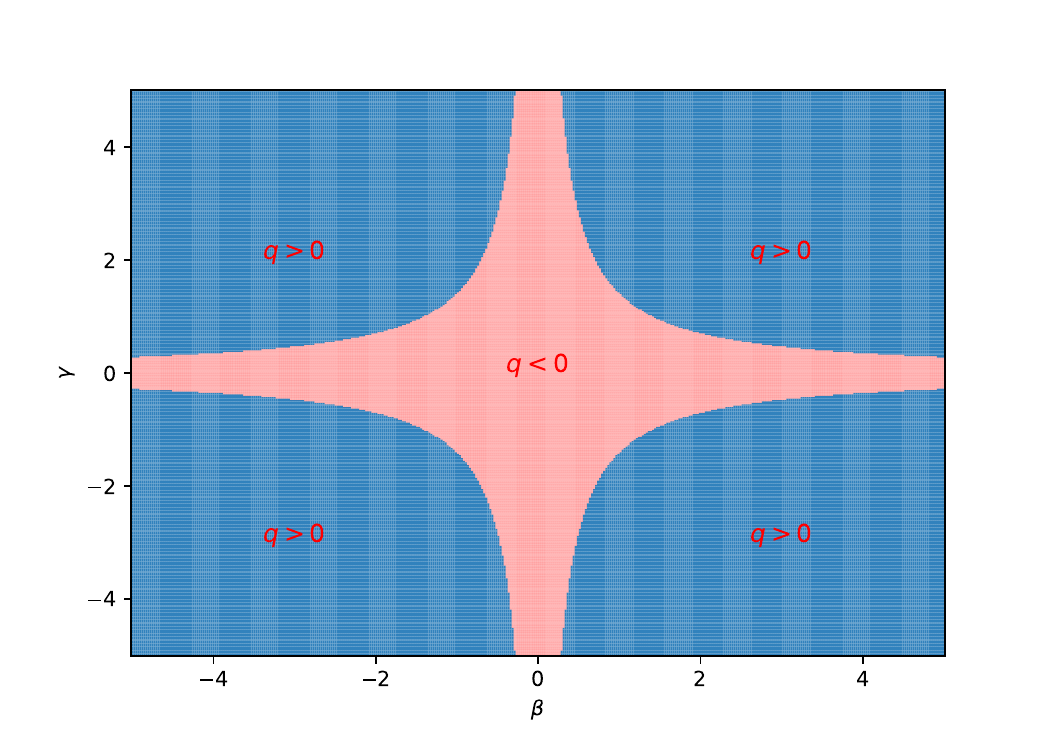}
         \caption{Region plot between the model parameters $\gamma$ and $\beta$ for the deceleration parameter}
         \label{ch6_fig:case3_q_region}
     \end{subfigure}
     \begin{subfigure}[b]{0.4\textwidth}
         \centering
         \includegraphics[width=60mm]{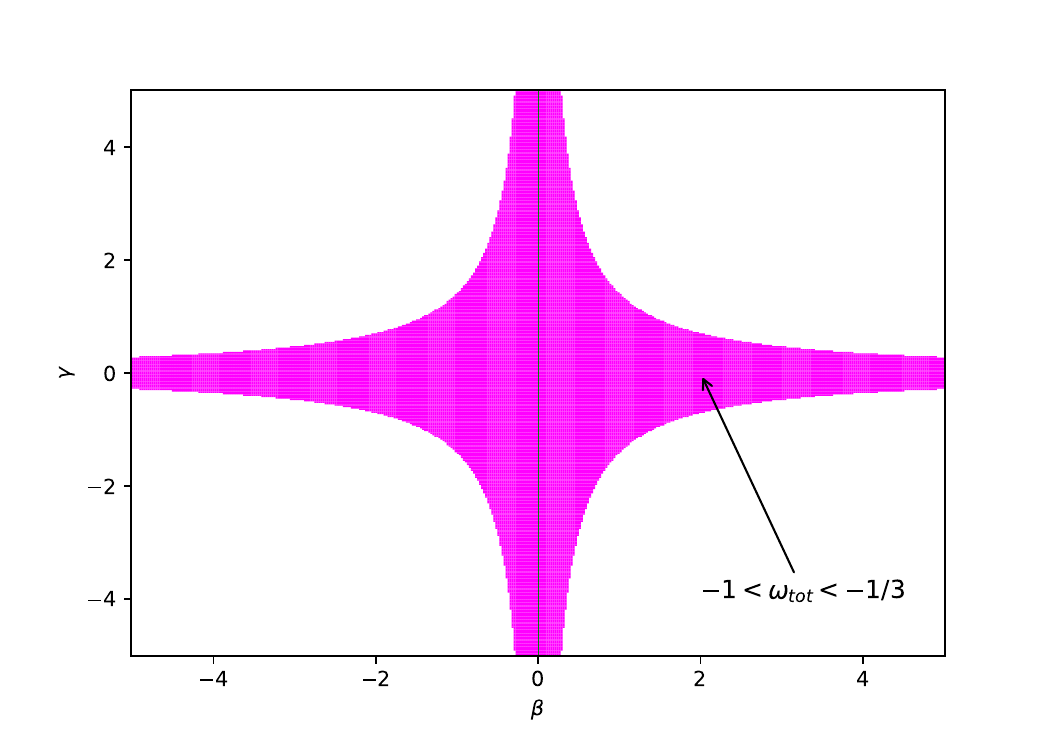}
         \caption{Region plot between the model parameters $\gamma$ and $\beta$ for the quintessence phase.  }
         \label{ch6_fig:case3_omega_region}
     \end{subfigure}
\caption{Region plot between the model parameters $\gamma$ and $\beta$ for the critical points $B_{9}$ and $B_{10}$ for Model-III.} 
\label{ch6_Figb9b10}
\end{figure}
\end{itemize}

{\bf{\large Stability Analysis:}}

\begin{itemize}
\item Eigenvalues of the critical point $B_{3\pm}$ and $B_{4\pm}$
\begin{eqnarray*}
\lambda_{1} = 3, \hspace{0.2cm} \lambda_{2} = 1, \hspace{0.2cm} \lambda_{3} = -6\pm\sqrt{6} \alpha , \hspace{0.2cm} \lambda_{4} = \pm2 \sqrt{6} \beta, \hspace{0.2cm} \lambda_{5} = 3\pm\sqrt{\frac{3}{2}} \beta  \gamma \,.  
\end{eqnarray*}
Sign($-$) in $\lambda_{3}$, $\lambda_{4}$ and $\lambda_{5}$ indicate the eigenvalues of the critical points $B_{3+}$ and $B_{4-}$, whereas Sign ($+$) in $\lambda_{3}$, $\lambda_{4}$ and $\lambda_{5}$ imply the eigenvalues of the critical points $B_{3-}$ and $B_{4+}$. The critical points $B_{3+}$ and $B_{4-}$ indicate the saddle behavior for the conditions $\alpha >-\sqrt{6}\land \gamma >0\land \beta >\frac{\sqrt{6}}{\gamma } $, nonetheless it is showing unstable node behaviour. The critical points $B_{3-}$ and $B_{4+}$ show the saddle behavior for the conditions $\alpha <\sqrt{6}\land \gamma >0\land \beta <-\frac{\sqrt{6}}{\gamma }$; else it is defined as an unstable node. 
    
\item Eigenvalues of the critical point $B_{5}$ and $B_{6}$
\begin{eqnarray*}
\lambda_{1} = -\frac{6}{\gamma }, \hspace{0.5cm} \lambda_{2} =-\frac{1}{2}, \hspace{0.5cm} \lambda_{3} =-3\pm\frac{3 \alpha }{\beta  \gamma }, \nonumber \\  \lambda_{4} =-\frac{3}{4}-\frac{3 \sqrt{24 \beta ^6 \gamma ^6-7 \beta ^8 \gamma ^8}}{4 \beta ^4 \gamma ^4}, \hspace{0.5cm}
\lambda_{5} =-\frac{3}{4}+\frac{3 \sqrt{24 \beta ^6 \gamma ^6-7 \beta ^8 \gamma ^8}}{4 \beta ^4 \gamma ^4} \,.
\end{eqnarray*}
Sign($-$) and ($+$) in the $\lambda_{3}$ imply the eigenvalues of the critical point $B_{5}$ and $B_{6}$ respectively. Both $B_{5}$ and $B_{6}$ are showing stable behavior for the conditions 
$\bigg(-2 \sqrt{\frac{6}{7}}<\alpha \leq -\sqrt{3}\land \gamma >0\land \bigg(-2 \sqrt{\frac{6}{7}} \sqrt{\frac{1}{\gamma ^2}}\leq \beta <\frac{\alpha }{\gamma }\lor -\frac{\alpha }{\gamma }<\beta \leq 2 \sqrt{\frac{6}{7}} \sqrt{\frac{1}{\gamma ^2}}\bigg)\bigg)\lor  \bigg(-\sqrt{3}<\alpha <\sqrt{3}\land \gamma >0\land \bigg(-2 \sqrt{\frac{6}{7}} \sqrt{\frac{1}{\gamma ^2}}\leq \beta <-\sqrt{3} \sqrt{\frac{1}{\gamma ^2}}\lor \sqrt{3} \sqrt{\frac{1}{\gamma ^2}}<\beta \leq 2 \sqrt{\frac{6}{7}} \sqrt{\frac{1}{\gamma ^2}}\bigg)\bigg)\lor \bigg(\sqrt{3}\leq \alpha <2 \sqrt{\frac{6}{7}}\land \gamma >0\land \bigg(-2 \sqrt{\frac{6}{7}} \sqrt{\frac{1}{\gamma ^2}}\leq \beta <-\frac{\alpha }{\gamma }\lor \frac{\alpha }{\gamma }<\beta \leq 2 \sqrt{\frac{6}{7}} \sqrt{\frac{1}{\gamma ^2}}\bigg)\bigg)$. It is saddle or unstable if the critical points do not satisfy the above conditions. 
 
\item Eigenvalues of the critical point $B_{7}$ and $B_{8}$
\begin{eqnarray*}
\lambda_{1} = -\frac{8}{\gamma }, \hspace{0.5cm} \lambda_{2} =1, \hspace{0.5cm} \lambda_{3} =-4\pm\frac{4 \alpha }{\beta  \gamma }, \nonumber \\  \lambda_{4} =-\frac{1}{2}-\frac{\sqrt{64 \beta ^6 \gamma ^6-15 \beta ^8 \gamma ^8}}{2 \beta ^4 \gamma ^4}, \hspace{0.5cm}
\lambda_{5} =-\frac{1}{2}+\frac{\sqrt{64 \beta ^6 \gamma ^6-15 \beta ^8 \gamma ^8}}{2 \beta ^4 \gamma ^4} \,.
\end{eqnarray*}
Indices ($-$) and ($+$) in $\lambda_{3}$ denote the eigenvalues of the critical points $B_{5}$ and $B_{6}$ respectively. The points $B_{5}$ and $B_{6}$ imply the saddle behavior for the condition $\bigg(-\frac{8}{\sqrt{15}}<\alpha \leq -2\land \bigg(\bigg(\beta <0\land \frac{\alpha }{\beta }<\gamma \leq \frac{8 \sqrt{\frac{1}{\beta ^2}}}{\sqrt{15}}\bigg)\lor \bigg(\beta >0\land -\frac{\alpha }{\beta }<\gamma \leq \frac{8 \sqrt{\frac{1}{\beta ^2}}}{\sqrt{15}}\bigg)\bigg)\bigg)\lor \bigg(-2<\alpha \leq 2\land \bigg(\bigg(\beta <0\land -\frac{2}{\beta }<\gamma \leq \frac{8 \sqrt{\frac{1}{\beta ^2}}}{\sqrt{15}}\bigg)\lor \bigg(\beta >0\land \frac{2}{\beta }<\gamma \leq \frac{8 \sqrt{\frac{1}{\beta ^2}}}{\sqrt{15}}\bigg)\bigg)\bigg)\lor \bigg(2<\alpha <\frac{8}{\sqrt{15}}\land \bigg(\bigg(\beta <0\land -\frac{\alpha }{\beta }<\gamma \leq \frac{8 \sqrt{\frac{1}{\beta ^2}}}{\sqrt{15}}\bigg)\lor \bigg(\beta >0\land \frac{\alpha }{\beta }<\gamma \leq \frac{8 \sqrt{\frac{1}{\beta ^2}}}{\sqrt{15}}\bigg)\bigg)\bigg)$. Except above conditions the points shows unstable node behavior.

\item Eigenvalues of the critical points $B_{9}$ and $B_{10}$\\
Both the critical points have same eigenvalues for $\lambda_{1}$, $\lambda_{2}$, $\lambda_{3}$ and $\lambda_{4}$ and only different is in the eigenvalue $\lambda_{5}$. 
\begin{eqnarray*}
\lambda_{1} = -2 \beta ^2 \gamma, \hspace{0.2cm} \lambda_{2} =-3+\frac{\beta ^2 \gamma ^2}{2}, \hspace{0.2cm} \lambda_{3} =-2+\frac{\beta ^2 \gamma ^2}{2} \,, \hspace{0.2cm} \lambda_{4} =-3+\beta ^2 \gamma ^2, \\
\lambda_{5} =-\beta  \gamma  (\alpha +\beta  \gamma ) \hspace{0.2cm} (B_{9}) \,, \hspace{0.2cm} \lambda_{5} =\beta  \gamma  (\alpha -\beta  \gamma ) \hspace{0.2cm} (B_{10})\,.
\end{eqnarray*}   
The points show stable node behavior for the conditions 
$\bigg(-\sqrt{3}<\alpha \leq 0\land \bigg(\bigg(\beta <0\land \frac{\alpha }{\beta }<\gamma <\sqrt{3} \sqrt{\frac{1}{\beta ^2}}\bigg)\lor \left(\beta >0\land -\frac{\alpha }{\beta }<\gamma <\sqrt{3} \sqrt{\frac{1}{\beta ^2}}\right)\bigg)\bigg)\lor \bigg(0<\alpha <\sqrt{3}\land \bigg(\bigg(\beta <0\land -\frac{\alpha }{\beta }<\gamma <\sqrt{3} \sqrt{\frac{1}{\beta ^2}}\bigg)\lor \bigg(\beta >0\land \frac{\alpha }{\beta }<\gamma <\sqrt{3} \sqrt{\frac{1}{\beta ^2}}\bigg)\bigg)\bigg)$. Points will demonstrate unstable node and saddle behavior if they fail to fulfill the above conditions. These points also demonstrate the DE dominance phase of the Universe.
\end{itemize}

{\bf{\large Numerical Solutions}:}\\

\begin{figure}
     \centering
     \begin{subfigure}[b]{0.3\textwidth}
         \centering
         \includegraphics[width=\linewidth]{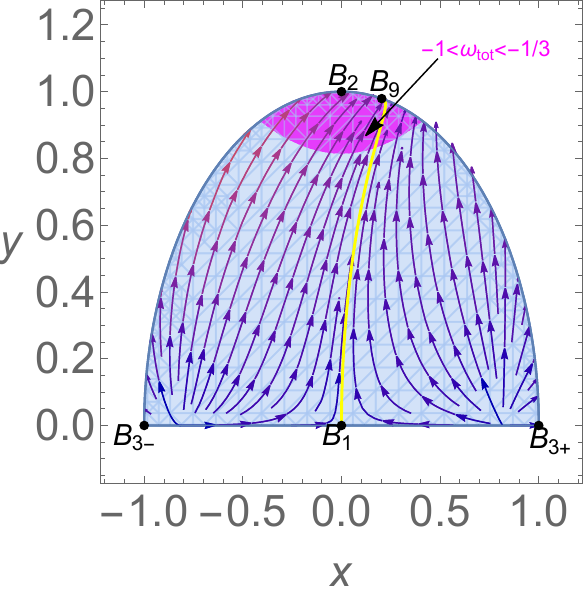}
         \caption{Phase space analysis with $\alpha-0.1$, $\gamma=0.5$ and $\beta=1$}
         \label{ch6_fig:case3_phaseportrait}
     \end{subfigure}
     \hfill
     \begin{subfigure}[b]{0.3\textwidth}
         \centering
         \includegraphics[width=\linewidth]{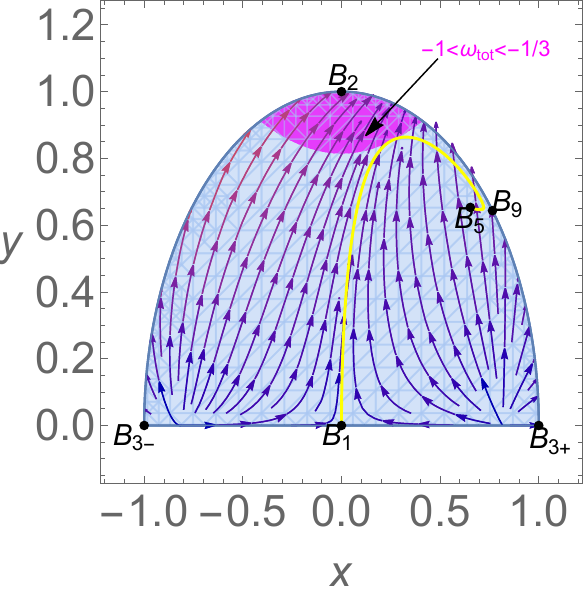}
         \caption{Phase space with $\alpha-0.1$, $\gamma=1.25$ and $\beta=1.5$}
         \label{ch6_fig:case3_phaseB5}
     \end{subfigure}
     \hfill
     \begin{subfigure}[b]{0.3\textwidth}
         \centering
         \includegraphics[width=\linewidth]{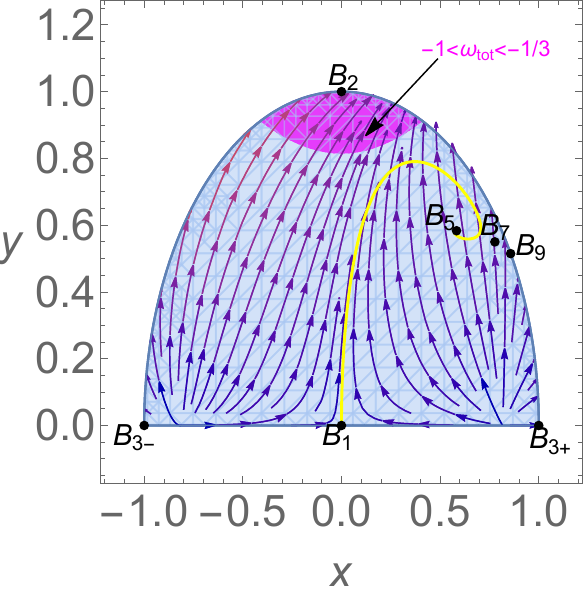}
         \caption{Phase space with $\alpha-0.1$, $\gamma=1.4$ and $\beta=1.5$}
         \label{ch6_fig:case3_phaseB7}
     \end{subfigure}
\caption{Model-III: 2D phase space portrait plot for the autonomous system [Eq.~\ref{ch6_autonomous-system1}-Eq.~\ref{ch6_autonomous-system6}]. The magenta/shaded region represents the portion of the phase space where the Universe undergoes accelerated expansion $(-1<\omega_{tot}<-\frac{1}{3})$. The initial conditions are: $x = 10^{-5}$, $y = 9 \times 10^{-13} $, $u=10^{-5}$, $\rho=\sqrt{0.999661}$, $\lambda=0.8$.} 
\label{ch6_Fig6phase_portrait}
\end{figure}

In Fig.~\ref{ch6_Fig6phase_portrait}, the phase space portrait for varying values of the model parameters $\gamma$ and $\beta$ has been presented. The phase space structure is divided into three distinct regions, determined by the specific ranges of $\gamma$ and $\beta$. These regions represent different dynamical behaviors of the system. The variation in the values of $\gamma$ and $\beta$ reveals the existence of an accelerating region of the critical points $B_9$ and $B_{10}$, indicated by the magenta (shaded) area and the existence of the critical points ($B_5-B_{8}$). The phase space portrait illustrates how variations in $\gamma$ and $\beta$ govern the onset of cosmic acceleration, emphasizing the crucial role these parameters play in the overall evolution of the Universe.

In Fig.~\ref{ch6_fig:case3_phaseportrait}, we present the phase portrait for $\gamma = 0.5$ and $\beta = 1$. For these values the critical points ($B_5$-$B_8$) do not exist since $\beta^2 \gamma^2 > 3$ and $\beta^2 \gamma^2 > 4$ are required for the appearance of the points ($B_5$-$B_6$) and ($B_7$-$B_8$) respectively. These conditions are consistent with the physical viability on the density parameters $(0 < \Omega_r < 1)$ and $(0 < \Omega_m < 1)$. In the accelerating region associated with the critical points $B_9$ and $B_{10}$, where $\beta^2 \gamma^2 < 2$, the critical points ($B_5$-$B_6$) and ($B_7$-$B_8$) do not exist. For $\gamma = 0.5$ and $\beta = 1$, the critical point lies within the accelerating region (magenta/shaded). The critical point $B_9$ corresponds to the accelerating phase when $\beta^2 \gamma^2 < 2$ also satisfying the stability conditions for $B_9$.

All the phase space trajectories are heteroclinic orbits starting from the point $B_{3\pm}$ and ending at $B_9$. Two such heteroclinic orbits $B_{3\pm} \to B_1 \to B_9$ provide a physical model for the transition from DM to DE. The total EoS parameter is given by $\omega_{tot} = -1 + \frac{\beta^2 \gamma^2}{3}$. However, at early times the model predicts stiff fluid domination represented by the points $B_{3\pm}$, which is not favored from a phenomenological standpoint. If $\beta^2 \gamma^2 > 2$, the point $B_9$ would lie outside the acceleration region (magenta/shaded) and would not represent an inflationary solution. 

For $\gamma = 0.5$ and $\beta = 1$, the critical point $B_9$ behaves as a stable node that shows the late-time cosmic acceleration of the Universe. The heteroclinic orbit solution (yellow line) is obtained from the numerical solution of the autonomous system [Eq.~\ref{ch6_autonomous-system1}-Eq.~\ref{ch6_autonomous-system6}) with the initial conditions $x = 10^{-5}$, $y = 9 \times 10^{-13}$, $u = 10^{-5}$, $\rho = \sqrt{0.999661}$ and $\lambda = 0.8$.

In Fig.~\ref{ch6_fig:case3_phaseB5}, we observe six critical points in the phase space for the condition $\beta^2 \gamma^2 > 3$. For $\gamma = 1.25$ and $\beta = 1.5$, the critical points $B_7$ and $B_8$ do not exist and the critical point $B_9$ lies outside the accelerating region (magenta/shaded). The critical points $B_5$, $B_6$ and $B_9$ exhibit saddle behavior (unstable) and correspond to the decelerating phase of the Universe. Notably, the point $B_9$ consistently lies outside the accelerating region (magenta/shaded) and thus cannot represent an inflationary solution. There are two heteroclinic orbits: $B_{3\pm} \to B_1 \to B_5$.

In Fig.~\ref{ch6_fig:case3_phaseB7}, we observe seven critical points in the phase space under $\beta^2 \gamma^2 > 4$. For $\gamma = 1.4$ and $\beta = 1.5$, the critical point $B_9$ lies outside the accelerating region (magenta/shaded). At these parameter values, the critical points $B_5$, $B_7$ and $B_9$ exhibit saddle-like (unstable) behavior, reflecting the decelerating phase of the Universe. Importantly, point $B_9$ consistently lies outside the accelerating region (magenta/shaded) and never corresponds to an inflationary solution. Two heteroclinic orbits are present: $B_{3\pm} \to B_1 \to B_5$.
 
\begin{figure}
     \centering
     \begin{subfigure}[b]{0.44\textwidth}
         \centering
         \includegraphics[width=70mm]{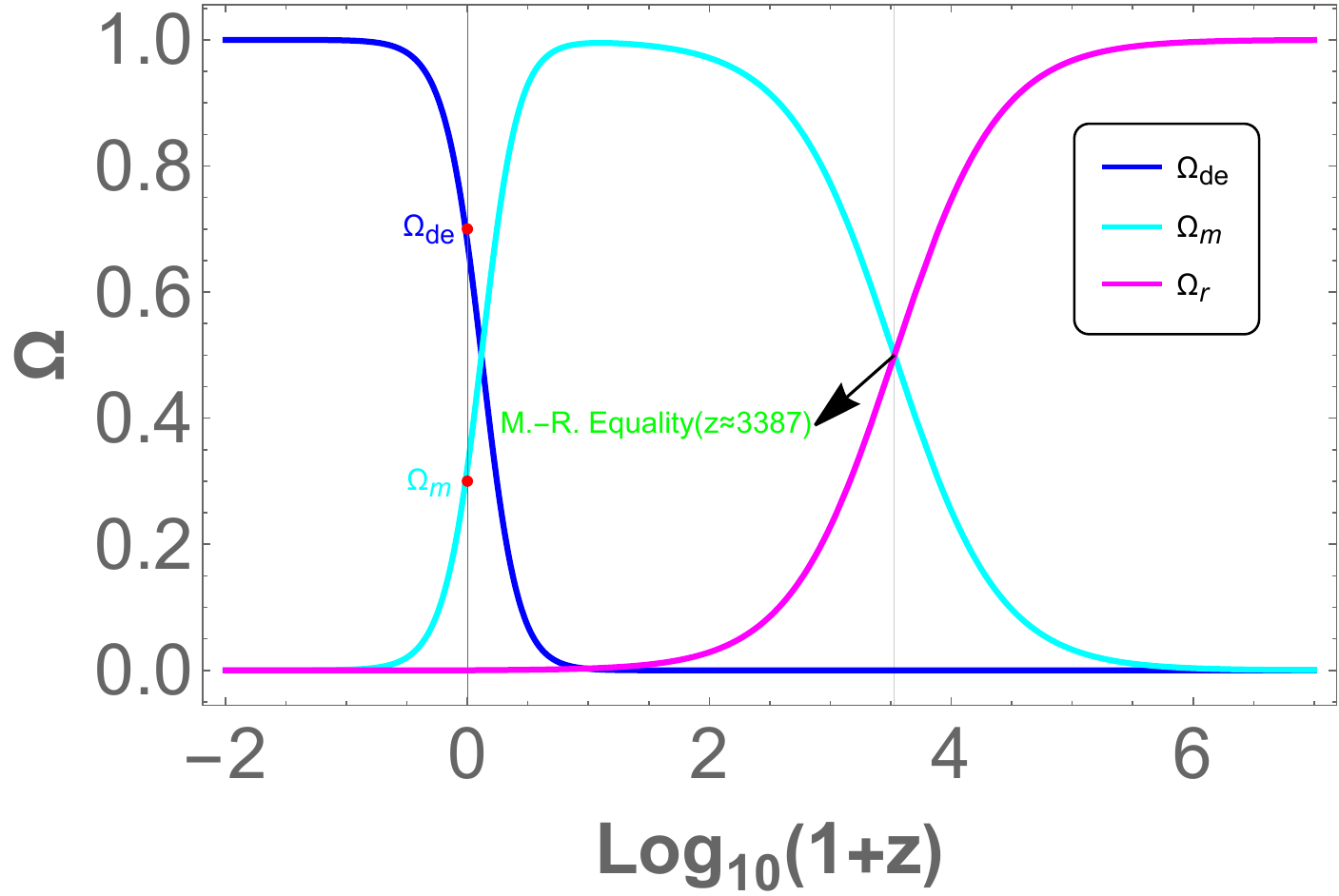}
         \caption{Evolution of density parameters.}
         \label{ch6_fig:case3_density}
     \end{subfigure}
     \begin{subfigure}[b]{0.44\textwidth}
         \centering
         \includegraphics[width=70mm]{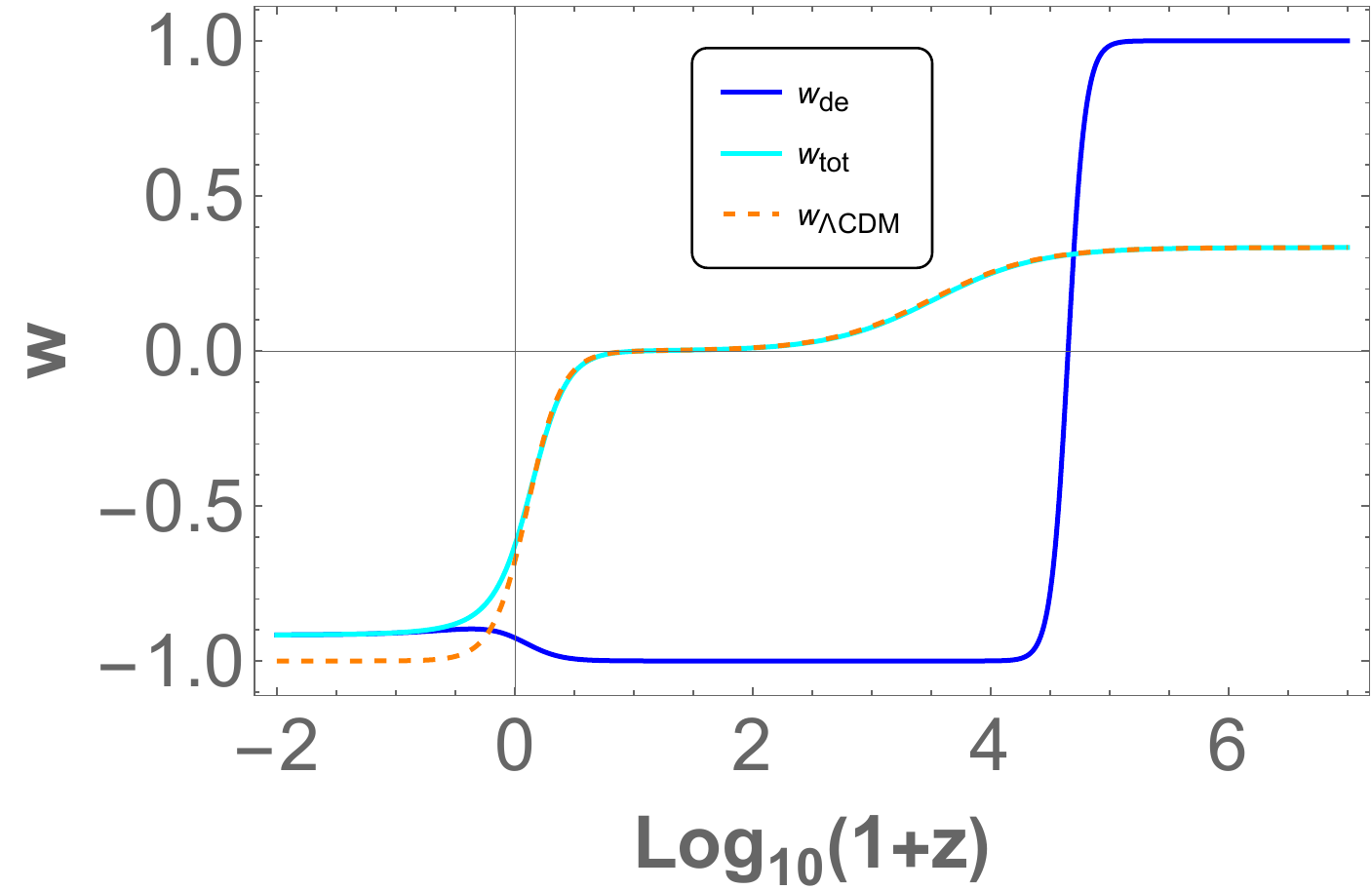}
         \caption{Evolution of EoS parameters.}
         \label{ch6_fig:case3_statepara}
     \end{subfigure}
\caption{In this figure, we set  $\alpha=-0.1$, $\gamma=0.5$ and $\beta=1$ with the initial conditions are the same as in Fig.~\ref{ch6_Fig6phase_portrait}.} 
\label{ch6_Fig5}
\end{figure}

\begin{figure}
     \centering
     \begin{subfigure}[b]{0.49\textwidth}
         \centering
         \includegraphics[width=70mm]{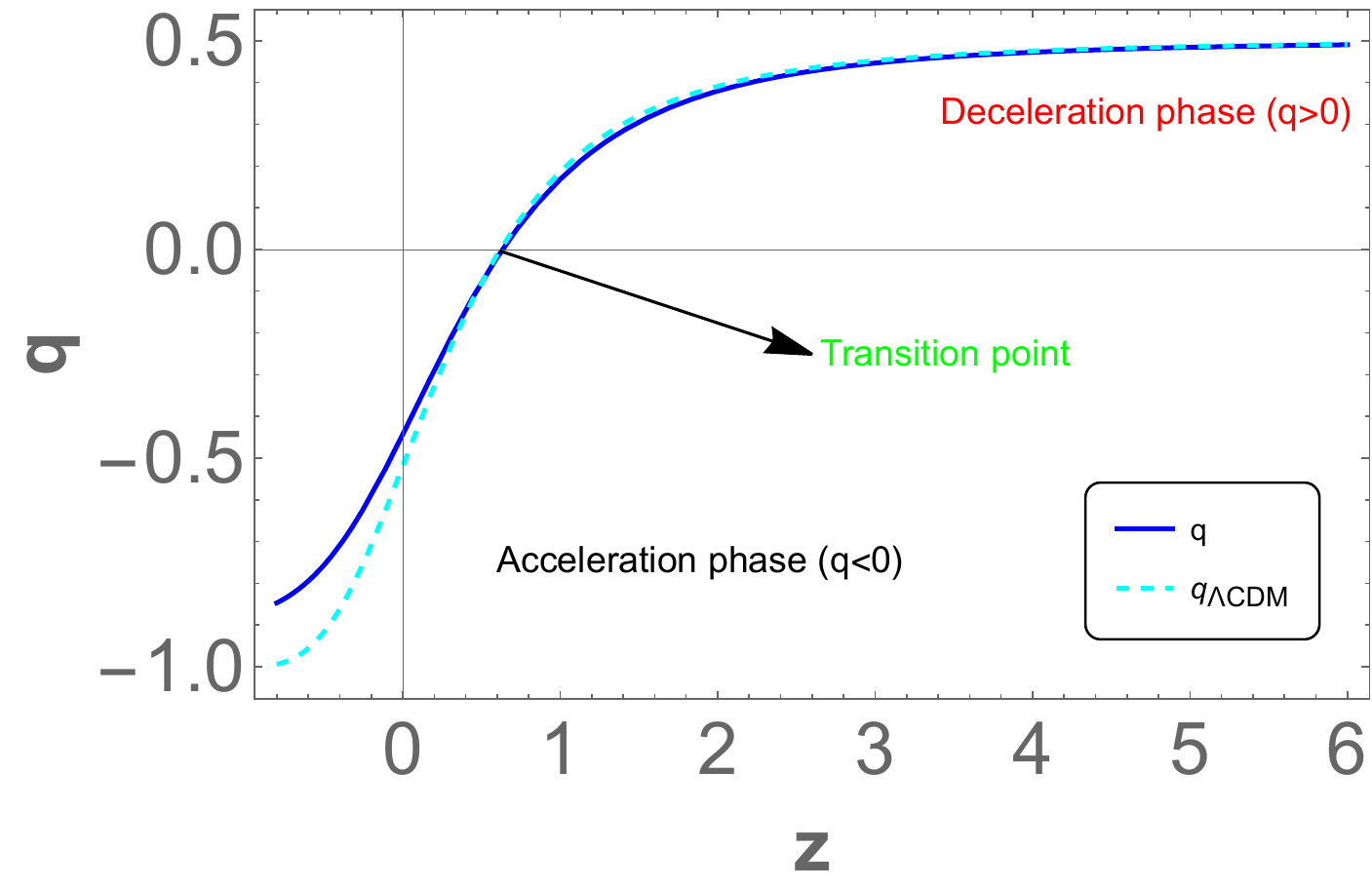}
         \caption{Evolution of the deceleration parameter $q$.}
         \label{ch6_fig:case3_qz}
     \end{subfigure}
     \begin{subfigure}[b]{0.49\textwidth}
         \centering
         \includegraphics[width=70mm]{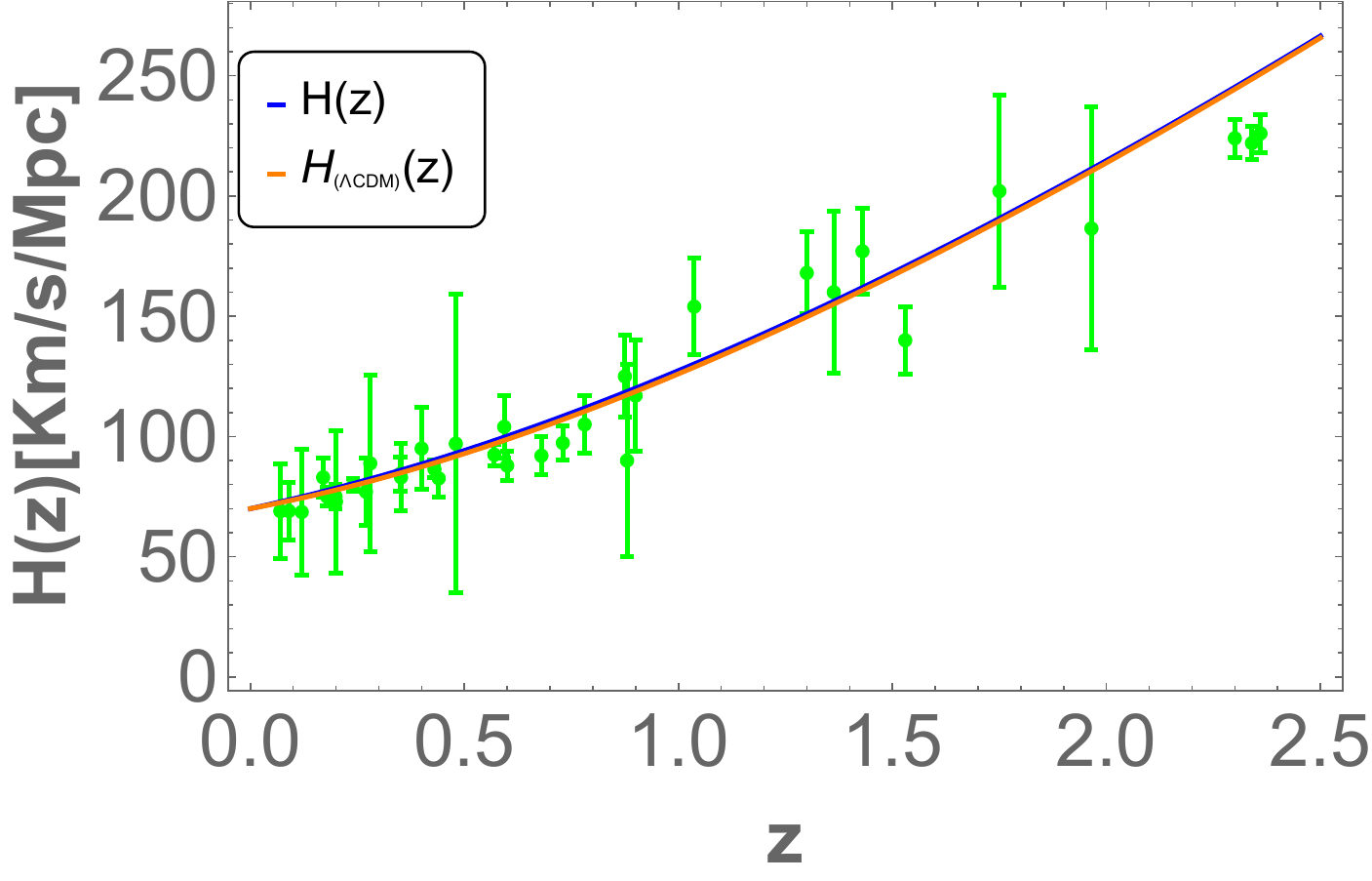}
         \caption{Evolution of the Hubble rate $H(z)$.}
         \label{ch6_fig:case3_Hz}
     \end{subfigure}
\caption{In this figure, we set  $\alpha=-0.1$, $\gamma=0.5$ and $\beta=1$ with the initial conditions are the same as in Fig.~\ref{ch6_Fig6phase_portrait}.} 
\label{ch6_Fig6}
\end{figure}

\begin{figure}[H]
 \centering
 \includegraphics[width=65mm]{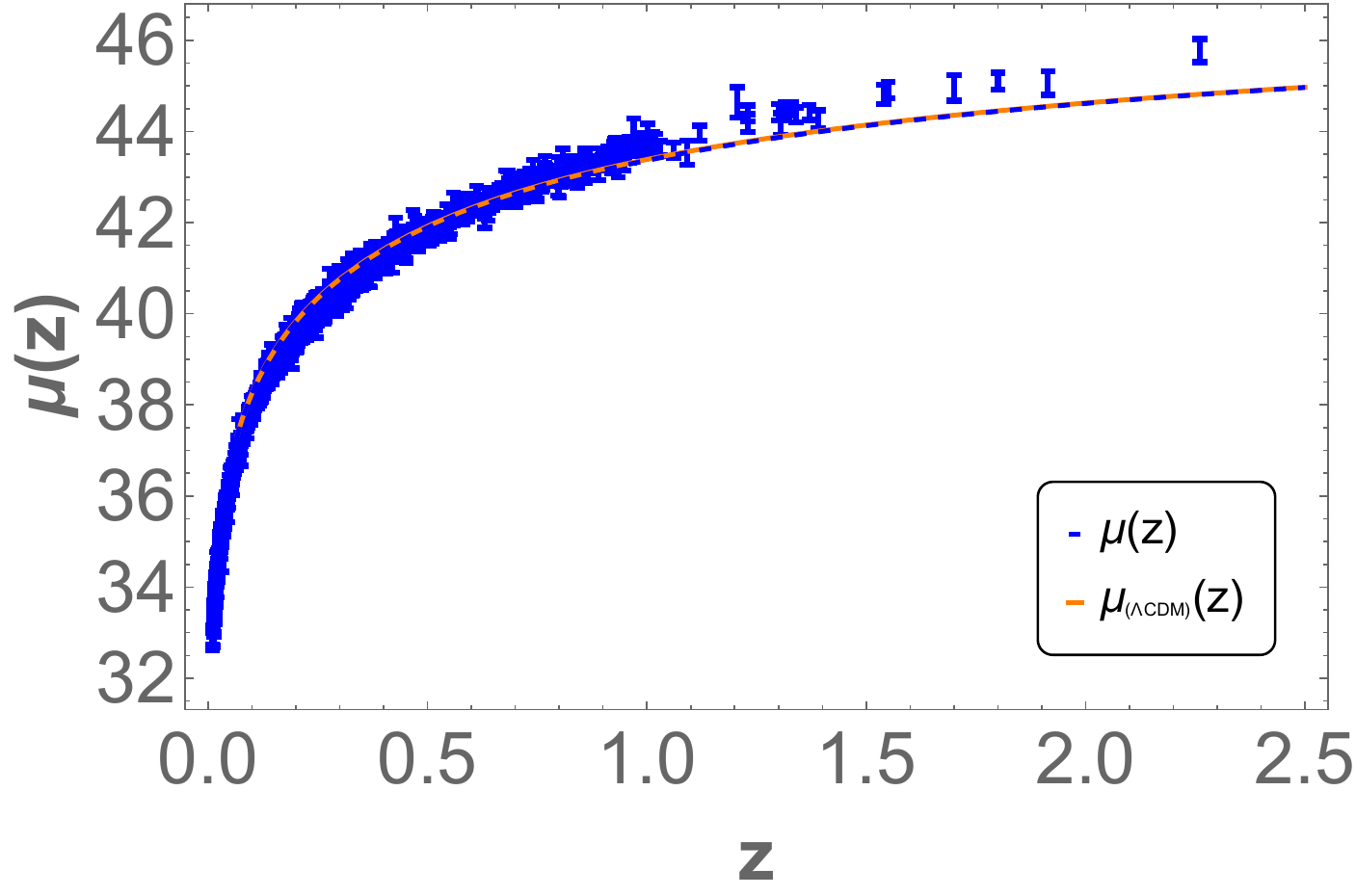} 
 \caption{We show the evolution of the distance modulus function $\mu(z)$ (dashed blue) and the $\Lambda$CDM model distance modulus function $\mu_{\Lambda CDM}(z)$ along with the 1048 SNIa  data points \cite{Scolnic_2018}. In this figure, we set  $\alpha=-0.1$, $\gamma=0.5$ and $\beta=1$ with the initial conditions are the same as in Fig.~\ref{ch6_Fig6phase_portrait}.} \label{ch6_Fig6pan}
 \end{figure}

For these solutions to be cosmologically viable, it is essential to have a sufficiently flat potential ($\beta^2 \gamma^2 < 2$) and to finely tune the initial conditions to ensure the persistence of DE domination. The solution should follow the sequence: $B_{3\pm} \to B_1 \to B_9$ ( Fig.~\ref{ch6_fig:case3_phaseportrait}). Moreover, at early times, the only viable solutions correspond to non-physical stiff fluid Universe, which are inadequate for describing the Universe. The EoS parameters, energy densities, deceleration parameters, Hubble rate and modulus function for a solution that mirrors the heteroclinic sequence $B_{3\pm} \to B_1 \to B_9$ are illustrated in Fig.~\ref{ch6_Fig5}, Fig.~\ref{ch6_Fig6} and Fig.~\ref{ch6_Fig6pan}.

From Fig.~\ref{ch6_fig:case3_density}, one can observe the radiation at the early epoch followed by dominance of DM for a while and finally the DE phase. At present, $\Omega_{m}\approx 0.32$ and $\Omega_{DE}\approx 0.68$, the matter-radiation equality observed at $z\approx 3387$. Fig.~\ref{ch6_fig:case3_statepara} explains the EoS parameter $\omega_{tot}$ (cyan), which begins at $\frac{1}{3}$ for radiation falls to $0$ during the matter-dominated period and finally reaches approximately $-1$.  We observed the current value of $\omega_{DE}(z=0)=-0.93$. The deceleration parameters [Fig.~\ref{ch6_fig:case3_qz}] show the transition from deceleration to acceleration at $z\approx 0.60$. At present, the deceleration parameter becomes $\approx -0.44$. The Hubble rate evolution as a function of redshift $z$, the Hubble rate $H_{\Lambda CDM}(z)$ and the Hubble data points \cite{Moresco_2022_H0} are displayed in Fig.~\ref{ch6_fig:case3_Hz}. It has been observed that the model presented here closely resembles the standard $\Lambda$CDM model. The $\Lambda$CDM model modulus function $\lambda_{\Lambda CDM}$, 1048 pantheon data points and the evolution of the modulus function $\mu(z)$ are shown in Fig.~\ref{ch6_Fig6pan}.
\section{Conclusion} \label{ch6_SECIV}
The dynamical systems technique is an important approach to investigating background cosmology.  Hence, in this chapter, we have presented the dynamic behavior of the modified Galileon cosmology which is based on higher derivatives in action and requires maintaining second-order equations of motion and imposition of Galilean symmetry. On the other hand,  the constant coefficients of the various action terms in the basic Galileon formulation are further expanded into any functions of the scalar field. We performed the phase space analysis for the exponential form of $F(\phi)$ with three well-motivated functions of potential $V(\phi)$. The cosmological parameters, such as EoS parameters, density parameters and deceleration parameters are obtained through the dynamical variables to analyze the cosmological features of the models.

For all three models [\ref{ch6_case1exp}, \ref{ch6_case2power}, \ref{ch6_case3sinh}], we have obtained the stable critical points that can describe the late-time cosmic accelerated phase of the Universe. We have also obtained scaling solutions for critical points. The points show that non-standard matter and radiation dominated eras of the Universe. We observe that the results obtained are similar to the standard quintessence model. We can conclude that the Galileon term has no contribution to the DE sector of the Universe. But there is the contribution of the Galileon term to the inflationary behavior of the Universe \cite{Burrage_2011, Renaux_Petel_2013, Gonzalez_Espinoza_2019, Choudhury_2024}. For Model-I, we have obtained the value of the density parameters of the matter and DE sectors at present ($z=0$), which is $\Omega_{m}^{0}\approx 0.3$, $\Omega_{DE}^{0}\approx 0.7$ and also obtained the value of the matter-radiation equality at $z\approx 3387$. Similarly, for Model-II and Model-III, we have $\Omega_{m}^{0}\approx 0.33$, $\Omega_{m}^{0}\approx 0.32$, $\omega_{DE}^{0}\approx 0.67$ and $\omega_{DE}^{0}\approx 0.68$ respectively.  At present, we have obtained the value of the DE dominated EoS parameter for Model-I, Model-II and Model-III as $-1$, $-0.92$ and $-0.93$, respectively. The values are compatible with the current observation \cite{Aghanim:2018eyx}.\\

For Model-\ref{ch6_case1exp}, through the behavior of Fig.~\ref{ch6_fig:case1_qz}, we can say that the deceleration parameter shows the transition from deceleration to acceleration phase at $z=0.66$ and the present value of the deceleration parameter is $q(z=0)=-0.55$. For Model-(\ref{ch6_case2power}), we got the transition point at $z=0.62$ and the present value of the deceleration parameter is $q(z=0)=-0.45$ Fig.~\ref{ch6_fig:case2_qz}. For Model-(\ref{ch6_case3sinh}), we obtain the transition point at $z=0.60$ and the present value of the deceleration parameter is $q(z=0)=-0.44$ Fig.~\ref{ch6_fig:case3_qz}. In all three models, transition point and current deceleration parameter values matched the cosmological observations \cite{PhysRevD.90.044016a, Camarena:2020prr}. In all three models [\ref{ch6_case1exp}, \ref{ch6_case2power}, \ref{ch6_case3sinh}], we also compared our results with the Hubble 31 data points \cite{Moresco_2022_H0} and the SNIa data 1048 data points \cite{Scolnic_2018}. Through the behavior of the Figs.~[\ref{ch6_fig:case1_Hz}, \ref{ch6_fig:case2_Hz}, \ref{ch6_fig:case3_Hz}], we can say that the results of our model are very close to the standard $\Lambda$CDM model. In Figs.~[\ref{ch6_Fig2pan}, \ref{ch6_Fig4pan}, \ref{ch6_Fig6pan}], we plotted the modulus function of our models with the standard $\Lambda$CDM model modulus function along with the 1048 SNIa data points. The results closely resemble the standard $\Lambda$CDM model.

The results of this analysis do not favor any particular form of potential. It shows that class potentials can describe the accelerated expansion of the Universe. As a consequence, the choice of potentials remains arbitrarily made. Even though this analysis is done by choosing three different potentials, it can be expanded to include more potentials. We have restricted ourselves to these three potential forms to keep the analysis from becoming unnecessarily lengthy. Furthermore, this general parameterization of $f(\lambda)=\lambda^2 (\Gamma-1)$ does not depend on the particular scalar field DE model but on the definitions of $\Gamma$ and $\lambda$. Furthermore, this general parametrization of $f$ can also be used with other scalar field DE models, like a quintessence, phantom, coupled DE, modified gravity theories, k-essence, and tracker solutions, etc. In all three models,  the scaling solution is an attractor during the radiation and matter eras but it does not exit the epoch of cosmic acceleration.
\chapter{Final Remarks and Future Scope} 

\label{Chapter7} 

\lhead{Chapter 7. \emph{Final Remarks and Future Scope}} 



\newpage

This thesis addresses late-time cosmic phenomena in teleparallel-based gravity through cosmological observation and dynamical system analysis. The dynamical system approach and the mathematical framework for cosmological observations are established in Chapter~\ref{Chapter1}. Additionally, this chapter outlines the mathematical formulations of GR and TEGR, along with several modifications to the TEGR framework.

In Chapter~\ref{Chapter2}, we explored analyzing dynamical systems within the $f(T)$ gravity framework. We addressed the non-linear differential equations associated with dynamic variables. This framework illustrates the evolution of the Universe by examining the critical points of autonomous systems. With these considerations, we analyzed the dynamical system in $f(T)$ gravity at both the background and perturbation levels in this work. In this study, we conducted a dynamical system analysis on three well-founded $f(T)$ models. Our findings indicate that each model predicts a Universe dominated by DE, leading to late-time cosmic acceleration. In contrast, the early Universe is characterized by a matter-dominated phase displaying unstable dynamics. We identified a critical point across all models delineating the DE phase at both background and perturbation levels. Additionally, we examined the growth and decay processes within the matter phase, providing insights into the evolutionary dynamics of these cosmological models.

In Chapter~\ref{Chapter3}, we conducted the dynamical system analysis within the framework of $f(T, \mathcal{T})$ gravity, focusing on two distinct forms of the theory. Our investigation revealed that these models exhibit a transition from an early-time matter-dominated phase to a late-time DE-dominated phase. We generated plots for key background cosmological parameters, including the deceleration parameter $q$, the total EoS $\omega_{tot}$ and the DE EoS $\omega_{DE}$, all of which illustrate the cosmic acceleration observed in the late Universe. Furthermore, we examined a specific instance of $f(T, \mathcal{T})$ against a variety of cosmological datasets, including the CC, PN$^{+}$ (without SH0ES), PN$^{+}$\& SH0ES (with SH0ES) and BAOs combined with $H_0$ priors from TRGB and R21. We performed comparative analyses of the results derived from the PN$^{+}$ dataset and the PN$^{+}$\& SH0ES dataset to assess the impact of including SH0ES observations. Our chosen model was also critiqued against the standard $\Lambda$CDM model. To evaluate the statistical robustness of our models, we employed the AIC and the BIC for goodness-of-fit assessments. In addition, we constructed Whisker plots to visually elucidate the variation of model parameters across different dataset combinations.  

In Chapter~\ref{Chapter4}, we conducted a detailed dynamical system analysis within the $f(T, \phi)$ gravity framework, exploring two distinct functional forms. Our investigation into the critical points of the autonomous system revealed various cosmological phases, including radiation, matter, stiff-matter and DE-dominated regimes, which correspond to the behavior of these critical points. The dynamical system analysis indicates that the selected $f(T, \phi)$ models exhibit early-time deceleration transitioning to late-time cosmic acceleration, providing insights into the evolution of the Universe across different epochs.

In Chapter~\ref{Chapter5}, we have examined cosmological observations within quintessence DE frameworks, utilizing three distinct potential functions \( V(\phi) \). It evaluates datasets from CC, PN$^+$\&SH0ES and BAOs to contrast these models with the conventional \( \Lambda \)CDM framework, represented through a whisker plot. The research also integrates \( H_0 \) priors R21 and F21 to assess their influence on the findings and thoroughly compares the quintessence models to evaluate their relative effectiveness.

In Chapter~\ref{Chapter6}, we have conducted a phase space analysis within the framework of modified Galileon cosmology, where the Galileon term is treated as a coupled scalar field, denoted as $F(\phi)$. We focus on the exponential form of $F(\phi)$ and three well-founded potential functions, $V(\phi)$. We derive the critical points of the autonomous system, stability criteria and cosmological characteristics. Our analysis identified the scaling solution for critical points to understand the eras dominated by matter-DE and radiation-DE. In these scaling solutions, DE is generally introduced along with another component, such as radiation or matter, which aids in elucidating the transition between different cosmological eras. The critical points dominated by DE display stable behavior, suggesting the late-time cosmic acceleration of the Universe. Additionally, we examine the results using the Hubble rate $H(z)$ and data sets from SNIa. We concluded that our analysis does not support any specific type of potential. The findings indicate that class potentials can explain the accelerated expansion of the Universe.

In future investigations, the problem can be extended by conducting a comprehensive phase space analysis at both the background and perturbation levels, focusing on the dynamics of matter perturbations. A pivotal area of research will be the examination of the $H_0$ and $S_8$ tensions within modified gravity frameworks, utilizing a diverse array of datasets including CMB measurements, Large-Scale Structure (LSS) observations, SNIa data, BAO and gravitational lensing surveys. Further analysis of cosmological perturbations is necessary to understand how modified gravity influences structure formation, specifically regarding scalar, vector and tensor perturbations. Notably, gravitational waves in teleparallel gravity, especially in models with boundary terms, may exhibit unique polarization signatures, offering valuable insights into the nature of gravity. These signatures could yield testable predictions for upcoming detectors such as LIGO-India, LISA and the Einstein Telescope. Additionally, investigating the effects of parity violation in gravitational waves and their relation to cosmic birefringence could enhance our understanding of fundamental physics. Extending our analysis to the early Universe, particularly examining inflationary models within the teleparallel framework, will help elucidate how torsion-based modifications affect the primordial power spectrum and reheating processes. Moreover, exploring interactions between DE and DM in these modified gravity scenarios might offer alternative explanations for cosmic acceleration. These research directions may establish a robust framework for assessing gravitational modifications across various cosmic epochs and observational modalities.




\addtocontents{toc}{\vspace{1em}} 

\backmatter


\label{References}
\lhead{\emph{References}}

\cleardoublepage
\pagestyle{fancy}
\lhead{\emph{Appendices}}
\chapter{Appendices}
\section*{Center Manifold Theory (CMT) for the non-hyperbolic critical point $B_{2}$}\label{ch5_Sec-app}
The CMT theory is discussed in the Section~\ref{Centralmanifoltheory}. The Jacobian matrix at the critical point $B_{2}$ for the autonomous system [Eq.~\ref{ch6_autonomous-system1}-Eq.~\ref{ch6_autonomous-system6}] is as follows:
\[
J(B_{2}) = 
\left( \begin{array}{ccccc}
 -3 & 0  & 0  & 0 & \sqrt{\frac{3}{2}} \\
  0 & -3 & 0 & 0 & 0 \\
  0 & 0 & -3 & 0 & 0 \\
  0 & 0 & 0 & -2 & 0 \\
  0 & 0 & 0 & 0 & 0 
\end{array} \right)
\]

The eigenvalues of Jacobian matrix $J(B_{2})$ are $-3$, $-3$, $-3$, $-2$ and $0$. The \begin{math}\left[1, 0, 0, 0, 0 \right]^T \end{math}, \begin{math}\left[0, 1, 0, 0, 0 \right]^T \end{math}, \begin{math}\left[0, 0, 1, 0, 0 \right]^T \end{math} are the eigenvector to the corresponding eigenvalues $-3$ and \begin{math} \left[0,0,0,1,0\right]^{T}\end{math} be the eigenvector corresponding to the eigenvalue $-2$ and \begin{math} \left[0,0,0,0,0\right]^{T}\end{math} be the eigenvector corresponding the eigenvalue $0$.

Using center manifold theory, we examine the stability of the critical point $B_{2}$ because of its non-hyperbolic nature. To apply CMT to this critical point, we must shift it to the origin using a shifting transformation. we have followed these transformations: $X=x$, $Y=1-y$, $Z=u$, $R=\rho$ and $ S=\lambda$ and then we can write equations in the new coordinate system as 
\begin{eqnarray}
\left( \begin{array}{c}
\dot{X} \\ 
\dot{Y} \\ 
\dot{Z} \\ 
\dot{R} \\ 
\dot{S} 
\end{array} \right) &=& 
\left( \begin{array}{ccccc}
-3 & 0 & 0 & 0 & 0 \\
0 & -3 & 0 & 0 & 0 \\
0 & 0 & -3 & 0 & 0 \\
0 & 0 & 0 & -2 & 0 \\
0 & 0 & 0 & 0 & 0
\end{array} \right) 
\left( \begin{array}{c}
X \\ 
Y \\ 
Z \\ 
R \\ 
S    
\end{array} \right) +
\left( \begin{array}{c}
non \\ 
linear \\ 
term   
\end{array} \right)
\end{eqnarray}

Comparing this diagonal matrix with the general form (\ref{CMT1}). After that, we can say that here $X$, $Y$, $Z$ and $R$ are the stable variables and $S$ is the central variable. At this critical point, the $A$ and $B$ matrix appears as 
\[
A =
\left( \begin{array}{cccc}
 -3 & 0 & 0 & 0 \\
  0 & -3 & 0 & 0 \\
  0 & 0 & -3 & 0 \\
  0 & 0 & 0 & -2
\end{array} \right)
\hspace{0.5cm}
B = 
\left( \begin{array}{c}
 0    
\end{array} \right)
\]

According to CMT, the manifold can be defined by a continuous differential function. we have assumed the following functions for the stable variables $X=h_{1}(S), Y=h_{2}(S), Z=h_{3}(S),$ and  $R=h_{4}(S)$. We have obtained the following zeroth approximation of the manifold functions 
\begin{eqnarray}
\mathcal{N}(h_1(S)) = -\sqrt{\frac{3}{2}}S\,, \quad \mathcal{N}(h_2(S)) =0\,, \quad  \mathcal{N}(h_3(S)) =0 \,, \quad \mathcal{N}(h_4(S)) =0.
\end{eqnarray}
As of this moment, the center manifold acquired by 
\begin{equation}\label{CMT-manifold}
\dot{S}= 3 S^{3}(\Gamma-1)+ higher \hspace{0.2cm} order \hspace{0.2cm} term   
\end{equation}

According to the CMT, this critical point shows stable behavior for  $\Gamma<1$ and unstable for $\Gamma>1$, where $\Gamma$ depends on the potential function $V(\phi)$.
\subsubsection*{\bf{Cosmological observation on the $\Lambda$CDM model}}\label{LCDMmodelresult}
\begin{figure}[H]
     \centering
         \includegraphics[width=65mm]{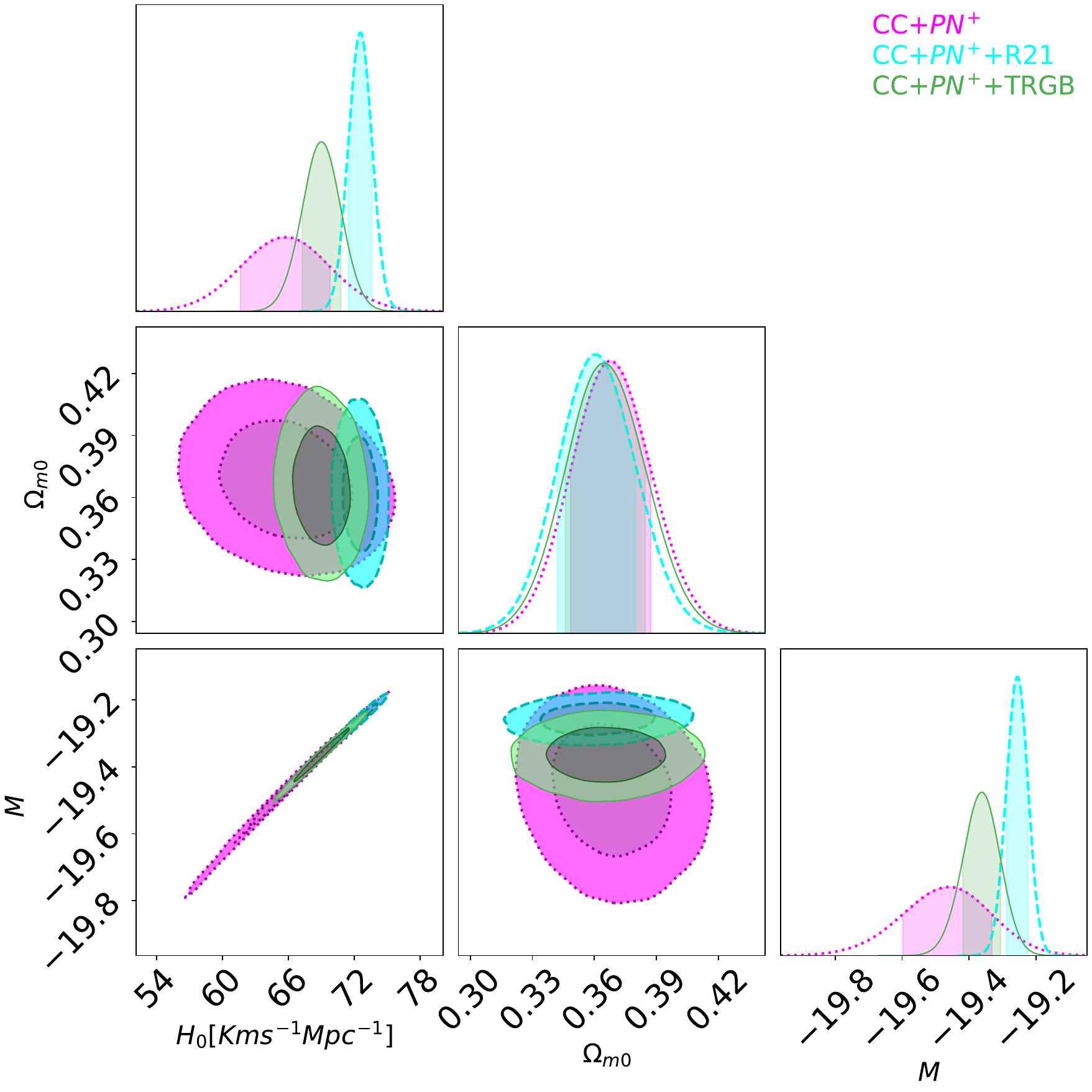}
    \includegraphics[width=65mm]{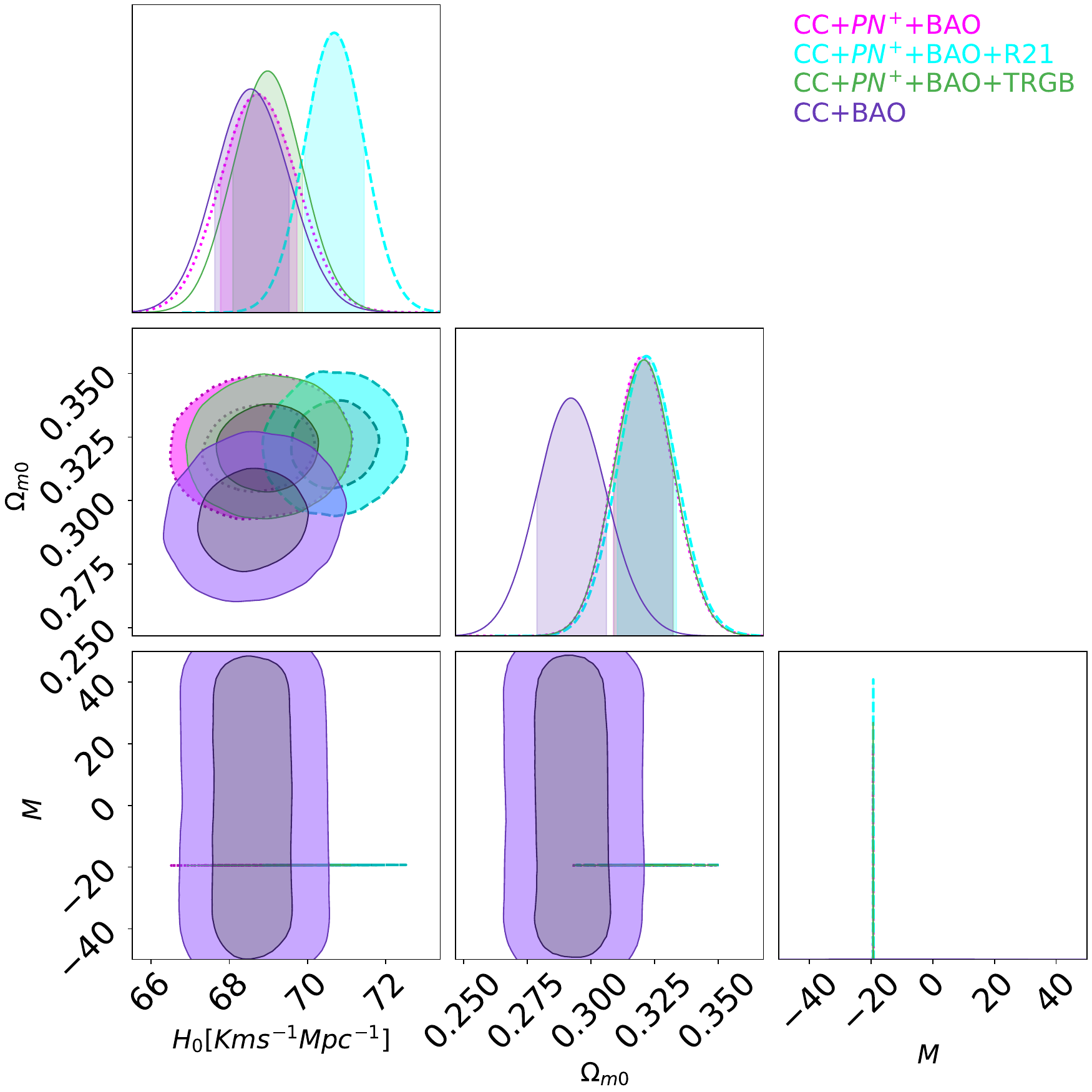} 
        \includegraphics[width=65mm]{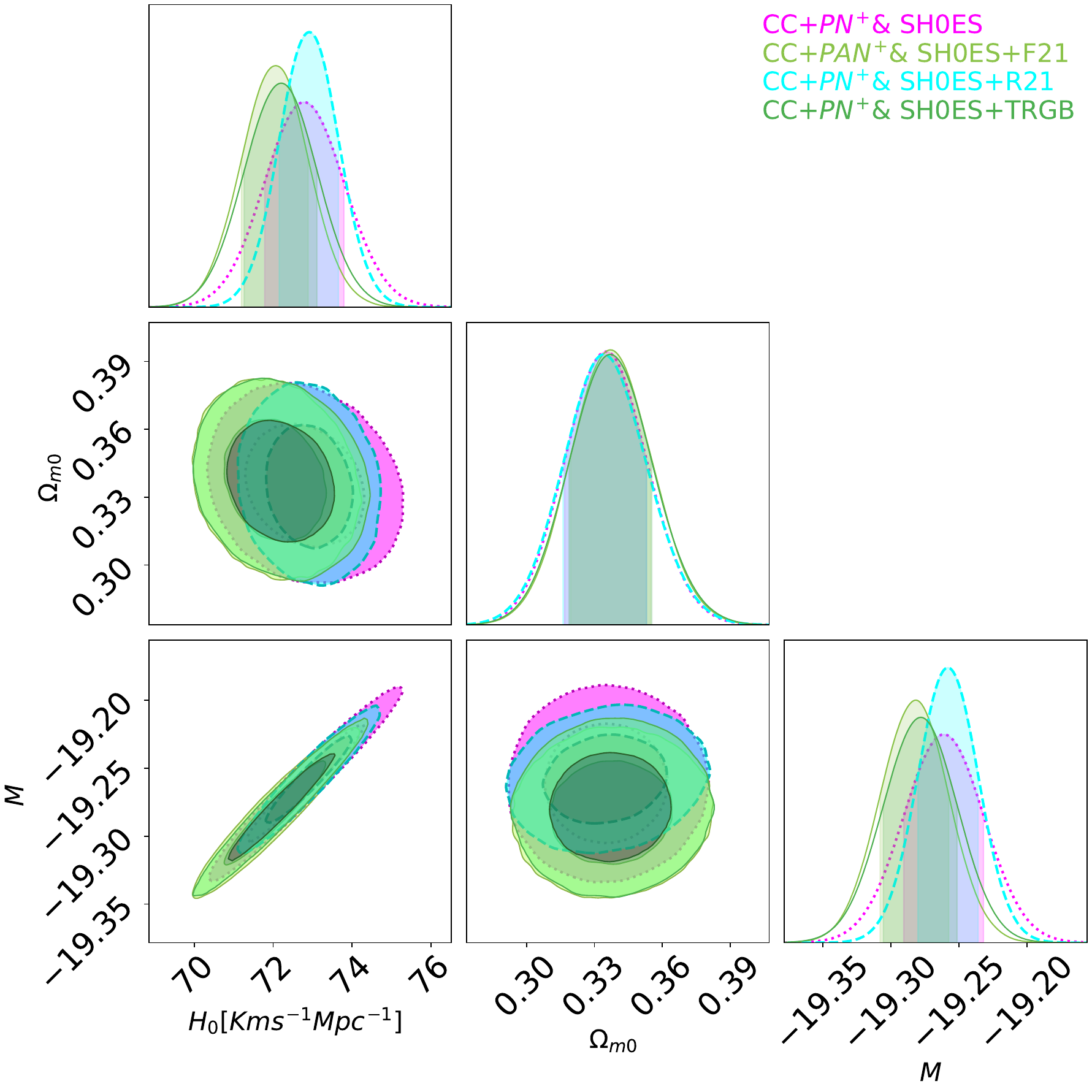}
          \includegraphics[width=65mm]{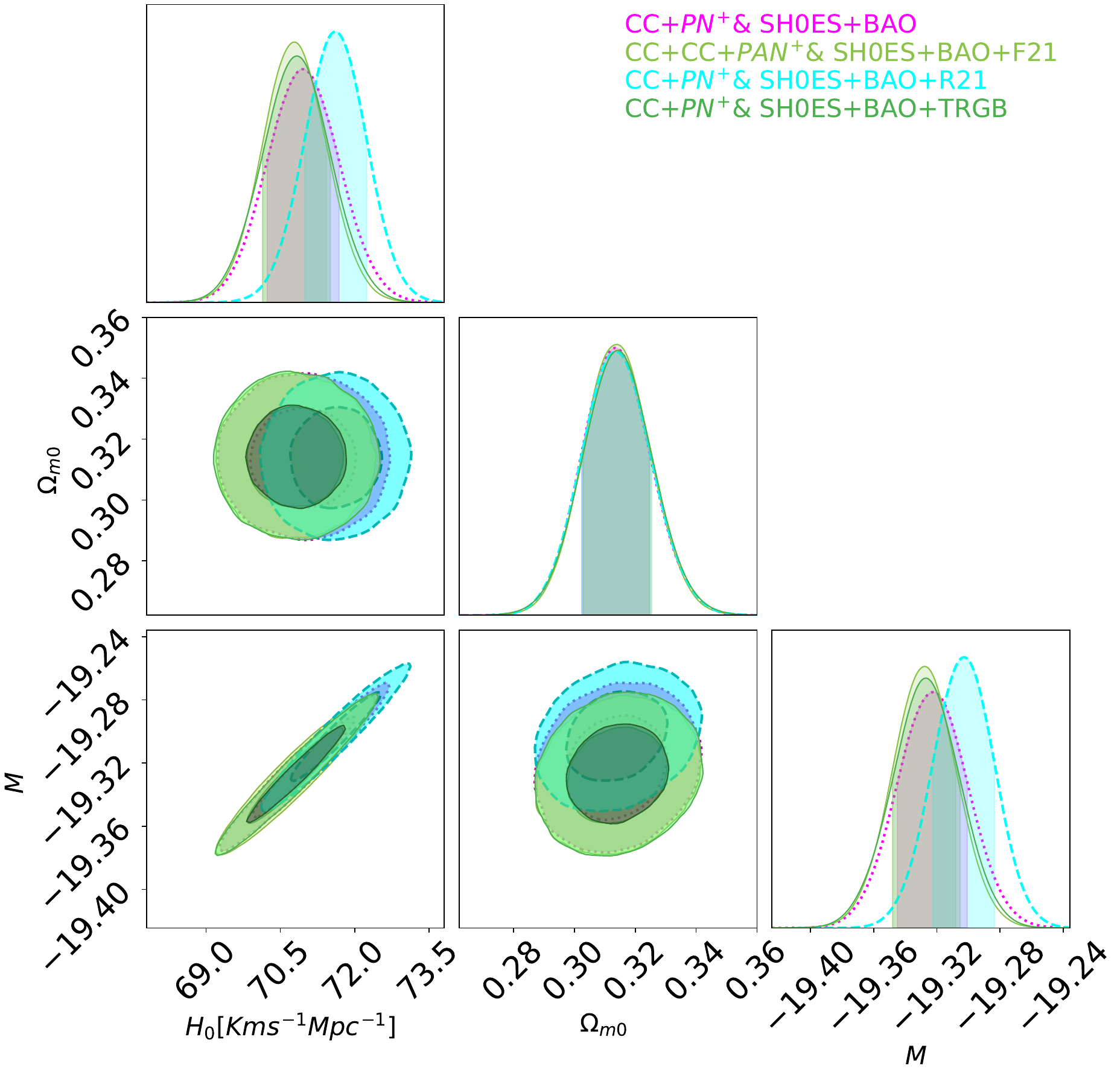} 
        \caption{The contour plot of $1\sigma$ and $2\sigma$ uncertainty regions and posterior distribution for the model parameters with the combination of data sets CC, PN$^{+}$, PN$^{+}$\& SH0ES, BAOs and the $H_0$ priors R21, F21 and TRGB. } 
\label{LCDMMCMCplus}
\end{figure}

We present the results of the $\Lambda$CDM model highlighting the posterior distributions in Fig.~\ref{LCDMMCMCplus}. These figures include the $1\sigma$ and $2\sigma$ confidence intervals for various combinations of datasets, providing a comprehensive examination of parameter constraints. Detailed results for each dataset combination are summarized in Table~\ref{tab:model_outputsLCDMplus} and Table~\ref{tab:model_outputsLCDM}. This methodology emphasizes the consistency of our findings with the standard predictions of the $\Lambda$CDM paradigm, facilitating a rigorous assessment of the performance of the model across the selected datasets.
\renewcommand{\arraystretch}{1} 

\begin{table}[H]
    \centering
    \caption{The results for the $\Lambda$CDM model. The first column identifies the data set combinations and the applied $H_{0}$ priors. The second and third columns present the derived values for $H_{0}$ and $\Omega_{m,0}$ respectively, while the fourth column displays the value of the nuisance parameter. The fifth column provides the minimized $\chi^2_{min}$ values, with the sixth and seventh columns showing the AIC and BIC values respectively. {\bf(Without SH0ES data points)}}
    \label{tab:model_outputsLCDMplus}
    \begin{tabular}{ccccccc}
        \hline
		$\Lambda$CDM & $H_0$ & $\Omega_{m,0}$ & $M$& $\chi^{2}_{min}$& AIC &BIC \\ 
		\hline
		CC+PN$^{+}$ & $65.7\pm 4.1$ & $0.368^{+0.019}_{-0.020}$ & $-19.46^{+0.13}_{-0.14}$ &1792.30 &  1798.30 &  1802.02  \\ 
		CC+PN$^{+}$ +R21 & $72.6^{+1.0}_{-1.1}$ & $0.361\pm 0.019$ & $-19.256^{+0.032}_{-0.033}$ &1795.64 & 1801.64 &  1805.35\\ 
		CC+PN$^{+}$+TRGB& $69.0^{+1.8}_{-1.7}$ & $0.365^{+0.020}_{-0.019}$ & $-19.362^{+0.055}_{-0.057}$ &1793.20 & 1799.20 &  1802.92 \\ 
		\cline{1-7}
        CC+PN$^{+}$+BAO & $68.73^{+1.00}_{-0.95}$ & $0.320^{+0.012}_{-0.011}$ & $-19.387^{+0.033}_{-0.030}$&1817.75 & 1823.75 &  1827.36 \\ 
		CC+PN$^{+}$+BAO+R21 & $70.68^{+0.76}_{-0.75}$ & $0.322^{+0.012}_{-0.011}$ & $-19.326\pm 0.024$ &1827.16 & 1833.16 &  1836.78\\ 
		 CC+PN$^{+}$+BAO+TRGB & $68.99^{+0.88}_{-0.90}$ & $0.321^{+0.011}_{-0.012}$ & $-19.379\pm 0.029$ &1818.00 & 1824.00 &  1827.61\\
          CC+BAO & $68.56^{+0.97}_{-0.93}$ & $0.292^{+0.014}_{-0.013}$ & - &20.14 & 26.14 &    29.76\\
		\hline
    \end{tabular}
\end{table}
\begin{table}[H]
    \centering
    \caption{The results for the $\Lambda$CDM model. The first column identifies the data set combinations and the applied $H_{0}$ priors. The second and third columns present the derived values for $H_{0}$ and $\Omega_{m,0}$ respectively, while the fourth column displays the value of the nuisance parameter. The fifth column provides the minimized $\chi^2_{min}$ values, with the sixth and seventh columns showing the AIC and BIC values respectively. {\bf(With SH0ES data points)}} Where A=CC+PN$^{+}\&$ SH0ES
    \label{tab:model_outputsLCDM}
    \begin{tabular}{ccccccc}
        \hline
		$\Lambda$CDM  & $H_0$ & $\Omega_{m,0}$ & $M$& $\chi^{2}_{min}$& AIC &BIC \\ 
		\hline
		A & $72.8\pm 1.0$ & $0.335\pm 0.018$ & $-19.261^{+0.029}_{-0.030}$  &1539.22 &  1545.22 &  1548.93  \\ 
        A+ F21 & $71.99^{+0.92}_{-0.78}$ & $0.337^{+0.018}_{-0.017}$ & $-19.284^{+0.027}_{-0.023}$&1541.55 & 1547.55 &  1551.26 \\ 
		 A+ R21 & $72.91\pm 0.75$ & $0.334^{+0.019}_{-0.018}$ & $-19.258\pm 0.022$ &1539.25 & 1545.25 &  1548.96\\ 
		 A+ TRGB& $72.22^{+0.89}_{-0.95}$ & $0.337^{+0.019}_{-0.018}$ & $-19.278\pm 0.027$ &1541.18 & 1547.18 &  1550.89 \\ 
		\cline{1-7}
         A+ BAO & $70.95^{+0.74}_{-0.71}$ & $0.313\pm 0.011$ & $-19.322^{+0.021}_{-0.023}$&1567.17 & 1573.17 &  1576.90 \\ 
        A+ BAO+F21 & $70.83^{+0.59}_{-0.69}$ & $0.313^{+0.012}_{-0.010}$ & $-19.327^{+0.019}_{-0.021}$& 1567.57& 1573.57 &  1577.29 \\ 
		A+ BAO+R21& $71.60^{+0.64}_{-0.61}$ & $0.314\pm 0.011$ & $-19.303^{+0.019}_{-0.020}$ &1569.96 & 1575.96 &  1579.67\\ 
		A+ BAO+TRGB& $70.82^{+0.69}_{-0.68}$ & $0.315^{+0.011}_{-0.012}$ & $-19.327^{+0.022}_{-0.021}$ &1567.50 & 1573.50 &  1577.23 \\ 
		\hline
    \end{tabular}
\end{table}

\cleardoublepage
\pagestyle{fancy}
\lhead{\emph{List of publications and presentations}}
\chapter{List of publications and presentations}
\justifying
\section*{Publications included in this thesis}
\begin{enumerate}[itemsep=2pt]

 \item \textbf{L K Duchaniya}, K Gandhi and B. Mishra, “Attractor behavior of f (T) modified gravity and the cosmic acceleration", \textit{Physics of the Dark Universe}, \textbf{44}, 101461 (2024).

 \item \textbf{L K Duchaniya}, Santosh V. Lohakare and B. Mishra, “Cosmological models in $f(T, \mathcal{T})$ gravity and the dynamical system analysis", \textit{Physics of the Dark Universe }, \textbf{43}, 101402 (2024).

  \item \textbf{L K Duchaniya}, B Mishra, 
    “Late Time Phenomena in $f(T,\mathcal{T})$ Gravity Framework: Role of $H_0$ Priors". \textbf{(Accepted for publication in European Physical Journal C)}

    \item \textbf{L K Duchaniya}, S. A. Kadam, J. L. Said and  B. Mishra, “Dynamical systems analysis in $f (T, \phi)$ gravity", \textit{European Physical Journal C}, \textbf{83}, 27 (2023).

   \item \textbf{L K Duchaniya}, J. L. Said, B Mishra, “Quintessence dark energy models". \textbf{(Communicated)}

   \item \textbf{L K Duchaniya}, B Mishra, IV Fomin, SV Chervon, “Dynamical system analysis in modified Galileon cosmology", \textit{Classical and Quantum Gravity}, \textbf{41}, 253016 (2024).
\end{enumerate}
\section*{Other Publications}
\begin{enumerate}[itemsep=4pt, topsep=5pt]

    \item \textbf{L K Duchaniya}, Santosh V. Lohakare, B. Mishra and S. K. Tripathy, “Dynamical Stability Analysis  of Accelerating $f(T)$ Gravity Models". \textit{European Physical 
Journal C }, \textbf{82}, 448 (2022).

     \item \textbf{L K Duchaniya}, B. Mishra and  
J. L. Said, “Noether symmetry approach in scalar-torsion $f(T,\phi)$  gravity", \textit{European Physical Journal C }, \textbf{83}, 613 (2023).

    \item S. A. Kadam, \textbf{L. K. Duchaniya}, B. Mishra, "Teleparallel Gravity and Quintessence: The Role of Nonminimal Boundary Couplings", \textit{Annals of Physics}, \textbf{470}, 169808 (2024).

    \item IV Fomin, SV Chervon, \textbf{L K Duchaniya}, B Mishra, “The scalar-torsion gravity corrections in the first-order inflationary models", \textit{Physics of the Dark Universe}, \textbf{48}, 101895 (2025).

     \item \textbf{L K Duchaniya}, B Mishra, G. Otalora, M. Gonzalez-Espinoza, “Late-time acceleration and structure formation in interacting $\alpha$-attractor dark energy". \textbf{(Communicated)}
\end{enumerate}

\section*{Conferences and talks}  
\begin{enumerate}[itemsep=4pt, topsep=5pt]
     \item Presented a paper entitled “Dynamical systems analysis in $f (T, \phi)$ gravity” in the \textbf{international 
conference Physical Interpretations of Relativity Theory (PIRT-2023)} organized by \textbf{Bauman Moscow 
State Technical University, Moscow, Russia}. (3rd July-6th July 2023). 
\item  Presented a paper entitled “Cosmological implications of $f(T)$ gravity models through phase space 
analysis” in the \textbf{Workshop on Tensions in Cosmology, (CORFU-2023)} organized by \textbf{National Technical University of Athens.} (6th September-13th September 2023). 
\item  Presented a paper entitled “Dynamical system analysis in $f(T)$ gravity at both background and perturbation levels” at the \textbf{89th annual conference of Indian Mathematical Society, (IMS-2023)} organized by \textbf{Department of Mathematics, BITS-Pilani, Hyderabad Campus}. (22nd December-25th December 2023).  
\item  Presented a paper entitled “Attractor behavior of the $f(T)$ modified gravity at both background and perturbation levels” at the \textbf{International Joint Meeting on Cosmology and Gravitation, (IJMCG-2024)} organized by \textbf{University of Tarapacá, Campus Velásquez, Arica-Chile}. (15th October-18th October 2024). 
\item  Presented a paper entitled “Dynamical system analysis in modified Galileon cosmology” at the \textbf{Space. Time. Civilization, (STC-2024)} organized by \textbf{Bauman Moscow State Technical University together with Birla Institute of Technology and Science, Pilani, Hyderabad Campus and with Egyptian Russian University}. (2nd November-7th November 2024). 
\item Participated in the \textbf{27th International Conference of the International Academy of Physical 
Sciences on Advances in Relativity and Cosmology} organized by the \textbf{Department of Mathematics, BITS– Pilani, Hyderabad Campus}. (26th October to 28th October 2021).

    \item Participated in the \textbf{workshop on General Relativity and Cosmology}, organized by \textbf{GLA University, Mathura (U.P.)}. (24th November 2022 to 26th November 2022).
    
    \item Participated in the \textbf{Workshop on Teacher's Enrichment Workshop}, organized by \textbf{Department of Mathematics, BITS-Pilani, Hyderabad Campus}. (9th January 2023 to 14th January 2023).

    
\end{enumerate}
\cleardoublepage

\pagestyle{fancy}
\lhead{\emph{Biography}}
\chapter{Biography}
\textbf{Brief Biography of Candidate}\\
\textbf{Mr. Lokesh Kumar Duchaniya} completed his B.Sc. in 2016 from the University of Rajasthan and M.Sc. in 2018 from Malaviya National Institute of Technology (MNIT), Jaipur. He achieved a UGC NET-JRF in December 2019 and has published nine research papers during his Ph.D. in national and international journals. His other two articles have been communicated to the journal for publication. He presented his work at various renowned national and international conferences. He was selected to present his work at ``CORFU-2023" in Greece, held in September 2023, with financial support from DST-SERB India. He also has an academic visit to IUCAA Pune in April 2024 for collaborative research.

\textbf{Brief Biography of Supervisor}\\
\textbf{Prof. Bivudutta Mishra} received his Ph.D. degree from Sambalpur University, Odisha, India, in 2003. His main research areas are Geometrically Modified Theories of Gravity, Theoretical Aspects of Dark Energy and Wormhole Geometry. He has published over 169 research papers in national and international journals, presented papers at conferences in India and abroad, supervised six Ph.D. students and is currently guiding six more. He has also organized academic and scientific events in the department. He has become a member of the scientific advisory committee of national and international academic events. He has completed multiple sponsored projects funded by Government Funding agencies and is at present working on three projects funded by CSIR, SERB-DST (MATRICS) and SERB-DST(CRG-ANRF).  He is also an awardee of DAAD-RISE, 2019,2022. He has also reviewed several research papers in highly reputed journals, is a Ph.D. examiner and is a BoS member of several universities. He has been invited by many foreign universities to share his research in scientific events, some of which are Canada, Germany, the Republic of China, Russia, Australia, Switzerland, Japan, the UK and Poland. As an academic administrator, he was Head of the Department of Mathematics from September 2012 to October 2016 and was Associate Dean of International Programmes and Collaborations from August 2018 to September 2024. He is also a visiting professor at Bauman Moscow State Technical University, Moscow, a visiting associate at Inter-University Centre for Astronomy and Astrophysics, Pune, a Fellow of the Royal Astronomical Society, UK and a Fellow of the Institute of Mathematics and Applications, UK.
He is a foreign member of the Russian Gravitational Society in Moscow and is among the top 2\% scientists in the world.
                         
\end{document}